%% file: planck_lensing.tex
\documentclass[longauth,twocolumn,traditabstract]{aa}
\usepackage[table,usenames,dvipsnames]{xcolor}
\usepackage{ulem}
\usepackage{enumerate}
\usepackage{ifthen}
\usepackage{perpage}
\usepackage{multirow}
\usepackage{upgreek}
\usepackage{overpic}
\usepackage{graphicx,amsmath}
\usepackage{txfonts}
\usepackage[breaklinks, citecolor=blue, colorlinks=true, linkcolor=blue]{hyperref}
\usepackage{natbib}
\bibpunct{(}{)}{;}{a}{}{,} 
\usepackage{pdfcolmk}
\usepackage{morefloats}
\usepackage{float}

\pdfmapfile{+txfonts.map}

\input Planck.tex

\newcommand{\be}{\begin{equation}}
\newcommand{\ee}{\end{equation}}

\newcommand{\phihatapo}{\widetilde{\phi}}
\newcommand{\maskapo}{\widetilde{M}}

\newcommand{\binner}{ {\cal B} }
\newcommand{\apod}{{\sc iso}}
\newcommand{\metis}{{\sc metis}}
\newcommand{\patches}{{\sc patches}}

\newcommand{\threej}[6]{\left(
    \begin{array}{ccc}
        \! #1\! & #2\!  & #3\!  \\
        \! #4\! & #5\!  & #6\!
      \end{array}
    \right)}

\newcommand{\Q}{ \bar{x} }
\newcommand{\norm}{{\cal N}}
\newcommand{\resp}{ {\cal R} }

\newcommand{\WMAP}{{WMAP}\ }

\newcommand{\HealpixPixelization}{{HEALPix}}

\newcommand{\muKarcmin}{\mu {\rm K}\, {\rm arcmin}}
\newcommand{\Jy}{ {\rm Jy}}

\newcommand{\hatn}{\hat{\vec{n}}}
\newcommand{\vecd}{\vec{d}}

\newcommand{\sylm}[3]{{}^{}_{#1} Y_{#2}^{#3}}

\newcommand{\elt}{\ell}
\newcommand{\elp}{L}

\newcommand{\WF}{{\rm WF}}
\newcommand{\MF}{{\rm MF}}

\newcommand{\lcdm}{$\Lambda$CDM}
\providecommand{\planck}{\Planck}
\def\simlt{\lower.5ex\hbox{\ltsima}}
\def\ltsima{$\; \buildrel < \over \sim \;$}
\newcommand{\As}{A_{\rm s}}
\newcommand{\WP}{WP}
\newcommand{\highL}{highL}
\newcommand{\lensing}{lensing}

\newcommand{\Aphiphi}{A_{\rm L}^{\phi\phi}}

\providecommand{\Omk}{\Omega_K}
\providecommand{\Oml}{\Omega_{\Lambda}}

\providecommand{\Omm}{\Omega_{\mathrm{m}}}

\newcommand{\NILC}{{\sc NILC}}
\newcommand{\SEVEM}{{\sc SEVEM}} 
\newcommand{\SMICA}{{\sc SMICA}}
\newcommand{\CR}{{\sc RULER}}

\begin{document}

\onecolumn

\title{\Planck\ 2013 results. XVII. Gravitational lensing by large-scale structure}
\input{AuthorList_P12_Lensing_authors_and_institutes.tex}

\abstract{
On the arcminute angular scales probed by \Planck, the cosmic microwave background (CMB) anisotropies are gently perturbed by gravitational lensing.
Here we present a detailed study of this effect, detecting lensing independently in the 100, 143, and 217\,GHz frequency bands with an overall significance of greater than $25\sigma$.
We use the temperature-gradient correlations induced by lensing to reconstruct a (noisy) map of the CMB lensing potential,  which provides an integrated measure of the mass distribution back to the CMB last-scattering surface.
Our lensing potential map is significantly correlated with other tracers of mass, a fact which we demonstrate using several representative tracers of large-scale structure.
We estimate the power spectrum of the lensing potential, finding generally good agreement with expectations from the best-fitting $\Lambda$CDM model for the \Planck\ temperature power spectrum, showing that this measurement at $z=1100$ correctly predicts the properties of the lower-redshift, later-time structures which source the lensing potential.
When combined with the temperature power spectrum, our measurement provides degeneracy-breaking power for parameter constraints; it improves CMB-alone constraints on curvature by a factor of two and also partly breaks the degeneracy between the amplitude of the primordial perturbation power spectrum and the optical depth to reionization, allowing a measurement of the optical depth to reionization which is independent of large-scale polarization data.
Discarding scale information, our measurement corresponds to a $4\%$ constraint on the amplitude of the lensing potential power spectrum, or a $2\%$ constraint on the root-mean-squared amplitude of matter fluctuations at $z \sim 2$.
}
   
\date{Arxiv v2: 30 July 2014}
   
\titlerunning{Gravitational lensing by large-scale structures with \textit{Planck}}
\authorrunning{Planck Collaboration}
\maketitle
\vspace{-0.15in}
\addtocounter{footnote}{2}
\twocolumn
\newpage
\section{Introduction}
\label{sec:intro}

This paper, one of a set of papers associated with the 2013 release of data from the \Planck\footnote{\Planck\ (\url{http://www.esa.int/Planck}) is a project of the European Space Agency (ESA) with instruments provided by two scientific consortia funded by ESA member states (in particular the lead countries France and Italy), with contributions from NASA (USA) and telescope reflectors provided by a collaboration between ESA and a scientific consortium led and funded by Denmark.} mission~\citep{planck2013-p01}, describes our reconstruction of the cosmic microwave background (CMB) lensing potential based on 15 months of data, estimation of the lensing potential power spectrum, and a first set of associated science results. \all2013resultspapers

When Blanchard and Schneider first considered the effect of gravitational lensing on the CMB anisotropies in 1987, they wrote with guarded optimism that although ``such an observation is far from present possibilities [...] such an effect will not be impossible to find and to identify in the future...''~\citep{1987A&A...184....1B}.
In the proceeding years, and with the emergence of the concordance $\Lambda$CDM cosmology, a standard theoretical picture has emerged, in which the large-scale, linear structures of the Universe which intercede between ourselves and the CMB last-scattering surface induce small but coherent \citep{1989MNRAS.239..195C} deflections of the observed CMB temperature and polarisation anisotropies, with a typical magnitude of $2\arcm$.
These deflections blur the acoustic peaks \citep{Seljak:1995ve}, generate small-scale power \citep{1990MNRAS.243..353L,Metcalf:1997ih}, non-Gaussianity \citep{Bernardeau:1996aa}, and convert a portion of the dominant $E$-mode polarisation to $B$-mode \citep{Zaldarriaga:1998ar}.
Gravitational lensing of the CMB is both a nuisance, in that it obscures the primordial fluctuations \citep{Knox:2002pe}, as well as a potentially useful source of information; the characteristic signatures of lensing provide a measure of the distribution of mass in the Universe at intermediate redshifts (typically \mbox{$0.1 < z < 5$}).
In the $\Lambda$CDM framework, there exist accurate methods to calculate the effects of lensing on the CMB power spectra \citep{Challinor:2005jy}, as well as optimal estimators for the distinct statistical signatures of lensing \citep{Hu:2001kj,Hirata:2002jy}.

In recent years there have been a number of increasingly sensitive experimental measurements of CMB lensing.
Lensing has been measured in the data of the \WMAP satellite both in cross-correlation with large-scale-structure probed by galaxy surveys \citep{Hirata:2004rp, Smith:2007rg, Hirata:2008cb,Feng:2012uf}, as well as internally at lower signal-to-noise \citep{Smidt:2010by,Feng:2011jx}.
The current generation of low-noise, high-resolution ground-based experiments has done even better;
the Atacama Cosmology Telescope (ACT) has provided an internal detection of lensing at $4.6\sigma$ \citep{Das:2011ak,Das:2013zf}, and the South Pole Telescope detects lensing at $8.1\sigma$ in the temperature power spectrum, and $6.3\sigma$ from a direct reconstruction of the lensing potential \citep{Keisler:2011aw,vanEngelen:2012va,Story:2012wx}.
Significant measurements of the correlation between the reconstructed lensing potential and other tracers of large-scale structure have also been observed \citep{Bleem:2012gm,Sherwin:2012mr}.

\Planck\ enters this field with unique full-sky, multi-frequency coverage.
Nominal map noise levels for the first data release (approximately $105$, $45$, and $60\, \muKarcmin$ for the three CMB channels at 100, 143, and 217\,GHz respectively) are approximately five times lower than those of \WMAP (or twenty five times lower in power), 
and the \Planck\ beams (approximately $10\arcm$, $7\arcm$ and $5\arcm$ at 100, 143, and 217\,GHz), are small enough to probe the $2\parcm4$ deflections typical of lensing.
Full sky coverage is particularly beneficial for the statistical analysis of lensing effects, as much of the ``noise'' in temperature lens reconstruction comes from CMB fluctuations themselves, which can only be beaten down by averaging over many modes.

Lensing performs a remapping of the CMB fluctuations, such that the observed temperature anisotropy in direction $\hatn$ is given in terms of the unlensed, ``primordial'' temperature anisotropy as (e.g.~\citealt{Lewis:2006fu})
\begin{align}
T(\hatn) 
&= T^{\rm unl}( \hatn + \nabla \phi(\hatn) ), \nonumber \\
&= T^{\rm unl}(\hatn) + \sum_i \nabla^{i} \phi(\hatn) \nabla_{i} T(\hatn) + {\cal O}(\phi^2),
\label{eqn:lensing_remapping}
\end{align}
where $\phi(\hatn)$ is the CMB lensing potential, defined by
\be
\phi(\hatn) = -2 \int_0^{\chi_*} d\chi
\frac{ f_K( \chi_* - \chi) }{ f_K(\chi_*) f_K(\chi)}
\Psi(\chi \hatn; \eta_0 - \chi ).
\label{eqn:lensing_deflection}
\ee
Here $\chi$ is conformal distance (with \mbox{$\chi_* \approx 14000\,\mathrm{Mpc}$} denoting the distance to the CMB last-scattering surface) and $\Psi( \chi \hatn, \eta )$ is the (Weyl) gravitational potential at conformal distance $\chi$ along the direction $\hatn$ at conformal time $\eta$ (the conformal time today is denoted as $\eta_0$).
The angular-diameter distance $f_K(\chi)$ depends on the curvature of the Universe, and is given by
\be
f_K(\chi)=\begin{cases}
 K^{-1/2} \sin (K^{1/2} \chi) & \text{for $K>0$ (closed)}  , \\
\chi & \text{for $K=0$ (flat)}  , \\
|K|^{-1/2} \sinh (|K|^{1/2} \chi) & \text{for $K<0$ (open)}. \\
\end{cases}
\label{f_K_def}
\ee
The lensing potential is a measure of the integrated mass distribution back to the last-scattering surface.
To first order, its effect on the CMB is to introduce a correlation between the lensed temperature and the gradient of the unlensed temperature, a property which can be exploited to make a (noisy) reconstruction of the lensing potential itself.

\begin{figure}[!htpb]
\begin{center}
\includegraphics[width=0.9\columnwidth]{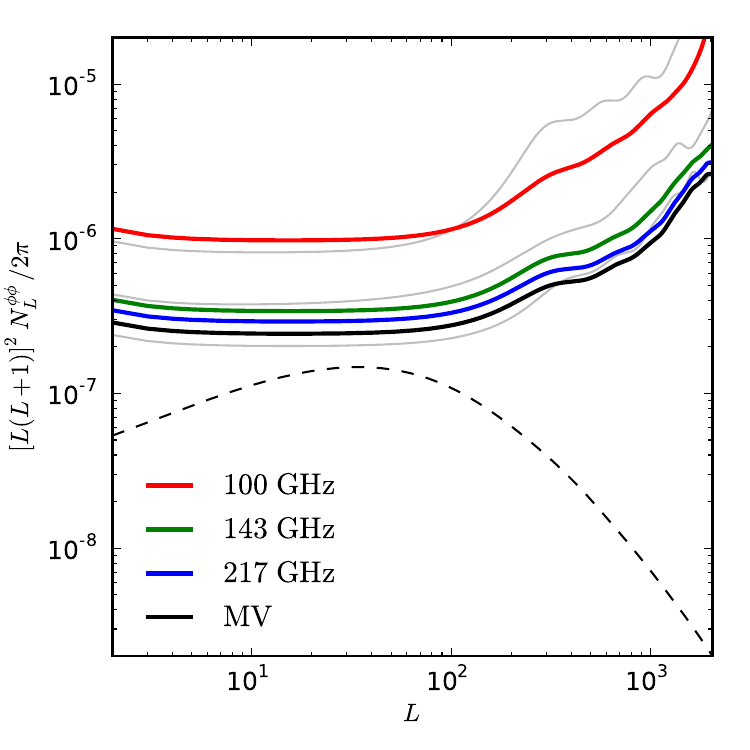}
\vspace{-0.1in}
\caption{Sky-averaged lens reconstruction noise levels for the 100, 143, and 217\,GHz \Planck\ channels (red, green, and blue solid, respectively), as well as for experiments that are cosmic-variance limited to a maximum multipole  $\elt_{\rm max}=1000$, $1500$, and $1750$ (upper to lower solid grey lines). 
A fiducial $\Lambda$CDM lensing potential using best-fit parameters to the temperature power spectrum from~\citet{planck2013-p11}
is shown in dashed black.
The noise level for a minimum-variance (``MV'') combination of $143+217$\,GHz is shown in black (the gain from adding 100\,GHz is negligible).
\label{fig:nlpp_hfi_and_cv}
}
\end{center}

\begin{center}
\includegraphics[width=0.9\columnwidth]{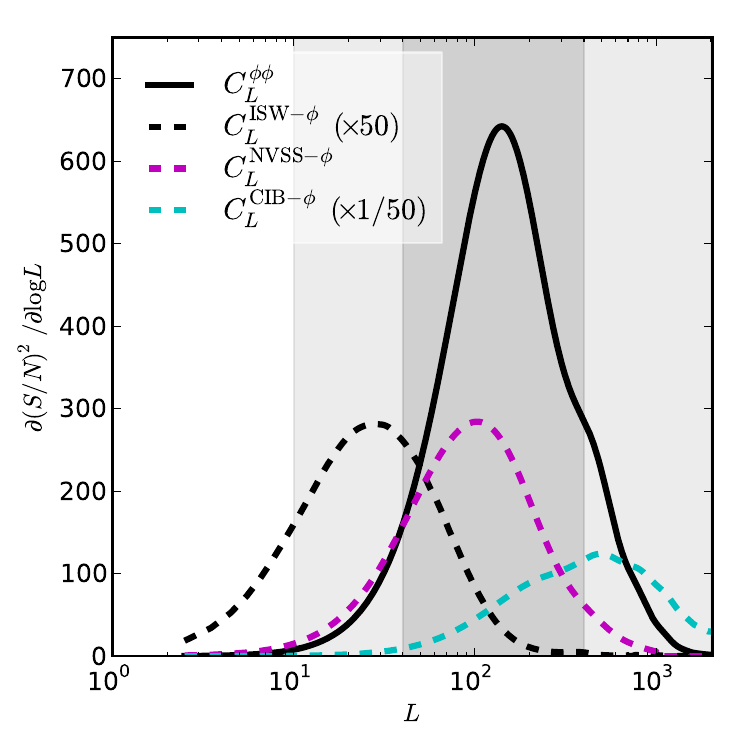}
\vspace{-0.1in}
\caption{
Overview of forecasted contributions to the detection significance as a function of lensing multipole $L$ for the $C_L^{\phi\phi}$ power spectrum (solid black), as well as for several other mass tracers, 
at the noise levels of our MV lens reconstruction.
Our measurement of the power spectrum $C_L^{\phi\phi}$ is presented in Sect.~\ref{sec:results},
The ISW-$\phi$ correlation believed to be induced by dark energy is studied in Sect.~\ref{sec:results:iswlensing}.
The NVSS-$\phi$ correlation is studied (along with other galaxy correlations) in Sect.~\ref{subsec:xcorr}.
The CIB-$\phi$ prediction (dashed cyan) uses the linear SSED model of \cite{Hall:2009rv}, assuming no noise or foreground contamination.
A full analysis and interpretation of the CIB-$\phi$ correlation is performed in \cite{planck2013-p13}. 
\label{fig:clpp_cltp_sn}
}
\end{center}
\end{figure}

In Fig.~\ref{fig:nlpp_hfi_and_cv} we plot the noise power spectrum $N_L^{\phi\phi}$ for reconstruction of the lensing potential using the three
\Planck\ frequencies which are most sensitive to the CMB anisotropies on the arcminute angular scales at which lensing effects become apparent.
The angular size of the \Planck\ beams ($5\arcm$ FWHM and greater) does not allow a high signal-to-noise ratio ($S/N$) reconstruction of the lensing potential for any individual mode (our highest $S/N$ on an individual mode is approximately $2/3$ for the 143 and 217\,GHz channels, or $3/4$ for a minimum-variance combination of both channels), however with full-sky coverage the large number of modes that are probed provides considerable statistical power.
To provide a feeling for the statistical weight of different regions of the lensing measurement, in Fig.~\ref{fig:clpp_cltp_sn} we plot (forecasted) contributions to the total detection significance for the potential power spectrum $C_L^{\phi\phi}$ as a function of lensing multipole $L$.
In addition to the power spectrum of the lensing potential, there is tremendous statistical power in cross-correlation of the \Planck\ lensing potential with other tracers of the matter distribution.
In Fig.~\ref{fig:clpp_cltp_sn} we also plot forecasted $S/N$ contributions for several representative tracers.

This paper describes the production, characterization, and first science results for two \Planck-derived lensing products:
\begin{enumerate}[(I)]
\item A map of the CMB lensing potential $\phi(\hatn)$ over a large fraction of the sky (approximately $70\%$). 
This represents an integrated measure of mass in the entire visible Universe, with a peak sensitivity to redshifts of $z \sim 2$.
At the resolution of \mbox{\Planck,} this map provides an estimate of the lensing potential down to angular scales of $5\arcm$ at $L=2048$, 
corresponding to structures on the order of $3\, \mathrm{Mpc}$ in size at $z=2$.
\item An estimate of the lensing potential power spectrum $C_L^{\phi\phi}$ and an associated likelihood, which is used in the cosmological parameter analysis of \mbox{\cite{planck2013-p11}}.
Our likelihood is based on the lensing multipole range \mbox{$40 \le L \le 400$}. 
This multipole range (highlighted as a dark grey band in Fig.~\ref{fig:clpp_cltp_sn}), was chosen as the range in which \Planck\ has the greatest sensitivity to lensing power, encapsulating over $90\%$ of the anticipated signal-to-noise, while conservatively avoiding the low-$L$ multipoles where mean-field corrections due to survey anisotropy (discussed in Appendix~\ref{sec:errorbudget:meanfields}) are large, and the high-$L$ multipoles where there are large corrections to the power spectra from Gaussian (disconnected) noise bias.
Distilled to a single amplitude, our likelihood corresponds to a $4\%$ measurement of the amplitude of the fiducial $\Lambda$CDM lensing power spectrum, or a $2\%$ measurement of the amplitude of the matter fluctuations (neglecting parameter degeneracies).
\end{enumerate}
Our efforts to validate these products are aided by the frequency coverage of the three \Planck\ channels that we employ, which span a wide range of foreground, beam, and noise properties.
For the mask levels that we use, the root-mean-squared (RMS) foreground contamination predicted by the \Planck\ sky model~\citep{delabrouille2012} has an amplitude of $14$, $22$, and $70\, \mu \mathrm{K}$ at 100, 143, and 217\,GHz, which can be compared to a CMB RMS for the \Planck\ best-fitting $\Lambda$CDM power spectrum of approximately $110\, \mu \mathrm{K}$.
The dominant foreground component at all three CMB frequencies is dust emission, both from our Galaxy as well as the cosmic infrared background (CIB), although at 100\,GHz free-free emission is thought to constitute approximately $15\%$ of the foreground RMS.
Contamination from the thermal Sunyaev-Zeldovich (tSZ) effect is a potential worry at 100 and 143\,GHz, but negligible at 217\,GHz \citep{SunyaevZeldovich1980}.
On the instrumental side, these frequency channels also span a wide range of beam asymmetry, with typical ellipticities of  
$19\%$, $4\%$, and $18\%$ at 100, 143, and 217\,GHz.
The magnitude of correlated noise on small scales (due to deconvolution of the bolometer time response) also varies significantly.
The ratio of the noise power (before beam deconvolution) at $\ell=1500$ to that at $\ell=500$ is a factor of $1.5$, $1.1$, and $1.0$ at 100, 143, and 217\,GHz.
The agreement of lens reconstructions based on combinations of these three channels allows a powerful suite of consistency tests for both foreground and instrumental biases. We will further validate the robustness of our result to foreground contamination using the component-separated maps from the \planck\ consortium \citep{planck2013-p06}.

At face value, the $4\%$ measurement of $C_L^{\phi\phi}$ in our fiducial likelihood corresponds to a $25\sigma$ detection of gravitational lensing effects.
In fact, a significant fraction (approximately $25\%$ of our error bar) is due to sample variance of the lenses themselves, and so the actual ``detection'' of lensing effects (under the null hypothesis of no lensing) is significantly higher.
We have also been conservative in terms of mask and multipole range in the construction of our fiducial lensing likelihood.
As we will show in Sect.~\ref{sec:consistency:comp_sep}, we obtain consistent results on sky fractions larger than our fiducial $70\%$ sky mask.

The \Planck\ lensing potential is part of a significant shift for CMB lensing science from the detection regime to that of precision cosmological probe.
The NVSS quasar catalogue, for example, has been a focus of previous lensing cross-correlation studies with \WMAP~\citep{Hirata:2004rp, Smith:2007rg, Hirata:2008cb}, where evidence for cross-correlation was found at approximately $3.5\sigma$. 
As we will see in Sect.~\ref{subsec:xcorr}, the significance for this correlation with \Planck\ is now $20\sigma$.
Notably, this is less than the significance with which lensing may be detected internally with \Planck.
The lensing potential measured by \Planck\ now has sufficient signal-to-noise that shot noise of the NVSS quasar catalogue is the limiting source of noise in the cross-correlation.

The majority of this paper is dedicated to the production and testing of the \Planck\ lensing map and power spectrum estimate.
Our focus here is on extracting the non-Gaussian signatures of lensing, although we note that lensing effects are also apparent at high significance ($10\sigma$) as a smoothing effect in the \Planck\ temperature power spectra \citep{planck2013-p08}.
We begin in Sect.~\ref{sec:methodology}, where we describe and motivate our methodology for producing unbiased estimates of the lensing potential and its power spectrum.
The \Planck\ maps and data cuts that are used for this purpose are described in Sect.~\ref{sec:data}, and the simulations that we use to characterize our reconstruction and its uncertainties are described in Sect.~\ref{sec:simulations}.
In Sect.~\ref{sec:error_budget} we give an overview of our error budget, and discuss the various sources of systematic and statistical uncertainty for our lensing estimates.
In Sect.~\ref{sec:results} we present our main results: the first \Planck\ lensing map and a corresponding estimate of the lensing potential power spectrum.
The likelihood based on this power spectrum is combined with the likelihood for the temperature anisotropy power spectrum~\citep{planck2013-p08} to derive parameter constraints in~\citet{planck2013-p11}. 
In Sect.~\ref{sec:results:cosmo} we highlight a subset of parameter results where the information provided by the lensing likelihood has proven particularly useful.
In the concordance $\Lambda$CDM cosmology, there is believed to be a correlation between the CMB lensing potential and the low-$\elt$ temperature anisotropies, driven by the effects of dark energy. 
We also present a measurement of this correlation in Sect.~\ref{sec:results:iswlensing}. 
Finally, we connect our lensing potential map to other tracers of large-scale structure with several illustrative cross-correlations using galaxy, quasar, cluster and infrared source catalogues in Sect.~\ref{subsec:xcorr}.
These main results are followed in Sect.~\ref{sec:consistency} by a large suite of systematic and consistency tests, where we perform null tests against a variety of different data cuts and processing.
We conclude in Sect.~\ref{sec:conclusions}. A series of appendices provide further details on some technical aspects of our methodology and lensing potential estimates.

\label{fiducialcosmology}
Throughout this paper, when we refer to the concordance or fiducial $\Lambda$CDM cosmology we are referring to a model with baryon density
$\omega_{\rm b} = \Omega_{\rm b}h^2 = 0.0221$, 
cold dark matter density $\omega_{\rm c} =\Omega_{\rm c}h^2 = 0.1199$, 
density parameter for the cosmological constant
$\Omega_{\Lambda} = 0.6910$, 
Hubble parameter $H_0 = 100 h \,\mathrm{km}\,\mathrm{s}^{-1}\,\mathrm{Mpc}^{-1}$ with $h=0.6778$, 
spectral index of the power spectrum of the primordial curvature perturbation
$n_{\rm s} = 0.96$, amplitude of the primordial power spectrum (at $k=0.05\,\mathrm{Mpc}^{-1}$) $A_{\rm s} = 2.21\times 10^{-9}$, and Thomson optical depth through reionization $\tau=0.093$. These values were determined from a pre-publication analysis of the \Planck\ temperature power spectrum, but are consistent with the best-fit values quoted in~\citet{planck2013-p11}.
Throughout this work we shall frequently quote lensing bandpower amplitudes relative to this fiducial model for ease of comparison, although as discussed in Sect.~\ref{sec:methodology} our lensing likelihood itself is designed to be insensitive to the choice of fiducial model.

\section{Methodology}
\label{sec:methodology}

In this section, we detail our methodology for reconstructing the lensing potential and estimating its angular power spectrum.
These are both obtained by exploiting the distinctive statistical properties of the lensed CMB.
\begin{enumerate}[(I)]
\item If we consider a fixed lensing potential applied to multiple realizations of the CMB temperature anisotropies, then lensing introduces \textit{statistical anisotropy} into the observed CMB; the fluctuations are still Gaussian, however the covariance varies as a function of position and orientation on the sky.
We use this idea to obtain a (noisy) estimate of $\phi(\hatn)$.
The noise of this map is a combination of instrumental noise and statistical noise due to the fact that we only have a single realization of the CMB to observe, analogous to shape noise in galaxy lensing.

\item If we consider averaging over realizations of both the lensing potential and the CMB fluctuations, then lensing introduces \textit{non-Gaussianity} into the observed CMB. This appears at lowest order in the connected part of the CMB 4-point function, or trispectrum\footnote{The ISW-lensing correlation also introduces a non-zero bispectrum. When correlating the reconstructed $\phi(\hatn)$ with the large-angle temperature anisotropies in Sect.~\ref{sec:results:iswlensing}, we are probing this bispectrum.}.
We use this to measure the lensing power spectrum $C_L^{\phi\phi}$.
\end{enumerate}
The estimators that we use are derived from maximizing the likelihood function of the lensed CMB 
under the assumption that the instrumental noise is Gaussian and the lensed CMB is perturbatively Gaussian and statistically isotropic.
These estimators are optimal (in the minimum-variance sense).
In cases where we have made suboptimal analysis choices, we provide estimates of the loss of signal-to-noise.

\subsection{Lens reconstruction}
\label{sec:methodology:lens_reconstruction}
To gain intuition for the process of lens reconstruction, it is useful to consider the effect of lensing on a small patch of the sky.
Lensing remaps the temperature fluctuations by a deflection field $\nabla \phi(\hatn)$.
The part of $\nabla \phi(\hatn)$ that is constant over our patch is not an observable effect; it describes only a re-centering of the map.
The variation of the deflection field across the patch \textit{is} observable, however. 
This can be usefully decomposed into convergence ($\kappa$) and shear modes ($\gamma_{+}$, $\gamma_{-}$) as
\be
-\nabla_{\!i} \nabla_{\!j}\phi(\hatn) = 
\left[ \! \begin{array}{cc} 
\kappa + \gamma_+ & \gamma_{-}   \\
\gamma_{-}  & \kappa - \gamma_+ 
\end{array} \! \right](\hatn).
\ee
If we observe a patch that is small enough that these quantities can be taken as constant, then the observational consequences are simple.
The convergence mode causes a local change of scale, either magnifying or demagnifying the fluctuations.
Taking the local power spectrum of our small patch, we would find that the CMB peaks would shift to larger or smaller scales, relative to the full-sky average.
The shear modes also describe changes of scale, however they are now orientation dependent. 
On a small patch, convergence and shear estimators can be constructed from local estimates of the (orientation-dependent) power spectrum and then stitched together to recover the lensing potential $\phi$ \citep{Zaldarriaga:1998ar,Bucher:2010iv}.
This procedure describes a \textit{quadratic estimator} for the local convergence and shear.

From the description above, it is not immediately clear how to go about stitching together estimates of convergence and shear in different regions of the sky, or what weight to give the local power spectrum estimates as a function of scale.
These questions can be resolved by considering a generic form for the quadratic estimator, and optimizing its weight function for sensitivity to lensing \citep{Okamoto:2003zw}.
To first order in the lensing potential, the statistical anisotropy introduced by lensing appears as an off-diagonal contribution to the covariance matrix of the CMB:
\be
\Delta \langle T_{\elt_1 m_1} T_{\elt_2 m_2} \rangle =
\sum_{LM}
(-1)^M
\threej{\elt_1}{\elt_2}{L}{m_1}{m_2}{-M} W_{\elt_1 \elt_2 L}^{\phi} \phi_{LM},
\label{eqn:tt_p_form}
\ee
where the average $\langle \rangle$ is taken over CMB realizations with a fixed lensing potential.
Here the bracketed term is a Wigner $3j$ symbol, $\phi_{LM} = \int d^2\hatn Y_{LM}^{*}(\hatn) \phi(\hatn)$ is the harmonic transform of the lensing potential, and the weight function $W_{\elt_1 \elt_2 L}^{\phi}$ is given by
\begin{multline}
W_{\elt_1 \elt_2 L}^{\phi} 
= -  \sqrt{\frac{(2\elt_1+1)(2\elt_2+1)(2L+1)}{4\pi}}  \sqrt{L(L+1) \elt_1 (\elt_1 +1) } \\ \times
C_{\elt_1}^{TT} 
\left( \frac{1 + (-1)^{\elt_1 + \elt_2 + L}}{2} \right) \threej{\elt_1}{\elt_2}{L}{1}{0}{-1} + (\elt_1 \leftrightarrow \elt_2).
\label{eqn:qe_weight_lensing}
\end{multline}
Here $C_{{\elt}}^{TT}$ is the ensemble-average power spectrum of the lensed CMB.
In our analysis, we will use the fiducial model described at the end of Sect.~\ref{sec:intro} to determine $C_{{\elt}}^{TT}$, however our lensing likelihood can be renormalized to account for uncertainties in this model.
In this approach, errors in the fiducial model do not bias our lensing bandpower estimates, they only result in slight sub-optimality.
Note that we use the lensed power spectrum here, rather than the unlensed spectrum that is sometimes used in the literature, as this is accurate to higher order in $\phi$ \citep{Lewis:2011fk}, an improvement which is necessary at \Planck\ sensitivity \citep{Hanson:2010rp}.
Use of the unlensed spectrum would lead to biases on the order of $15\%$ at $L<200$.

Now we construct a quadratic estimator to search for the covariance which is introduced by lensing.
We will use several different estimators for the lensing potential, as well as to probe possible point-source contamination, and so it will be useful to keep this discussion as general as possible.
A completely generic quadratic estimator for the lensing potential can be written as
\be
\hat{\phi}^{x}_{LM} = \sum_{L'M'} \left[ {\cal R}^{x \phi} \right]^{-1}_{LM, L'M'} \left[ \Q_{L'M'} - \Q_{L'M'}^{\MF} \right],
\ee
where ${\cal R}^{x \phi}$ is a normalization matrix, and $\bar{x}_{LM}$ is a quadratic ``building block'' which takes in a pair of filtered sky maps $\bar{T}_{{\elt}m}^{(1)}$ and
$\bar{T}_{{\elt}m}^{(2)}$, and sums over their empirical covariance matrix with a weight function $W_{\elt_1 \elt_2 L}^{x}$:
\be
\Q_{LM} = 
\frac{1}{2} \sum_{\elt_1 m_1, \elt_2 m_2} 
(-1)^{M}
\threej{\elt_1}{\elt_2}{L}{m_1}{m_2}{-M} W_{\elt_1 \elt_2 L}^{x} 
\bar{T}^{(1)}_{\elt_1 m_1} \bar{T}^{(2)}_{\elt_2 m_2}.
\label{eqn:qe_block}
\ee
The ``mean-field'' term $\Q^{\MF}_{LM}$ accounts for all \textit{known} sources of statistical anisotropy in the map, which could otherwise bias the lensing estimate.
It is given by
\be
\Q^{\MF}_{LM} = 
\frac{1}{2} \sum_{\elt_1 m_1, \elt_2 m_2} 
(-1)^{M}
\threej{\elt_1}{\elt_2}{L}{m_1}{m_2}{-M} W_{\elt_1 \elt_2 L}^{x} 
\langle \bar{T}^{(1)}_{\elt_1 m_1} \bar{T}^{(2)}_{\elt_2 m_2} \rangle,
\label{eqn:lensing_meanfield}
\ee
where the ensemble average here is taken over realizations of the CMB and noise.

We may now optimize the generic quadratic estimator above.
If the primordial CMB fluctuations and instrumental noise are Gaussian and the lensing potential is fixed, then the likelihood for the observed CMB fluctuations is still a Gaussian, which may be maximized with respect to the lensing potential modes $\phi_{LM}$ \citep{Hirata:2002jy}.
The optimal quadratic estimator is the first step of an iterative maximization of this likelihood, and it has been shown that additional iterations of the estimator are not necessary for temperature lens reconstruction \citep{Hirata:2002jy,Okamoto:2003zw}.
The optimal quadratic estimator has the following choices for the weight function and filtering.
\begin{enumerate}[(I)]
\item The weight function $W^{x}$ should be a matched filter for the covariance induced by lensing (i.e., one should use $\bar{\phi}$, with weight function given by Eq.~\ref{eqn:qe_weight_lensing}). 
We shall use this weight function for all of our fiducial results, although for consistency tests we will also use ``bias-hardened'' estimators, which have weight functions constructed to be orthogonal to certain systematic effects \citep{Namikawa:2012pe}.
This is discussed further in Sect.~\ref{sec:biashardened}.

\item The filtered temperature multipoles $\bar{T}_{{\elt}m}$  should be given by $\bar{T}_{{\elt}m} = (C^{-1} T)_{{\elt}m}$, where $T$ is a beam-deconvolved sky map and $C$ is its total signal+noise covariance matrix.
We describe our approximate implementation of this filtering in Appendix~\ref{app:ivf}. 
When combining multiple frequencies for our minimum-variance estimator, all of the available data are combined into a single map which is then filtered and used for both input multipoles of the quadratic estimator.
It can be desirable to use different pairs of maps however, and we use this for several consistency tests.
For example, we feed maps with independent noise realizations into the quadratic estimator to avoid possible noise biases in Sect.~\ref{sec:consistency:noisebias}.
\end{enumerate}
In the quadratic maximum-likelihood estimator, the mean-field correction emerges from the determinant term in the likelihood function, and it can be seen that the normalization matrix ${\cal R}$ is the Fisher matrix for the $\phi_{LM}$; this means that the normalization is the same as the covariance matrix of the lens reconstruction, and so the \textit{unnormalized} lensing estimate $\bar{\phi}= \bar{x}-\bar{x}^{\MF}$ is equivalent to an inverse-variance-weighted lens reconstruction, which is precisely the quantity needed for most statistical analysis.
This is why we have denoted it with an overbar, in analogy to $\bar{T}$. 

We choose to treat the map noise as if it were homogeneous when constructing the filtered $\bar{T}_{{\elt}m}$, and do not account for variation with hit count across the sky.
This is a slightly suboptimal filtering choice; in Appendix.~\ref{app:ivf} we estimate that it leads to a $5\%$ loss of total signal-to-noise when constraining the power spectrum of the lensing potential.
The advantage of this approach, however, is that far from the mask boundaries our filtering asymptotes to a simple form, given by
\be
\bar{T}_{{\elt}m} \approx \left[C_{\elt}^{TT} + C_{\elt}^{NN} \right]^{-1} T_{{\elt}m} \equiv F_{\elt} T_{{\elt}m},
\label{eqn:fl}
\ee
where $C_{\elt}^{TT}$ is the temperature power spectrum and $C_{\elt}^{NN}$ is the power spectrum of the homogeneous noise level that we use in our filtering.
For the purposes of compact notation, in the following equations we combine both of these elements in the ``filter function'' $F_{\elt}$.
The asymptotic form of our filtering, Eq.~\eqref{eqn:fl}, will prove useful, as it means that the normalization of our estimator, as well as its variance and response to various systematic effects, may be accurately modelled analytically.
It allows us to propagate uncertainties in the beam transfer function and CMB power spectrum, for example, directly to our lens reconstruction.
This filtering choice also means that the normalization does not vary as a function of position on the sky, which simplifies the analysis of cross-correlations between the lensing potential map and external tracers.
Under the approximation of Eq.~\eqref{eqn:fl}, the estimator normalization is given by 
\be
{\cal R}^{x\phi}_{LM, L'M'} = \delta_{L L'} \delta_{M M'} {\cal R}_L^{x\phi},
\ee
where the response function ${\cal R}_L^{x\phi}$ for filtered maps $\bar{T}^{(1)}$ and
$\bar{T}^{(2)}$ is
\be
\resp_L^{x \phi, (1)(2)} =  \frac{1}{(2L+1)} \sum_{\elt_1 \elt_2} \frac{1}{2} W_{\elt_1 \elt_2 L}^{x} W_{\elt_1 \elt_2 L}^{\phi} F_{\elt_1}^{(1)} F_{\elt_2}^{(2)}.
\label{eqn:respxp}
\ee
This can be read as ``the response of estimator $x$ to lensing on scale $L$''. 
The filter functions $F_{{\elt}}$ are those used for $\bar{T}^{(1)}$ and $\bar{T}^{(2)}$ respectively.
In cases where the filter functions are obvious, we will drop the indices above.

Putting all of the above together, for a chosen quadratic estimator $\Q$ we obtain normalized, mean-field-debiased estimates of the lensing potential $\phi$ as
\be
\hat{\phi}^{x}_{LM} = \frac{1}{\resp^{x\phi}_{L}} 
\left( \Q_{LM} - \Q^{\MF}_{LM} \right).
\label{eqn:phihat}
\ee
Note that our normalization function $\resp^{x\phi}_L$ is only approximate, but we will verify its accuracy in Sect.~\ref{sec:simulations}.
For the standard lensing estimator of \cite{Okamoto:2003zw} (which uses the weight function of Eq.~\ref{eqn:qe_weight_lensing}), we use $x=\phi$. 
This estimator is denoted simply as $\hat{\phi}_{LM}$.

\subsection{Lensing power spectrum estimation}
\label{sec:methodology:power_spectrum_reconstruction}
We form estimates for the power spectrum of the lensing potential by taking spectra of the lensing estimates from Sect.~\ref{sec:methodology:lens_reconstruction}, using a simple pseudo-$C_{\elt}$ estimator.
In order to reduce mode coupling, as well as to downweight regions near the analysis boundary where the mean-field due to masking can be large, we take the power spectrum from an apodized version of our lensing estimate, given by
\be
\phihatapo^x_{LM} = P_L^{-1} \int d\hatn Y_{LM}^*(\hatn) \maskapo(\hatn) \left[ \sum_{L' M'} Y_{L'M'}(\hatn) P_{L'} \hat{\phi}^x_{L'M'} \right],
\label{eqn:apophi}
\ee
where $\maskapo(\hatn)$ is an apodized version of the analysis mask $M(\hatn)$ used in our filtering and $P_L \equiv L(L+1)$ is an approximate pre-whitening operation.
The construction of $\maskapo(\hatn)$ is described in Sect.~\ref{sec:data}.
Our fiducial apodization occurs over a band of approximately $5\deg$, and effectively reduces the sky fraction by $9\%$.

The power spectrum of $\phihatapo$ probes the 4-point function of the observed CMB, which contains both disconnected and connected parts.
We model it as being due to a combination of Gaussian CMB fluctuations, lensing effects and unresolved point-source shot noise, and estimate the power spectrum of the lensing potential with
\begin{multline}
\hat{C}_{L, x}^{\phi\phi} = \frac{f_{\rm sky, 2}^{-1}}{2L+1} \sum_{M} | \phihatapo^x_{LM} |^2 
- \left. \Delta C_L^{\phi\phi} \right|_{\mathsc{N0}}
\\
- \left. \Delta C_L^{\phi\phi} \right|_{\mathsc{N1}}
- \left. \Delta C_L^{\phi\phi} \right|_{\mathsc{PS}}
- \left. \Delta C_L^{\phi\phi} \right|_{\mathsc{MC}},
\label{eqn:clppest}
\end{multline}
where $f_{\rm sky, 2} = \int d\hatn \maskapo^2(\hatn) / 4\pi$ is the average value of the square of the apodizing mask.
The first line of Eq.~\eqref{eqn:clppest} isolates the connected part of the CMB 4-point function, or trispectrum, which would be zero for Gaussian fluctuations.
The second line contains corrections which isolate the part of the trispectrum which is directly proportional to the non-Gaussianity induced by $C_L^{\phi\phi}$.
In the following paragraphs, we explain these terms in more detail.

The first correction term $\left. \Delta C_L^{\phi\phi} \right|_{\mathsc{N0}}$ subtracts the (large) disconnected contribution to the power spectrum of $\phihatapo$.
To determine this term, we use the data-dependent subtraction which emerges for maximum-likelihood estimators of the CMB trispectrum (\citealt{Regan:2010cn}; see also Appendix~\ref{app:likelihood}).
For lensing, this procedure has the additional advantage of reducing the correlation between different multipoles $L \ne L'$ of the lens reconstruction 
to a level which is negligible at $\Planck$ resolution and noise levels \citep{Schmittfull:2013uea}, 
as well as reducing sensitivity to uncertainties in our model of the CMB and noise covariance matrices \citep{Namikawa:2012pe}.
Writing the power spectrum of $\phihatapo_{LM}$ explicitly as a function of the four inverse-variance filtered temperature maps
\be
 C_{L, x}^{\phihatapo \phihatapo}[ \bar{T}^{(1)}, \bar{T}^{(2)}, \bar{T}^{(3)}, \bar{T}^{ (4)} ] \equiv \frac{f^{-1}_{\rm sky,2}}{2L+1} \sum_M  | \phihatapo^x_{LM} |^2,
\ee
the disconnected contribution reads 
\begin{multline}
\left. \Delta C_{L, x}^{\phi\phi} \right|_{{\mathsc N0}} = 
\Bigg<  - C_{L, x}^{\phihatapo \phihatapo}\left[ \bar{T}^{(1)}_{\mathsc{mc}}, \bar{T}^{(2)}_{\mathsc{mc}'}, \bar{T}^{(3)}_{\mathsc{mc}'}, \bar{T}^{ (4)}_{\mathsc{mc}} \right] \\
+ C_{L, x}^{\phihatapo\phihatapo}\left[ \bar{T}^{(1)}_{\mathsc{mc}}, \bar{T}^{(2)} , \bar{T}^{ (3)}_{\mathsc{mc}}, \bar{T}^{(4)} \right] 
+ C_{L, x}^{\phihatapo \phihatapo} \left[ \bar{T}^{(1)}_{\mathsc{mc}}, \bar{T}^{(2)} , \bar{T}^{ (3) }, \bar{T}^{ (4)}_{\mathsc{mc}} \right]  \\
+ C_{L, x}^{\phihatapo \phihatapo}\left[ \bar{T}^{(1)}, \bar{T}^{(2)}_{\mathsc{mc}} , \bar{T}^{(3)}, \bar{T}^{(4)}_{\mathsc{mc}} \right]  
+ C_{L, x}^{\phihatapo \phihatapo}\left[ \bar{T}^{(1)}, \bar{T}^{(2)}_{\mathsc{mc}} , \bar{T}^{(3)}_{\mathsc{mc}}, \bar{T}^{(4)} \right]  \\
- C_{L, x}^{\phihatapo \phihatapo}\left[ \bar{T}^{(1)}_{\mathsc{mc}}, \bar{T}^{(2)}_{\mathsc{mc}'}, \bar{T}^{(3)}_{\mathsc{mc}}, \bar{T}^{ (4)}_{\mathsc{mc}'} \right]
\Bigg>_{\mathsc{mc}, \mathsc{mc}'} ,
\label{eqn:cln0}
\end{multline}
where $\bar{T}_{\mathsc{mc}}$ indicates a Monte-Carlo simulation of the corresponding map. The ensemble average is taken over two sets of independent realizations $\mathsc{mc}$ and $\mathsc{mc'}$.
Note that because of the way we have used pairs of Monte-Carlo simulations and data with independent CMB and noise realizations, the mean-field correction is zero for all of the terms above.

The term $\Delta C_L^{\phi\phi} |_{\mathsc{N1}}$ corrects for the ``$N^{(1)}$'' bias due to secondary contractions of the lensing trispectrum \nopagebreak\citep{Hu:2001fa,Kesden:2003cc}.
It is only a large effect at $L>100$, and so we calculate it using a flat-sky expression as
\begin{multline}
\left. \Delta C_{|\vec{L}|, x}^{\phi\phi} \right|_{\mathsc{N1}} = 
\frac{1}{ \resp_{|\vec{L}|}^{x\phi, (1)(2)} \resp_{|\vec{L}|}^{x\phi, (3)(4)} }
\int \frac{d^2 \vec{l}_1}{(2\pi)^2}  
\int \frac{d^2 \vec{l}_3}{(2\pi)^2} \\
F^{(1)}_{ | \vec{l}_1 |} F^{(2)}_{ | \vec{l}_2 |} F^{(3)}_{ | \vec{l}_3 |} F^{(4)}_{ | \vec{l}_4 |}
W^{x}( \vec{l}_1, \vec{l}_2 ) W^{x}( \vec{l}_3, \vec{l}_4 ) 
\\ \times \bigg[
C_{ | \vec{l}_1 - \vec{l}_3 | }^{\phi\phi, {\rm fid.}} W^{\phi}( -\vec{l}_1, \vec{l}_3 ) W^{\phi}( -\vec{l}_2, \vec{l}_4 ) \\ + 
C_{ | \vec{l}_1 - \vec{l}_4 | }^{\phi\phi, {\rm fid.}} W^{\phi}( -\vec{l}_1, \vec{l}_4 ) W^{\phi}( -\vec{l}_2, \vec{l}_3 ) \bigg],
\label{eqn:n1}
\end{multline}
where $\vec{l}_1 + \vec{l}_2 = \vec{l}_3 + \vec{l}_4 = \vec{L}$ and 
$C_{\elp}^{\phi\phi,\,{\rm fid.}}$ is a fiducial model for the lensing potential power spectrum.
The $W(\vec{l}, \vec{l}')$ are flat-sky analogues of the full-sky weight functions. 
The flat-sky lensing weight function, for example, is
\be
W^{\phi}(\vec{l}_1, \vec{l}_2) = 
C_{ | \vec{l}_1 | }^{TT} \vec{l}_1 \cdot \vec{L} + 
C_{ | \vec{l}_2 | }^{TT} \vec{l}_2 \cdot \vec{L}.
\label{eqn:flat_w}
\ee
The $N^{(1)}$ term is proportional to the lensing potential power spectrum, and so in principle it should be used to improve our constraints on $C_L^{\phi\phi}$ rather than subtracted as an additive bias. However the statistical power of this term is relatively small at \Planck\ noise levels.
From a Fisher matrix calculation, the trispectrum contractions which source the $N^{(1)}$ term are only detectable in the \Planck\ data at $4\sigma$ significance, compared to approximately $25\sigma$ for the primary contractions.
We choose simply to subtract the $N^{(1)}$ term from our power spectrum estimates.
There is a small cosmological uncertainty in the $N^{(1)}$ correction due to uncertainty in the $C_L^{\phi\phi}$ power spectrum, which we discuss in Sect.~\ref{sec:errorbudget:cosmological}.

The $\left. \Delta{C}_L^{\phi\phi} \right|_{\mathsc{PS}}$ term is a correction for the bias induced by the 
shot-noise 
of unresolved point sources 
(including SZ clusters), 
and will be discussed in more detail in Sect.~\ref{sec:pointsources}.

Finally, the $\left. \Delta C_L^{\phi\phi} \right|_{\mathsc{MC}}$ term is a small correction that we obtain by estimating $\hat{C}_L^{\phi\phi}$ following the procedure above on a number of lensed CMB realizations, and then subtracting the input power spectrum. This term can be non-zero due to pseudo-$C_{\elt}$ leakage effects from masking, which we have not accounted for other than apodization, errors in our calculation of the $N^{(1)}$ term, or errors in the normalization at the power spectrum level. We will find that $\left. \Delta C_L^{\phi\phi} \right|_{\mathsc{MC}}$ is sufficiently small that in practice it does not matter whether we account for it as a renormalization or an additive offset, and we choose to treat it as an offset for simplicity.

We will ultimately characterize the uncertainty of $\hat{C}_{\elp}^{\phi\phi}$ by Monte Carlo, however the following analytical expression is a useful approximation
\be
{\rm Var}( \hat{C}_{L, x}^{\phi\phi} ) 
\approx
V_L
\equiv
\frac{1}{f_{{\rm sky, 2}}} \frac{2}{2L+1}  \left[  C_L^{\phi\phi} + N_{L, x}^{\phi\phi} \right]^2 
\label{eqn:var_clpp_analytical}
\ee
where $N_{L, x}^{\phi\phi}$ is the reconstruction noise level.
We take it to be
\begin{multline}
N_{L, x}^{\phi\phi} =
\frac{1}{(2L+1)}
\frac{1}{ \resp_L^{x \phi, (1)(2)} \resp_L^{x \phi, (3)(4)} }
\sum_{\elt_1 \elt_2} \frac{1}{4} \left| W_{\elt_1 \elt_2 L}^{x} \right|^2 
\\ \times
\left( 
\hat{C}^{TT, (1)(3)}_{\elt_1}  \hat{C}^{TT, (2)(4)}_{\elt_2}
+
\hat{C}^{TT, (1)(4)}_{\elt_1}  \hat{C}^{TT, (2)(3)}_{\elt_2}
\right)
\label{eqn:nlpp_fullsky}
\end{multline}
where $\hat{C}^{(i)(j)}_L$ is the ensemble-average cross-spectrum between $\bar{T}^{(i)}$ and $\bar{T}^{(j)}$, given by
\be
\hat{C}^{TT, (i)(j)}_L = \frac{f_{{\rm sky}, {(i)(j)}}^{-1}}{2L+1} \sum_{M} \left< \bar{T}^{(i)}_{LM} \bar{T}^{(j)*}_{LM} \right>,
\ee
where $f_{{\rm sky}, {(i)(j)}}$ is the fraction of sky common to both $\bar{T}^{(i)}$ and $\bar{T}^{(j)}$.
We will use this analytical estimate of the variance for weighting our $C_L^{\phi\phi}$ estimates in statistical analyses.

\subsection{Lensing power spectrum likelihood}
\label{sec:methodology:likelihood}
Based on our measurements of the lensing potential power spectrum,
we construct a Gaussian likelihood based on bins in $C_L^{\phi\phi}$.
Our likelihood has the form
\be
-2 \ln {\cal L}_{\phi} (C^{\phi\phi}) =
\binner^{L}_{i} 
\left( \hat{C}^{\phi \phi}_L - C^{\phi \phi}_L \right)
\left[ \Sigma^{-1} \right]^{ij}
\binner^{L'}_{j} \
\left( \hat{C}^{\phi \phi}_{L'} - C_{L'}^{\phi \phi} \right),
\label{eqn:likelihood_form}
\ee
where $\binner$ represents a binning function, $\Sigma$ is the covariance matrix between bins, and sums are performed over paired upper/lower indices.
As the shape and amplitude of $C_{\elp}^{\phi\phi}$ are constrained strongly in the concordance $\Lambda$CDM model, with which our results are broadly consistent, we choose a binning function designed to maximize our sensitivity to small departures from the fiducial $\Lambda$CDM expectation. This is given by
\be
\binner^{L}_{i} =
\frac{
C_L^{\phi\phi,\,{\rm fid.}} V_L^{-1} 
}{
\sum_{L' = L_{\rm min}^{i}}^{L_{\rm max}^{i}} 
(C_{L'}^{\phi\phi,\,{\rm fid.}})^2 V^{-1}_{L'} 
}
\qquad  \mbox{if } 
L_{\rm min}^{i} \le L \le L_{\rm max}^{i},
\ee
where $V_L$ is defined in Eq.~\eqref{eqn:var_clpp_analytical} and $L_{\rm min}^{i}$ and $L_{\rm max}^{i}$ define the multipole range of the bin.
Our binned results  correspond to an estimate of the amplitude of a fiducial $C_L^{\phi\phi}$ power spectrum in a given multipole range, normalized to unity for the fiducial model.
We denote these amplitude estimates explicitly for a given lens reconstruction $\hat{C}_L^{\phi\phi}$ as
\be
\hat{A}_i = 
\binner_i^{L} \hat{C}^{\phi\phi}_L.
\label{eqn:amplitude_a_def}
\ee

In principle there are several reasons why a likelihood approach such as the one we have described above
could fail; the usual issues with a pseudo-$C_{\elt}$ likelihood,
such as the non-Gaussianity of the $C_L^{\phi\phi}$ are compounded
by the fact that the $\phi$ estimates themselves are non-Gaussian,
and derived from the temperature data itself, which means that the
measurement uncertainties on the lens reconstruction are potentially correlated
with those of the temperature power spectrum. 
In Appendix~\ref{app:likelihood}, we validate the above approach
to the likelihood by considering these issues in more detail.

\subsection{Unresolved point-source correction}
\label{sec:pointsources}
At the high multipoles ($\elt > 1000$) that provide most of the modes which are useful for lens reconstruction, the contribution from unresolved 
(and therefore undetected and unmasked) 
extragalactic foregrounds becomes apparent in the \Planck\ power spectra. 
For measurements of the lensing potential power spectrum, we should therefore be concerned with possible biases from the trispectrum non-Gaussianity of these unresolved sources.
The work of \cite{2013arXiv1310.7547O} indicates that at \Planck\ resolution and sensitivity, the contribution of unresolved foregrounds to lens reconstruction is small but potentially non-negligible, with the largest contribution coming from the shot noise trispectrum of unresolved radio and tSZ sources. 
Given uncertainties in modelling the non-Gaussianity of unresolved point sources, our approach is heavily data dependent.
In addition to tests for the consistency of lens reconstructions at 143 and 217\,GHz, we use two additional methods to measure and correct for point-source contamination in our analysis.

In our fiducial lensing power spectrum analysis,
we measure the amplitude of the shot-noise contribution to the data trispectrum and correct the measured lensing spectrum accordingly. 
We measure the shot-noise amplitude using the power spectrum of a quadratic estimator designed to detect the ``noise'' due to sources, similar to the approach advocated in \cite{Munshi:2009wy}.
The trispectrum (or connected four-point function) for point-source shot noise is defined in position space as
\begin{multline}
\langle T(\hatn_1) T(\hatn_2) T(\hatn_3) T(\hatn_4) \rangle = \\
S^4 \delta( \hatn_1 - \hatn_2 ) \delta( \hatn_2 - \hatn_3 ) \delta( \hatn_3 - \hatn_4 ),
\end{multline}
or in harmonic space:
\be
\langle T_{\elt_1 m_1} T_{\elt_2 m_2} T_{\elt_3 m_3} T_{\elt_4 m_4} \rangle = S^4 \int d\hatn Y_{\elt_1 m_1} Y_{\elt_2 m_2} Y_{\elt_3 m_3} Y_{\elt_4 m_4}.
\ee
The quadratic estimator that we use to measure the amplitude $S^4$ is denoted as $\bar{s}$, and is defined by the weight function
\be
W_{\elt_1 \elt_2 L}^{s} = 
 \sqrt{\frac{(2\elt_1+1)(2\elt_2+1)(2L+1)}{4\pi}}
\threej{\elt_1}{\elt_2}{L}{0}{0}{0}.
\label{eqn:qe_weight_ptsrc}
\ee
We measure the amplitude of the shot-noise trispectrum $S^4$ as
\be
\widehat{S^4} = 
\left( \sum_{L} \hat{C}^{ss}_{L} S^{-1}_L \right)
\Bigg/
\left( \sum_{L'} S^{-1}_{L'} \right)
,
\ee
where $\hat{C}^{ss}_\elp$ is defined analogously to Eq.~\eqref{eqn:clppest} and the sums are taken over \mbox{$\elp, {\elp}' \in [\elp_{\rm min}^s, \elp_{\rm max}^s]$}. 
The power spectrum weighting $S_{L}$ is given by 
\be
S_{L}
\equiv
\frac{1}{f_{{\rm sky, 2}}} \frac{2}{2L+1}  \frac{1}{ \resp^{ss, (1)(2)}_L \resp^{ss, (3)(4)}_L }.
\ee
In the limit that \mbox{$[L_{\rm min}^s , L_{\rm max}^s]=[0, \infty]$} this estimator is equivalent to measuring the shot-noise trispectrum using the 1-point kurtosis of the map. 
The advantage of the trispectrum-related power spectrum approach is that it allows us to separate out regions of the trispectrum measurement which are contaminated by lensing, as well as to look for evidence of clustering, which would appear as a deviation of the measured spectrum $\hat{C}_L^{ss}$ from the shape expected for unclustered sources.
We then calculate and remove an estimated bias to the measured lensing potential given by
\be
\Delta \hat{C}_{L, x}^{\phi\phi} |_{\rm PS} = 
\widehat{S^4 }
\resp_L^{x s, (1)(2)}  \resp_L^{x s, (3)(4)}.
\label{eqn:dclps}
\ee

The shot-noise correction described above does not take into account the correlation of sources with the dark matter distribution and hence the lensing potential. 
In Sect.~\ref{sec:biashardened} we therefore additionally perform tests using a ``point source bias-hardened estimator'', constructed using the weight function of Eq.~\eqref{eqn:qe_weight_ptsrc}. 
This bias-hardened estimator has zero response to both the point source shot-noise trispectrum $(S^4)$, as well as to the primary trispectrum contractions involving the correlation between point sources and the lensing potential $(S^2 \phi)$. 

\section{Data and cuts}
\label{sec:data}

\paragraph{\Planck\ sky maps:}
The majority of the results in this paper are based on the \Planck\ nominal-mission frequency
maps at 100, 143, and 217\,GHz,
built from the first 15.5 months of data.
These are in HEALPix\footnote{\url{http://healpix.jpl.nasa.gov}} format, with resolution parameter $N_{\rm side}=2048$, corresponding to pixels with a typical width of $1\parcm7$.
These have effective noise levels of approximately $105\,\muKarcmin$ at 100\,GHz, $45\,\muKarcmin$ at 143\,GHz, and $60\,\muKarcmin$ at 216\,GHz.
The beam widths shrink with frequency, and are $10\arcmin$ at 100\,GHz, $7\arcmin$ at 143\,GHz, and $5\arcmin$ at 217\,GHz.
Our primary products 
-- a map of the lensing potential and an associated power spectrum likelihood -- 
are based on a minimum-variance (hereafter MV) combination of the 143 and 217\,GHz maps.
Although lensing may be detected at a significance of approximately $10\sigma$ at 100\,GHz, the CMB modes used in this reconstruction are already sample-variance dominated at 143 and 217\,GHz, and so adding it to the MV combination would lead to negligible improvement.
In addition to these maps, we also use the 857\,GHz \Planck\ map as a dust template, which is projected out of both maps independently in our filtering procedure (described further in Appendix~\ref{app:ivf}).
This projection is primarily intended to remove diffuse Galactic dust contamination, although it also removes a portion of the CIB fluctuations, which have a similar spectral index to that of Galactic dust over these frequency bands \citep{planck2011-7.12,planck2011-6.6}.
As a simple approach, this template projection overlooks several potential difficulties, including variation of Galactic dust spectral indices across the sky as well as mismatch between the beams at 100/143/217 and 857\,GHz, although we find it is adequate for our purposes.
As lens reconstruction is most sensitive to small-scale modes, the coupling to large-scale Galactic foregrounds is relatively weak.
In Sect.~\ref{sec:consistency} we will also perform lens reconstruction using the more rigorously component-separated maps of \cite{planck2013-p06}, finding good agreement with our baseline results.

In analyzing these maps, we use the fiducial beam transfer functions described in \cite{planck2013-p03c} and \cite{planck2013-p28}.
There are uncertainties associated with these transfer functions, which we propagate to an uncertainty in the lensing estimator normalization in Sect.~\ref{sec:errorbudget:beamtransfer}.

\paragraph{Galaxy mask:} 
We avoid the majority of Galactic foreground power using the temperature analysis masks described in \cite{planck2013-p28}.
These are constructed using a combination of the $30\,$GHz and $353\,$GHz maps, corrected for an estimate of the CMB contribution, smoothed to $5'$ and thresholded until a desired sky fraction is obtained \citep{planck2013-p06}.
For our baseline results, we use the $70\%$ masks (which remove $30\%$ of the sky), although in Sect.~\ref{sec:consistency} we will show that we obtain consistent results with both larger and smaller masks.
When computing the power spectrum, we additionally multiply our $\phi$ estimates by an apodized version of the Galaxy mask. 
Each pixel outside of the masked region is multiplied by an apodized weight varying between zero at the mask boundary and unity at a distance greater than $5\deg$ from the closest masked pixel.
We use a sinusoid weight function, similar to the one used in \cite{Namikawa:2012pe} and \cite{BenoitLevy:2013bc}.
Note that due to this apodization, the effective sky fraction used in our power spectrum analysis is approximately $9\%$ lower than the sky fraction of our reconstruction.

\paragraph{CO and extended-object masks:} 
We mask regions believed to be contaminated by carbon-monoxide (CO) emission lines at 100 and 217\,GHz using the {\sc{type~2}} CO map described in \cite{planck2013-p03a}.
We mask all pixels above 
$60\,\muK$ in the map after degrading to $N_{\rm side} = 256$, and then restore by hand all isolated pixels that are removed by this cut but do not appear to be extended CO regions.
This low-resolution mask is prograded back to $N_{\rm side} = 2048$ and smoothed at $20\arcm$, slightly larger than the map resolution of $15\arcm$.
We additionally exclude extended nearby 
objects (the two Magellanic Clouds, M31, M33, and M81) by cutting out disks centred on their locations. 
The radii of these disks range from $250\arcm$ for the Large Magellanic Cloud to $30\arcm$ for M81.

\paragraph{Point-source masks:} 
We remove detected point sources (or otherwise compact objects) using a mask constructed from a combination of sources identified in
the \Planck\ Early Release Compact Source Catalogue (ERCSC; \citealt{planck2012-VII}),
the SZ clusters from the \Planck\ Cluster Catalogue (PCC; \citealt{planck2013-p05a}),
and the \Planck\ Catalogue of Compact Sources (PCCS; \citealt{planck2013-p05}). 
We produce individual masks for 100, 143, and 217\,GHz,
cutting out detected sources with disks having radii of either $3\sigma$ or $5\sigma$ depending on their flux level, 
where $\sigma$ describes the Gaussian beam-width of the given channel and is taken to be 
$4.1\arcm$ at 100\,GHz, $3.1\arcm$ at 143\,GHz, and $2.1\arcm$ at 217\,GHz.
SZ clusters are not masked at 217\,GHz. 
For the PCCS we mask all sources with $S/N \geq 5$ in a given band, or $S/N \geq 10$ in either adjacent frequency band, using a $3\sigma$ disc,
extending to a $5\sigma$ disk for any sources with $S/N \geq 10$ in the target band.
For the ERCSC, we make a $3\sigma$ cut for every source detected at the target frequency, as well as for every source which is detected with a flux greater than 
$1\,\Jy$ at 70\,GHz, $0.7\,\Jy$ at 100\,GHz, $0.4\,\Jy$ at 143\,GHz, $0.5\,\Jy$ at 217\,GHz, or $0.9\,\Jy$ at 353\,GHz.
For sources that satisfy the brightness criterion at the target frequency, we also enlarge the cut to $5\sigma$.
For SZ clusters we make a cut between $3\sigma$ and $5\sigma$ depending on the cluster angular size $\theta_{500}$ 
for all sources in the MMF1 catalog with $S/N \geq 5$.
We further mask cool cores using the Early Cold Cores catalog \citep{planck2011-1.10} in a similar fashion, using a maximum 15\arcmin excluding radius.

In Sect.~\ref{sec:consistency:pointsource} we will test our sensitivity to this S/N cut, varying it from $S/N \geq 5$ to $S/N \geq 4.2$ and $S/N \geq 10$.

\section{Simulations}
\label{sec:simulations}
\begin{figure}[!ht]
\begin{center}
\hspace{-0.2in} \includegraphics[width=0.95\columnwidth]{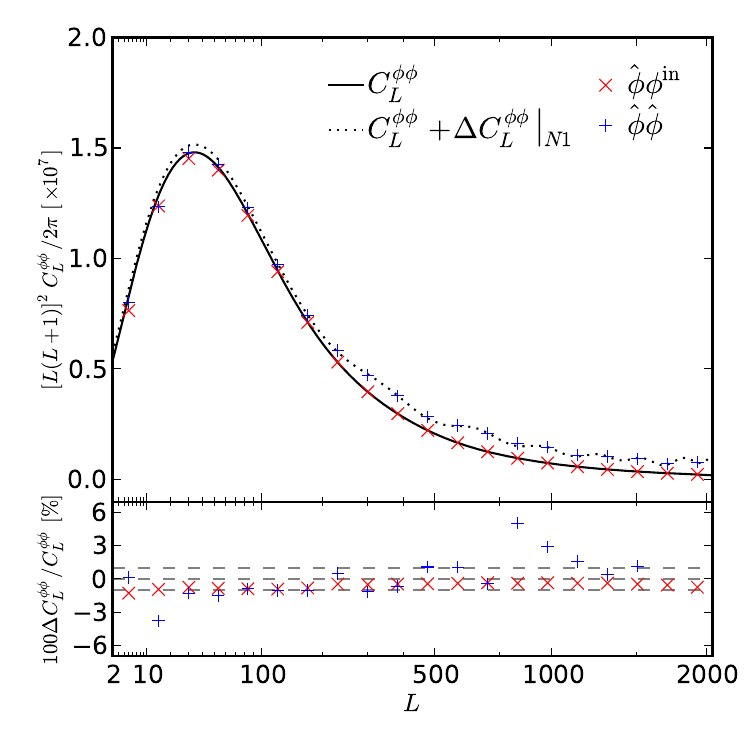}
\end{center}
\vspace{-0.25in}
\caption{
Validation of our estimator normalization for simulations of the MV reconstruction at the map and power spectrum levels.
The map normalization (plotted as $\hat{\phi} \phi^{\rm in}$) is tested by taking the cross-spectrum of the input $\phi$ with the reconstruction averaged over Monte-Carlo simulations, divided by an $f_{\rm sky}$ factor to account for missing power in the mask.
The power spectrum normalization (plotted as $\hat{\phi} \hat{\phi}$) is obtained by averaging the first line of Eq.~\eqref{eqn:clppest} over  simulations, and then comparing it to the expected value, which is $C_L^{\phi\phi} + \left. \Delta C_L^{\phi\phi} \right|_{\mathsc{N1}}$ because our simulations do not contain point-source non-Gaussianity.
\label{fig:normalizationtest}
}
\vspace{0.4in}

\centerline{
\begin{overpic}[width=0.45\columnwidth]{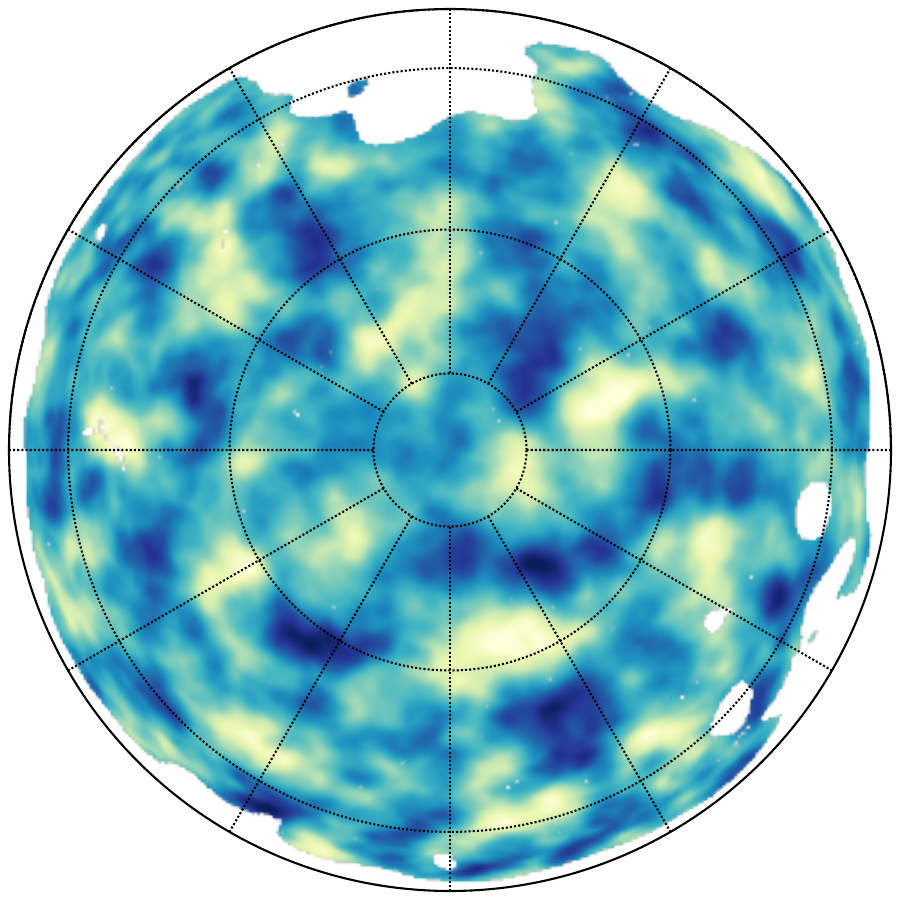}
\put(10,0){\tiny Sim}
\put(0,95){$\phi^{\WF}_{} (\hatn)$}
\end{overpic}
\hspace{1em}
\begin{overpic}[width=0.45\columnwidth]{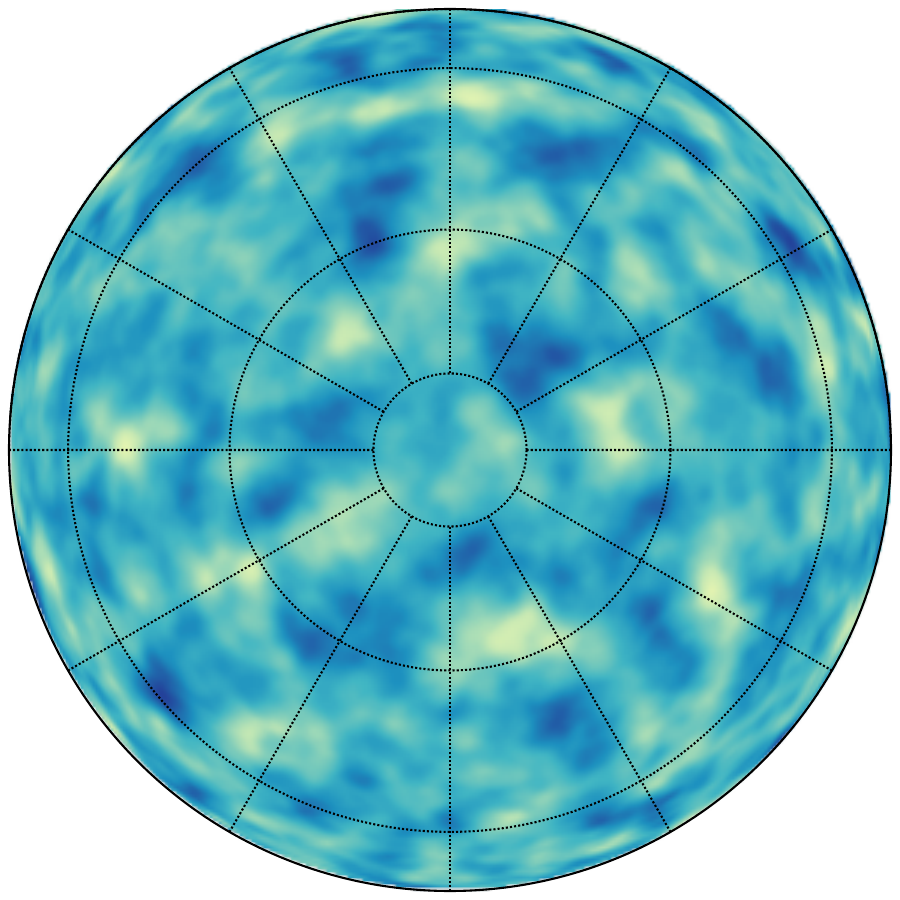}
\put(10,0){\tiny Input}
\end{overpic}
}
\vspace{0.125cm}
\caption{
Simulation of the Wiener-filtered lensing potential estimate \mbox{$\phi^{\WF}_{LM} \equiv C_L^{\phi\phi} (\bar{\phi}_{LM} - \bar{\phi}_{LM}^{\MF})$} for the MV reconstruction (left), and the input $\phi$ realization (right; filtered by $C_L^{\phi\phi} \resp_L^{\phi\phi}$ to be directly comparable to the Wiener estimate).
Both maps show the southern Galactic sky in orthographic projection. 
The lensing reconstruction is noise dominated on all scales, however correlations between the two maps can still be seen visually.
\label{fig:wiener_filtered_mv_reconstruction_sim}
}
\end{figure}

We require simulations of the data described in Sect.~\ref{sec:data} for several aspects of our analysis: 
to determine the lensing-mean field $\phi^{\MF}$, to determine the $\left. \Delta C_L^{\phi\phi}\right|_{\mathsc{N0}}$ and $\left. \Delta C_L^{\phi\phi}\right|_{\mathsc{MC}}$ correction terms of Eq.~\eqref{eqn:clppest}, and to determine our measurement error bars, as well as to validate our reconstruction methodology.
Here we outline the basic algorithm we use to generate these simulations.
\begin{enumerate}[(1)]
\item
Simulate the unlensed CMB and lensing potential. 
Using the fiducial $\Lambda$CDM theoretical unlensed power spectra for the cosmology given at the end of Sect.~1, 
we generate realizations ${T}^{\rm unl}_{{\elt}m}$ and $\phi_{LM}$ of the unlensed temperature and lensing potential with appropriate correlations. 
By default we simulate these fields up to $\elt_{\rm max} = 2560$.
In addition to these stochastic components, we also include a dipolar lensing mode in our simulations, to account for the Doppler aberration due to our motion with respect to the CMB \citep{Kamionkowski:2002nd,Challinor:2002zh}:
\be
\phi_{\beta}(\hatn) \approx \vec{\beta} \cdot \hatn,
\ee
where $\vec{\beta}$ is in the direction of Galactic coordinates $(l,b)=(264.4, 48.4)$, with amplitude $|\vec{\beta}| = 0.00123$, corresponding to a velocity of 369\kms \citep{Hinshaw:2008kr}. 
\item 
\label{sims:step:lens}
Lens the temperature field. We use a fast spherical harmonic transform to compute the temperature and deflection fields
\begin{align}
{T}^{\rm unl}(\hatn) &=\sum_{\elt m} Y_{\elt m}(\hatn) {T}^{\rm unl}_{{\elt}m}, \nonumber \\
\vec{d}(\hatn) &= \sum_M \mbox{\boldmath $\nabla$} Y_{LM}(\hatn) \phi_{LM}.
\end{align}
This computation is performed on an equicylindrical pixelization (ECP) grid with $(N_{\theta}, N_{\phi}) = (16384, 32768)$ equally-spaced points in $(\theta, \phi)$. We  determine the lensed temperature on this ECP grid as
\be
T(\hatn) = {T}^{\rm unl}(\hatn + \vec{d}(\hatn)).
\ee
This equation is evaluated following the same algorithm as the {\tt LENSPix}\ code \citep{Lewis:2005tp}, with 2D cubic Lagrange interpolation on the ECP grid, which has conveniently uniform pixel spacing.
We then obtain the lensed temperature multipoles $T^{\phi}_{{\elt}m}$ by a reverse harmonic transform.
The average power spectrum of these lensed simulations agrees with the implementation in CAMB \citep{Challinor:2005jy} to better than $0.1\%$ at $l<2048$. 
We have also used the algorithm of \cite{Basak:2008pq}, which gives consistent results.

In addition to the lensing dipole, our motion with respect to the CMB also causes a small modulation of the observed fluctuations, which we determine from the lensed temperature field as
\be
T^{\rm{mod}}(\hatn) = f_{\nu}  \vec{\beta} \cdot \hatn  {T}(\hatn),
\ee
where $f_{\nu}$ is a frequency dependent boost factor that we approximate as $1.5$, 2, and 3 at $100$, $143$, and $217$\,GHz, respectively.
There is an extended discussion of this factor in \cite{planck2013-pipaberration}.
\item 
Generate a Gaussian realization of extragalactic foreground power $T^{\rm fg}_{{\elt}m}$.
We simulate Gaussian foreground power at 100, 143, and 217\,GHz such that the auto- and cross-spectra of the foreground power at these frequencies has the form
\be
C_{\elt}^{A \times B} = A^{\rm PS}_{A \times B} C_{\elt}^{\rm PS} + A^{\rm CIB}_{A \times B} C_{\elt}^{\rm CIB},
\ee
where $A$ and $B$ label frequency bands.
The template power spectra are based on the modelling of  \cite{planck2013-p08,planck2013-p11}, and are given by
\begin{align}
C_{\elt}^{\rm PS} = \frac{2\pi}{3000^2}, \hspace{0.3in} C_{\elt}^{\rm CIB} = \frac{2\pi}{\elt(\elt+1)} \left( \frac{\elt}{3000} \right)^{0.8}.
\end{align}
We use the parameters
\begin{align}
A^{\rm PS}_{100 \times 100} &= 208\,\mu{\rm K}^2 & A^{\rm PS}_{100 \times 143} &= 72\,\mu {\rm K}^2 & A^{\rm PS}_{100 \times 217} &= 44\,\mu{\rm K}^2  \nonumber \\
A^{\rm PS}_{143 \times 143} &= 64\,\mu{\rm K}^2 & A^{\rm PS}_{143 \times 217} &= 43\,\mu{\rm K}^2 & \nonumber A^{\rm PS}_{217 \times 217} &= 57\,\mu{\rm K}^2 \\
A^{\rm CIB}_{143 \times 143} &= 4\, \mu{\rm K}^2 & A^{\rm CIB}_{143 \times 217} &= 14\, \mu {\rm K}^2 & \nonumber A^{\rm CIB}_{217 \times 217} &= 54\,\mu {\rm K}^2 
\end{align}
The parameters above were obtained by fitting the auto- and cross-spectra of our inverse-variance-filtered temperature maps to the templates above.
Our point source masks differ from those used in \cite{planck2013-p11}, and so these numbers cannot be directly compared to the values there.
The purpose of including a Gaussian foreground component in our simulations is primarily so that our measurement error bars will include the additional scatter due to the foreground power (we will see in Sect.~\ref{sec:error_budget} that foreground power constitutes a few percent of our total error budget).
Our lensing power spectrum results are otherwise unaffected by this component of our simulations. 
Our lensing estimates do include an overall correction for point-source shot noise, although we do not include the non-Gaussian covariance of the shot noise in our error budget, given that point sources are a small fraction of the total power at the multipoles probed by \Planck. We test this approach further in Sect.~\ref{sec:consistency:pointsource}.

\item 
Convolve $T^{\phi} + T^{\rm fg} + T^{\rm mod}$ with the (asymmetric) instrumental beam using Eq.~\eqref{eqn:map_beam_centered}. 
This results in a \HealpixPixelization\ map of the beam-convolved CMB $T^b(\hatn)$, from which we extract beam-convolved multipoles $T^{b}_{{\elt}m}$.
\item 
Make a final map of the observed sky $T(\hatn)$ with the \HealpixPixelization\ $N_{\rm side}=2048$ pixelization, with
\be
T(\hatn) = \sum_{{\elt}m} Y_{{\elt}m}(\hatn) B_{\elt}^{\rm pix} T_{{\elt}m} + n(\hatn),
\ee
where $B_{\elt}^{\rm pix}$ is the \HealpixPixelization\ $N_{\rm side}=2048$ pixel window function, which accounts for the effective smearing of the beam due to the binning of hits into pixels.
Although this smearing is an inadequate description of the pixelization-smearing effect for any individual pixel, it is a good approximation when averaged over all pixels on the sky.
We will discuss pixelization effects further in Sect.~\ref{sec:pixelization}.
The quantity $n(\hatn)$ is a simulated realization of the instrumental noise.
We use noise realizations from the FFP6 simulation set \citep{planck2013-p01}.
In order to make the power spectrum of our simulations more closely match the power spectrum of the data, we find that it is necessary to include an additional white noise component with an amplitude of $20$, $10$, and $10\,\muKarcmin$ at $100$, 143 and $217$\,GHz respectively. Indeed, the FFP6 simulations are produced on the basis of the half-ring estimates of the noise which are known to underestimate the map noise  \citep{planck2013-p03}.

\end{enumerate}
The simulations described above are used throughout this work both to debias our lensing estimates, as well as to characterize their uncertainty.
Here we briefly use them to validate the normalization of our lensing estimates.
There are two relevant normalizations: the map-level normalization which describes how our reconstruction traces the input $\phi$ realization, as well as the analogous normalization for the lensing potential power spectrum.
Throughout this work, we use the analytical normalization estimates of Eq.~\eqref{eqn:respxp} to normalize our map estimates.
In Fig.~\ref{fig:normalizationtest} we plot the average cross-correlation between the input and reconstructed $\phi_{LM}$ for our minimum-variance reconstruction, divided by a simple $f_{\rm sky}$ factor to account for missing power in the mask.
We can see that our analytical map-level normalization is accurate at the $1\%$ level.
This is small enough that we do not further renormalize our $\phi$ map.
At the power spectrum level, we find similar accuracy of our normalization in the $40<L<400$ region which forms the basis of our fiducial lensing likelihood.
The power spectrum estimates in Fig.~\ref{fig:normalizationtest} are obtained by averaging the first line of Eq.~\eqref{eqn:clppest} over a set of lensed but otherwise Gaussian simulations.
This $1\%$ error in the power spectrum normalization \textit{is} accounted for in our likelihood by the $\left. \Delta C_L^{\phi\phi} \right|_{\mathsc{MC}}$ term.
Given the good agreement of the $C_L^{\phi\phi}$ power spectrum used in our simulations with that measured from data, it does not matter whether we incorporate
the Monte-Carlo correction as a multiplicative or additive term.
We have chosen to incorporate it as additive for simplicity.
At higher multipoles, there is additional scatter in our power spectrum normalization test (up to $10\%$ for the plotted bins).
It is difficult to know from this test whether it is due to actual scatter in our power spectrum normalization (which seems unlikely, given that the map-level normalization is quite accurate), a small mis-estimation of the $N^{(1)}$ term (which begins to dominate over $C_L^{\phi\phi}$ above $L=500$), or simply pseudo-$C_{\elt}$ leakage issues.
In any case, it is still small, and is absorbed into our $\left. \Delta C_L \right|_{\rm MC}$ term when estimating power spectra at these multipoles.
Although we have only presented results for our minimum-variance reconstruction here, those for individual frequency results are similar.

As a visual illustration, and a preview of our data results in Sect.~\ref{sec:results}, in Fig.~\ref{fig:wiener_filtered_mv_reconstruction_sim}, we show a simulated lens reconstruction as well as the input $\phi$ map. This gives a visual impression of the signal-to-noise in our lens reconstruction.

\section{Error budget}
\label{sec:error_budget}
\begin{figure}[!htpb]
\begin{center}
\hspace{-0.2in} \includegraphics[width=0.95\columnwidth]{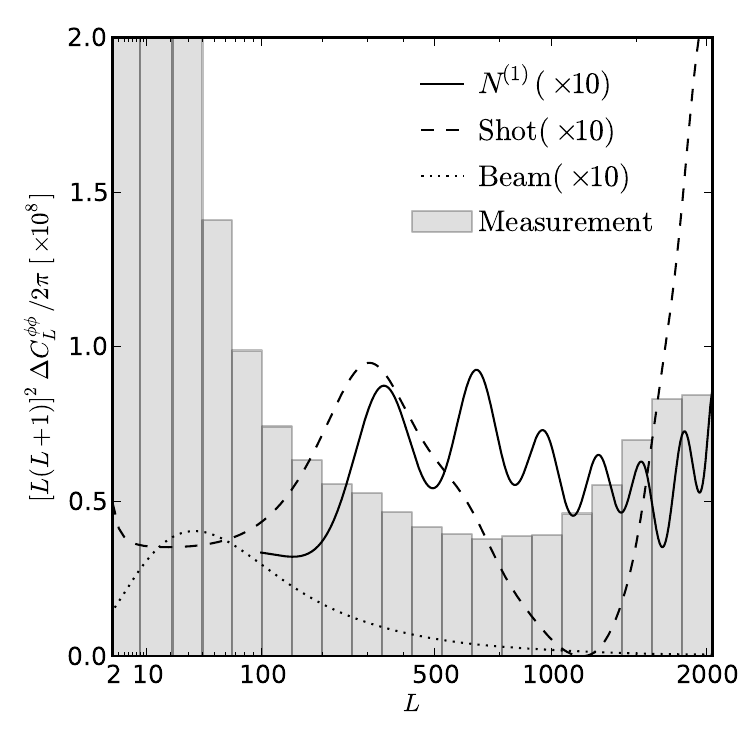}
\caption{Relative contribution of statistical measurement uncertainty (grey band, representing the $1\sigma$ uncertainty on $C_L^{\phi\phi}$ for the plotted bins) and the systematic errors we assign due to uncertainty in the $N^{(1)}$ bias correction, point-source shot noise correction, and the beam transfer function.
The contributions represent error eigenmodes in $C_L^{\phi\phi}$, and so are completely correlated between $L$.
Note that these contributions are much smaller than our measurement uncertainty, and they have been multiplied by a factor of 10 for clarity.
\label{fig:error_budget_mv}
}
\vspace{0.25in}
\hspace{-0.2in} \includegraphics[width=0.95\columnwidth]{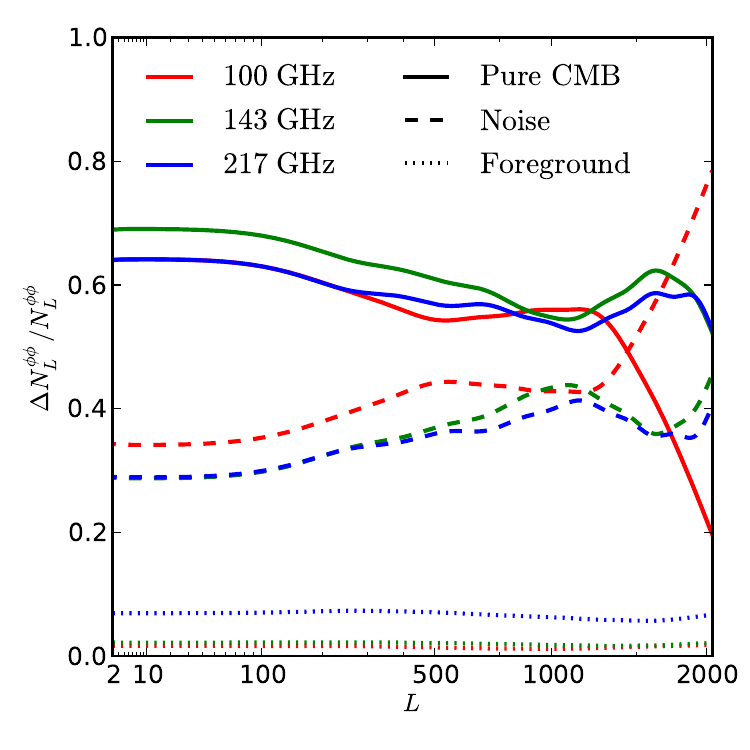}
\caption{
Relative contributions of CMB, instrumental noise, and foreground power terms discussed in Sect~\ref{sec:errorbudget:measurement} to the approximate lens reconstruction variance of Eq.~\eqref{eqn:nlpp_fullsky}, previously plotted in Fig.~\ref{fig:nlpp_hfi_and_cv}.
As discussed in Sect.~\ref{sec:errorbudget:measurement}, 
the finite number of CMB modes 
observed by \Planck
is the dominant source of variance for the lens reconstruction.
\label{fig:nlpp_hfi_and_cv_contributions}
}
\end{center}
\end{figure}

In this section, we describe the measurement and systematic error budget for our estimation of the lensing potential power spectrum.
This is broken down into three sections; in Sect.~\ref{sec:errorbudget:measurement} we describe our measurement (or ``statistical'') error bars, which are due to the fact that we have only a single noisy sky with a finite number of modes to observe.
In Sect.~\ref{sec:errorbudget:beamtransfer} we consider uncertainty in the instrumental beam transfer function, which we will see propagates to a normalization uncertainty for our lensing estimates.
Finally, in Sect.~\ref{sec:errorbudget:cosmological} we discuss the effect of cosmological uncertainty; possible errors in the fiducial model for $C_{\elt}^{TT}$ result in a normalization uncertainty for our lensing estimates, and uncertainties in the fiducial $C_L^{\phi\phi}$ power spectrum lead to uncertainties in the $N^{(1)}_L$ correction.
As a guide to the relative size and scale dependence of these terms, in Fig.~\ref{fig:error_budget_mv} we summarize the error budget for our fiducial minimum-variance lens reconstruction, based on 143 and 217\,GHz.
Individual frequency bands, as well as 100\,GHz are qualitatively similar.

\subsection{Measurement}
\label{sec:errorbudget:measurement}

Although our measurement uncertainties are ultimately assigned by Monte Carlo, we can use the analytical expression of Eq.~\eqref{eqn:nlpp_fullsky} to gain intuition for how they are sourced by various components.
Our simple model of the sky after masking and dust cleaning is that it consists of three uncorrelated signals: CMB, instrumental noise, and unresolved isotropic foreground power.
The noise variance of the lens reconstruction in Eq.~\eqref{eqn:nlpp_fullsky} involves two power spectra, and so we can think of the noise contribution as the sum of six possible terms involving pairs of the CMB, noise, and foreground power spectra.
In Fig.~\ref{fig:nlpp_hfi_and_cv_contributions} we combine these contributions into three representative contributions to the reconstruction noise: ``pure CMB'' in which both spectra are due to CMB fluctuations; the ``noise'' contribution in which either both spectra are those for noise power, or one is noise and one is CMB; and, finally, the ``foreground'' contribution in which either one or both of the spectra are due to unresolved foreground power.
We can see that for most reconstruction multipoles, the pure CMB contribution constitutes the largest part of the reconstruction noise, followed by noise.
The unresolved foreground power is a fairly small contribution to our measurement error.
Note that the dominant terms for both the ``noise'' and ``foreground'' contributions are the ones in which one of the spectra is a CMB fluctuation.
For this reason, we will focus less on the use of cross-spectra to avoid noise biases than is done for the usual CMB power spectra
\citep{planck2013-p08}, although we will perform consistency tests using cross-spectra of data to avoid noise biases.
Note that our realization-dependent method for removing the disconnected noise bias (Eq.~\ref{eqn:cln0}) means that the majority of this contribution is estimated directly from the data itself, reducing our sensitivity to uncertainty in the noise and foreground power.

\subsection{Beam transfer function}
\label{sec:errorbudget:beamtransfer}
Errors in the effective beam transfer function appear as an error in the normalization of our lensing estimates.
For simplicity here we will describe the case for a single standard quadratic lensing estimator that uses the same map for both of its inputs, although when dealing with combinations of channels for our actual results we account for differences in the beam transfer function and errors between ``legs'' of the estimator (or pair of estimators, in the case of the lensing potential power spectrum).

We model the effect on $\hat{\phi}$ from a discrepancy between the ``measured'' and ``true'' effective beam transfer function as
\be
\langle \hat{\phi}_{LM} \rangle = \frac{1}{{\cal R}_L^{\phi\phi}} \left[ \frac{1}{2L+1} \sum_{\elt_1 \elt_2} \frac{1}{2} \left| W_{\elt_1 \elt_2 L}^{\phi} \right|^2 
F_{\elt_1} F_{\elt_2}
\frac{B_{\elt_1}^{\rm true}}{B_{\elt_1}^{\rm meas}}  \frac{B_{\elt_2}^{\rm true}}{B_{\elt_2}^{\rm meas}} \right]  \phi_{LM} ,
\label{eqn:beam_normalization_error}
\ee
where the average is over CMB fluctuations for fixed lenses.
The statistics of $B^{\rm true}/B^{\rm meas}$ are discussed in \cite{planck2013-p03c}, where it is found that the beam uncertainty may be described by a small number
$N_{\rm eig}$ of eigenmodes (typically five),
\be
\ln \left( \frac{B_{\elt}^{\rm meas}}{B_{\elt}^{\rm true}} \right) = 
\sum_{i=1}^{N_{\rm eig}} g_i E_{\elt}^{b_i}.
\label{eqn:beam_error_eigenmodes}
\ee
Here, $E_{\elt}^{b_i}$ are the uncertainty eigenmodes (indexed by $i$) for the beam determination, and $g_{i}$ are independent Gaussian random variables with unit variance.
The frequency-band eigenmodes are small for 100--217\,GHz (no more than $1\%$ at $l<2048$),
and so we can equate
\be
\frac{B_{\elt}^{\rm true}}{ B_{\elt}^{\rm meas}} \approx 1 - \ln \left( \frac{B_{\elt}^{\rm meas}}{B_{\elt}^{\rm true}} \right).
\ee 
We can also ignore terms of order $E^2$ and propagate the independent beam uncertainties directly to a corresponding set of independent eigenmodes for $\phi$. 
These eigenmodes are given as
\be
E^{\phi_i}_{L} = -\frac{1}{ {\cal R}_L^{\phi\phi} } \left[ \frac{1}{2L+1} \sum_{\elt_1 \elt_2} \frac{1}{2} \left| W_{\elt_1 \elt_2 L}^{\phi} \right|^2 
F_{\elt_1} F_{\elt_2}
\sum_{i}
\left( E_{\elt_1}^{b_i} + E_{\elt_2}^{b_i} \right) \right],
\label{eqn:beam_error_lensing_eigenmodes}
\ee
with the model that
\be
\frac{\langle \hat{\phi}_{LM}\rangle }{\phi_{LM}} = 1 + \sum_{i=0}^{N_{\rm eig}} g_i E^{\phi_i}_{L}.
\ee
\begin{figure}[!ht]
\begin{center}
\hspace{-0.2in}
\includegraphics[width=0.95\columnwidth]{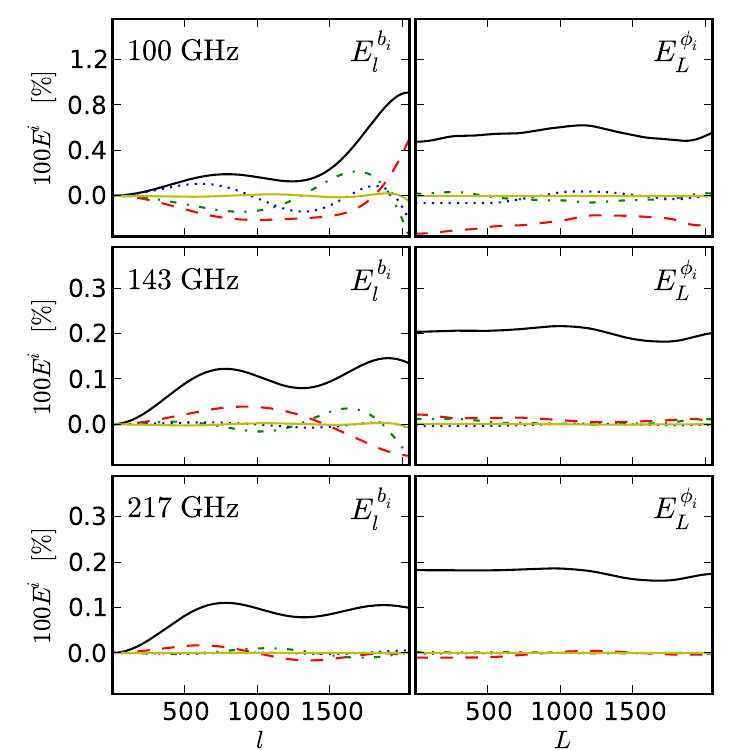}
\caption{
Propagation of eigenmodes for beam transfer function uncertainty (left-hand panels) to an uncertainty in the lensing normalization (right-hand panels).
The normalization eigenmodes are flat, with amplitudes given by approximately twice the beam uncertainty at the beam and noise cutoff scale (at $\elt\approx1000$ for 100\,GHz and $\elt\approx1500$ for 143 and 217\,GHz).
\label{fig:err_bl_al_eigs}
}
\end{center}
\end{figure}
In Fig.~\ref{fig:err_bl_al_eigs} we plot the beam uncertainty eigenmodes as well as the corresponding normalization eigenmodes.
The normalization eigenmodes are flatter than the beam uncertainty eigenmodes, and do not converge to zero at low-$L$, reflecting the fact that at all scales the lensing estimate takes most of its weight from modes of $\elt_1$ and $\elt_2$ close to the noise/beam cutoff (at $\elt\approx1000$ for 100\,GHz and $\elt\approx1500$ for 143 and 217\,GHz).
As can be seen in Fig.~\ref{fig:err_bl_al_eigs}, the beam normalization eigenmode is essentially twice the value of the beam uncertainty eigenmode on these scales.

Owing to the small size of the eigenmodes, the propagation to a fractional uncertainty on $C_L^{\phi\phi}$ is accomplished essentially by applying a factor of two to the eigenmodes of Eq.~\eqref{eqn:beam_error_lensing_eigenmodes}.
The normalization uncertainty due to beams for the lensing potential power spectrum is therefore approximately $1\%$ at 100\,GHz and $0.5\%$ at 143 and 217\,GHz.
When computing the beam uncertainty for our minimum-variance result, we take the 143 and 217\,GHz beam eigenmodes to be uncorrelated.

\subsection{Cosmological}
\label{sec:errorbudget:cosmological}
Similarly to the case with the beam transfer function of the previous subsection, uncertainty in the cosmological model may also be propagated to an uncertainty in the estimator normalization.
There are two aspects in which cosmological uncertainty can enter our analysis.
First, when estimating the $\left. \Delta C_L^{\phi\phi} \right|_{\mathsc{N1}}$ contribution of Eq.~\eqref{eqn:clppest}
we assume a fiducial $C_L^{\phi\phi}$, obtained from the parameters of the fiducial $\Lambda$CDM cosmology described in Sect.~\ref{sec:intro}.
We take a random subset of samples from \planck\ MCMC chains for the $\Lambda$CDM cosmology and calculate the corresponding $\left. \Delta C_L^{\phi\phi} \right|_{\mathsc{N1}}$ bias.
Compared to our fiducial correction, we see small-scale fluctuations with an amplitude of  $0.2\%$ and a typical scale of $\Delta L = 150$, superimposed on a larger overall amplitude scatter with a standard deviation of approximately  $6\%$.
This result is reasonable: in the $\Lambda$CDM model the \Planck\ $TT$ measurements have the power to detect the smoothing effects of gravitational lensing on the acoustic peaks at a significance level of $10\sigma$, corresponding to a $10\%$ constraint on the overall amplitude of $C_L^{\phi\phi}$, consistent with the result from analysis of individual samples above. 
To account for cosmological uncertainty in the $\left. \Delta C_L^{\phi\phi} \right|_{\mathsc{N1}}$ correction, we therefore assign it a $10\%$ overall amplitude uncertainty.
This correction is small for the $L$ range considered in our fiducial likelihood, and so this uncertainty is essentially negligible compared to our statistical error bars (although we include it in our error budget nevertheless).

The second point at which cosmological uncertainty enters our analysis is in the normalization of our estimates.
In Eq.~\eqref{eqn:respxp}, as well as in our tests of the normalization in Sect.~\ref{sec:simulations}, we have assumed that the statistically-anisotropic covariance induced by lensing is equal to that assumed when weighting the estimator, given by Eq.~\eqref{eqn:qe_weight_lensing}.
If the true cosmological power spectrum differs from the fiducial one used in Eq.~\eqref{eqn:qe_weight_lensing} then we will misestimate $\phi$ as
\be
\frac{ \langle\hat{\phi}_{LM} \rangle}{\phi_{LM}} = 1 + \Delta_L^{TT} = \frac{1}{ {\cal R}_L^{\phi \phi} } \left[ \frac{1}{2L+1} \sum_{\elt_1 \elt_2} \frac{1}{2} W_{\elt_1 \elt_2 L}^{\phi} 
W_{\elt_1 \elt_2 L}^{\phi,\, {\rm true}}
F_{\elt_1} F_{\elt_2} \right],
\label{eqn:cosmological_normalization_error}
\ee
where $W_{\elt_1 \elt_2 L}^{\phi, {\rm true}}$ is given by Eq.~\eqref{eqn:qe_weight_lensing}, with the modification that it is evaluated with the ``true'' ensemble-average lensed CMB power spectrum rather than the fiducial one $C_{\elt}^{TT}$.
We have chosen to account for this normalization uncertainty coherently with the temperature likelihood.
Given a proposed $C_{\elt}^{TT}$, $C_{L}^{\phi\phi}$ pair, we renormalize $C_L^{\phi\phi} \rightarrow (1+\Delta^{TT}_L)^2 C_L^{\phi\phi}$ in our likelihood (Eq.~\ref{eqn:likelihood_form}) so that it is directly comparable to our measurement. 

\section{Results}
\label{sec:results}
In this section, we present the primary results of this paper; a measurement of the CMB lensing potential over approximately $70\%$ of the sky and its associated power spectrum estimate $\hat{C}_L^{\phi\phi}$. 
Our lensing potential map and power spectrum have several science implications, which we will discuss in proceeding subsections.
\begin{enumerate}[(1)]
\item The lensing power spectrum probes the matter power spectrum integrated back to the last-scattering surface, 
and provides additional low-redshift leverage for CMB-alone parameter constraints with a broad maximum at $z\sim2$.
Highlights of the additional information that the \Planck\ lensing likelihood contributes to the parameter constraints of \cite{planck2013-p11} are presented in Sect.~\ref{sec:results:likelihood}.
\item Our lensing potential map correlates with the CMB temperature via the late-time integrated Sachs-Wolfe (ISW) effect sourced by dark energy.
We measure the cross-spectrum between our lens reconstruction and the low-$\elt$ temperature multipoles in Sect.~\ref{sec:results:iswlensing}, 
which we find is present at the level predicted in $\Lambda$CDM, and discrepant with the null hypothesis of no correlation at approximately $2.5\sigma$.
\item Our lensing potential map correlates with the fluctuations of the cosmic infrared background, which also trace large-scale structure and have  good redshift overlap with the CMB lensing potential. We do not discuss this correlation here, but instead refer to the detailed analysis of \cite{planck2013-p13}.
\item Finally, our lensing potential map correlates with external tracers of large-scale structure. In Sect.~\ref{subsec:xcorr} we present cross-correlations with several representative galaxy catalogues. 
\end{enumerate}

Our primary results are based on a minimum-variance (MV) combination of the 143 and 217\,GHz \Planck\ maps, although as an initial consistency check we will also present most of our results in this section for the individual frequency maps as well. 
In Sect.~\ref{sec:consistency} we will present a more extensive suite of consistency tests on our power spectrum estimates and lensing map. In particular in Sect.~\ref{sec:consistency:foregrounds}, we will compare with results obtained on different masks and on the foreground-cleaned \planck\ maps described in \cite{planck2013-p06}.

\begin{figure}[!htpb]
\begin{center}
\begin{overpic}[width=0.75\columnwidth]{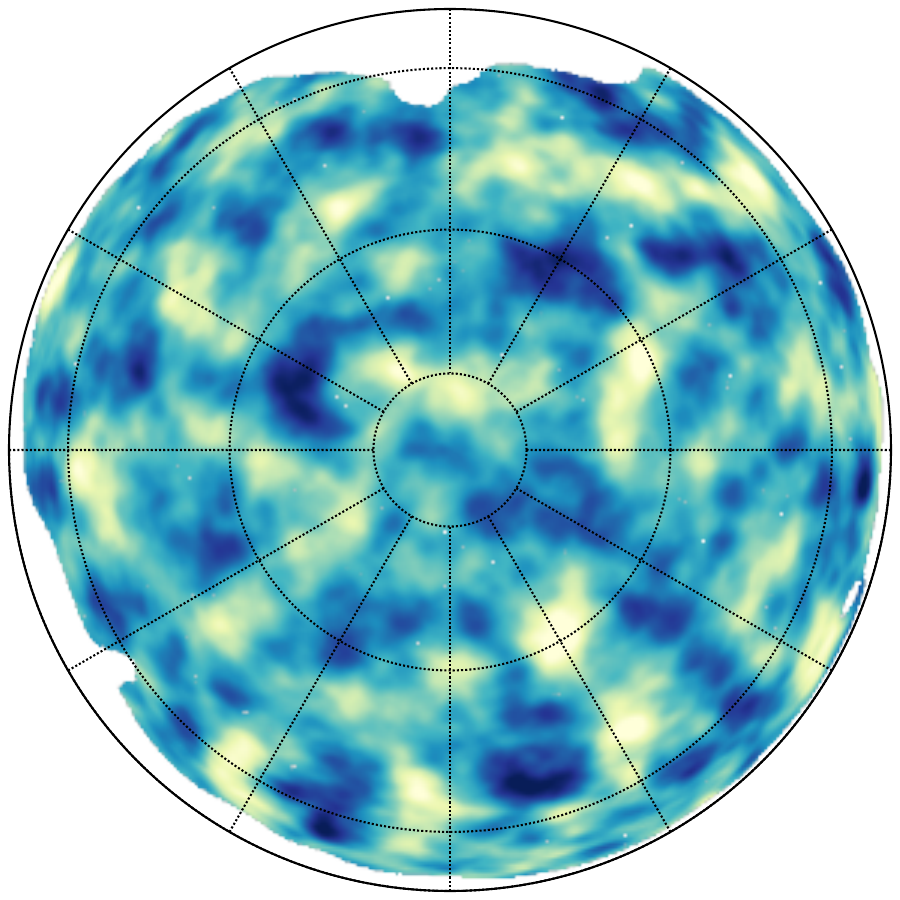} 
\put(0,90){\large $\phi^{\WF}_{} (\hatn)$}
\end{overpic}
\\ \large{Galactic north}
\vspace{0.1in}

\begin{overpic}[width=0.75\columnwidth]{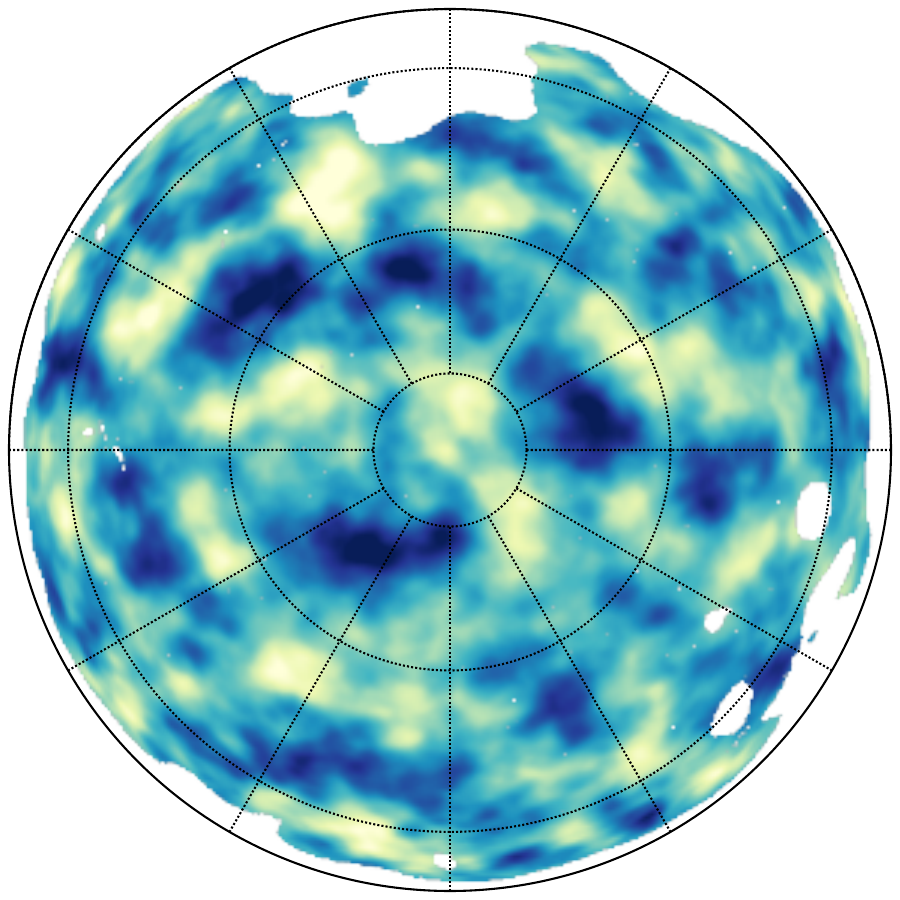} 
\put(0,90){\large $\phi^{\WF}_{} (\hatn)$}
\end{overpic}
\\ \large{Galactic south}
\vspace{0.1in}
\caption{
Wiener-filtered lensing potential estimate \mbox{$\phi^{\WF}_{LM} \equiv C_L^{\phi\phi} (\bar{\phi}_{LM} - \bar{\phi}_{LM}^{\MF})$} for our MV reconstruction, in Galactic coordinates using orthographic projection. 
The reconstruction is bandpass filtered to \mbox{$L \in [10, 2048]$}.
The \Planck\ lens reconstruction has $S/N \le 1$ for individual modes on all scales, so this map is noise dominated. 
Comparison between simulations of reconstructed and input $\phi$ in  Fig.~\ref{fig:wiener_filtered_mv_reconstruction_sim}
show the expected level of visible correlation between our reconstruction and the true lensing potential.
\label{fig:wiener_filtered_mv_reconstruction}
}
\end{center}

\centerline{
\begin{overpic}[width=0.45\columnwidth]{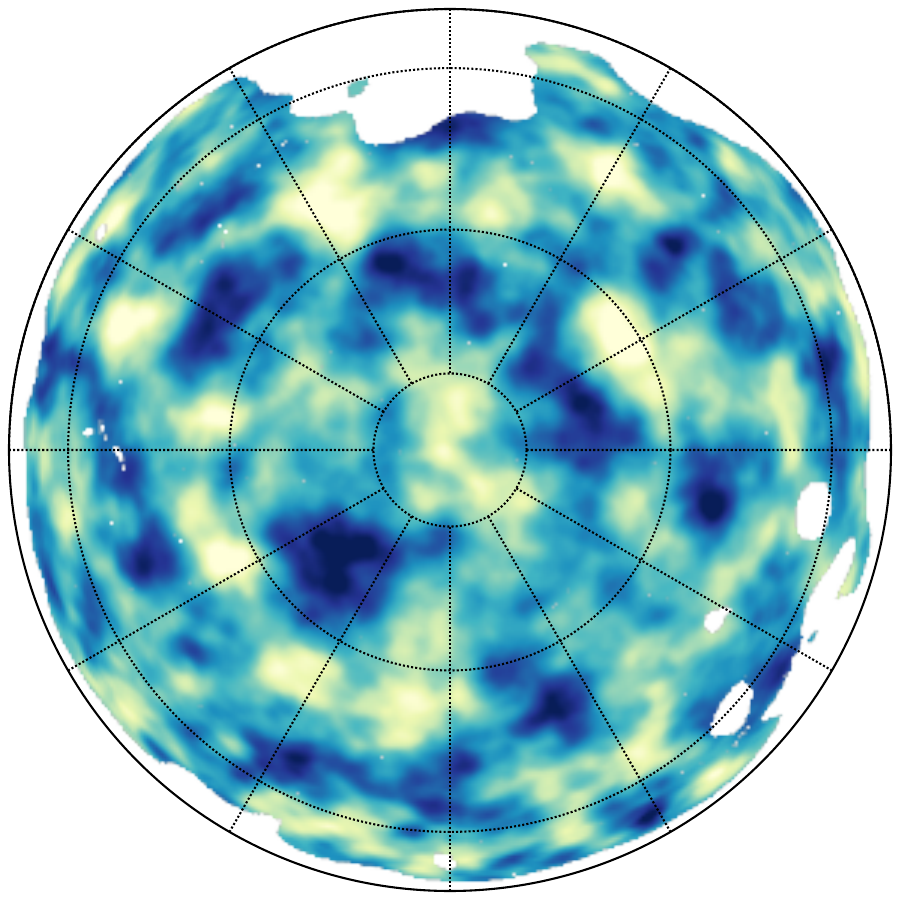}
\put(10,-8){\small{Galactic south - 143\,GHz}}
\end{overpic}
\hspace{1em}
\begin{overpic}[width=0.45\columnwidth]{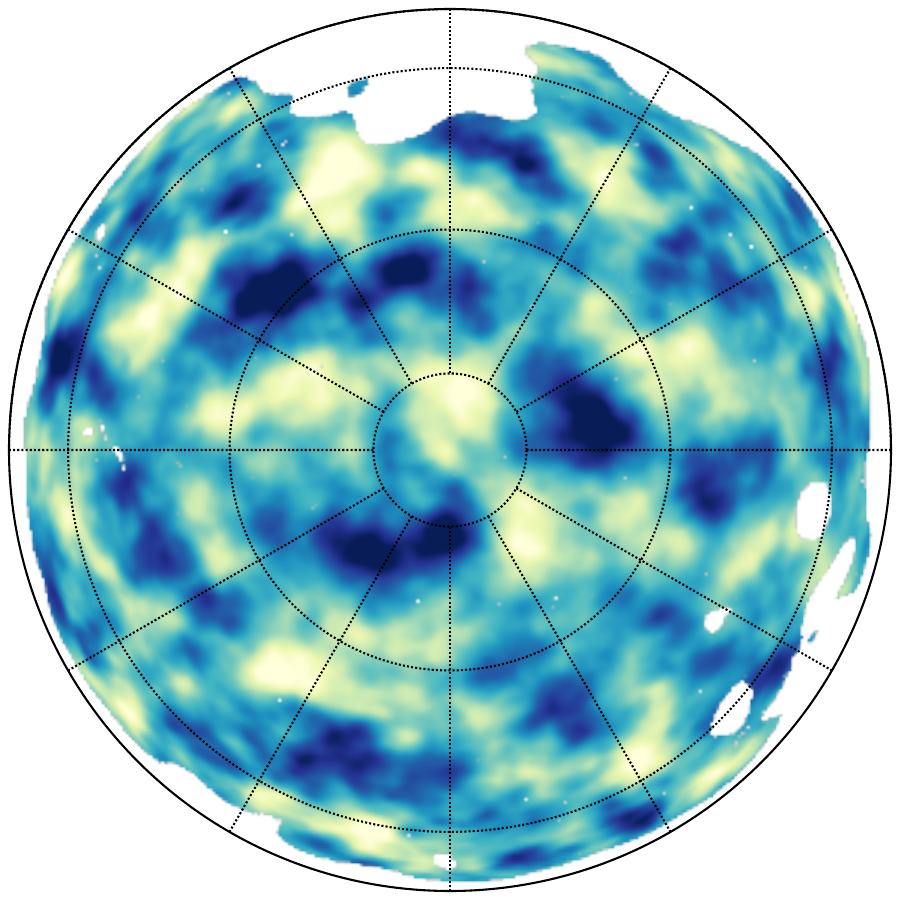}
\put(10,-8){\small{Galactic south - 217\,GHz}}
\end{overpic}
}
\vspace{0.15in}
\caption{
Wiener-filtered lensing potential estimates, as in Fig.~\ref{fig:wiener_filtered_mv_reconstruction}, for the individual 143 and 217\,GHz maps.
Note that the ``noise'' due to CMB fluctuations is correlated between these two estimates.
\label{fig:143217reconstructions}
}
\end{figure}

Our MV lensing map is shown in Fig.~\ref{fig:wiener_filtered_mv_reconstruction},  where we plot the Wiener-filtered reconstruction of the lensing potential.
At the noise levels of our reconstruction, approximately half of the modes in this map are noise.
At a visual level, the lensing map of Fig.~\ref{fig:wiener_filtered_mv_reconstruction} has comparable signal-to-noise to the \textit{COBE}-DMR maps of the CMB temperature anisotropy \citep{Bennett:1996ce}.
Future measurements, including the \Planck\ full mission release which contains approximately twice the amount of data used here, will improve on this first full-sky map of the CMB lensing potential.
As is illustrated in the simulated reconstruction of Fig.~\ref{fig:wiener_filtered_mv_reconstruction_sim}, there will be clear visual correlations between this map and future measurements.
For comparison with the full MV result, in Fig.~\ref{fig:143217reconstructions} we also plot the individual 143\, and 217\,GHz reconstructions for the southern hemisphere.

\begin{figure*}[!ht]
\vspace{0.1in}
\begin{center}
\begin{overpic}[width=\textwidth]{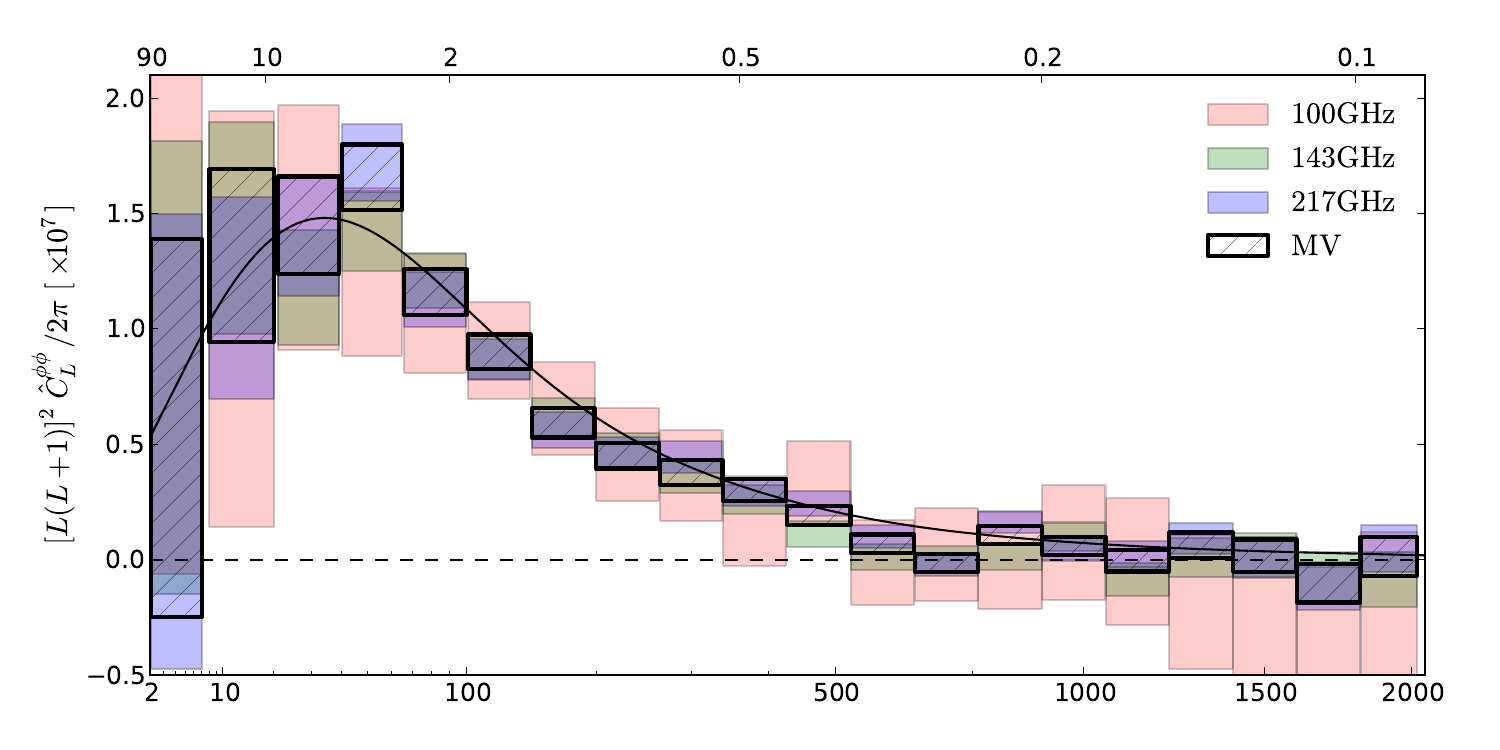}
\put(45, 0){ \large{Lensing multipole $L$} }
\put(45, 48.5){ \large{Angular scale [deg.]} }
\end{overpic}
\vspace{-0.1in}
\caption{Lensing potential power spectrum estimates based on the individual 100, 143, and 217\,GHz sky maps, as well our fiducial minimum-variance (MV) reconstruction which forms the basis for the \Planck\ lensing likelihood.
The black line is for the best-fit $\Lambda$CDM model of \cite{planck2013-p11}.
\label{fig:analysis_lens_clpp_pub}
}
\end{center}
\end{figure*}

\begin{figure}[htbp]
  \centering
  \begin{overpic}[width=\linewidth]{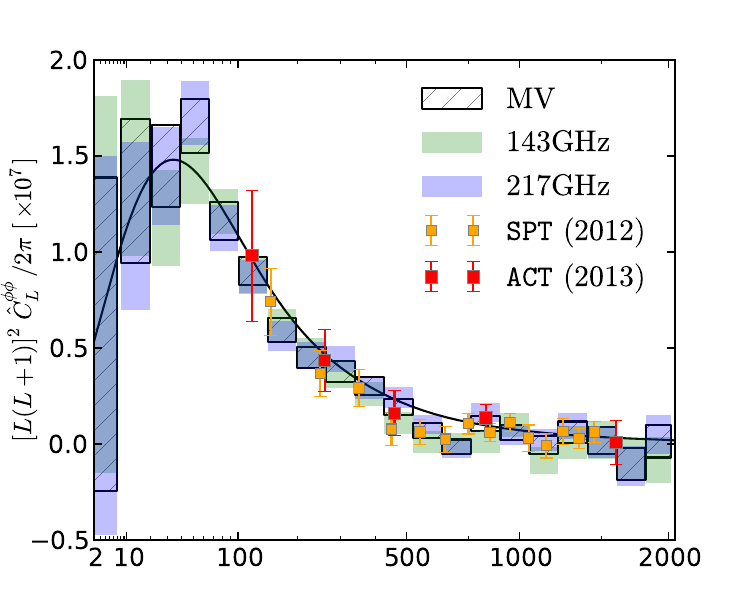} \put(50,0){$L$} \end{overpic}
  \includegraphics[width=\linewidth]{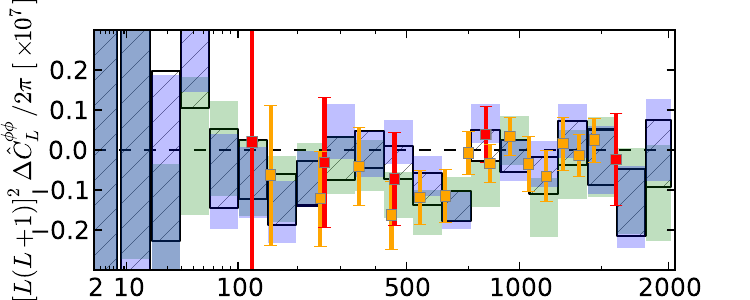}
  \caption{\label{fig:compactspt} 
  	Replotting of Fig.~\ref{fig:analysis_lens_clpp_pub}, removing 100\,GHz for easier comparison of 143 and 217\,GHz.
	Also plotted are the SPT bandpowers from \cite{vanEngelen:2012va}, 
	and the ACT bandpowers from \cite{Das:2013zf}. 
	All three experiments are very consistent.
	The lower panel shows the difference between the measured bandpowers and the fiducial best-fit $\Lambda$CDM model.
	    }
\end{figure}

In Fig.~\ref{fig:analysis_lens_clpp_pub} we plot the power spectra of our individual 
100, 143, and 217\,GHz reconstructions as well as the minimum-variance reconstruction.
The agreement of all four spectra is striking.
Overall, our power spectrum measurement is reasonably consistent with the $\Lambda$CDM prediction, given our measurement error bars.
Dividing the $L \in [1, 2048]$ multipole range into bins of 
$\Delta L = 64$ and 
binning uniformly in $[L(L+1)]^2 C_L^{\phi\phi}$, we obtain a reduced $\chi^2$ 
for the difference between our power spectrum estimate and the model of
$40.7$
with 32 degrees of freedom.
The associated probability to exceed is 
$14\%$.
On a detailed level, there are some discrepancies between the shape and amplitude of our power spectrum and the fiducial model however.
Our likelihood is based on the multipole range $40 \le L \le 400$, which captures $90\%$ of the 
forecasted ensemble-average 
signal-to-noise for an amplitude constraint on $C_L^{\phi\phi}$
in our fiducial cosmological model (see Fig.~\ref{fig:clpp_cltp_sn}).
This range was chosen as the region of our spectrum least likely to be contaminated by systematic effects (primarily uncertainties in the mean-field corrections at low-$L$, and uncertainties in the Gaussian and point-source bias corrections at high-$L$).
Estimating an average amplitude for the fiducial lensing power spectrum for a single bin over this multipole range using Eq.~\eqref{eqn:amplitude_a_def} 
we find an amplitude of 
$\hat{A}_{40 \rightarrow 400} = 0.94 \pm 0.04$ 
relative to the fiducial model (which has $A = 1$).
The power in this region is consistent with the fiducial model, although 
$1.5\sigma$ 
low (the corresponding probability-to-exceed for the $\chi^2$ of this difference is $15\%$). 
The low- and high-$L$ extent of our likelihood were deliberately chosen to have enough expected lensing signal to enable a $10\sigma$ detection of lensing on either side, bookending our likelihood with two additional consistency tests. 
On the low-$L$ side, we have a good agreement with the expected power. 
As will be discussed in Sect.~\ref{sec:biashardened}, our measurement at $L<10$ fails some consistency tests at a level comparable to the expected signal. 
The $L<10$ modes, which we suspect are somewhat contaminated by errors in the mean-field subtraction, are nevertheless consistent with the fiducial expectation, as can be seen in Fig.~\ref{fig:analysis_lens_clpp_pub}; we measure 
$\hat{A}_{1 \rightarrow 10} = 0.44 \pm 0.54$.
Extending to the lower limit of our likelihood, with a single bin from 
$10 \le L \le 40$ 
we measure 
$\hat{A}_{10 \rightarrow 40} = 1.02 \pm 0.12$.
On the high-$L$ side of our fiducial likelihood, there is tension however.
Extending from the final likelihood multipole at $L=400$ to the maximum multipole of our reconstruction, we find $\hat{A}_{400 \rightarrow 2048} = 0.68 \pm 0.13$, which is in tension with $A=1$ at a level of just over $2.4\sigma$.
The relatively low power in our reconstruction is driven by a dip relative to the
$\Lambda$CDM model spectrum between $500 < L < 750$, as can be seen in Fig.~\ref{fig:analysis_lens_clpp_pub}.
We show this feature more clearly in the residual plot of Fig.~\ref{fig:compactspt}.
This deficit of power is in turn driven by the 143\,GHz data. 
For an estimate of the power spectrum using only 143\,GHz, we measure 
$\hat{A}^{143}_{400 \rightarrow 2048} = 0.37 \pm 0.18$.
The 217\,GHz reconstruction is more consistent with the model, having
$\hat{A}^{217}_{400 \rightarrow 2048} = 0.82 \pm 0.17$.
These two measurements are in tension; we have
$\hat{A}^{217 - 143}_{400 \rightarrow 2048} = 0.45 \pm 0.18$, which is a $2.5\sigma$ discrepancy.
The error bar on this difference accounts for the expected correlation between the two channels due to the fact that they see the same CMB sky.
A larger set of consistency tests will be presented in Sect.~\ref{sec:consistency}.
We note for now that the bins from $40<L<400$ used in our likelihood pass all consistency tests, and show better agreement between 143 and 217\,GHz.
Although $L<40$ and $L>400$ are not included in our nominal likelihood, when discussing the use of the lensing likelihood for cosmological parameter constraints in the following section we will perform additional cross-checks using these bins to ascertain whether they would have any significant implications for cosmology.

In addition to the \Planck\ power spectrum measurements, in Fig.~\ref{fig:compactspt} we have overplotted the ACT and SPT measurements of the lensing potential power spectrum
\citep{Das:2013zf,vanEngelen:2012va}. It is clear that all are very consistent.
The \Planck\ data provides the largest signal-to-noise of these measurements; as we have already discussed the $40<L<400$ lensing likelihood provides a $4\%$ constraint on the amplitude of the lensing potential power spectrum, while the constraint from current ACT and SPT measurements are $32\%$ and $16\%$ respectively.
These measurements are nevertheless quite complementary.
As a function of angular scale, the full-sky \Planck\ power spectrum estimate has the smallest uncertainty per multipole of all three experiments at $L<500$, at which point the additional small-scale modes up to $\elt_{\rm max}=3000$ used in the SPT lensing analysis lead to smaller error bars.
The good agreement in these estimates of $C_L^{\phi\phi}$ is reassuring; in addition to the fact that the experiments and analyses are completely independent, these measurements are sourced from fairly independent angular scales in the temperature map, with $\ell \lea 1600$ in the case of \Planck, $\ell < 2300$ in the case of ACT, and $\ell < 3000$ in the case of SPT.
Cross-correlation of the \Planck\ lensing map with these independent measures of the lensing potential will provide an additional cross-check on their consistency, however at the power spectrum level they are already in good agreement.

\subsection{Parameters}
\label{sec:results:cosmo}
\label{sec:results:likelihood}

Weak gravitational lensing of the CMB provides sensitivity to cosmological parameters affecting the late-time growth of structure that are otherwise degenerate in the primary CMB anisotropies imprinted around recombination.
Examples include the dark energy density in models with spatial curvature and the mass of neutrinos that are light enough ($m_\nu < 0.5\, \mathrm{eV}$) still to have been relativistic at recombination. 

To connect our measurement of the lensing power spectrum to parameters, we construct a lensing likelihood nominally based on the multipole range $40 \le L \le 400$, cut into eight equal-width bins with $\Delta L = 45$ to maintain parameter leverage from shape information in addition to our overall amplitude constraint.
In Table~\ref{tab:bandpowers} we present bandpowers for these eight bins using the individual 100, 143, and 217\,GHz reconstructions as well as the MV reconstruction that is the basis for our nominal likelihood.
The bandpower estimates and their uncertainties are broken down into constituent parts as discussed in Sect.~\ref{sec:methodology}.
Based on these bandpowers, we form a likelihood following Eq.~\eqref{eqn:likelihood_form}.
The measurement errors on each bin are measured by Monte-Carlo using 1000 simulations, and the bins are sufficiently wide that we can neglect any small covariance between them (this is discussed further in Appendix~\ref{app:likelihood}).
We analytically marginalize over uncertainties that are correlated between bins, including them in the measurement covariance matrix. 
This includes beam transfer function uncertainties (as described in Sect~\ref{sec:errorbudget:beamtransfer}), uncertainties in the point source correction (Sect.~\ref{sec:consistency:pointsource}) and uncertainty in the $N^{(1)}$ correction. 

\begin{table*}[!t]
\begingroup
\caption{Reconstruction bandpower amplitudes and minimum-variance 
reconstruction for 100, 143, and 217\,GHz, computed following 
Eq.~\eqref{eqn:amplitude_a_def}.
The overall amplitude $A$ is evaluated as $\hat{A} = 
A^{\mathsc{raw}} - A^{\mathsc{n0}} - A^{\mathsc{ps}} - 
A^{\mathsc{n1}} - A^{\mathsc{mc}}$.
The bands are wide enough that the correlation of the measurement 
uncertainty between bins is negligible, as discussed in 
Appendix~\ref{app:likelihood}; however, most of the systematic errors 
are significantly correlated between bands.
The final column lists the physical bandpower estimates and their 
errors obtained by scaling flat averages of $L^4 
\hat{C}_L^{\phi\phi}$ across each bin in our fiducial $\Lambda$CDM 
model by the given $\hat{A}$ value.
\label{tab:bandpowers}}
\nointerlineskip
\vskip -3mm
\footnotesize
\setbox\tablebox=\vbox{
    \newdimen\digitwidth
    \setbox0=\hbox{\rm 0}
    \digitwidth=\wd0
    \catcode`*=\active
    \def*{\kern\digitwidth}
    \newdimen\signwidth
    \setbox0=\hbox{+}
    \signwidth=\wd0
    \catcode`!=\active
    \def!{\kern\signwidth}
\halign to \hsize{\hbox to 0.8in{#\leaderfil}\tabskip=1em&
    \hfil$#$\hfil&
    \hfil$#$\hfil&
    \hfil$#$\hfil&
    \hfil$#$\hfil&
    \hfil$#$\hfil&
    \hfil$#$\hfil&
    \hfil$#$\hfil\tabskip=0pt\cr
\noalign{\doubleline}
\omit&\multispan6\hfil Reconstruction Bandpower Amplitude\hfil&\cr
\noalign{\vskip -3pt}
\omit&\multispan6\hrulefill\cr
\omit\hfil$L_{\rm min}$--$L_{\rm max}$\hfil& A^{\mathsc{raw}} - 
A^{\mathsc{n0}}& A^{\mathsc{ps}}& A^{\mathsc{n1}}& A^{\mathsc{mc}}& 
A^{\mathsc{beam}}&\hat{A}& L^4 C_L^{\phi\phi} \times 10^{7}\cr
\noalign{\vskip 3pt\hrule\vskip 5pt}
\omit{\bf 100\,GHz}\hfil\cr
\noalign{\vskip 4pt}
\hglue 0.6em*40--*84& (*7.883-*6.948) \pm 0.191& +0.005 \pm 0.003& 
+0.035 \pm 0.003& !0.007& \pm 0.011& !0.889 \pm 0.191& 7.297 \pm 
1.568\cr
\hglue 0.6em*85--129& (11.077-*9.999) \pm 0.197& +0.008 \pm 0.004& 
+0.040 \pm 0.004& -0.010& \pm 0.011& !1.040 \pm 0.197& 6.711 \pm 
1.272\cr
\hglue 0.6em130--174& (16.360-15.411) \pm 0.256& +0.014 \pm 0.007& 
+0.045 \pm 0.004& !0.025& \pm 0.011& !0.866 \pm 0.257& 4.294 \pm 
1.273\cr
\hglue 0.6em175--219& (25.031-23.896) \pm 0.337& +0.024 \pm 0.013& 
+0.057 \pm 0.006& !0.043& \pm 0.011& !1.011 \pm 0.337& 3.947 \pm 
1.316\cr
\hglue 0.6em220--264& (36.642-35.881) \pm 0.442& +0.040 \pm 0.021& 
+0.104 \pm 0.010& !0.009& \pm 0.010& !0.607 \pm 0.443& 1.920 \pm 
1.400\cr
\hglue 0.6em265--309& (51.002-50.170) \pm 0.556& +0.058 \pm 0.030& 
+0.205 \pm 0.020& -0.076& \pm 0.010& !0.645 \pm 0.557& 1.693 \pm 
1.462\cr
\hglue 0.6em310--354& (66.348-65.291) \pm 0.707& +0.071 \pm 0.037& 
+0.304 \pm 0.030& -0.010& \pm 0.010& !0.692 \pm 0.708& 1.534 \pm 
1.571\cr
\hglue 0.6em355--400& (81.994-81.084) \pm 0.793& +0.077 \pm 0.040& 
+0.332 \pm 0.033& -0.121& \pm 0.010& !0.621 \pm 0.795& 1.177 \pm 
1.507\cr
\hglue 0.6em*40--400& (17.725-16.731) \pm 0.112& +0.016 \pm 0.008& 
+0.062 \pm 0.006& !0.004& \pm 0.011& !0.912 \pm 0.113&\dots\cr
\noalign{\vskip 8pt}
\omit{\bf 143\,GHz}\hfil\cr
\noalign{\vskip 4pt}
\hglue 0.6em*40--*84& (*3.936-*2.841) \pm 0.089& +0.005 \pm 0.002& 
+0.026 \pm 0.003& -0.021& \pm 0.004& !1.086 \pm 0.090& 8.917 \pm 
0.736\cr
\hglue 0.6em*85--129& (*5.008-*4.058) \pm 0.090& +0.007 \pm 0.004& 
+0.032 \pm 0.003& -0.016& \pm 0.004& !0.928 \pm 0.090& 5.991 \pm 
0.582\cr
\hglue 0.6em130--174& (*7.076-*6.088) \pm 0.109& +0.012 \pm 0.006& 
+0.040 \pm 0.004& -0.020& \pm 0.004& !0.956 \pm 0.109& 4.743 \pm 
0.540\cr
\hglue 0.6em175--219& (10.092-*9.207) \pm 0.137& +0.021 \pm 0.010& 
+0.058 \pm 0.006& !0.013& \pm 0.004& !0.793 \pm 0.138& 3.094 \pm 
0.538\cr
\hglue 0.6em220--264& (14.254-13.279) \pm 0.174& +0.032 \pm 0.016& 
+0.107 \pm 0.011& -0.010& \pm 0.004& !0.845 \pm 0.175& 2.671 \pm 
0.553\cr
\hglue 0.6em265--309& (19.049-18.063) \pm 0.211& +0.045 \pm 0.022& 
+0.191 \pm 0.019& !0.008& \pm 0.004& !0.742 \pm 0.213& 1.948 \pm 
0.558\cr
\hglue 0.6em310--354& (24.524-22.862) \pm 0.244& +0.055 \pm 0.026& 
+0.271 \pm 0.027& -0.011& \pm 0.004& !1.347 \pm 0.246& 2.988 \pm 
0.547\cr
\hglue 0.6em355--400& (28.355-27.576) \pm 0.273& +0.060 \pm 0.029& 
+0.298 \pm 0.030& -0.006& \pm 0.004& !0.427 \pm 0.276& 0.808 \pm 
0.524\cr
\hglue 0.6em*40--400& (*8.403-*7.401) \pm 0.048& +0.016 \pm 0.008& 
+0.064 \pm 0.006& -0.013& \pm 0.004& !0.935 \pm 0.050&\dots\cr
\noalign{\vskip 8pt}
\omit{\bf 217\,GHz}\hfil\cr
\noalign{\vskip 4pt}
\hglue 0.6em*40--*84& (*3.884-*2.742) \pm 0.088& +0.003 \pm 0.003& 
+0.026 \pm 0.003& -0.007& \pm 0.004& !1.120 \pm 0.088& 9.200 \pm 
0.721\cr
\hglue 0.6em*85--129& (*4.817-*3.866) \pm 0.088& +0.005 \pm 0.005& 
+0.032 \pm 0.003& -0.012& \pm 0.004& !0.927 \pm 0.089& 5.982 \pm 
0.572\cr
\hglue 0.6em130--174& (*6.601-*5.788) \pm 0.103& +0.007 \pm 0.009& 
+0.040 \pm 0.004& -0.000& \pm 0.004& !0.766 \pm 0.103& 3.800 \pm 
0.512\cr
\hglue 0.6em175--219& (*9.544-*8.637) \pm 0.131& +0.012 \pm 0.014& 
+0.058 \pm 0.006& !0.012& \pm 0.004& !0.824 \pm 0.132& 3.217 \pm 
0.516\cr
\hglue 0.6em220--264& (13.409-12.317) \pm 0.162& +0.018 \pm 0.021& 
+0.103 \pm 0.010& !0.004& \pm 0.004& !0.967 \pm 0.163& 3.057 \pm 
0.516\cr
\hglue 0.6em265--309& (17.825-16.583) \pm 0.202& +0.025 \pm 0.028& 
+0.179 \pm 0.018& -0.003& \pm 0.004& !1.041 \pm 0.204& 2.733 \pm 
0.536\cr
\hglue 0.6em310--354& (22.391-20.915) \pm 0.227& +0.030 \pm 0.034& 
+0.251 \pm 0.025& -0.014& \pm 0.004& !1.210 \pm 0.231& 2.683 \pm 
0.513\cr
\hglue 0.6em355--400& (26.119-25.068) \pm 0.265& +0.032 \pm 0.037& 
+0.279 \pm 0.028& -0.026& \pm 0.004& !0.766 \pm 0.269& 1.451 \pm 
0.510\cr
\hglue 0.6em*40--400& (*8.122-*7.111) \pm 0.046& +0.009 \pm 0.011& 
+0.064 \pm 0.006& -0.004& \pm 0.004& !0.942 \pm 0.048&\dots\cr
\noalign{\vskip 8pt}
\omit{\bf MV}\hfil\cr
\noalign{\vskip 4pt}
\hglue 0.6em*40--*84& (*3.297-*2.170) \pm 0.073& +0.006 \pm 0.003& 
+0.026 \pm 0.003& -0.011& \pm 0.003& !1.105 \pm 0.073& 9.077 \pm 
0.601\cr
\hglue 0.6em*85--129& (*4.062-*3.062) \pm 0.074& +0.009 \pm 0.004& 
+0.032 \pm 0.003& -0.014& \pm 0.003& !0.972 \pm 0.074& 6.276 \pm 
0.476\cr
\hglue 0.6em130--174& (*5.480-*4.542) \pm 0.084& +0.015 \pm 0.007& 
+0.040 \pm 0.004& -0.012& \pm 0.003& !0.895 \pm 0.085& 4.442 \pm 
0.420\cr
\hglue 0.6em175--219& (*7.616-*6.758) \pm 0.103& +0.024 \pm 0.011& 
+0.058 \pm 0.006& !0.009& \pm 0.003& !0.766 \pm 0.104& 2.992 \pm 
0.406\cr
\hglue 0.6em220--264& (10.587-*9.576) \pm 0.131& +0.036 \pm 0.017& 
+0.101 \pm 0.010& -0.003& \pm 0.003& !0.876 \pm 0.133& 2.771 \pm 
0.419\cr
\hglue 0.6em265--309& (14.004-12.903) \pm 0.156& +0.048 \pm 0.022& 
+0.175 \pm 0.017& !0.000& \pm 0.003& !0.878 \pm 0.158& 2.304 \pm 
0.415\cr
\hglue 0.6em310--354& (17.677-16.285) \pm 0.181& +0.057 \pm 0.026& 
+0.246 \pm 0.025& -0.009& \pm 0.003& !1.099 \pm 0.185& 2.438 \pm 
0.409\cr
\hglue 0.6em355--400& (20.647-19.583) \pm 0.208& +0.061 \pm 0.028& 
+0.274 \pm 0.027& -0.024& \pm 0.003& !0.753 \pm 0.212& 1.427 \pm 
0.401\cr
\hglue 0.6em*40--400& (*6.788-*5.768) \pm 0.039& +0.020 \pm 0.009& 
+0.066 \pm 0.007& -0.009& \pm 0.003& !0.943 \pm 0.040&\dots\cr
\noalign{\vskip 5pt\hrule\vskip 3pt}}}
\endPlancktablewide
\endgroup
\end{table*}

As the lensing likelihood is always used in conjunction with the \Planck\ $TT$ power spectrum likelihood, we coherently account for uncertainty in $C_{\elt}^{TT}$ by renormalizing our lensing potential measurement for each sample, as described in Sect.~\ref{sec:errorbudget:cosmological}.

The lensing likelihood is combined with the main \Planck\ $TT$ likelihood~\citep{planck2013-p08} -- constructed from the temperature (pseudo) cross-spectra between detector sets at intermediate and high multipoles, and an exact approach for Gaussian temperature anisotropies at low multipoles -- in~\mbox{\citet{planck2013-p11}} to derive parameter constraints for the six-parameter \lcdm\ model and well-motivated extensions. Lensing also affects the power spectrum, or 2-point function, of the CMB anisotropies, and this effect is accounted for routinely in all \Planck\ results. On the angular scales relevant for \planck, the main effect is a smoothing of the acoustic peaks and this is detected at around $10\sigma$ in the \Planck\ temperature power spectrum~\citep{planck2013-p11}. The information about $C_L^{\phi\phi}$ that is contained in the lensed temperature power spectrum for multipoles $\elt \simlt
3000$ is limited to the amplitude of a single eigenmode~\citep{Smith:2006nk}. In extensions of \lcdm\ with a single additional late-time parameter, lensing of the power spectrum itself can therefore break the geometric degeneracy~\citep{1999MNRAS.302..735S,Sherwin:2011gv,vanEngelen:2012va,planck2013-p11}. As discussed in Appendix~\ref{app:likelihood} and~\citet{Schmittfull:2013uea}, cosmic variance of the lenses produces weak correlations between the CMB 2-point function and our estimates of $C_L^{\phi\phi}$, but they are small enough that ignoring the correlations in combining the two likelihoods should produce only sub-percent underestimates of the errors in physical cosmological parameters.

In the following, we illustrate the additional constraining power of our $C_L^{\phi\phi}$ measurements in \lcdm\ models and one-parameter extensions, highlighting those results from~\citet{planck2013-p11} where the lensing likelihood is influential.

\subsubsection{Six-parameter \lcdm\ model}

\begin{figure*}
\centering
\includegraphics[width=85mm,angle=0]{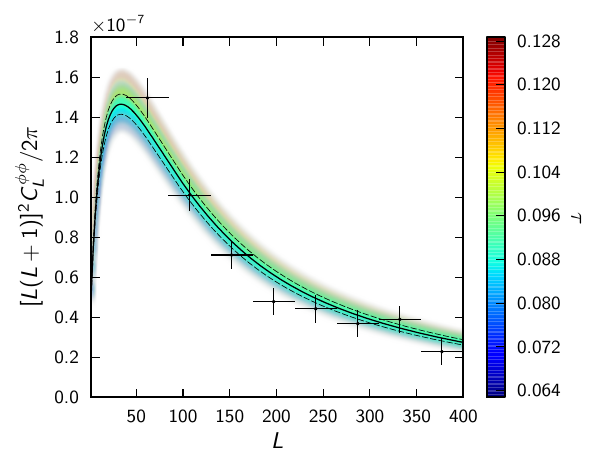}
\includegraphics[width=85mm,angle=0]{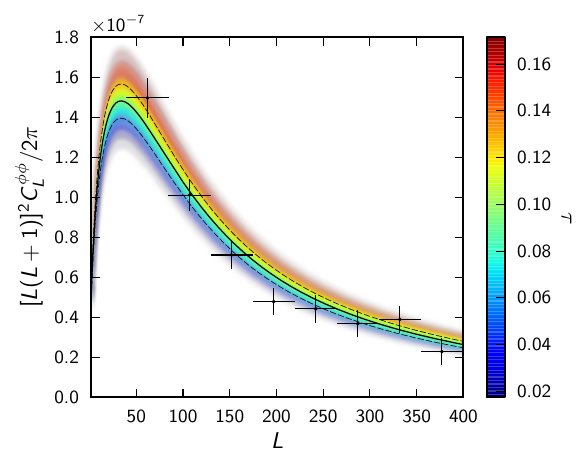}
\\
\includegraphics[width=85mm,angle=0]{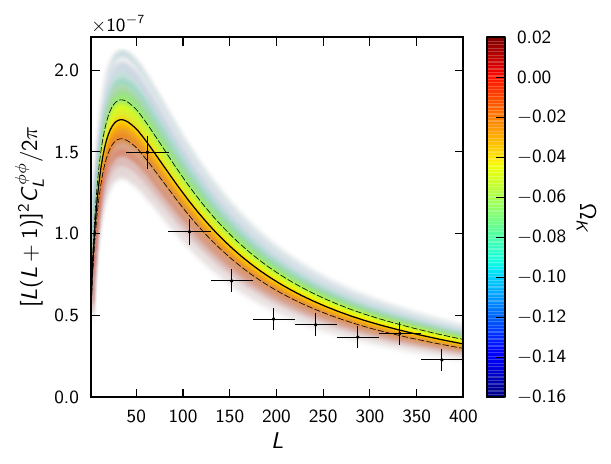}
\includegraphics[width=85mm,angle=0]{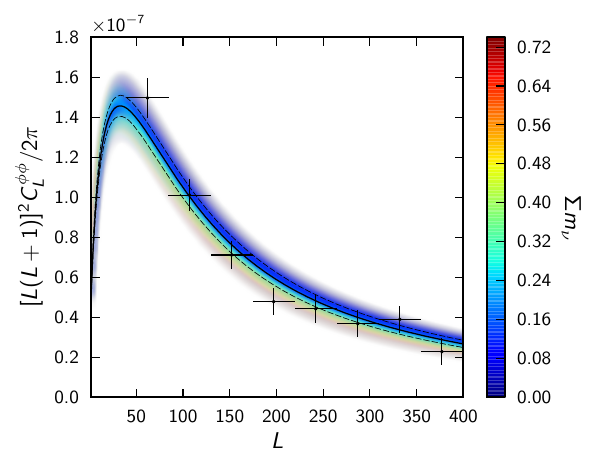}
\caption{\emph{Upper left}: \Planck\ measurements of the lensing power spectrum compared to
the \lcdm\ mean prediction and 68\% confidence interval (dashed lines) for models fit to \Planck+\WP+\highL\ (see text).
The eight bandpowers are those used in the \planck\ lensing likelihood; they are
renormalized, along with their errors, to account for the small differences between the lensed $C_\elt^{TT}$ in the best-fit model and the fiducial model used throughout this paper. The error bars are the $\pm 1\sigma$ errors from the diagonal of the covariance matrix. The colour coding shows how $C_\elp^{\phi\phi}$ varies with the optical depth $\tau$ across samples from the
\lcdm\ posterior distribution. \emph{Upper right}: as upper-left but using only the temperature power spectrum from \planck. \emph{Lower left}: as upper-left panel but in models with spatial curvature. The colour coding is for $\Omega_K$. \emph{Lower right}: as upper-left but in models with three massive neutrinos (of equal mass). The colour coding is for the summed neutrino mass $\sum m_\nu$.}
\label{fig:CMBlensing_LCDM}
\end{figure*}

In the six-parameter \lcdm\ model, the matter densities, Hubble constant and spectral
index of the primordial curvature perturbations are tightly constrained by the \planck\ temperature power spectrum alone.
However, in the absence of lensing the amplitude $\As$ of the primordial power spectrum and the reionization optical depth $\tau$ are degenerate, with only the combination $\As e^{-2\tau}$, which directly controls the amplitude of the  anisotropy power spectrum on intermediate and small scales, being well determined.
This degeneracy is broken by large-angle polarization since
the power from scattering at reionization depends on the combination $\As \tau^2$.
In this first release of \planck\ data, we use the \WMAP\ nine-year polarization maps~\citep{Bennett:2012zja} in combination with \planck\ temperature data.
With this data combination, $C_L^{\phi\phi}$ is rather tightly constrained in the \lcdm\ model (see Fig.~\ref{fig:CMBlensing_LCDM}) and the direct measurements reported here provide a non-trivial consistency test of the model.

The eight $C_L^{\phi\phi}$ bandpowers used in the lensing likelihood are compared to the expected spectrum in Fig.~\ref{fig:CMBlensing_LCDM} (upper-left panel).
For the latter, we have used parameter values  determined from the main \planck\ likelihood in combination with \WMAP\ polarization (hereafter denoted \WP) and
small-scale power spectrum measurements (hereafter \highL) from ACT~\citep{Das:2013zf} and SPT~\citep{2012ApJ...755...70R}\footnote{As discussed in detail in~\citet{planck2013-p11}, the primary role of the ACT and SPT data in these parameter fits is to constrain more accurately the contribution of extragalactic foregrounds, which must be carefully modelled to interpret the \planck\ power spectra on small scales. For \lcdm, the foreground parameters are sufficiently decoupled from the cosmological parameters that the inclusion of the ACT and SPT data has very little effect on the cosmological constraints.}.
In this plot, we have renormalized the \emph{measurements} and their error bars (rather than the theory) using the best-fit model with a variant of the procedure described in Sect.~\ref{sec:errorbudget:cosmological}. Since
the lensed temperature power spectrum in the best-fit model is very close to that in the fiducial model used to normalise the power spectrum estimates throughout this paper, the power spectrum renormalisation factor $(1+\Delta^{TT}_L)^2$ of Eq.~(\ref{eqn:cosmological_normalization_error}) differs from unity by less than $0.5\%$. The predicted $C_L^{\phi\phi}$ in the best-fit model differs from the fiducial model by less than $2.5\%$ for $L < 1000$. The best-fit model
is a good fit to the measurements, with $\chi^2=10.9$ and the corresponding probability to exceed equal to $21\%$\footnote{%
For comparison, the $\chi^2$ of the eight lensing bandpowers in the range $L=40$--400 relative to the joint best-fit $\Lambda$CDM model to the main \Planck+\WP+\highL\ likelihood is $9.5$, and relative to the fiducial model $\chi^2=11.6$.
}. Significantly, we see that the \lcdm\ model, calibrated with the CMB fluctuations imprinted around $z=1100$, correctly predicts the evolution of structure and geometry at much lower redshifts.
The 68\% uncertainty in the \lcdm\ prediction of $C_L^{\phi\phi}$ is shown by the dashed lines in the upper-left panel of Fig.~\ref{fig:CMBlensing_LCDM}. We can assess consistency with the direct measurements, properly accounting for this uncertainty, by introducing an additional parameter $\Aphiphi$ that scales the theory $C_L^{\phi\phi}$ in the lensing likelihood. Note that we choose not to alter the lensing effect in $C_\elt^{TT}$. As reported in~\cite{planck2013-p11}, we find
\[
\Aphiphi  = 0.99\pm 0.05 \quad (\mbox{68\%; \Planck+\lensing+\WP+\highL}),
\]
in excellent agreement with $\Aphiphi=1$. This analysis differs in detail from the simple one-parameter analysis with the fiducial template, reported in Table~\ref{tab:bandpowers} as $A=0.943 \pm 0.04$ for the MV reconstruction. 
The latter does not account for uncertainty in the amplitude or shape of the template between bandpowers due to variations in the six $\Lambda$CDM parameters, nor small variations in the lensed temperature power spectrum (which affect the normalisation of the reconstruction power spectrum). 

\begin{figure}
\centering
\includegraphics[width=88mm,angle=0]{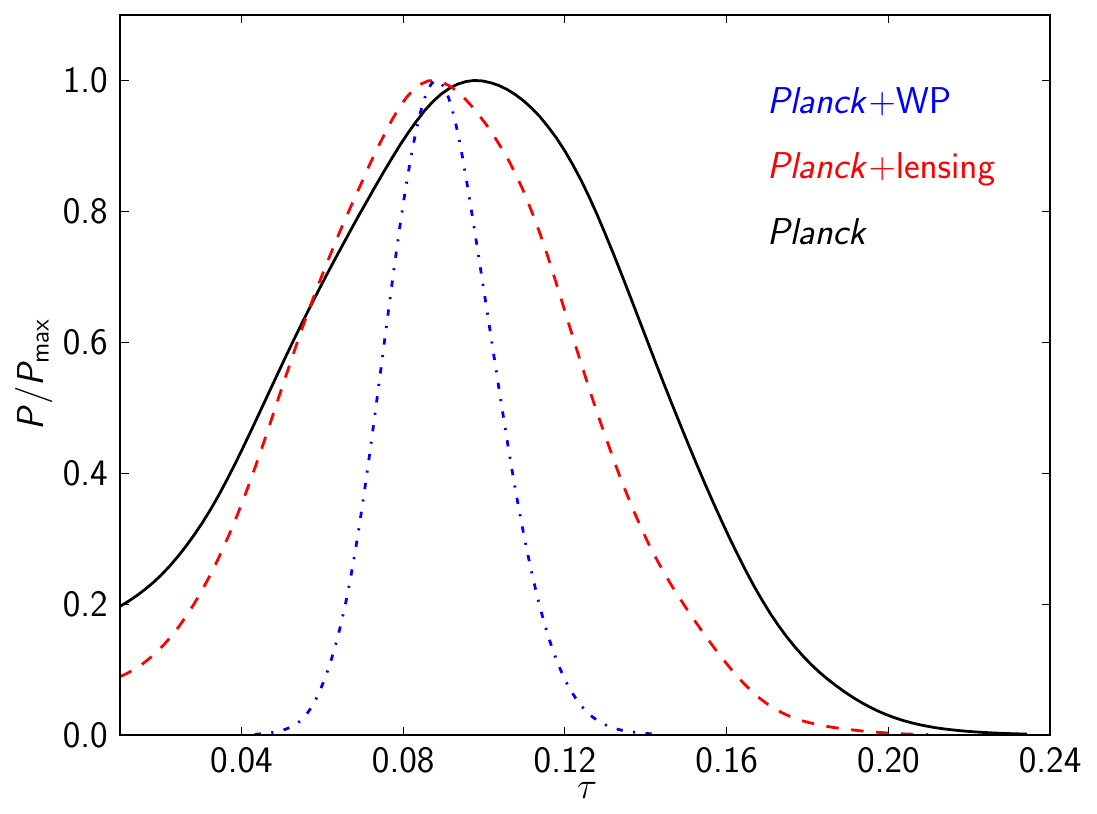}
\caption{Marginalised constraints on the optical depth in \lcdm\ models from the \planck\ temperature power spectrum (\planck; solid black), and additionally including the lensing likelihood (\planck+\lensing; dashed red) or \textit{WMAP} polarization (\planck+\WP; dashed-dotted blue). We use a prior $\tau > 0.01$ in all cases.}
\label{fig:lenstau_LCDM}
\end{figure}

An alternative route to breaking the $\As$-$\tau$ degeneracy is possible for the first time with \planck. Since $C_L^{\phi\phi}$ is directly proportional to $\As$, the lensing power spectrum measurements and the smoothing effect of lensing in $C_\elt^{TT}$ (which at leading order varies as $\As^2 e^{-2\tau}$) can separately constrain $\As$ and $\tau$ \emph{without} large-angle polarization data. The variation of $C_L^{\phi\phi}$ with $\tau$ in \lcdm\ models constrained \emph{only} by the \planck\ temperature power spectrum is illustrated in the upper-right panel of Fig.~\ref{fig:CMBlensing_LCDM}, and suggests that the direct $C_L^{\phi\phi}$ measurements may be able to improve constraints on $\tau$ further. This is indeed the case, as shown in Fig.~\ref{fig:lenstau_LCDM} where we compare the posterior distribution of $\tau$ for the \planck\ temperature likelihood alone with that including the lensing likelihood. We find
\begin{eqnarray*}
\tau &=& 0.097 \pm 0.038 \quad(\mbox{68\%; \planck}) \nonumber \\
\tau &=& 0.089 \pm 0.032 \quad(\mbox{68\%; \planck+\lensing}) . \nonumber 
\end{eqnarray*}
At 95\% confidence, we can place a lower limit on the optical depth of $0.04$
(\planck+\lensing). This is very close to the optical depth for instantaneous reionization at $z=6$, providing further support for reionization being an extended process.

The $\tau$ constraints via the lensing route are consistent with, though weaker, than those from \WMAP\ polarization. However, since the latter measurement requires very aggressive cleaning of Galactic emission (see e.g.,\ Fig.~17 of~\citealt{2007ApJS..170..335P}), the lensing constraints are an important cross-check.

\subsubsection{Effect of the large and small scales on the six-parameter \lcdm\ model}

Before exploring the further parameters that can be constrained with the lensing likelihood, we test the effect on the \lcdm\ model of adding the large-scale ($10 \le L \le 40$) and small-scale ($400 \le L \le 2048$) lensing data to our likelihood. Adding additional data will produce random shifts in the posterior distributions of parameters, but these should be small here since the multipole range $40 \le L \le 400$ is designed to capture over 90\%\ of the signal-to-noise (on an amplitude measurement). If the additional data are expected to have little statistical power, i.e., the error bars on parameters do not change greatly, but their addition produces large shifts in the posteriors, this would be symptomatic either of internal tensions between the data or an incorrect model.

\begin{figure}
\centering
\includegraphics[width=\columnwidth]{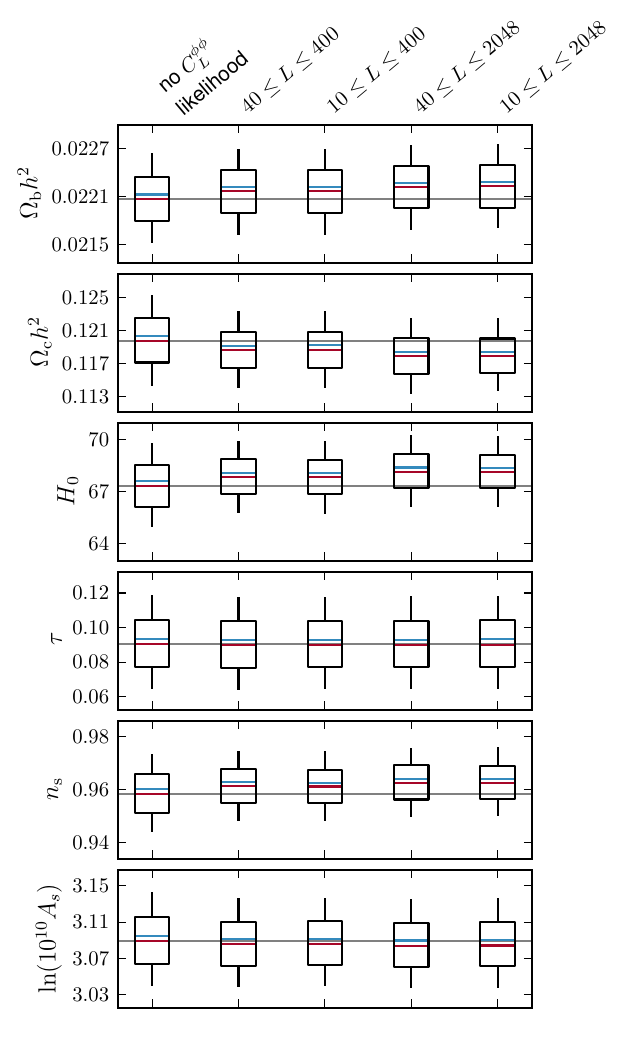}
\caption{Marginalised posteriors for the six-parameter \lcdm\ model, shown as box plots, for \planck+\WP+\highL\ with various lensing likelihoods.
The red and blue lines are the median and mean, respectively.
The box and bar correspond to $68\%$ and $95\%$ of the probability density, both centered on the median.  The left-most column is without the lensing likelihood and the median of these constraints is shown by the grey line. The remaining columns show the effect of adding in the fiducial lensing likelihood (second column), and further adding a low-$L$ bin (third column), high-$L$ bins (fourth column), or both (final column).}
\label{fig:6parbincomp}
\end{figure}

In Fig.~\ref{fig:6parbincomp}, we compare the posterior distributions of the $\Lambda$CDM parameters for \planck+\WP+\highL\ alone with those after combining with various lensing likelihoods. 
Adding our fiducial lensing likelihood (second column) reduces the errors on parameters by a small amount and the median values shift by rather less than $1\sigma$ for all parameters. The largest gain is for $\Omega_{\rm c}h^2$ (and $H_0$) where the errors improve by 20\%.
Adding further large- and small-scale data produces no significant reduction in error bars, as expected.  Parameter medians also change very little except for $\Omega_{\rm c} h^2$, which is dragged low by a further $0.3\sigma$ on adding the small-scale lensing information\footnote{The shift in $H_0$ is due to the anti-correlation between $H_0$ and $\Omega_c h^2$ caused by the acoustic-scale degeneracy in the temperature power spectrum; see~\citealt{planck2013-p11}.}. These findings are consistent with the power spectrum amplitude measurements discussed in Sect.~\ref{sec:results}: we can lower the lensing power by reducing the matter density, and this is favoured by the lower amplitudes measured from the small-scale lensing power spectrum.

The tension between the small-scale power and the power over the $L=40$--$400$ range included in our fiducial likelihood, coupled with our lower confidence in the accuracy of the bias removal on small scales, is the reason that we do not include these smaller scales at this stage in the \planck\ lensing likelihood.

\subsubsection{Spatial curvature and dark energy}

Inflation models with sufficient number of $e$-folds of expansion naturally predict that the Universe should be very close to spatially flat. Constraining any departures from flatness is therefore a critical test of inflationary cosmology. However, the primary CMB anisotropies alone suffer from a geometric degeneracy, whereby models with identical primordial power spectra, physical matter densities, and angular-diameter distance to last-scattering have almost identical power spectra~\citep{1999MNRAS.304...75E}. The degeneracy is partly broken by lensing~\citep{1999MNRAS.302..735S}, with small additional contributions from the late-ISW effect (on large scales) and by projection effects in curved models~\citep{2012JCAP...04..027H}. In \lcdm\ models with curvature, the geometric degeneracy is two-dimensional, involving the curvature and dark energy density, and this limits the precision with which either can be determined from the CMB alone.

\begin{figure*}
\centering
\includegraphics[height=65mm,angle=0]{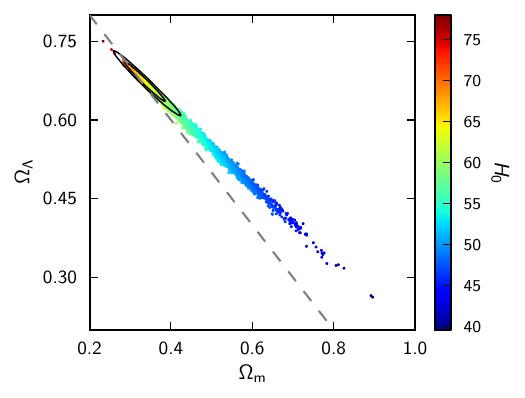}
\includegraphics[height=65mm,angle=0]{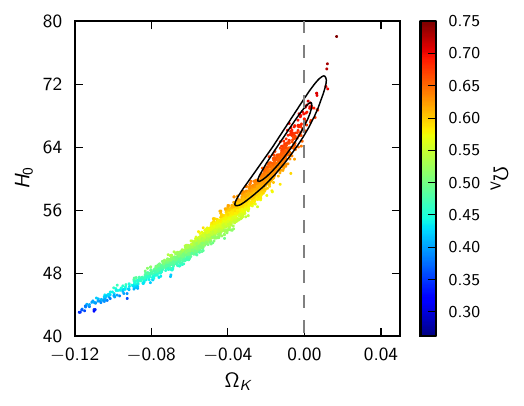}
\caption{Two views of the geometric degeneracy in curved \lcdm\ models which is partially broken by lensing. \emph{Left}: the degeneracy in the $\Omm$-$\Oml$ plane, with samples from \planck+\WP+\highL\ colour coded by the value of $H_0$. The contours delimit the 68\% and 95\% confidence regions, showing the further improvement from including the lensing likelihood. \emph{Right}: the degeneracy in the $\Omk$-$H_0$ plane, with samples colour coded by $\Oml$. Spatially-flat models lie along the grey dashed lines.}
\label{fig:lensOLCDM}
\end{figure*}

With the high-significance detection of lensing by \planck\ in the temperature power spectrum~\citep{planck2013-p11}, and via the lens reconstruction reported here, the geometric degeneracy is partially broken, as shown in Fig.~\ref{fig:lensOLCDM}. The long tail of closed models with low dark energy density (and expansion rate at low redshift) allowed by the geometric degeneracy have too much lensing power to be consistent with \planck's measured temperature and lensing power spectra (see also
Fig.~\ref{fig:CMBlensing_LCDM}).
We find marginalised constraints on the curvature parameter of 
\begin{eqnarray*}
\Omega_K &=& -0.042^{+0.027}_{-0.018} \quad \mbox{(68\%; \planck+\WP+\highL)} \\
\Omega_K &=& -0.0096^{+0.010}_{-0.0082} \quad \mbox{(68\%; \planck+\lensing+\WP+\highL)} ,
\end{eqnarray*}
so that lensing reconstruction reduces the uncertainty on $\Omega_K$ by more than a factor of two over limits driven by the smoothing effect on the acoustic peaks of $C_\elt^{TT}$. This improvement is consistent with the spread in $C_L^{\phi\phi}$ in curved models constrained by the temperature power spectrum, relative to the errors on the reconstruction power spectrum; see Fig.~\ref{fig:CMBlensing_LCDM}. Note that the mean value of $\Omega_K$ also moves towards zero with the inclusion of the $C_L^{\phi\phi}$ measurements. Adding the high-$L$ and low-$L$ data to the likelihood brings no more than a percent-level improvement on the constraint.
We see that the CMB \emph{alone} now constrains the geometry to be flat at the percent level. Previous constraints on curvature via CMB lensing have been reported by SPT in combination with the \textit{WMAP}-7 data: $\Omega_K = -0.003^{+0.014}_{-0.018}$ (68\%;~\citealt{Story:2012wx}). This constraint is consistent, though almost a factor of two weaker, than that from \Planck. Tighter constraints on curvature result from combining the \planck\ data with other astrophysical data, such as baryon acoustic oscillations, as discussed in~\citet{planck2013-p11}.

Lensing effects provide evidence for dark energy from the CMB alone, independent of other astrophysical data \citep{Sherwin:2011gv,vanEngelen:2012va}. 
In curved \lcdm\ models, we find marginalised constraints on $\Omega_\Lambda$ of
\begin{eqnarray*}
\Omega_\Lambda &=& 0.57^{+0.073}_{-0.055} \quad \mbox{(68\%; \planck+\WP+\highL)} \\
\Omega_\Lambda &=& 0.67^{+0.027}_{-0.023} \quad \mbox{(68\%; \planck+\lensing+\WP+\highL)} .
\end{eqnarray*}
Again, lensing reconstruction improves the errors by more than a factor of two over those from the temperature power spectrum alone. 
Note that part of the preference for closed models from the $\mbox{\planck+\WP+\highL}$ data combination (i.e.,\ without the 4-point function) is related to the tendency of the temperature power spectrum, considered over the full
multipole range, to favour models with more lensing power than in the best-fit $\Lambda$CDM model. This effect is discussed in some detail in~\citet{planck2013-p11}.

\subsubsection{Neutrino masses}

The unique effect in the unlensed temperature power spectrum of massive neutrinos that are still relativistic at recombination is small. With the angular scale of the acoustic peaks fixed from measurements of the temperature power spectrum, neutrino masses increase the expansion rate at $z >1$ where dark energy is negligible\footnote{With all other physical densities held fixed to preserve the acoustic physics before recombination, increasing the neutrino mass increases the expansion rate at all times after the neutrinos become non-relativistic (which is after recombination for the light masses considered here). However, to preserve the angular scale of the sound horizon and hence the location of the acoustic peaks and troughs, one must \emph{reduce} the expansion rate at low redshift. This is done by reducing the physical density in dark energy.} and so suppress clustering on scales smaller than the horizon size at the non-relativistic transition~\citep{Kaplinghat:2003bh}. This effect reduces $C_\elp^{\phi\phi}$ for $L >10$ (see Fig.~\ref{fig:CMBlensing_LCDM}) and gives less smoothing of the acoustic peaks in $C_\elt^{TT}$. As discussed in~\cite{planck2013-p11}, the constraint on $\sum m_\nu$ from the \planck\ temperature power spectrum (and \WMAP\ low-$\elt$ polarization) is driven by the smoothing effect of lensing: $\sum m_\nu < 0.66\,\mathrm{eV}$ (95\%; \planck+\WP+\highL). Curiously, this constraint is weakened by additionally including the lensing likelihood to
\begin{equation}
\sum m_\nu < 0.85\,\mathrm{eV}, \quad \mbox{(95\%; \planck+\lensing+\WP+\highL)},
\nonumber
\end{equation}
reflecting mild tensions between the measured lensing and temperature power spectra, with the former preferring larger neutrino masses than the latter.
Possible origins of this tension are explored further in~\cite{planck2013-p11} and are thought to involve both the $C_{\elp}^{\phi\phi}$ measurements and features in the measured $C_{\elt}^{TT}$ on large scales ($\elt < 40$) and small scales $\elt > 2000$ that are not fit well by the \lcdm+foreground model. In regard to $C_{\elp}^{\phi\phi}$, Fisher estimates show that the bandpowers in the range $130 < \elp < 309$ carry most of the statistical weight in determining the marginal error on $\sum m_\nu$, and Fig.~\ref{fig:CMBlensing_LCDM} reveals a preference for high $\sum m_\nu$ from this part of the spectrum. We have checked that removing the first bandpower from the lensing likelihood, which is the least stable to data cuts and the details of foreground cleaning as discussed in Sect.~\ref{sec:consistency}, has little impact on our neutrino mass constraints. We also note that a similar trend for lower lensing power than the \lcdm\ expectation on intermediate scales is seen in the ACT and SPT measurements (Fig.~\ref{fig:compactspt}).
Adding the high-$L$ information to the likelihood weakens the constraint further, pushing the 95\% limit to $1.07\,\mathrm{eV}$. This is consistent with our small-scale measurement having a significantly lower amplitude. At this stage it is unclear what to make of this mild tension between neutrino mass constraints from the 4-point function and those from the 2-point function, and we caution over-interpreting the results. We expect to be able to say more on this issue with the further data that will be made available in future \planck\ data releases.

\subsection{Correlation with the ISW Effect}
\label{sec:results:iswlensing}

As CMB photons travel to us from the last-scattering surface, the gravitational potentials they traverse may undergo a non-negligible amount of evolution.
This produces a net redshift or blueshift of the photons concerned, as they fall into and then escape from the evolving potentials.
The overall result is a contribution to the CMB temperature anisotropy known as the late-time ISW effect, or the Rees-Sciama (R-S) effect depending on whether the evolution of the potentials concerned is in the linear (ISW) or non-linear (R-S) regime of structure formation \citep{SachsWolfe1967,ReesSciama1968}.
In the epoch of dark energy domination, which occurs after $z \sim 0.5$ for the concordance $\Lambda$CDM cosmology, large-scale potentials tend to decay over time as space expands, resulting in a net blueshifting of the CMB photons which traverse these potentials.

In the concordance $\Lambda$CDM model, there is significant overlap between the large-scale structure which sources the CMB lensing potential $\phi$ and the ISW effect (greater than $90\%$ at $L<100$), although it should be kept in mind that we cannot observe the ISW component by itself, and so the effective correlation with the total CMB temperature is much smaller, on the order of $20\%$.

The correlation between the lensing potential and the ISW effect results in a non-zero bispectrum or three-point function for the observed CMB fluctuations.
This bispectrum is peaked for ``squeezed'' configurations, in which one short leg at low-$\elt$ supported by the ISW contribution is matched to the lensing-induced correlation between two small-scale modes at high-$\elt$.
Constraints on the amplitude of the lensing-ISW bispectrum using several different estimators are presented in \cite{planck2013-p09a}.
Here we will present an additional constraint, in which the bispectrum measurement is recast as an estimate for the amplitude of the cross-spectrum $C_L^{T\phi}$, using the filtering and frequency map combinations of our baseline lensing reconstruction.
Our measurements are in good agreement with those made in \cite{planck2013-p09a}; a detailed comparison of several lensing-ISW bispectrum estimators, including the one used here, is presented in \cite{planck2013-p14}.

\begin{figure}[!ht]
\centerline{\includegraphics[width=0.95\columnwidth]{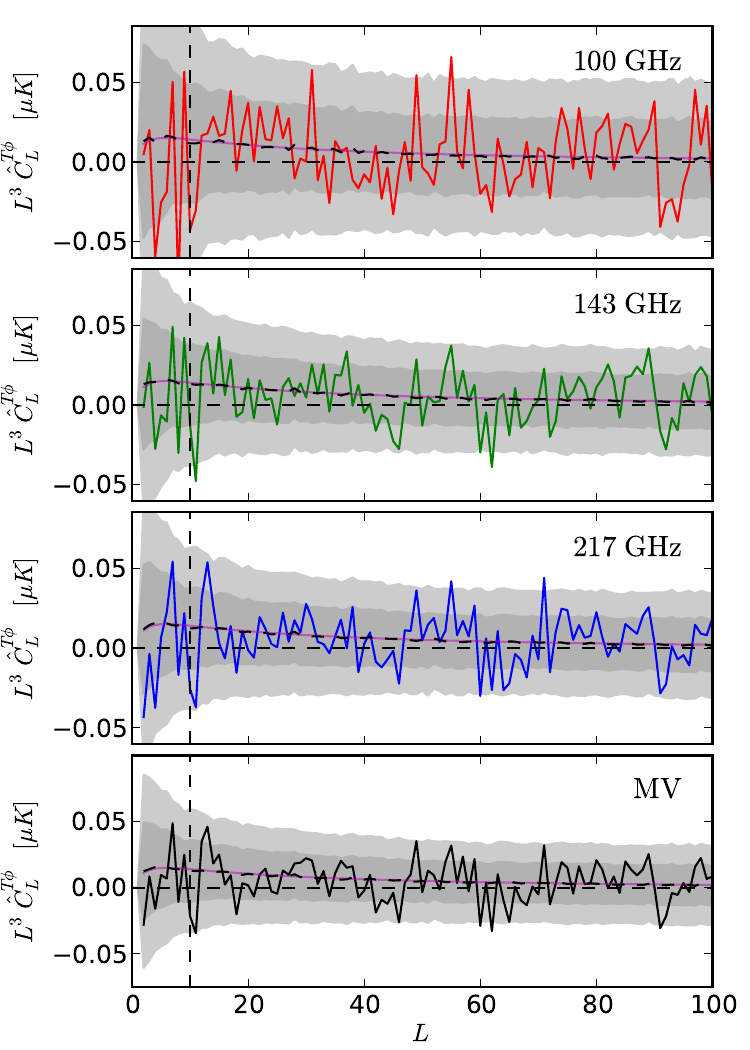}}
\caption{
Lensing-ISW bispectrum-related cross spectra computed from 
Eq.~\eqref{eqn:cltp_hat}.
Black dashed lines indicate the average value for simulations, while dark/light grey filled regions indicate the expected one/two standard deviation scatter, also measured from simulations.
The thin magenta line gives the expected $C_{\elp}^{T\phi}$ cross-spectrum for our fiducial model. 
The agreement of this curve with the simulation average illustrates that our estimator is accurately normalized.
In all the quantitative analysis of this section we ignore $L<10$, although we have plotted the cross-spectra at these multipoles for interest.
\label{fig:analysis_lens_isw_phi_triplot}
}
\end{figure}

Following \cite{Lewis:2011fk}, we begin with an estimator for the cross-spectrum of the lensing potential and the ISW effect as
\be
\hat{C}_L^{T\phi} = \frac{f_{\rm sky}^{-1}}{2L+1} \sum_{M} \hat{T}_{LM} \hat{\phi}_{LM}^*,
\label{eqn:cltp_hat}
\ee
where $\hat{T}_{{\elt}m} = C_{{\elt}}^{TT} \bar{T}_{{\elt}m}$ is the Wiener-filtered temperature map and $\hat{\phi}$ is given in Eq.~\eqref{eqn:phihat}.
In Fig.~\ref{fig:analysis_lens_isw_phi_triplot} we plot the measured cross-spectra for our individual frequency reconstructions at 100, 143, and 217\,GHz as well as the MV reconstruction. 
We also plot the mean and scatter expected in the fiducial $\Lambda$CDM model.

To compare quantitatively the overall level of the measured $C_L^{T\phi}$ correlation to the value in $\Lambda$CDM, we estimate an overall amplitude for the cross-spectrum as
\be
\hat{A}^{T\phi} 
= 
\norm^{T\phi}
\sum_{L=L_{\rm min}}^{L_{\rm max}} (2L+1) C_L^{T \phi, {\rm fid.}} \hat{C}_L^{T \phi} / (C_L^{TT} N_{L}^{\phi \phi}).
\label{eqn:cltp_estimator}
\ee
The overall normalization $\norm^{T\phi}$ is determined from Monte-Carlo simulations.
For our processing of the data, we find that it is well approximated (at the $5\%$ level) by the analytical approximation
\be
\norm^{T\phi} \approx \left[ \sum_{L=L_{\rm min}}^{L_{\rm max}} (2L+1) \left( C_{L}^{T \phi, {\rm fid.}} \right)^2 / (C_{L}^{TT} N_{L}^{\phi \phi}) \right]^{-1}.
\ee
The estimator above is equivalent to the KSW and skew-$C_{\elt}$ estimators of \cite{Komatsu:2003iq,Munshi:2009fr} for the lensing-ISW bispectrum that are used in \mbox{\cite{planck2013-p09a}} (up to implementation details such as filtering).
The mean-field subtraction performed when computing $\hat{\phi}_{LM}$ can be identified with the linear term of \cite{Creminelli:2005hu}, which is necessary to minimize the estimator variance.
The contribution to the total $S/N$ of this estimator as a function of the short leg $L$ is plotted in Fig.~\ref{fig:clpp_cltp_sn}, where it can be seen that the constraining power for the fiducial correlation is almost entirely at $L<100$.

In Table~\ref{table:lensing_isw_bispectrum_amps} we present measured values for the amplitude of the lensing-ISW bispectrum using Eq.~\eqref{eqn:cltp_hat}. 
The uncertainties on $\hat{A}^{T\phi}$ are determined by Monte-Carlo. 
We use the multipole range $10 < L < 100$, given some of the potential systematic issues with $L<10$ identified in Sect.~\ref{sec:biashardened}, although as can be seen from Fig.~\ref{fig:analysis_lens_isw_phi_triplot}, the inclusion of lower multipoles does not significantly affect our results. 
Note that for the ISW-lensing measurements, inaccuracies in the mean-field subtraction do not bias the estimator although they may degrade the statistical errors on large scales.
The differences between the different amplitude fits are well within the expected scatter, as we show in Table~\ref{table:lensing_isw_bispectrum_diffs}.
\begin{table}[!tbh]
\begingroup
\newdimen\tblskip \tblskip=5pt
\caption{
Best-fit amplitudes of the lensing-ISW bispectrum $\hat{A}^{T\phi}$ given by Eq.~\eqref{eqn:cltp_estimator} with $L_{\rm min}=10$ and $L_{\rm max}=100$, as well as split into even/odd-$L$ contributions.
All of the fits are consistent (within $2\sigma$) with the amplitude of unity expected in our fiducial $\Lambda$CDM model.
\label{table:lensing_isw_bispectrum_amps}
}
\nointerlineskip
\vskip -3mm
\footnotesize
\setbox\tablebox=\vbox{
 \newdimen\digitwidth
 \setbox0=\hbox{\rm 0}
  \digitwidth=\wd0
  \catcode`*=\active
  \def*{\kern\digitwidth}
  \newdimen\signwidth
  \setbox0=\hbox{+}
  \signwidth=\wd0
  \catcode`!=\active
  \def!{\kern\signwidth}
\halign{\hbox to 2.0cm{$#$\leaderfil}\tabskip=0.5em&
  \hfil$#$\hfil\tabskip=1.0em&
  \hfil$#$\hfil&
  \hfil$#$\hfil\tabskip=0pt\cr
\noalign{\doubleline}
\omit&\multispan3\hfil Lensing-ISW amplitudes\hfil\cr
\noalign{\vskip -4pt}
\omit&\multispan3\hrulefill \cr
\noalign{\vskip 2pt}
\omit&\hat{A}^{T\phi}$ (all L)$&\hat{A}^{T\phi}$ (even L)$&\hat{A}^{T\phi}$ (odd L)$\cr
\noalign{\vskip 4pt\hrule\vskip 5pt}
100\,{\rm GHz}&0.93\pm0.52&0.45\pm0.72&1.44\pm0.73\cr
143\,{\rm GHz}&0.81\pm0.36&0.27\pm0.48&1.37\pm0.52\cr
217\,{\rm GHz}&0.87\pm0.35&0.54\pm0.49&1.22\pm0.49\cr
{\rm MV}&0.78\pm0.32&0.25\pm0.45&1.32\pm0.46\cr
\noalign{\vskip 4pt\hrule\vskip 3pt}}}
\endPlancktable
\endgroup
\end{table}
\begin{table}[!tbh]
\begingroup
\newdimen\tblskip \tblskip=5pt
\caption{
Differences between the lensing-ISW amplitude fits of Table~\ref{table:lensing_isw_bispectrum_amps}, along with $1\sigma$ scatter determined from Monte-Carlo simulations.
\label{table:lensing_isw_bispectrum_diffs}
}
\nointerlineskip
\vskip -3mm
\footnotesize
\setbox\tablebox=\vbox{
 \newdimen\digitwidth
 \setbox0=\hbox{\rm 0}
  \digitwidth=\wd0
  \catcode`*=\active
  \def*{\kern\digitwidth}
  \newdimen\signwidth
  \setbox0=\hbox{+}
  \signwidth=\wd0
  \catcode`!=\active
  \def!{\kern\signwidth}
\halign{\hbox to 2.0cm{$#$\leaderfil}\tabskip=0.5em&
  \hfil$#$\hfil\tabskip=1.0em&
  \hfil$#$\hfil&
  \hfil$#$\hfil\tabskip=0pt\cr
\noalign{\doubleline}
\omit&\multispan3\hfil Lensing-ISW pair differences\hfil\cr
\noalign{\vskip -4pt}
\omit&\multispan3\hrulefill \cr
\noalign{\vskip 2pt}
\omit&143\,{\rm GHz}&217\,{\rm GHz}&$MV$ \cr
\noalign{\vskip 4pt\hrule\vskip 5pt}
100\,{\rm GHz}&+0.13 \pm 0.38 &  +0.06 \pm 0.42 &  +0.16 \pm 0.41\cr
143\,{\rm GHz}&   &  -0.07 \pm 0.26 &  +0.03 \pm 0.18\cr
217\,{\rm GHz}&   &  &  +0.10 \pm 0.13\cr
$Even$-$Odd$&-1.10\pm0.69 & -0.68 \pm 0.69 &  -1.07 \pm 0.64\cr
\noalign{\vskip 4pt\hrule\vskip 3pt}}}
\endPlancktable
\endgroup
\end{table}

As a point of interest, we have also split our amplitude constraint into the contribution from even and odd multipoles.
There are well known odd/even-multipole power asymmetries in the temperature anisotropies on large angular scales, the study of which is somewhat limited by the small number of available modes \citep{Land:2005jq,Kim:2010gf,Gruppuso:2010nd,Bennett:2010jb}.
The lensing potential gives a potentially new window on these power asymmetries, as a third somewhat independent measurement of power on large angular scales.
As we can see in Table~\ref{table:lensing_isw_bispectrum_amps}, 
there is a large difference between the odd/even-$L$ contributions in our lensing-ISW bispectrum estimate, related to the odd/even temperature asymmetry.

In Table~\ref{table:lensing_isw_bispectrum_diffs} we present differences between the amplitude measurements of Table~\ref{table:lensing_isw_bispectrum_amps}, as well as the expected scatter, accounting for correlations between the estimates due to common CMB and (in the case of MV vs. 143 or 217\,GHz) noise fluctuations.
We see that our estimates are all very consistent.
The difference in measured odd/even multipole power, while striking, is not statistically significant at greater than $2\sigma$.

To conclude, we see a correlation between our lensing potential measurement and the large-scale temperature anisotropies that is consistent with the level expected due to the present-day dark energy domination in our fiducial $\Lambda$CDM cosmology.
Our amplitude measurements are in agreement with those presented in \cite{planck2013-p09a}; these independent measurements are compared in detail in \cite{planck2013-p14}.

\subsection{Correlation with Galaxy Catalogues}
\label{subsec:xcorr}
\begin{figure*}[!ht]
\begin{center}
\begin{overpic}[width=\textwidth]{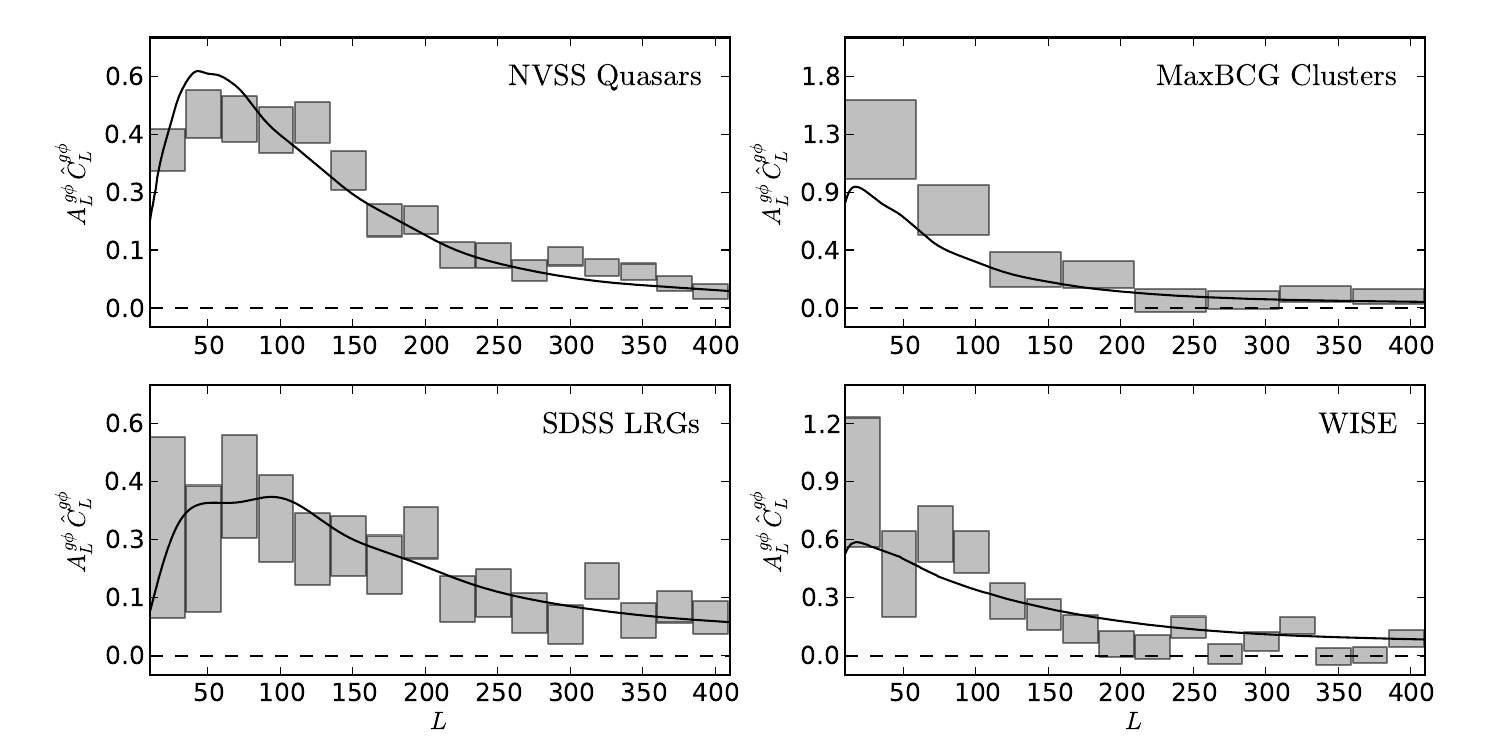} 
\end{overpic}
\end{center}
\vspace{-0.2in}
\caption{
Cross-spectra of the \Planck\ MV lensing potential with several galaxy catalogs, scaled by the signal-to-noise weighting factor $A_L^{g\phi}$ defined in Eq.~\eqref{eqn:algp}.
Cross-correlations are detected at approximately $20\sigma$ significance for the NVSS quasar catalog, $10\sigma$ for SDSS LRGs, and $7\sigma$ for both MaxBCG and WISE. 
\label{fig:analysis_lens_clgp_pub}
}
\end{figure*}
The CMB lensing potential is an integrated measure of all matter in the Universe back to the last-scattering surface, with a broad kernel which peaks at $z \sim 2$, but has significant contributions from both lower and higher redshifts. 
There are therefore significant correlations between the CMB lensing potential and other tracers of large-scale structure. 
Such correlations have already been observed
between the lensing potential reconstructed from WMAP data and the NVSS quasar catalogue \citep[on the order of $3\sigma$ in significance]{Smith:2007rg,Hirata:2008cb},
as well as between
the lensing potential from SPT with optical and infrared catalogues from BCS, WISE, and \textit{Spitzer}/IRAC \citep[between $4$ and $5\sigma$]{Bleem:2012gm},
and the ACT lensing potential with quasars from SDSS \citep[at $3.8\sigma$]{Sherwin:2012mr}.
Here we will show several representative examples of correlations with the \Planck\ lensing potential, between $7\sigma$ and $20\sigma$ in significance.
Our goal is not to perform an in depth study, but rather to demonstrate the power of our public lensing map.
In addition to the correlations presented here, there is a powerful correlation with the cosmic infrared background (CIB).
Correlation of the \Planck\ lensing potential with the CIB fluctuations as probed by the highest frequency \Planck\ channels is observed at greater than $40\sigma$ in significance,
and has been subjected to a more detailed analysis and modelling, which is presented in an accompanying paper, \cite{planck2013-p13}.

To predict expected levels of correlation for a given galaxy catalogue, we use the Limber approximation  \citep{Limber1954} with a simple linear bias model, in which the cross-spectrum between two mass tracers is given by
\be
C_{\elp}^{g \phi} = \int d\chi K_L^{g} (\chi) K_L^{\phi}(\chi) P(k=L/\chi, \chi),
\label{eqn:limber}
\ee
where $\chi$ is conformal distance and $P(k, \chi)$ is the 3D matter power spectrum for wavenumber $k$ at conformal lookback time $\chi$. 
$K^\phi$ and $K^g$ are kernels associated with lensing and with the galaxy catalogue of interest respectively, and are given by
\begin{align}
K_L^{\phi} (\chi)  &= -\frac{3 \Omega_m H_0^2}{L^2} \frac{\chi}{a} \left( \frac{\chi_{\ast}-\chi}{\chi_{\ast} \chi} \right), \nonumber \\
K_L^{g}(\chi)   &= \frac{dN}{dz} \frac{dz}{d\chi} \frac{b(z)}{\chi}.
\end{align}
Here $b(z)$ is a redshift-dependent linear bias parameter, and $dN/dz$ describes the redshift distribution of the galaxy population.
As with the lensing-ISW bispectrum, we can construct a simple pseudo-$C_{L}$ estimator for the cross correlation using
\be
\hat{C}_L^{g \phi} = \frac{ f_{ {\rm sky}, {\rm j} }^{-1} }{ 2L+1} \sum_{M} g_{LM} \hat{\phi}_{LM}^*,
\ee
where $f_{ {\rm sky},{\rm j}}$ is the sky fraction common to both the lens reconstruction and the catalogue, and $g_{LM}$ is the harmonic transform of the  galaxy fractional overdensity.
Denoting the positions of the objects $p$ in the catalogue as $\hat{n}_p$, the transform is given by
\be
g_{LM} =  \frac{1}{N^{gg}}  \sum_{p} Y_{LM}^*(\hat{n}_p),
\ee
where $N^{gg}$ is the surface density of objects in ${\rm gal}/{\rm steradian}$.
With a fiducial model $C_L^{g \phi,  {\rm fid.}}$ for the cross spectrum obtained using Eq.~\eqref{eqn:limber}, the minimum-variance estimator for its overall amplitude is
\be
\hat{A}^{g\phi} = \norm^{g\phi} \sum_{L} \frac{ (2L+1) C_L^{g\phi, {\rm fid.}} \hat{C}_L^{g \phi} }{ (C_L^{gg} + N_{\vphantom{L}}^{gg})( C_L^{\phi\phi} + N_L^{\phi \phi} ) } 
\equiv \sum_L A_L^{g\phi} \hat{C}_L^{g\phi},
\label{eqn:algp}
\ee
where $C_L^{gg}$ is the signal power spectrum of the catalogue, which can be estimated using Eq.~\eqref{eqn:limber} with $K^{g}$ for both weight functions.
Here we have defined the spectrum $A_L^{g\phi}$ as a scaling at each multipole $L$ of the cross-correlation power spectrum.
The normalization $\norm^{g\phi}$ is given by
\be
\norm^{g\phi} = \left[ \sum_{L} \frac{ (2L+1) \left( C_L^{g\phi, {\rm fid.}} \right)^2 }{ (C_L^{gg} + N_{\vphantom{L}}^{gg})( C_L^{\phi\phi} + N_L^{\phi \phi} ) } \right]^{-1}.
\ee

In Fig.~\ref{fig:analysis_lens_clgp_pub} we plot the contributions to $\hat{A}^{g\phi}$ as a function of $L$ for several surveys that have significant correlations with the \Planck\ MV lensing potential: 
the NVSS quasar catalogue, the MaxBCG cluster catalogue, an SDSS LRG catalogue, and an infrared catalogue from the WISE satellite.
The error bars for each correlation are measured from the scatter of simulated lens reconstructions correlated with each catalogue map, and are in generally good agreement (at the $20\%$ level) with analytical expectations.
These catalogues are discussed in more detail below.
\begin{enumerate}
\item NVSS Quasars: 
The NRAO VLA Sky Survey \citep[NVSS;][]{NVSS1998} is a catalogue of approximately two million sources north of $\delta = -40\deg$ which is $50\%$ complete at $2.5\,$mJy. Most of the bright sources are AGN-powered radio galaxies and quasars. 
We process this catalogue following \cite{Smith:2007rg}, pixelizing the catalogue at \HealpixPixelization\ $N_{\rm side}=256$ and projecting out the azimuthally-symmetric modes of the galaxy distribution in ecliptic coordinates to avoid systematic striping effects in the NVSS dataset.
We model the expected cross-correlation for this catalogue using a constant $b(z)=1.7$ and a redshift distribution centered at $z_0=1.1$ given by
\be
\frac{dN}{dz} \propto \left\{ \begin{array}{cl}
\exp\left(-\frac{(z-z_0)^2}{2 (0.8)^2}\right) & \qquad (z \le z_0) \\
\exp\left(-\frac{(z-z_0)^2}{2 (0.3)^2}\right) & \qquad (z \ge z_0).
\end{array} \right.  \label{eq:dnvssdz}
\ee
For this model, in the correlation with the MV lens reconstruction we measure an amplitude of 
$\hat{A}^{g\phi}_{\rm NVSS} = 1.03 \pm 0.05$.

\item SDSS LRGs:
We use the LRG catalogue of \cite{Ross:2011cz,Ho:2012vy} based on  Sloan Digital Sky Survey Data Release 8 (SDSS DR8), which covers $25\%$ of the sky.
After cutting to select all sources with photometric redshift $0.4 \le z \le 0.8$, and $p_{\rm gal} > 0.2$, we are left with approximately
$1.4 \times 10^6$ objects with a mean redshift of $z = 0.55$ and a scatter of $\pm 0.07$.
Apart from the cut above, we do not perform any additional weighting on $p_{\rm gal}$.
We model this catalogue using $dN/dz$ taken from the histogram of photometric redshifts, and take $b(z) = 2$.
We measure
$\hat{A}^{g\phi}_{{\rm LRGs}} = 0.96 \pm 0.10$, very consistent with expectation.

\item MaxBCG Clusters: 
The MaxBCG cluster catalogue \citep{Koester:2007bg} is a collection of $13,823$ clusters over approximately 
$20\%$ of the sky selected from the SDSS photometric data, covering a redshift range $0.1 \le z \le 0.3$.
It is believed to be $90\%$ pure and more than $85\%$ complete for clusters with $M \ge 1 \times 10^{14} \Msolar$.
To simplify the sky coverage, we have discarded the three southern SDSS stripes in the catalogue, which reduces the overall sky coverage to approximately $17\%$.
There are accurate photometric redshifts ($\Delta_z\! \sim\! 0.01$) for all objects in the catalogue, and so we can construct $dN/dz$ directly from the histogram of the redshift distribution.
Although these clusters are at very low redshift compared to the typical structures which source the CMB lensing potential, 
they are strong tracers of dark matter, with an effective bias parameter of $b(z)=3$ \citep{Huetsi:2009zq}.
We obtain a similar average bias parameter $\langle b(M,z) \rangle$ for the MaxBCG clusters if we combine the mass-richness relation of \mbox{\cite{Bauer:2012rs}} and the halo bias prescription of \mbox{\cite{Tinker:2010my}}.
Here we measure a correlation with the \Planck\ lensing potential of
$\hat{A}^{g\phi}_{\rm MaxBCG} = 1.54 \pm 0.21$.
This is significantly larger than expected given the simple model above, although as can be seen in Fig.~\ref{fig:analysis_lens_clgp_pub} the shape of the correlation is in reasonable agreement.

\item WISE Catalogue:
The Wide Field Survey Infrared Explorer (WISE) satellite \citep{Wright:2010qw} has mapped the full sky in four frequency bands $W1$---$W4$ at 
$3.4$, $4.6$, $12$, and $22\,\mu m$ respectively.
We start from the full mission catalogue, which contains over $560 \times 10^6$ objects.
To obtain a catalogue with roughly uniform sensitivity over the full sky and to eliminate stellar contamination we follow \cite{Kovacs:2013rs}, selecting all sources with W1 magnitudes less than $15.2$ at galactic latitudes greater than $10^{\deg}$, and require $W1-W2 > 0.2$ and $W2 - W3 > 2.9$.
We cut all sources that are flagged as potentially spurious, or for which more than $30\%$ of frames observing a given source are marked as contaminated by moon-glow.
These cuts leave us with $2,308,751$ sources.
We additionally remove all sources that lie outside an $f_{\rm sky}\!=\!0.6$ Galaxy mask.
Following \cite{Goto:2012yc}, we model the redshift distribution of these sources using
\be
\frac{dN}{dz} \propto z \exp\left( -\frac{(z-z_0)^2 }{2 \sigma_z^2} \right),
\ee
with $z_0=0.1$, $\sigma_z = 0.1$, 
and where the proportionality constant is chosen such that $\int dz dN/dz = 1$.
The bias is taken to be $b=1$.
For this model we measure an amplitude of
$\hat{A}^{g \phi}_{\rm WISE} = 0.97 \pm 0.13$.
Note that the cuts we have used above are fairly conservative-- in particular, relaxing the magnitude cut above we can obtain much larger cross-correlation signals (on the order of $40\sigma$ in significance).
\end{enumerate}
It can be seen from Fig.~\ref{fig:analysis_lens_clgp_pub} that we have significant correlations with the objects in all four of these catalogues, with amplitudes and shapes roughly at the expected level.
The correlations here are only a demonstrative sample of what is possible with the \Planck\ lensing potential map.
Note that although the lensing potential has a peak sensitivity to high-redshift tracers (such as NVSS), significant correlations may also be observed with relatively low redshift catalogs.
From the perspective of cross-correlation, the \Planck\ lensing potential represents a mass survey with a well understood redshift kernel over the full-sky, and 
we anticipate that correlations with it will prove useful as a direct probe of the correlation between luminous and dark matter.

\section{Consistency and systematic tests}
\label{sec:consistency}

\begin{figure*}[!htpb]
\begin{center}
\def\diffaxscal{\put(8,15.5){ \colorbox{white}{$\times 10^8$}} }
\def\diffspacer{\vspace{-0.025in}}
\def\diffwidth{\textwidth}
\def\startdiff{ \hspace{-0.1in} }
\def\enddiff{\diffaxscal \end{overpic} \diffspacer}

\startdiff\begin{overpic}[width=\diffwidth]{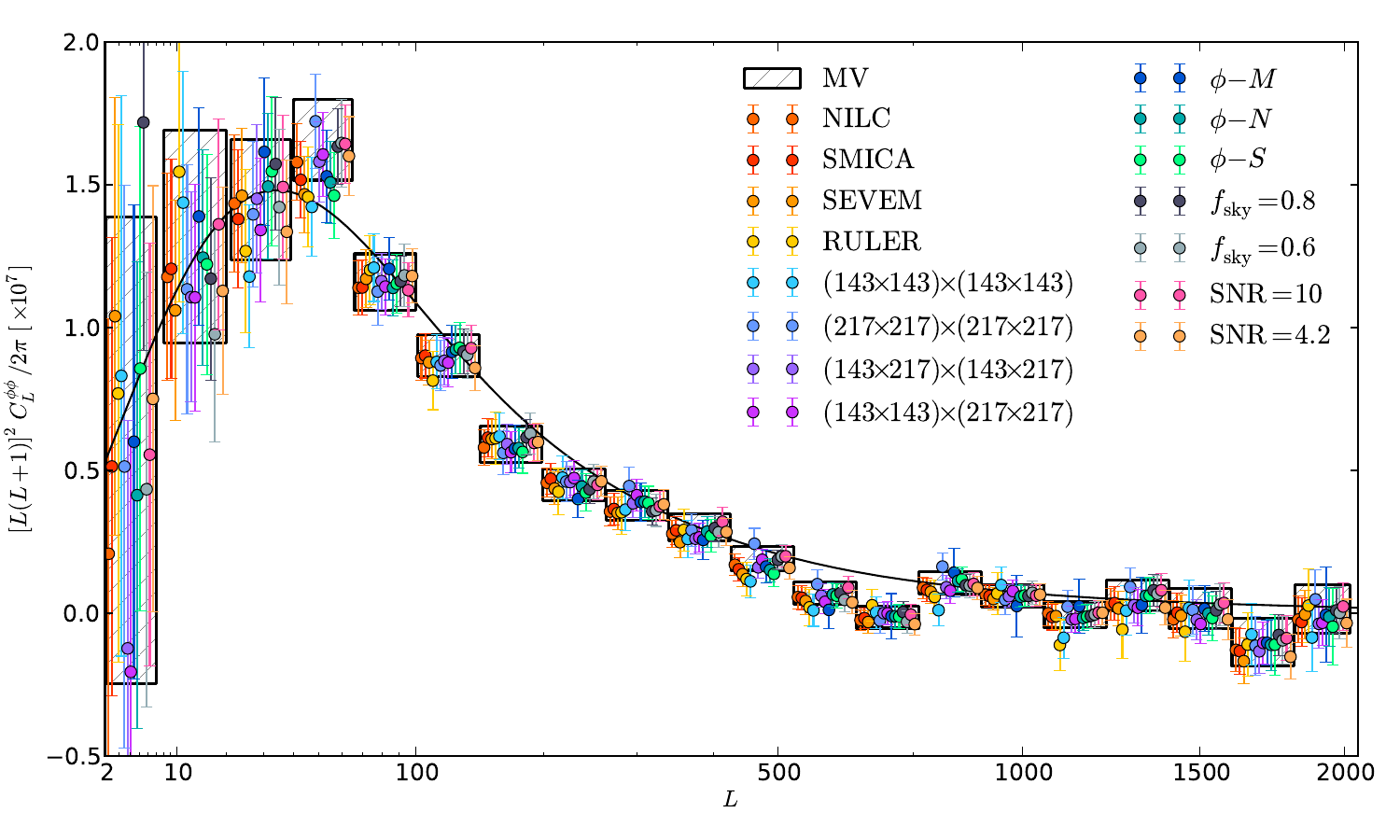} 
\end{overpic}

\startdiff\begin{overpic}[width=\diffwidth]{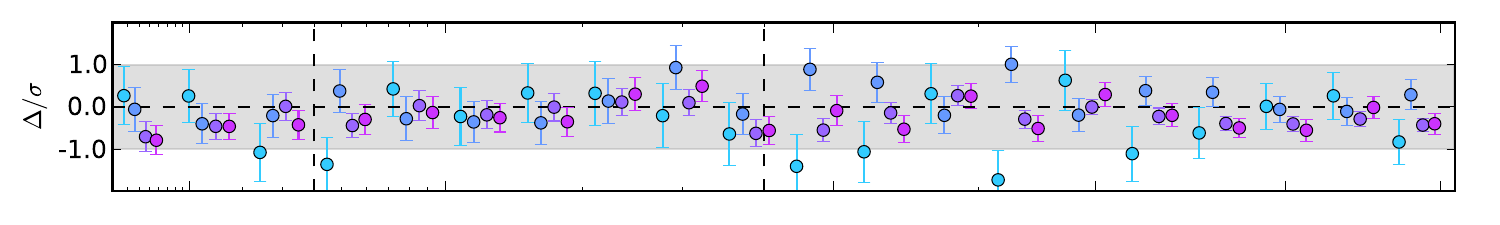} 
\put(8,14.2){Frequency Combinations:}
\end{overpic} 
\diffspacer

\startdiff\begin{overpic}[width=\diffwidth]{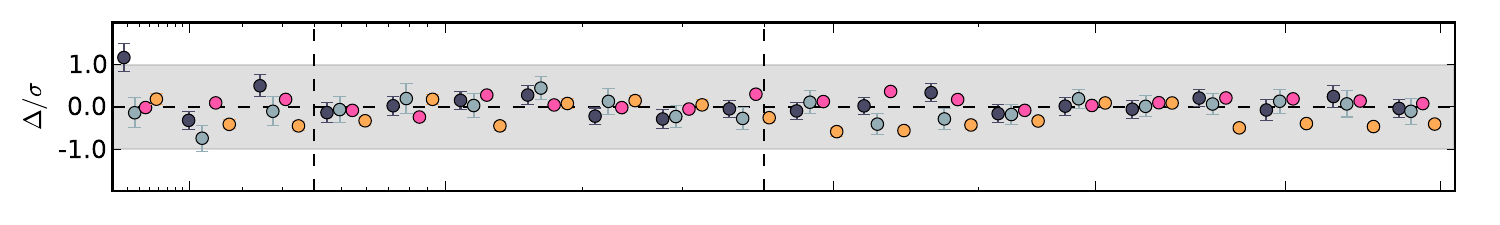}
\put(8,14.2){Mask Variation:}
\end{overpic}
\diffspacer

\startdiff\begin{overpic}[width=\diffwidth]{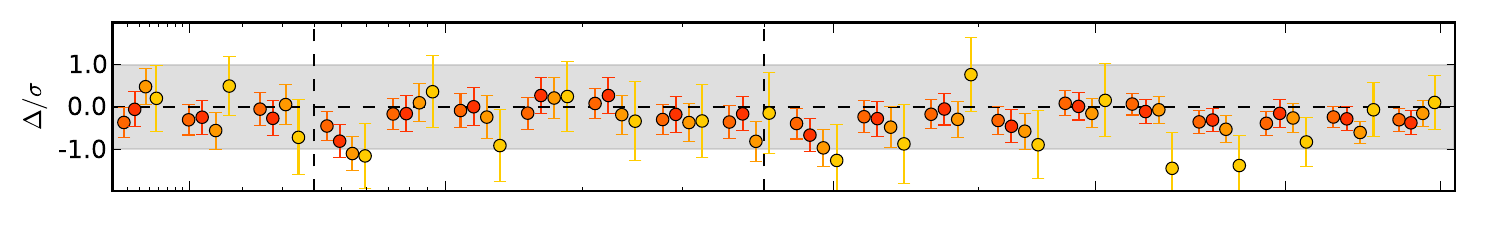}
\put(8,14.2){Component Separated Maps:}
\end{overpic}
\diffspacer

\startdiff\begin{overpic}[width=\diffwidth]{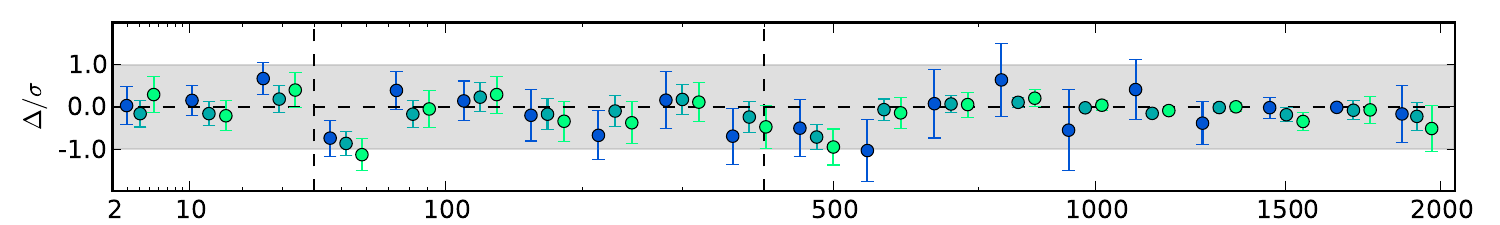}
\put(8,14.2){Bias-hardened Estimators:}
\end{overpic} 
\end{center}
\vspace{0.05in}
\caption{
Summary of internal consistency tests between our fiducial minimum-variance (MV) reconstruction and a set of alternatives designed to test sensitivity to potential issues.
The top panel shows $C_L^{\phi\phi}$ estimates, with measurement error bars.
The bottom panels show the residual with respect to the MV reconstruction in units of the MV measurement uncertainty.
The grey band marks the $1\sigma$ deviation uncertainty of the MV reconstruction.
The error bar on each data point in the lower panels gives the standard deviation of the scatter between each result and the MV, determined from Monte-Carlo simulations that account for the correlated CMB, noise and foreground power between estimators.
Comparison of the uncertainty on the scatter points and the grey band gives an indication of how constraining each test is.
The various tests are described in more detail in subsections of Sect.~\ref{sec:consistency}.
}
\label{fig:consistency_panel_mv}
\end{figure*}

In this section we perform a number of consistency and systematic tests.
Most of our tests will focus on the robustness of the MV reconstruction that forms the basis for our fiducial lensing likelihood.
In Fig.~\ref{fig:consistency_panel_mv} we summarize several of our main consistency checks:
foreground tests obtained by comparing individual frequencies, adjusting the Galactic mask level, or comparing to results obtained analyzing component-separated maps (discussed in Sect.~\ref{sec:consistency:foregrounds}),
tests of our correction at the power spectrum level for contamination by the shot-noise from unresolved point-source 
(in Sect.~\ref{sec:consistency:pointsource}),
tests for possible noise bias obtained by taking appropriate correlations of maps with independent noise realizations
(discussed further in Sect.~\ref{sec:consistency:noisebias}),
and tests using alternative estimators that are bias hardened against various systematics (Sect.~\ref{sec:biashardened}). 
In all cases, we find that our MV power spectrum is generally consistent with these alternative reconstructions.

Our baseline reconstruction method uses the methodology outlined in Sect.~\ref{sec:methodology}, as well as a specific choice of filtering to produce the $\bar{T}$ that are the inputs to our lensing estimators. 
We test our robustness to these choices in Sect.~\ref{sec:consistency:filtering}, as well as the implementation of our main reconstruction pipeline, by comparing to three alternative and largely independent pipelines.

In addition to these targeted tests, we have also performed several generic null tests for unexpected signals in our data.
These are deferred to Appendix~\ref{app:nulltests}, where we study survey-to-survey discrepancies, lensing curl modes, 
and the map-level statistics of our lens reconstruction.


\subsection{Foreground contamination}
\label{sec:consistency:foregrounds}
\label{sec:consistency:comp_sep}

In this section, we test the consistency of our results with respect to the treatment of Galactic foreground contamination.
We use several different tests to gain perspective on possible issues:
\begin{enumerate}[(1)]
\item
We compare individual frequency reconstructions.
The agreement of the lensing power spectra estimated at 100, 143, and 217\,GHz is apparent within the error bars plotted in Fig.~\ref{fig:analysis_lens_clpp_pub}.
However, because a large fraction of the uncertainty in these estimates comes from CMB fluctuations themselves they are significantly correlated.
In Fig.~\ref{fig:consistency_panel_mv} we plot the difference between the 143 and 217\,GHz reconstructions (the scatter for 100\,GHz is too large to provide a useful test of the MV reconstruction so we have not plotted it, although it is also consistent).
Both are in reasonable agreement, although we can see that the somewhat large amplitude for our first likelihood bin, just above $L=40$ is being driven by 217\,GHz.
\item
We perform lens reconstruction using both more aggressive and more conservative Galaxy masks than our fiducial $f_{\rm sky}=0.7$ analysis mask, 
constructed following the description in Sec.~\ref{sec:data}.
The results of this test are also shown in Fig.~\ref{fig:consistency_panel_mv}, for masks with 
\mbox{$f_{\rm sky}=0.6$} and 
\mbox{$f_{\rm sky}=0.8$}.
We can see a highly significant discrepancy at very low-$L$ between our fiducial 
\mbox{$f_{\rm sky} =0.7$}
mask and the more aggressive 
\mbox{$f_{\rm sky} = 0.8$} mask, however other bins (in particular in the $40<L<400$ range) are untouched. Our fiducial results are completely consistent with the more conservative $f_{\rm sky}=0.6$ results.
\item 
We perform lens reconstruction on the  component-separated CMB maps from \cite{planck2013-p06}.
These are the product of four different foreground-removal algorithms: \CR, \NILC, \SEVEM, and \SMICA.
Each of these methods combines the full set of nine \Planck\ frequency maps from $30 \rightarrow 857$\,GHz to obtain a best-estimate CMB map.
This makes a detailed characterization of these maps more difficult.
However analysis of these maps does provide a valuable consistency test. 
In Fig.~\ref{fig:consistency_panel_mv} we compare the reconstructions obtained on these component-separated maps to our MV reconstruction, analyzed using the same Galaxy mask.
The foreground-cleaned maps are generally in good agreement with our fiducial reconstruction, although the first bin above $L=40$ is consistently about one standard-deviation lower than our MV result, which as we have already pointed out is driven high by 217\,GHz.
The confidence masks produced in \cite{planck2013-p06} cover significantly more sky than the $f_{\rm sky}=0.7$ mask used in our baseline analysis.
\end{enumerate}
The remainder of this subsection describes our analysis of the component-separated maps in more detail.
For the power spectrum tests of Fig.~\ref{fig:consistency_panel_mv}, we analyze these maps following the same methodology as for our fiducial results.
As a mask, we use the union of our fiducial $f_{\rm sky}=0.7$ Galaxy mask, point source masks at 100, 143, and 217\,GHz, as well as the ``method'' masks provided for each component separation algorithm.
We characterize the bias terms of Eq.~\eqref{eqn:clppest} as well as the statistical uncertainties of the lens reconstruction on each map using the same lensed CMB and FFP6 noise realizations that are discussed in Sect.~\ref{sec:simulations}.
These have been run through each of the component-separation algorithms above, resulting in a set of simulated foreground-cleaned maps 
(note that no foreground signal was included in the simulations which were run through the separation algorithms, and so these simulations only reproduce the mixing employed by each algorithm of the individual frequency maps).
To match the power spectrum of these simulations to the power spectrum of the data maps, we find it is necessary to add extragalactic foreground power following the model in Sect.~\ref{sec:simulations}, with $A_{\rm cib} = 18\, \muK^2$ and $A_{\rm src} = 28\, \muK^2$.
The resulting simulations have a power spectrum that agrees with that of the CMB map estimate based on the data to better than $2\%$ at $l < 2048$.
This could be improved slightly by tailoring a specific correction for each map.
We also add homogeneous pixel noise with a level of $12\, \muKarcmin$.
If we neglected this power, the agreement would be only at the $8\%$ level, primarily due to the noise term (the $A_{\rm cib}$ and $A_{\rm src}$ contributions are each at the level of $1-2\%$).
Due to the procedure that we use to subtract the disconnected noise bias (Eq.~\ref{eqn:cln0}) from our lensing power spectrum estimates, the inclusion of these components does not significantly affect our results, but comparison with the values used for our single-frequency simulations in Sect.~\ref{sec:simulations} are a useful indicator of the extent to which the component-separation algorithms are able to remove extragalactic foreground power in the high-$\elt$ regime.

As already discussed, our results on the component-separated CMB maps are presented in Fig.~\ref{fig:consistency_panel_mv}.
Because the CMB and FFP6 noise components of the foreground-cleaned map simulations are the same as those used to characterize our fiducial lens reconstruction, we can measure the expected scatter between the component-separated maps and our fiducial reconstruction.
This scatter will be slightly overestimated because we have not attempted to model coherently the contribution to the reconstruction noise from residual diffuse extragalactic foreground power.
For the eight bins in $40 \le L \le 400$ on which our fiducial likelihood is based, we measure a $\chi^2$ for the difference between our fiducial reconstruction and the corresponding foreground-cleaned reconstruction of $\chi^2 = (3.14, 4.3, 2.5, 14.7)$ for \NILC, \SMICA, \SEVEM, and \CR\  respectively. 
These $\chi^2$ values associated have probability-to-exceed (PTE) values of $(79\%, 64\%, 86\%, 2\%)$.
At the level which we are able to test, the \NILC, \SMICA, and \SEVEM\  foreground-cleaned maps give results that are quantitatively consistent with our fiducial reconstruction.
There is more scatter between our fiducial reconstruction and the \CR\  map than expected from simulations, as evidenced by a very high $\chi^2$ for the difference, however as can be seen in Fig.~\ref{fig:consistency_panel_mv}, there are not any clear systematic differences.
Indeed, the discrepancy for the bins plotted in Fig.~\ref{fig:consistency_panel_mv} (which differ somewhat from the linear bins used in our likelihood) is much less significant than for the bins of our fiducial likelihood.

\begin{figure}[!ht]
\centering
\vfill
\begin{overpic}[trim=0 48 0 45, clip, width=\columnwidth]{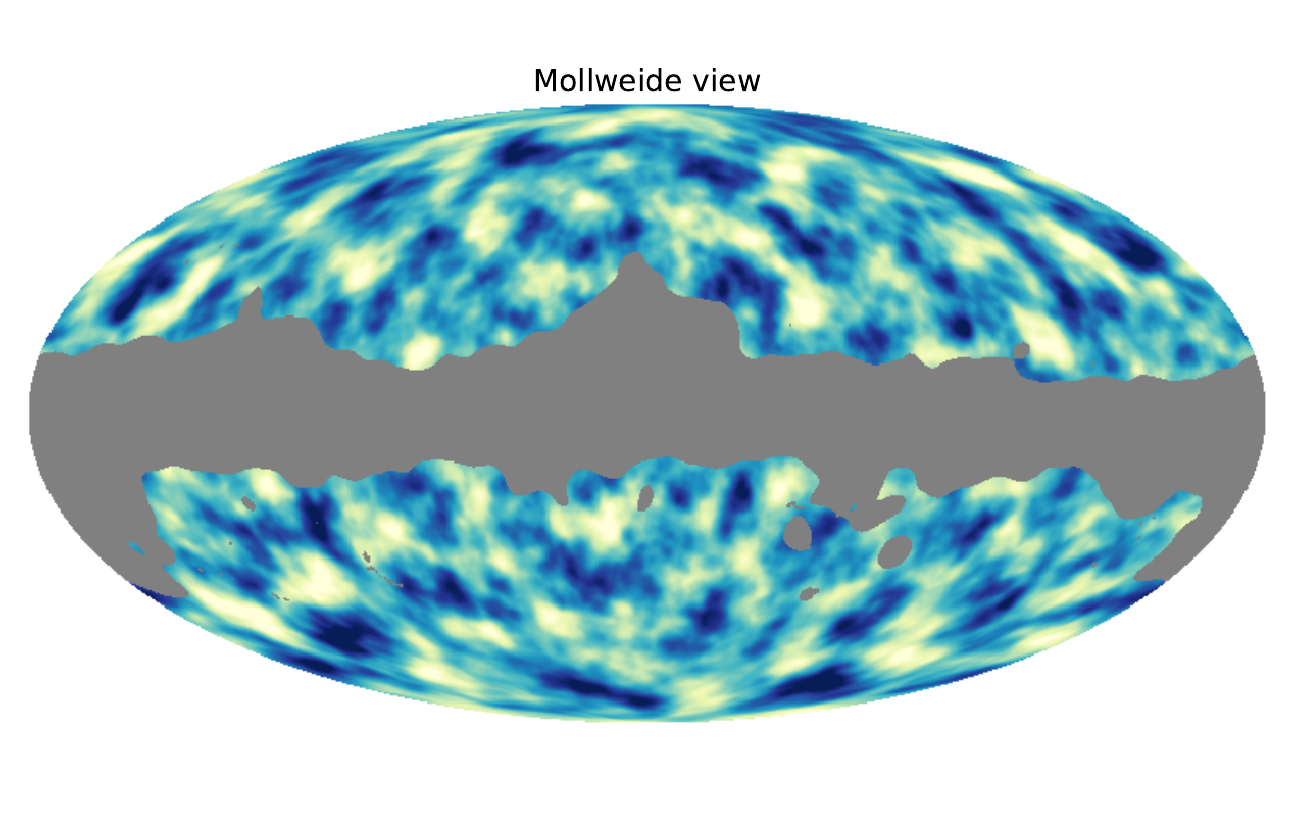}
\put(78,0){\tiny {\sc MV}, $f_{\rm{sky}} = 0.70$}
\end{overpic}
\vfill
\vspace{0.125cm}
\begin{overpic}[trim=0 48 0 45, clip, width=\columnwidth]{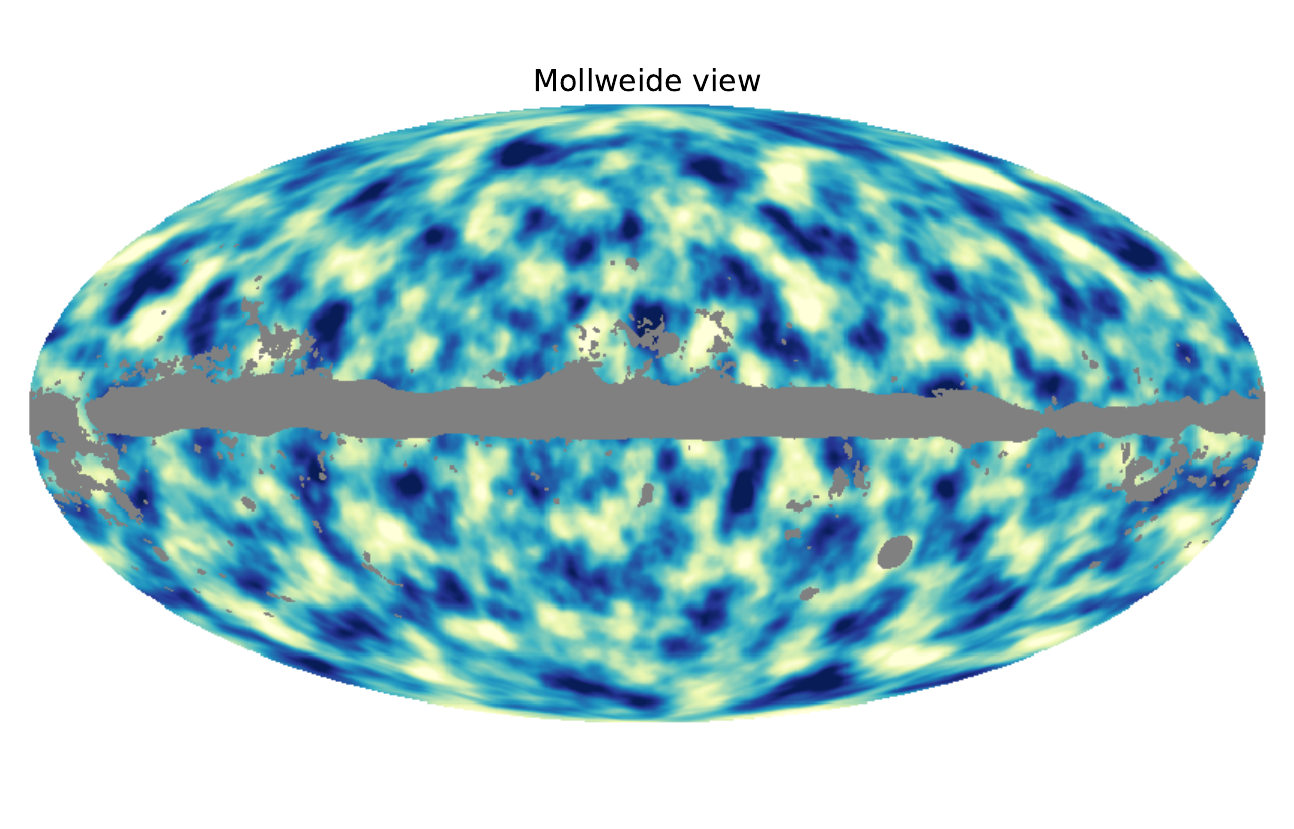}
\put(75,0){\tiny \NILC, $f_{\rm{sky}} = 0.87$}
\end{overpic}
\vfill
\vspace{0.125cm}
\caption{
Wiener-filtered potential maps in Galactic coordinates, as in Fig.~\ref{fig:wiener_filtered_mv_reconstruction}, plotted here in Mollweide projection. 
 {\em Top}: MV reconstruction; {\em Bottom}: extended reconstruction on the \NILC\ component-separated map. 
}
\label{fig:nilc_wiener_filtered_d}
\end{figure}

When using the component-separated maps above, we have used the same $f_{\rm sky}=0.7$ Galactic mask as for our MV result, 
although the confidence regions associated with each foreground-cleaned map allow more sky, ranging up to $f_{\rm sky}=0.94$ for the \NILC\ method.
We have used the \metis\  pipeline (described later in Sect.~\ref{sec:consistency:filtering}) to test whether this improved sky coverage could benefit our lens reconstruction.
The same method has been used in \cite{planck2013-p06} to evaluate possible biases to lens reconstruction induced by these methods using the FFP6 simulated CMB realization, described in \cite{planck2013-p01}, 
indicating that the different component-separation algorithms do not alter significantly the lensing signal (at the level which can be tested on a single simulation).
Analyzing the \NILC\ map, which has the largest confidence region, we find that we can increase the usable sky area up to $f_{\rm sky}=0.87$ without encountering significant Galactic contamination. In Fig.~\ref{fig:nilc_wiener_filtered_d} we show the striking improvement in sky coverage on the \NILC\ map. \SMICA\ and \SEVEM\ are very similar; we have not considered \CR\ because of its larger noise level.

Power spectrum estimates at this mask level show consistency with the MV reconstruction within two standard deviations of the measurement uncertainty. 
The increased sky coverage does not bring significant improvements in the error-bars of the power spectrum, however. 
Using Eq.~\ref{eqn:var_clpp_analytical} as an estimate of the power spectrum variance, the larger sky coverage  yields only a $3.5\%$ improvement at $L<40$ over the MV result, decreasing down to 0 at $L=400$. 
This could be due to the different weighting used in the component separation compared to the one of the MV map, which results in slightly noisier maps for our purpose.
While the component-separated maps allow for a reduced mask, they lead to a marginal improvement of the power spectrum uncertainties. 
Nevertheless, their agreement with the MV result is reassuring.


\subsection{Point source correction}
\label{sec:consistency:pointsource}
As can be seen in Table~\ref{tab:bandpowers}, the unresolved point-source shot-noise correction in any individual band for our MV likelihood is on the order of a few percent, reaching up to $6\%$ for the highest multipole bands.
Averaged over the $40 \le L \le 400$ band, the shot noise correction amounts to a $2\%$ shift in the amplitude of $\hat{C}_L^{\phi\phi}$, which is small but non-negligible compared to our statistical uncertainty of $4\%$.
Physically, the amplitudes of our source corrections are reasonable; at 143\,GHz we measure 
$\hat{S}^4_{143} = (1.3 \pm 0.6)\times 10^{-12}\, \mu K^4$.
From the radio point-source model of \cite{DeZotti:2009an}, this corresponds to an effective flux cut of approximately $150\,$mJy at this frequency, roughly comparable to that expected for the $S/N > 5$ cut we make when masking sources in our fiducial analysis \citep{planck2013-p05}.
The shot noise measured at 217\,GHz is lower, as expected given the smaller contribution from radio sources, with 
$\hat{S}^{4}_{217} = (0.4 \pm 0.4) \times 10^{-12}\, \mu K^4$.
The shot-noise level that we measure for the MV reconstruction lies between these two, with 
$\hat{S}^{4}_{\rm MV} = (0.6 \pm 0.3) \times 10^{-12}\, \mu K^4$.

As a consistency test of our $S/N\!>\!5$ source masks, 
we construct two new sets of source masks that only exclude $S/N >10$ sources 
(to check the stability of our reconstruction under significantly higher source contamination), 
or which exclude all detected sources with $S/N\!>\!4.2$ (to test the conservatism of our fiducial flux cut).
The results of this test are shown in the masks panel of Fig.~\ref{fig:consistency_panel_mv}.
Although not shown explicitly in the figure, the source correction term with the $S/N > 10$ mask is considerably larger than our fiducial correction.
For this mask we measure 
$\hat{S}^{4}_{\rm MV} = (2.8 \pm 0.3)\times 10^{-12}\, \mu K^4$, 
now a $9\sigma$ detection of shot noise in the trispectrum.
The importance of the correction grows correspondingly, becoming a $9\%$ overall correction to the $C_L^{\phi\phi}$ amplitude in $40 \le L \le 400$.
With the source correction, we measure 
$\hat{A}^{\rm MV}_{40 \rightarrow 400} = 0.95 \pm 0.04$, 
consistent with the 
$\hat{A}^{\rm MV}_{40 \rightarrow 400} = 0.94 \pm 0.04$ measurement for our fiducial source mask.
Without the source correction we would have measured 
$\hat{A}^{\rm MV}_{40 \rightarrow 400} =1.04 \pm 0.04$,
significantly discrepant.
We can see that even for this extremely permissive source mask, our correction for the point-source shot-noise trispectrum does a good job of rendering 
$\hat{C}_L^{\phi\phi}$ 
insensitive to source contamination.
Turning to the $S/N\!>\!4.2$ mask, we measure 
$\hat{S}^{4}_{{\rm MV}, S/N>4} = (-0.1 \pm 0.3)\times 10^{-12}\, \mu K^4$, 
reduced from a $2\sigma$ measurement of shot noise to one very consistent with zero.
For this reconstruction we measure
$\hat{A}_{40 \rightarrow 400} = 0.93 \pm 0.04$,
in good agreement with the MV result,
although there does appear to be an overall trend for lower power above $L=400$ in the $S/N\!>\!4.2$ mask.
At a detailed level, both of these results do fail $\chi^2$ tests on the difference with the MV reconstruction.
As the change in sky fraction between these three reconstructions is small (the difference between the fiducial and $S/N\!>\!4.2$ mask is 
$1\%$, 
and even smaller for the $S/N\!>\!10$ mask), even the small differences quoted above are significantly larger than they should be if the source correction were perfect.
This can be seen in Fig.~\ref{fig:consistency_panel_mv}, where the error bars on the difference between each of these reconstructions and the MV result is too small to be visible for the plotted points.
For the 
$40 \le L \le 400$
range on which our likelihood is based, we believe that the discrepancy between the power spectra estimated with these source masks is acceptable.
Particularly given the results of the $S/N>10$ cut, the source correction that we perform appears to behave reasonably, and reduces the sensitivity of our estimates to the shot-noise trispectrum of unmasked point sources.
In Sect. 7.4, we will also use a modified lensing estimator that has zero response to the shot-noise trispectrum and therefore does not require a source correction.
As with the tests above, we will find good overall agreement with the standard estimator used for the MV reconstruction.


\subsection{Instrumental noise bias}
\label{sec:consistency:noisebias}

Errors in the instrumental noise model are a source of concern for our lensing power spectrum estimates, in which all four legs of the trispectrum estimate come from a single map.
In this situation there are three Wick contractions of the noise which contribute to the power spectrum estimate in two distinct ways, depending on whether they couple between the two quadratic estimators in the power spectrum estimate or not:
\begin{enumerate}[(1)]
\item The Wick contraction that couples noise contributions inside the same estimator leads to a mean-field term as in Eq.~\eqref{eqn:lensing_meanfield}.
This term is sourced by the statistical anisotropy of the noise (due to the uneven hit distribution across the sky, or the correlations along the scan direction induced by deconvolution of the bolometer time constant).
The shape of this mean-field is dictated by the \Planck\ scan strategy, which is coherent over large scales, and so this mean-field is primarily at low-$L$, below the $L_{\rm min}=40$ cutoff of our likelihood.
It can be avoided by forming a quadratic estimator in which the two input maps have different noise realizations.
We perform a null test against this contribution in Fig.~\ref{fig:consistency_panel_mv}, where the 
$(143 \times 217) \times (143 \times 217)$ panel shows a comparison of our MV power spectrum estimate to that obtained using a quadratic estimator in which one leg comes from 143\,GHz and one from 217\,GHz.
We see generally good agreement with the MV result. For the eight bins in 
$40 \le L \le 400$
of our fiducial likelihood, we find a $\chi^2$ for the difference (accounting for correlations between the two estimates) of 
$4.2$, 
which has a PTE of
$83\%$.
The noise mean-field contribution can also be reduced by using an appropriate bias-hardened estimator. 
We will perform additional tests with such estimators in Sect.~\ref{sec:biashardened}.
\item 
The two Wick contractions that couple noise terms across-estimators contribute to the disconnected bias $\left. \Delta C_L^{\phi\phi} \right|_{\mathsc{N0}}$. 
This term  describes the noise contribution to the reconstruction variance. 
Our use of the data covariance matrix in Eq.~\eqref{eqn:cln0} makes our estimate of the disconnected noise bias insensitive to small errors in the noise model.
The noise contribution can also be avoided by taking the cross-spectrum of two quadratic estimators with independent noise.
We show the result of such a correlation compared to our MV reconstruction in Fig.~\ref{fig:consistency_panel_mv}, where the
$(143 \times 143) \times (217 \times 217)$ panel shows the cross-spectrum between the individual 143 and 217\,GHz reconstructions.
Again, we see generally good agreement with the MV result. For the eight bins in $40 \le L \le 400$ of our fiducial likelihood, we find a $\chi^2$ for the difference (accounting for correlations between the two estimates) of 
$6.5$,
which has a PTE of
$59\%$.
\end{enumerate}
Based on these tests, we believe that noise bias cannot be a significant contaminant for our power spectrum estimates.


 \subsection{Bias-hardened estimators}
 \label{sec:biashardened}
 
\begin{figure}[!htpb]
\hspace{-0.1in}
\includegraphics[width=\columnwidth]{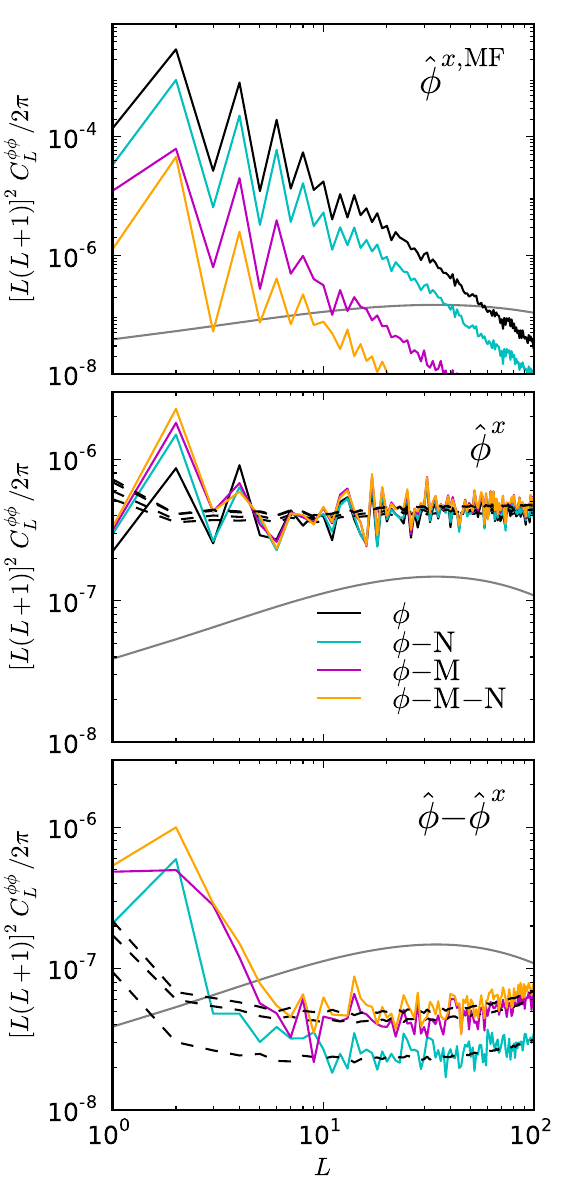}
\caption{
Bias-hardened estimator consistency tests for our MV reconstruction.
Following the discussion in Sect.~\ref{sec:biashardened} we form estimators that are bias-hardened against the mean-fields due to inhomogeneous noise levels 
($\phi\!-\!N$, \textcolor{cyan}{cyan}), 
masking 
($\phi\!-\!M$, \textcolor{magenta}{magenta}), 
and both of these effects simultaneously ($\phi\!-\!M\!-\!N$, \mbox{\textcolor{YellowOrange}{orange}}).
The top panel shows the raw power spectra of the mean-field for each of these estimators.
The middle panel shows the power spectra of the reconstructions themselves, compared to the average expected from signal+noise (dashed black). 
The lower panel shows the power spectrum of the difference between each bias-hardened estimator and the standard result. 
Dashed black lines give the expected average of this difference (measured on simulations). 
The grey line is the fiducial $C_L^{\phi\phi}$ power spectrum.
We can see that the differences are consistent with the expected scatter for all estimators at $L \ge 10$.
\label{fig:bh_lowl}
}
\end{figure}

Most of the results in this paper use the standard weight function $W^{\phi}_{\elt_1 \elt_2 L}$ to form quadratic estimates for the lensing potential.
This weight function is a matched filter for lensing, and results in estimators with minimal variance.
As we discuss in Appendix~\ref{sec:errorbudget:meanfields}, however, the standard estimator has large mean-field contributions at low-$L$, primary due to masking and noise inhomogeneity.
As pointed out in \cite{Namikawa:2012pe}, these mean-fields can be mitigated with the use of ``bias-hardened'' lensing estimators, which use weight functions $W^{x}_{\elt_1 \elt_2 L}$ specially constructed to project out these effects.
Following the notation of Appendix~\ref{sec:errorbudget:meanfields}, we consider an uncertain bias field $z_{LM}$ with associated weight function $W^{z}_{\elt_1 \elt_2 L}$ that gives a contribution to the off-diagonal elements of the CMB covariance matrix (and therefore the estimator mean-field) of
\be
\Delta \langle T_{\elt_1 m_2} T_{\elt_2 m_2} \rangle =
\sum_{LM} 
(-1)^M 
\threej{\elt_1}{\elt_2}{L}{m_1}{m_2}{M}
W^{z}_{\elt_1 \elt_2 L} z_{LM}.
\label{eqn:tcov_aniso3}
\ee
We can then form an estimator which is insensitive to $z_{LM}$ simply as
\be
W^{(\phi - z)}_{\elt_1 \elt_2 L} = W^{\phi}_{\elt_1 \elt_2 L} - \frac{ \resp_L^{\phi z} }{ \resp_L^{z z} } W^{z}_{\elt_1 \elt_2 L}.
\ee
The quadratic estimator $(\phi\!-\!z)$ that uses the weight function above has a significantly reduced mean-field contribution from $z_{LM}$.
This bias-hardening procedure may be repeated iteratively to produce weight functions that are insensitive to several sources of mean-field simultaneously. 
In Fig.~\ref{fig:bh_lowl} we plot a comparison at low-$L$ between estimates with the standard estimator and estimators that are bias hardened against noise and mask mean-fields.
We see generally good agreement between these sets of estimates for $L \ge 10$.
Below $L=10$, the pseudo-spectrum of the difference between the standard and bias-hardened estimators is unexpectedly large.
For this reason, we band-limit our lensing potential map to $L\ge10$, and do not consider $L<10$ for any of the quantitative results in this paper.

In addition to low-$L$ consistency tests, we may also use these bias-hardened estimators to estimate $C_L^{\phi\phi}$.
The resulting power spectra are plotted in Fig.~\ref{fig:consistency_panel_mv}.
Taking the difference with the eight bins in $40 \le L \le 400$ of our fiducial MV reconstruction, with scatter estimated from simulations, we find 
$\chi^2$ values of 
$6.6$ and $4.3$ 
for the mask/noise-hardened estimators respectively.
These have corresponding PTE values of $58\%$ and $83\%$. 
The error bars on $C_L^{\phi\phi}$ obtained with the bias hardened estimators used here are generally between 10 and 20\% percent
larger than the error bars of the standard estimator.
We also construct an estimator that is bias hardened against the point source weight function 
$W_{\elt_1 \elt_2 L}^{s}$ of Eq.~\eqref{eqn:qe_weight_ptsrc}.
This estimator has the distinction of having zero response to point-source shot noise, 
as well as any correlation between 
the point-source shot-noise power and the lensing potential.
Again, we find consistent results with the standard estimator, with $\chi^2 = 5$ and a PTE of $76\%$.

 
\subsection{Alternative methods}
\label{sec:consistency:filtering}

\begin{figure}[!ht]
\centerline{\includegraphics[width=\columnwidth]{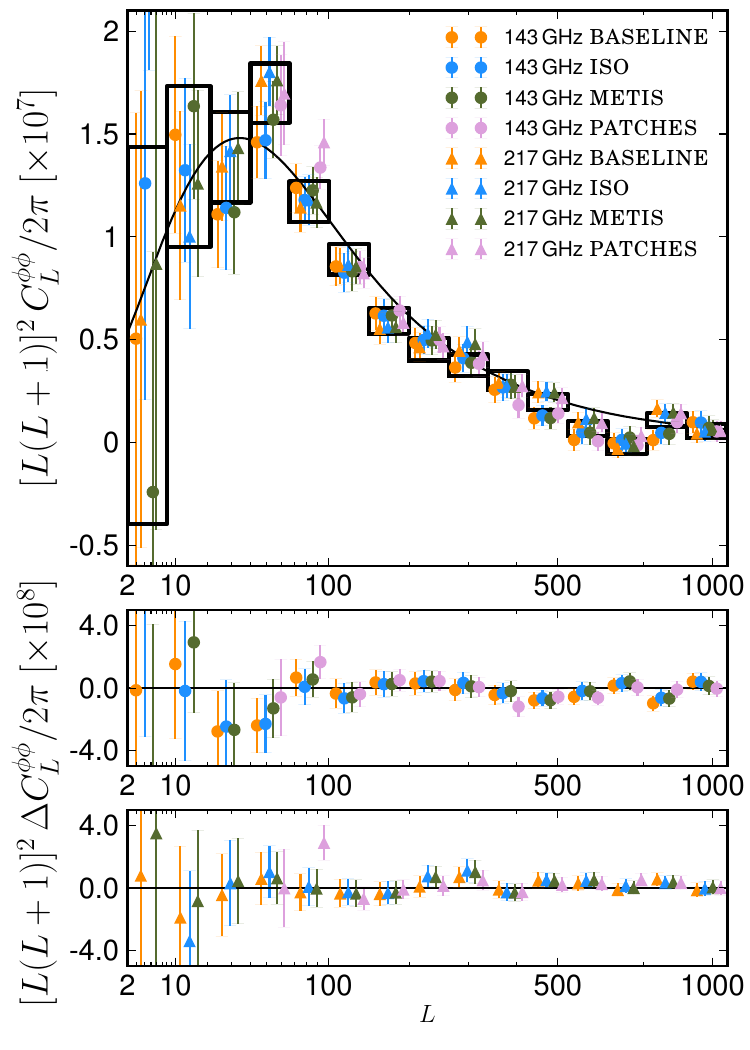}}
\caption{Comparison of alternative lensing pipelines. The {\sc BASELINE} results use the methodology of Sect.~\ref{sec:methodology}. 
Boxes are for the MV reconstruction, circles show the 143\,GHz results, and triangles show the 217\,GHz results. 
The two bottom panels show differences relative to the MV result for 143 and 217\,GHz.}
\label{fig:compalt}
\end{figure}

All of the primary results in this paper use a lens reconstruction pipeline based on the methodology outlined in Sect.~\ref{sec:methodology}.
As a robustness test, both of this methodology and of its implementation, we have implemented three independent reconstruction pipelines.
These independent pipelines make significantly different choices than our baseline approach, 
primarily in the choice of the inverse-variance filter function and the calculation of the correction terms in Eq.~\eqref{eqn:clppest}.
Our main result of this section is Fig.~\ref{fig:compalt}, where we compare the lens reconstruction power spectra for these alternative pipelines to our baseline results at 143 and 217\,GHz, as well as the MV reconstruction.
The agreement is excellent.
The only aspect of implementation common to these four results (apart from the data that was analyzed) is the set of simulations, described in Sect.~\ref{sec:simulations}, which are used to estimate the various bias terms and characterize the estimator scatter.

We now proceed to describe these alternative pipelines in somewhat more detail.
We note first two common aspects of all three alternative pipelines, which differ from our baseline results.
First, 
rather than accounting for the Galaxy and point-source mask in the filter function as is done for our baseline method,
these methods use a completely diagonal filter, which is the same as the form (given in Eq.~\ref{eqn:fl}) that
our baseline filtering asymptotes to far from the mask boundaries.
Instead, as we will discuss, the alternative pipelines deal with the mask using different choices of data preparation and selection.
A second difference common to all three pipelines with our baseline approach is that 
when computing the disconnected noise bias $\left. \Delta C_L^{\phi\phi} \right|_{\mathsc{N0}}$, these
methods do not use the two-point expression of Eq.~\eqref{eqn:cln0},
but rather the approximation to it based on Eq.~\eqref{eqn:nlpp_fullsky}, 
evaluated using an estimate of the data power spectrum.
Because this correction is quite large, the agreement of results that use this alternative method to calculate it is reassuring.
We now proceed to describe the individual methods in more detail; 
the common motivation in the development of each has been the reduction of the sharp gradients induced when masking, which can induce a mean-field several orders of magnitude larger than the lensing signal at low multipoles (as discussed in Appendix~\ref{sec:errorbudget:meanfields}).
Each method takes a different approach to mitigating this mask effect, as discussed below:
\begin{enumerate}[(1)]
\item 
The method \apod\ consists of applying the standard quadratic lensing estimator to the sky map after multiplying by an \textit{apodized} Galactic mask, 
and filling point source holes using local constrained Gaussian realizations of the CMB signal+noise.
The mask mean-field is proportional to the power spectrum of the mask, and so 
as apodization smooths the mask boundary (suppressing its power spectrum on small scales), it correspondingly reduces the mean-field significantly.
The combination of apodization and source filling makes this estimator very fast to apply to simulations but does require an involved set of correction terms and $f_{\rm sky}$ factors.
Our implementation and calculation of this method is described in detail in \cite{BenoitLevy:2013bc}.
\item 
The \metis\ method consists of \textit{inpainting} the Galactic mask as well as the point-source holes, 
using the sparse-inpainting algorithm described in \cite{inpainting:abrial06,Abrial:2008}.
The resulting map resembles a full-sky CMB map, and therefore has no mask mean-field contribution. The inhomogeneous noise and beam-induced mean-fields do still have to be corrected however. 
Our implementation is based on that described in \cite{Perotto:2009tv}, with several improvements.
In \cite{Perotto:2009tv} lens reconstruction was performed on the inpainted map and then analyzed on the full sky. 
However, further inspection has revealed that there are some spurious features in the lens reconstruction,
localized to the inpainted region inside the Galactic mask. 
This is likely due to the inhomogeneous noise in \Planck\ that was ignored in previous work and cannot be reproduced by the inpainter.
We therefore \textit{remask} the full-sky lens
reconstruction with an apodized Galactic mask (as in Eq.~\ref{eqn:apophi}) to remove these regions
from our analysis.
We follow the same procedure when evaluating the analytical expression for the $\left. \Delta C_L^{\phi\phi} \right|_{\mathsc{N0}}$ bias, prewhitening and then
applying an apodized Galactic mask to the inpainted temperature multipoles to estimate their power 
spectrum.
Small residual biases are corrected using the same $\left. \Delta C_L^{\phi\phi} \right|_{\mathsc{MC}}$ procedure used in the main method. 
\item 
The \patches\ method avoids the Galactic mask completely by cutting the sky into a collection of 410 small overlapping $10\deg \times 10\deg$ patches centered on the locations of \mbox{$N_{\rm side}\!=\!8$} \HealpixPixelization\ pixels, which are then analyzed under the flat-sky approximation.
Our implementation of this method is described in \cite{plaszczynski}. 
As with the \apod\ method, point source holes are filled using constrained Gaussian realizations.
The patches are extracted from a pre-whitened CMB map, and apodized with a Kaiser-Bessel window function. 
The Fourier modes in each patch are fitted in real space using a fast Fourier-Toeplitz algorithm. No mean-field correction is applied. Residual biases due to noise inhomogeneity are removed using a $\left. \Delta C_L^{\phi\phi} \right|_{\mathsc{MC}}$ correction that is found to be small.
The patches method has been particularly useful in the early stages of our analysis, to identify outliers caused by unmasked point sources.
\end{enumerate}
As can be seen in Fig.~\ref{fig:compalt}, all three of these methods are in good agreement with the results of our baseline method,
providing reassurance that our results are insensitive to the precise details of our data filtering and reconstruction methodology.

\section{Conclusions}
\label{sec:conclusions}

The \Planck\ maps have unprecedented sensitivity to gravitational lensing effects.
We see significant and consistent measurements of lensing for each of the high-resolution CMB channels at 100, 143 and 217\,GHz.
Even the noisiest channel that we have considered, 100\,GHz, provides a $10\sigma$ detection of lensing, which is greater than all previous detections.
Our fiducial lens reconstruction, based on a minimum-variance combination of the 143 and 217\,GHz channels does even better, with a detection of lensing (relative to the null hypothesis of no lensing) at a significance of greater than $25\sigma$. 
Notably, the noise on our reconstruction is low enough that it is no longer the limiting source of noise for many correlations with large-scale structure catalogs (several examples of which we have given in Sect.~\ref{subsec:xcorr}).
This marks a shift for CMB lensing, from the detection regime into that of standard cosmological probe.
Our lensing potential map is publicly available, and we look forward to the uses which may be found for it.

The percent-level \Planck\ lensing potential measurement pushes into the realm of precision cosmology, and requires careful validation tests that we have performed in Sect.~\ref{sec:consistency}.
Our fiducial likelihood, based on the $40 \le L \le 400$ range that is most sensitive to lensing, passes all of the tests that we have performed at an acceptable level.
Most importantly, it is consistent with individual frequency reconstructions, with more aggressive or more conservative masking, and with results obtained on more rigorously component-separated maps.
Our measurement of the lensing potential power spectrum $C_L^{\phi\phi}$ is broadly consistent with the $\Lambda$CDM expectations of \cite{planck2013-p11}, 
although there are some shape and amplitude tensions that lead to smaller-than-forecasted improvements on parameters such as the sum of neutrino masses $\sum m_{\nu}$.
As we saw in Sect.~\ref{sec:results:cosmo}, our estimate of the potential power spectrum has direct cosmological implications, where it significantly improves CMB  power-spectrum-only constraints on curvature (or alternatively, dark energy).
Our lensing likelihood also breaks the $A_s\!-\!\tau$ degeneracy, allowing a measurement of the optical depth to reionization which is a useful independent cross-check on the values determined with large angular-scale polarization measurements.

CMB lensing science has undergone a remarkable development in the previous two years \citep{Das:2011ak,Sherwin:2011gv,vanEngelen:2012va}.
The next year is expected to bring even greater improvements.
The $2500{\rm deg}^2$ lensing survey of the South Pole Telescope is expected have comparable statistical power
to the one reported here, with a highly complementary weighting in $L-$space that is more sensitive to smaller angular
scale lenses.
There is also additional \Planck\ data, which will improve on the measurement reported here considerably.
The results of this paper are based on the nominal-mission \Planck\ maps, collected during the first fifteen months of science operations.
The cryogens used to maintain the 100\,mK cooling stage for the \Planck\ HFI instrument lasted considerably longer than the nominal-mission lifetime, 
to a full-mission duration of approximately thirty months.
This additional data, set to be released in 2014, results in map noise levels that are roughly a factor two lower in power than those used here.
As instrumental noise constitutes approximately half of our error budget, we expect a corresponding decrease of approximately 25\% 
in the variance of our lensing potential map as well as the uncertainties on $C_L^{\phi\phi}$.
The \Planck\ polarization data, also set for release in 2014, will provide an additional powerful probe of lensing.

\begin{acknowledgements}
The development of Planck has been supported by: ESA; CNES and CNRS/INSU-IN2P3-INP (France); ASI, CNR, and INAF (Italy); NASA and DoE (USA); STFC and UKSA (UK); CSIC, MICINN, JA and RES (Spain); Tekes, AoF and CSC (Finland); DLR and MPG (Germany); CSA (Canada); DTU Space (Denmark); SER/SSO (Switzerland); RCN (Norway); SFI (Ireland); FCT/MCTES (Portugal); and PRACE (EU). A description of the Planck Collaboration and a list of its members, including the technical or scientific activities in which they have been involved, can be found at \url{http://www.sciops.esa.int/index.php?project=planck&page=Planck_Collaboration}.
Some of the results in this paper have been derived using the HEALPix package \citep{gorski2005}.
This research used resources of the National Energy Research Scientific Computing Center, which is supported by the Office of Science of the U.S. Department of Energy under Contract No. DE-AC02-05CH11231, as well as of the IN2P3 Computer Center (http://cc.in2p3.fr) and the Planck-HFI data processing center infrastructures hosted at the Institut d'Astrophysique de Paris (France) and financially supported by CNES. We acknowledge support from the Science and Technology Facilities Council [grant number ST/I000976/1].
\end{acknowledgements}

\appendix

\section{Null tests}
\label{app:nulltests}
In this appendix we present several generic null tests of our lens reconstruction.
In Appendix~\ref{sec:consistency:survey} we perform tests using lens reconstructions on the two individual six-month full-sky surveys contained in the first \Planck\ data release.
In Appendix.~\ref{sec:consistency:curl} we perform the curl-mode null test which is standard practice in Galaxy lensing.
Finally, in Appendix.~\ref{sec:consistency:statest} we will perform several tests on the isotropy and Gaussianity of our MV lensing potential map.


\subsection{Survey consistency}
\label{sec:consistency:survey}
\begin{figure}[!t]
\centerline{
\begin{overpic}[width=\columnwidth]{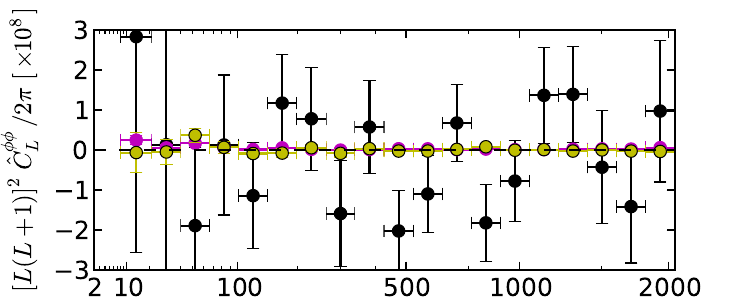}
\put(13,37){\tiny 143\,GHz}
\end{overpic}
}
\centerline{
\begin{overpic}[width=\columnwidth]{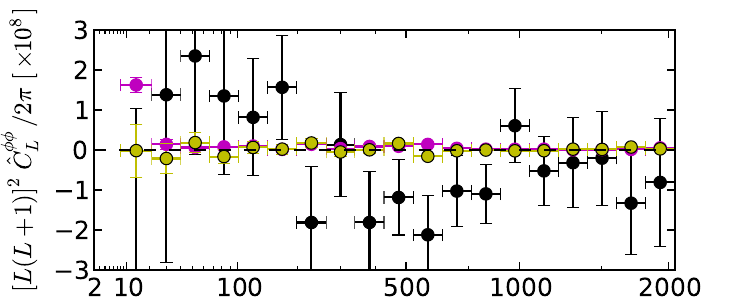}
\put(13,37){\tiny 217\,GHz}
\put(50,-4){$L$}
\end{overpic}
}
\vspace{0.15in}
\caption{
Survey consistency null tests at 143\,GHz 
 and 217\,GHz. 
Solid black points show difference of $\hat{C}_L^{\phi\phi}$ estimated separately on the two six-month surveys of the nominal mission.
Magenta points show the auto-spectrum of $\hat{\phi}$ obtained from maps made by taking the half-difference of the two surveys $({\rm S1} - {\rm S2}) \times ({\rm S1} - {\rm S2})$.
Yellow points show the cross-spectrum of $\hat{\phi}$ obtained on half-difference and half-sum maps $({\rm S1} - {\rm S2}) \times ({\rm S1} + {\rm S2})$.
The scatter on the tests that contain difference maps is very small because (as can be seen in Fig.~\ref{fig:nlpp_hfi_and_cv_contributions}) most of the lens reconstruction ``noise'' is due to primary CMB fluctuations, which is nulled in these tests.
\label{fig:analysis_lens_survey_diff}
}
\end{figure}

The \Planck\ nominal mission consists of approximately two full-sky six-month surveys.
Our fiducial analysis is based on maps made from the full data range, however we may also analyze the first two surveys separately to increase our sensitivity to effects such as 
beam asymmetry, 
correlated noise along the scan direction,
and
zodiacal light.
In Fig.~\ref{fig:analysis_lens_survey_diff} we plot several consistency tests exploiting the breakdown of the data into separate surveys, 
for both 143\,GHz and 217\,GHz.
We test for gross contamination of either survey by estimating the lensing power spectrum on each individually and then difference the resulting $\hat{C}_L^{\phi\phi}$, finding results consistent with the scatter expected from simple signal+noise simulations.
To test the noise model more thoroughly, we also run lens reconstruction on a map constructed by taking the half-difference of the two surveys, filtering it in the same way as the full nominal-mission map.
As can be seen in Fig.~\ref{fig:analysis_lens_survey_diff}, the power spectrum of this reconstruction null test is small compared to the measurement errors on $\hat{C}_L^{\phi\phi}$.
Finally, to isolate the effect of systematics which couple to the CMB and which flip sign between surveys (such as odd moments of the beam asymmetry), we take the cross-spectrum of lens reconstruction estimates obtained from the half-difference of the two survey maps with those obtained from the half-sum.
Again, this null test shows residuals that are negligibly small compared to our measurement errors.
\begin{figure}[!ht]

\vspace{0.2in}
\centerline{\begin{overpic}[width=\columnwidth]{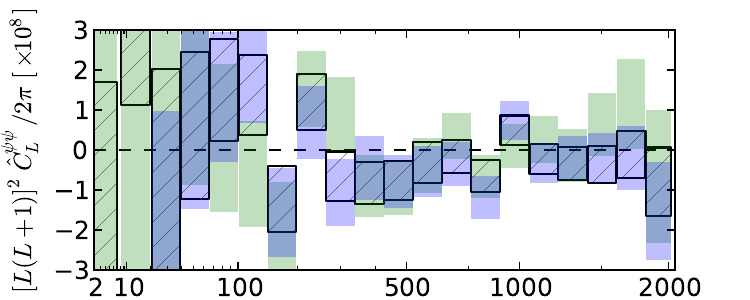}
\put(50,-4){$L$}
\end{overpic}}
\vspace{0.1in}
\caption{
Lensing curl power spectrum estimates for the MV reconstruction (black hatched), as well as the individual 143 (green filled) and 217\,GHz (blue filled) frequency reconstructions.
}
\label{fig:analysis_lens_clxx_pub}
\end{figure}


 \subsection{Curl-mode reconstruction}
\label{sec:consistency:curl}
In addition to the standard gradient lensing potential $\phi$ sourced by large-scale structure, we may also use the \Planck\ maps to reconstruct a curl-mode lensing field $\psi$.
The weight function associated with the curl field is given by \citep{Namikawa:2011cs}
\begin{multline}
W_{\elt_1 \elt_2 L}^{\psi} 
= -  \sqrt{\frac{(2\elt_1+1)(2\elt_2+1)(2L+1)}{4\pi}} \sqrt{L(L+1)  \elt_1 (\elt_1+1) } \\ \times 
C_{\elt_1}^{TT}
  \left( \frac{1 - (-1)^{\elt_1 + \elt_2 + L}}{2} \right) \threej{\elt_1}{\elt_2}{L}{1}{0}{-1} + (\elt_1 \leftrightarrow \elt_2).
\label{eqn:qe_weight_lensing_curl}
\end{multline}
Using this weight function, we may construct an estimator for the curl-mode power spectrum, completely analogous to that for $\phi$.
The response function of the curl estimator to point source shot noise is zero, and so we neglect this correction.
Consistency of the reconstructed curl mode with zero is a standard systematic test in weak lensing, as the curl field due to gravitational lensing should be negligible \citep{Hirata:2003ka}.
One caveat for this test in the CMB lensing context is that the variance of the curl mode reconstruction is still affected by the presence of gravitational lenses, by means of an ``$N^{(1)}$''-type \citep{Kesden:2003cc} contraction in the trispectrum \citep{vanEngelen:2012va}.
Our curl-mode power spectrum estimates are plotted in Fig.~\ref{fig:analysis_lens_clxx_pub}, after correction for the ``$N^{(1)}$''-type bias using Eq.~\eqref{eqn:n1}.
This bias is also given explicitly for the curl mode in \cite{BenoitLevy:2013bc}.
We do not see evidence for anomalous power in the curl reconstruction, although we do find in Sect.~\ref{sec:pixelization} that the curl estimate is contaminated by pixelization effects at a level too small to be seen in this plot.


\subsection{Statistical tests at the map level}
\label{sec:consistency:statest}
In this section we present several sanity tests of our lensing estimates at the map level.
\begin{figure}[!htpb]
\centering
\begin{overpic}[width=0.9\columnwidth, trim=80 0 60 0, clip]{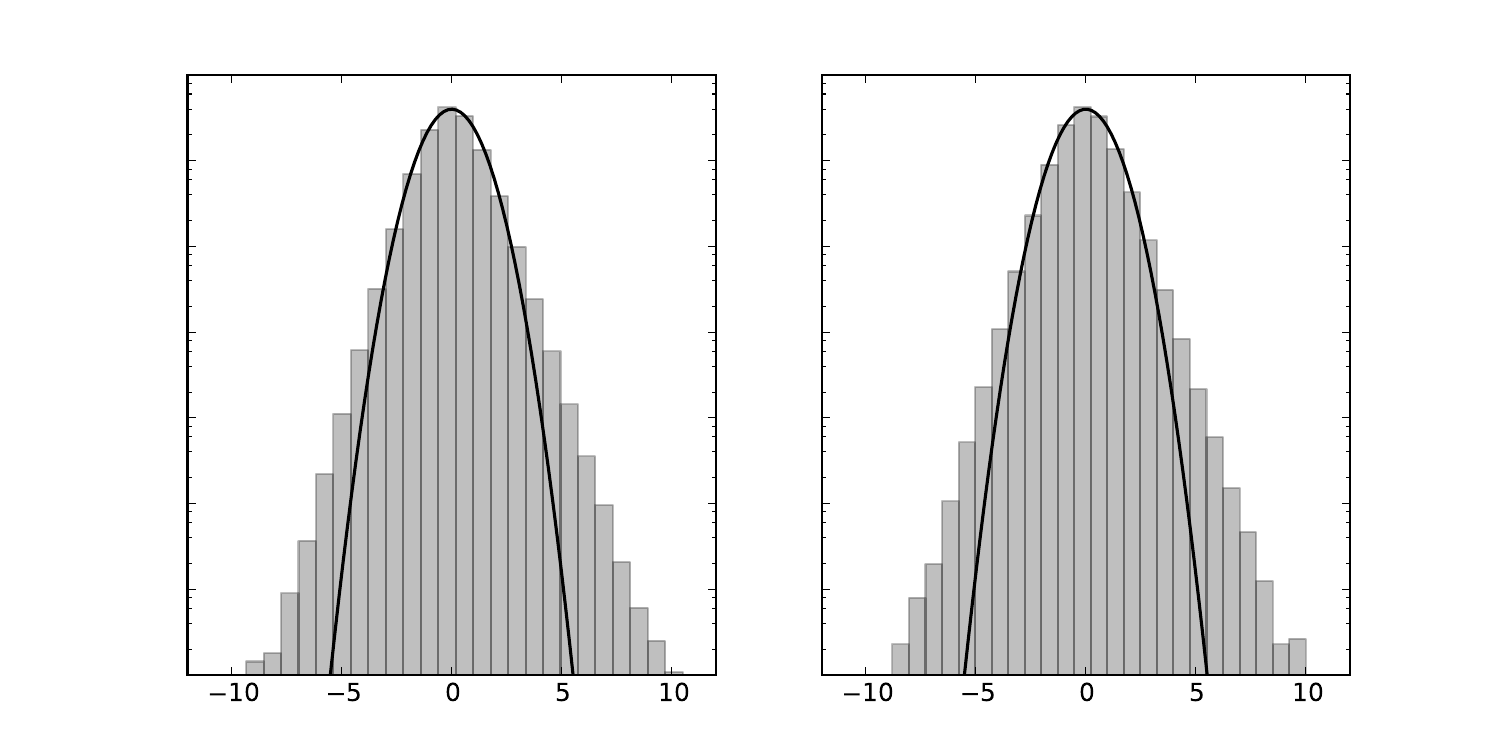}
\put(5,51){\tiny MV Data}
\put(57,51){\tiny MV Sim}
\put(22,0){$\sigma$}
\put(76,0){$\sigma$}
\end{overpic}
\caption{\label{fig:gaussmap}
Histogram of pixel values for the unnormalized lensing estimate $\bar{\phi}(\hatn)$ for the MV reconstruction, in units of the standard deviation of the map.
A Gaussian with $\sigma = 1$ is overplotted in black.
Note that the $y$-axis is logarithmic.
The left column shows the histogram for the data itself, while the right column shows a simulated reconstruction.
The reconstructed map is slightly non-Gaussian, but in a way that is expected from simulations given the non-Gaussian nature of the quadratic lensing estimator.
}
\vspace{0.2in}

{\includegraphics[width=0.8\columnwidth]{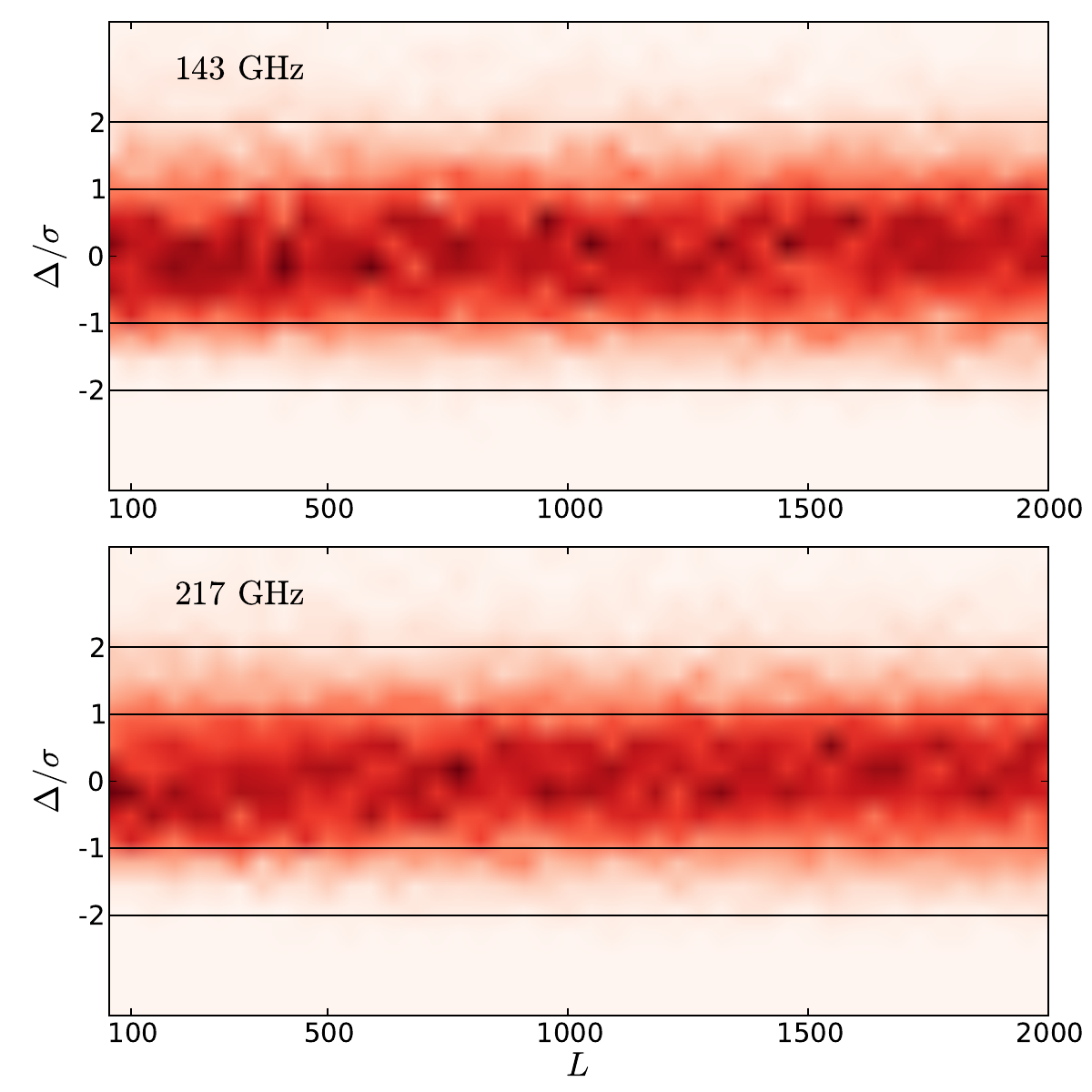}}
  \caption{\label{fig:patches_multispectra} Density plot of the individual deflection power spectra
    constructed on approximately 400 $10\deg\times10\deg$ patches for the 143\,GHz
    and 217\,GHz channels. The lines are the theoretical $1\sigma$ and $2\sigma$ error for this patch size.
    No strong outliers are seen.
    }
     \vspace{0.1in}

\begin{overpic}[width=\columnwidth, trim=20 2 20 20, clip]{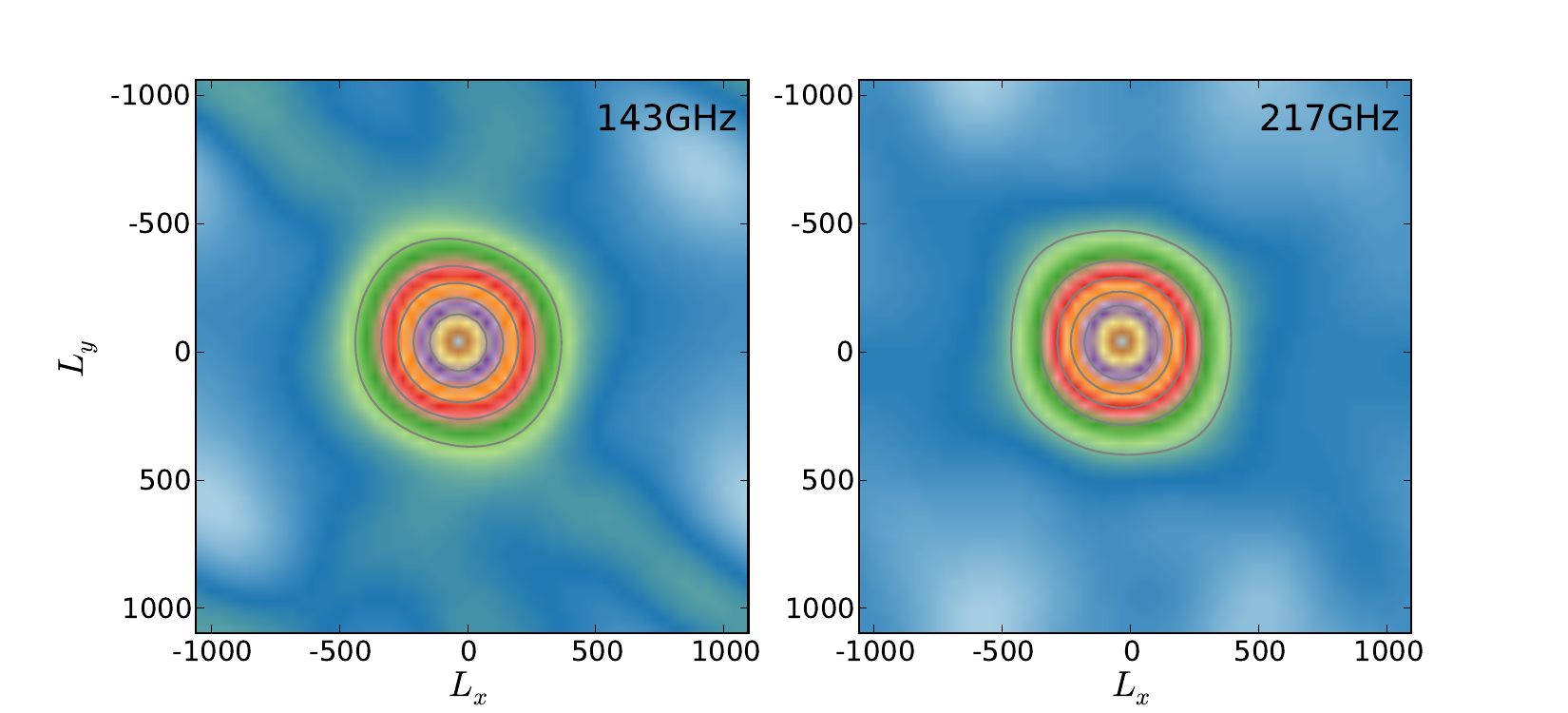} \end{overpic}
\vspace{-0.1in}
  \caption{\label{fig:patches_2Dspectra} 
    Bi-dimensional deflection power spectrum obtained by stacking the
    individual Fourier amplitude obtained on patches, after $\left. \Delta C_L^{\phi\phi} \right|_{\mathsc{n0}}$ subtraction and
    smoothing for the 143 and 217\,GHz channels.
    }

\end{figure}

\paragraph{1-point PDF of $\bar{\phi}$:}
In the left panel of Fig.~\ref{fig:gaussmap} we plot the histogram, or 1-point function, of our unnormalized potential estimate $\bar{\phi}$ for the MV reconstruction, bandlimited to 
$10 \le L \le 2048$.
As can be seen from the histogram, the map is slightly non-Gaussian, as might be expected for an estimator that is based on the product of two Gaussian random fields.
In the right panel of Fig.~\ref{fig:gaussmap} we plot the same histogram for a simulated MV reconstruction, which shows a comparable level of non-Gaussian structure.
In Appendix~\ref{app:likelihood} we show that this small non-Gaussianity is negligible for our lensing likelihood.

\paragraph{Power Spectra on Patches:}
To test for gross contamination of our reconstruction from any localized region of the sky, 
we use the \patches\  method (described in Sect~\ref{sec:consistency:filtering}) to build an estimate of the lensing power spectra on a collection of
approximately 400 $10\deg\times10\deg$ tiles.
We fill point-source holes with local constrained Gaussian realizations of the CMB signal.
The patches were chosen to have no overlap with the Galaxy mask.

The distribution of the power spectra estimated for each patch can be used as a diagnostic for deficiency in the mask and dust correction.
Significantly contaminated but localized regions would appear as outliers.
This procedure successfully detected several point sources that were overlooked in early versions of our source masks.
In Fig.~\ref{fig:patches_multispectra} we present results of this analysis using our final Galaxy and point source masks for 143\,GHz and 217\,GHz.

\paragraph{Isotropy on Patches:}
We can also use the same data to build a coarse test of the isotropy of the reconstructed lensing map.
We perform lens reconstruction on patches which are oriented along the local spherical basis ($\vec{e_\theta},\vec{e_\phi}$), 
and then stack the corresponding 2D power spectrum estimates to build a 
\emph{bi-dimensional} power spectrum for the 143\,GHz and 217\,GHz maps.
This is shown in Fig.~\ref{fig:patches_2Dspectra}, where we have additionally applied smoothing based on an undecimated wavelet transform to reduce the visual noise.
The overplotted contour lines are placed at levels given by the azimuthally averaged power spectrum for $L=100$, $200$, $300$ and $400$.

\section{Filtering}
\label{app:ivf}

In this section, we discuss our filtering methodology, which takes us from a set of
maps $T^{\rm obs, \nu}_p$ indexed by frequency $\nu$ and pixel $p$ to the multipoles $\bar{T}_{{\elt}m}$
that are fed into the quadratic estimators used to measure lensing and possible contaminants.
For an observed CMB that consists of Gaussian signal and noise, 
the optimal filtering (in the minimum-variance sense) for the construction of any quadratic anisotropy estimator is ``$C^{-1}$'' filtering, 
in which the observed map is beam-deconvolved and then multiplied by the inverse of a signal+noise covariance matrix.
For the purposes of lens reconstruction, the Gaussian assumption above is quite accurate for \Planck.
The \Planck\ map noise is well approximated as Gaussian \citep{planck2013-p03}.
The observed CMB signal is less-obviously well approximated as Gaussian, with the non-Gaussianity due to lensing detected at high significance, 
however for lens reconstruction at \Planck\ noise levels the difference in signal-to-noise using a
filter constructed in the Gaussian limit versus a more complicated ``delensing'' filter that properly accounts 
for lensing non-Gaussianity has been shown to be negligible \citep{Hirata:2002jy}.
Other sources of non-Gaussianity, such as unresolved point sources, are completely subdominant to either primary CMB 
or noise at all of the multipoles probed by our lens reconstruction, and so we neglect them in the construction of our filters.

For a single map, with a symmetric beam transfer function, full-sky coverage and homogeneous map noise levels, the $C^{-1}$ filter is given by
\begin{align}
\bar{T}_{{\elt}m} = \frac{F_l}{B_l} \sum_{p=1}^{n_{\rm pix}} \frac{4\pi}{n_{\rm pix}} {Y}_{{\elt}m}^{*}(\hatn_p) T^{\rm obs}_p,
\label{eqn:filtfullsky}
\end{align}
where $Y_{{\elt}m}(\hatn_p)$ gives the value of the spherical harmonic at the center of pixel $p$, $B_l$ gives the beam transfer function, and $F_l$ is given by
\be
F_l = \frac{1}{C_{\elt}^{TT} + \left(  \frac{\pi}{180 \times 60}  \sigma B_{\elt}^{-1} \right)^2 },
\label{eqn:flnlev}
\ee
where $\sigma$ is the map noise level in units of $\muKarcmin$.
The maps we use must always be masked to remove bright Galactic foregrounds and point sources, and so we do not ever use the full-sky filter above, although as we will discuss the filter that we do use asymptotes to this form in regions far from the mask, and the simple form $F_l$ of the filter function given in Eq.~\eqref{eqn:flnlev} will prove useful in many analytical calculations.

For the more realistic case of inhomogeneous map noise levels, with pixel-space noise correlations and a sky cut, 
the construction of $\bar{T}_{{\elt}m}$ is more involved.
To account for the sky cut, we can take the noise covariance to be infinite for masked pixels.
Reformulating the $C^{-1}$ filter using the inverse of the noise covariance matrix $N^{-1}_{p p'}$ in pixel space
(defined to be zero if either $p$ or $p'$ is a masked pixel),
and allowing for the joint-filtering of multiple input maps (with no noise correlations between them) we have
\be
\bar{T}_{{\elt}m} = \left( C_{\elt}^{TT} \right)^{-1}
\sum_{{\elt}' m'} \sum_{p p'} \sum_{\nu} 
{\cal C}_{{\elt}m, {\elt}'m'}^{-1}
{\cal Y}_{{\elt}'m'}^{* p, \nu} N_{p p'}^{-1, \nu} T^{\rm obs, \nu}_{p'}.
\label{eqn:cinvfilt}
\ee
where $\nu$ indexes the maps which are being combined and the pointing matrix 
${\cal Y}_{{\elt}m}^{p, \nu} \equiv B^{\nu}_{{\elt}} Y_{{\elt}m} (\hatn_p) 4\pi/n_{\rm pix}$ gives the value of the spherical harmonic at the center of pixel $p$, convolved with the appropriate beam+pixel transfer function $B_{{\elt}}^{\nu}$.
The matrix ${\cal C}_{{\elt}m, {\elt}'m'}$ is given by
\be
{\cal C}_{{\elt}m, {\elt}'m'} \equiv
\left( C_{\elt}^{TT} \right)^{-1} \delta_{{\elt} {\elt}'} \delta_{m m'} \\ +
\sum_{p, p'} \sum_{\nu}
 {\cal Y}_{{\elt}m}^{* p, \nu} N_{p p'}^{-1, \nu} {\cal Y}_{{\elt}'m'}^{p', \nu}.
\label{eqn:cinvmat}
\ee
The filter of Eq.~\eqref{eqn:cinvfilt} is the one that we will use, 
with a slightly suboptimal choice for the noise covariance matrix $N_{p p'}$ that is described below.

The \Planck\ map noise has several sources of pixel-pixel correlations for $p \ne p'$.
On small scales, there are correlations along the scan direction induced by the deconvolution of the bolometer time response \citep{planck2013-p03c}.
On large scales, there are correlations between scan rings, due to residual $1/f$ noise fluctuations not cancelled by the baseline subtraction procedure \citep{planck2013-p03}.
Accounting for these pixel correlations would be computationally intractable, as the full covariance matrix $N_{p p'}$ (an $n_{\rm pix}^4$ sized object) is too large to work with on a modern computer, and so we neglect them in our noise filter.
The remaining structure in the noise matrix is due to the variable coverage in different regions of the sky.
Noise levels are lowest near the Ecliptic poles, which are visited more frequently than the region around the Ecliptic plane due to the \Planck\ scan strategy.
Although the variable noise level is tractable to include in our filtering procedure (and is indeed included in the $C^{-1}$ filtering of e.g., \cite{planck2013-p09a}), we have chosen to neglect this aspect of the noise, using instead a fixed effective noise level. 
We use an inverse-noise map in pixel space given by 
\begin{equation}
N_{p}^{-1, \nu} = \left( \frac{180 \times 60 \times M_p}{\pi \times \sigma_{\nu}} \right)^2 \left( \frac{4\pi}{n_{\rm pix}} \right),
\label{eqn:ncov}
\end{equation}
where $M_p$ is a sky mask and $\sigma_{\nu}$ is an average map noise level in units of $\muKarcmin$.
In addition to this noise weighting, we also project out the five modes corresponding to the monopole, dipole, and 857GHz \Planck\ map (as a simple dust template).
The noise level for each of these modes is taken to infinity using the Woodbury formula; putting each of these modes into a map $T^{i}_p$ indexed by $i$,
our total noise covariance matrix is given in terms of the inverse noise map above by
\be
N_{pp'}^{-1, \nu} = \delta_{p p'} N_{p}^{-1, \nu}  - \sum_{ij} {\cal T}_{ij}^{-1} T_{p}^{i} T^{j}_{p'}  N_{p}^{-1, \nu}  N_{p'}^{-1, \nu},
\ee
where ${\cal T}_{ij}$ is the overlap matrix between templates given by
\be
{\cal T}_{ij} = \sum_{p}  T_{p}^{i} T^{j}_{p} N_{p}^{-1, \nu}.
\ee
Our motivation for taking a fixed noise level is that with this approach, in regions sufficiently far from the mask boundary, our filter asymptotes to the diagonal form of Eq.~\eqref{eqn:filtfullsky}. 
This means that the normalization of our lensing estimates can be well-approximated analytically, which is very useful for the propagation of systematic effects, and also that the normalization of our lensing estimates does not vary across the sky with noise level, which simplifies cross-correlation analysis.
Our $C^{-1}$ filter is therefore optimally accounting for masking effects, but not for noise correlations and inhomogeneity.
We estimate the suboptimality of neglecting these noise properties by calculating the quantity
\be
\frac{ (S/N)^{\mathsc{use}}}{  (S/N)^{\mathsc{opt}}  } =
 \frac{ \left(R_L^{\phi\phi, {\mathsc{use}}} \right)^2}{ \left( R_L^{\phi\phi, {\mathsc{opt}}} \right)^{\phantom{2}} } 
\left(
\sum_{\elt_1 \elt_2} 
\frac{1}{2} \left| W_{\elt_1 \elt_2 L}^{\phi} \right|^2
\frac{ \left( F_{\elt_1}^{ {\mathsc{use}} } F_{\elt_2}^{ {\mathsc{use}} } \right)^2 }{ F_{\elt_1}^{ {\mathsc{opt}} } F_{\elt_2}^{ {\mathsc{opt}} } }
\right)^{-1},
\label{eqn:snuseopt}
\ee
where $F_{{\elt}}^{ {\mathsc{opt}} }$ is the optimal filter and $F_{{\elt}}^{ {\mathsc{use}} }$ is the suboptimal filter that we have actually used.
This equation gives the $S/N$ loss as a function of lens multipole $L$, however in practice we find that the $L$ dependence is small enough that it suffices to quote a single average loss.
To estimate the degradation due to ignoring noise correlations we set
\be
F_{{\elt}}^{ {\mathsc{opt}} } = \frac{1}{C_{\elt}^{TT} + B_{\elt}^{-2, \nu} N_L^{TT}},
\ee
where $N_L^{TT}$ is the power spectrum of the map noise.
We find that the degradation due to neglect of noise correlations is small;
less than $2\%$ for all $L \le 2048$ at 100\,GHz, and 
less than $0.1\%$ at 143 and 217\,GHz.
To calculate the degradation due to ignoring noise inhomogeneity, 
we determine the map noise level in the 3072 regions corresponding to $N_{\rm side}=16$ \HealpixPixelization\ pixels, take $F_{\elt}^{\mathsc{opt}}$ using Eq.~\eqref{eqn:flnlev} with the local noise level, and estimate a resulting $S/N$ degradation using Eq.~\eqref{eqn:snuseopt}.
The neglect of noise inhomogeneity is the dominant suboptimality of our filtering, although it is still small.
We find an average $S/N$ loss (averaged over the entire sky) of approximately $4\%$ at 100, 143, and 217\,GHz, consistent with the simulation-based results of \cite{Hanson:2009dr}.
We take this loss as justified, given the simpler normalization properties of our lensing estimates when neglecting variations in the map noise level.

\section{Mean-fields}
\label{sec:errorbudget:meanfields}

\begin{figure}[!htbp]
\begin{overpic}[width=0.95\columnwidth]{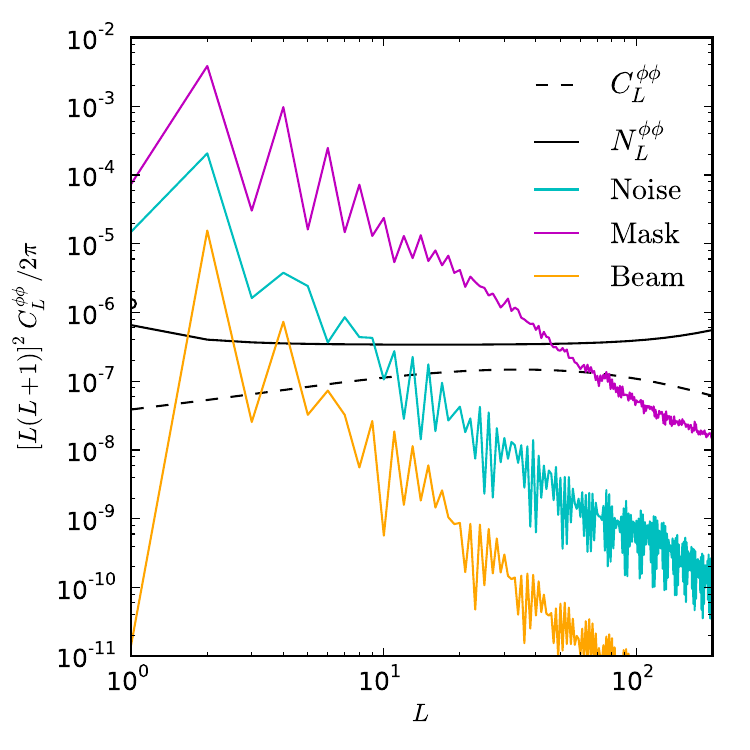} \put(22,18){ 143\,GHz } \end{overpic}
\caption{Analytical estimates for the power spectra of the largest low-$L$ mean-fields at 143\,GHz.
The various components are discussed in more detail in
Sect.~\ref{sec:mf:mask} (\textcolor{Purple}{mask}), 
Sect.~\ref{sec:mf:noise} (\textcolor{cyan}{noise}), and 
Sect.~\ref{sec:mf:beam} (\textcolor{orange}{beams}).
The mean-fields all couple most strongly to even modes of the lens reconstruction, due to the approximate north/south symmetry of the scan strategy and Galactic mask.
\label{fig:clpp_mf_beams_mask_noise}
}
\end{figure}

As discussed in Sect.~\ref{sec:methodology}, the quadratic lensing estimators that we use are designed to detect  \textit{statistical anisotropy} induced by lensing.
There are a number of non-lensing sources of statistical anisotropy that can mimic the lensing signal to some extent.
In our analysis, the effects we consider are
\begin{enumerate}[(1)]
\item The application of a sky mask, which introduces sharp gradients that may be misinterpreted as lensing.
\item Noise inhomogeneity, which causes the overall power to fluctuate across the sky and can resemble the convergence component of lensing.
\item Beam asymmetry, which smears the fluctuations more along one direction than another and can mimic the shear component of lensing.
\item Pixelization, in which detector samples are accumulated into pixels, introduces a spurious deflection field on the pixel scale because the centroid of the hit distribution in each pixel does not necessarily lie at the pixel center.
\end{enumerate}
In our analysis, we account for most of these effects with a corrective mean-field term, given by Eq.~\eqref{eqn:lensing_meanfield}, which is determined using Monte Carlo simulations.
In this appendix, we will break this mean-field down into its constituent parts and discuss each in more detail.
As an overview of the results in this section, in Fig.~\ref{fig:clpp_mf_beams_mask_noise} we plot estimates for the three largest mean-fields, due to masking, noise inhomogeneity, and beam asymmetry at 143\,GHz (100 and 217\,GHz are qualitatively similar).
These mean-fields all have most of their contributions on very large scales, dictated by the coherency of the scan strategy in the case of beam asymmetry and noise inhomogeneity, and of the large-scale nature of the Galactic foregrounds in the case of the sky mask.
We note also that the mask mean-field is concentrated near the edges of the Galaxy cut, and so for our power spectrum analysis (where we apodize the estimated $\phi$ map, removing the contributions near the edges of the mask), the effective mean-field is significantly reduced.

Our discussion will focus on constructing simple models for each source of mean-field.
Following \cite{Hanson:2010gu}, we will identify each of the individual contributions to the mean-field with a tracer $z_{LM}$ that sources a contribution to the CMB covariance matrix given by
\be
\Delta \langle T_{\elt_1 m_2} T^*_{\elt_2 m_2} \rangle =
\sum_{LM} z_{LM} 
(-1)^M
\threej{\elt_1}{\elt_2}{L}{m_1}{m_2}{-M}
W^{z}_{\elt_1 \elt_2 L},
\label{eqn:tcov_aniso2}
\ee
where $W^{z}_{\elt_1 \elt_2 L}$ is a weight function describing how $z_{LM}$ couples multipoles.
Such a contaminant leads to a bias for the standard lensing estimator $\hat{\phi}_{LM}$ given by
\be
\hat{\phi}^{\MF}_{LM} = \frac{ \resp^{\phi z}_{LM}}{ \resp^{ \phi \phi}_L } z_{LM},
\label{eqn:mf_analytical2}
\ee
where the response function $\resp^{\phi z}_L$ is defined in Eq.~\eqref{eqn:respxp}. 
The analytical forms for the mean-fields that we present here are used in Sect.~\ref{sec:biashardened} to construct ``bias hardened'' estimators that have smaller mean-field corrections, as proposed by \cite{Namikawa:2012pe}.

\subsection{Mask}
\label{sec:mf:mask}
The application of a mask $M(\hatn)$ to the beam-convolved sky, followed by subsequent beam deconvolution, can be modelled as a statistical anisotropy ``$M$'', 
with  $M_{LM} = \int d\hatn Y_{LM}^*(\hatn) M(\hatn)$ and weight function \citep{Namikawa:2012pe}
\begin{multline}
W_{\elt_1 \elt_2 L}^{M} = 
\sqrt{ \frac{ (2\elt_1 +1) (2 \elt_2 +1) (2L+1) }{4\pi} }
\threej{\elt_1}{\elt_2}{L}{0}{0}{0} \\ \times
\left(
C_{\elt_1}^{TT} \frac{ B_{\elt_1}^{(2)} }{ B_{\elt_2}^{(2)} } +
C_{\elt_2}^{TT} \frac{ B_{\elt_2}^{(1)} }{ B_{\elt_1}^{(1)} }
\right).
\end{multline}
The mask mean-field using our fiducial Galaxy and point-source analysis mask (described in Sect.~\ref{sec:data}) is plotted in Fig.~\ref{fig:clpp_mf_beams_mask_noise}.
At very low-$L$, the mean field induced by this mask can be much larger than the reconstruction noise level. At $L=2$ for example, it can be up to $10^4$ times greater than the reconstruction noise level, although the amplitude of the mean-field is a very steep function of $L$.

The mask mean-field is proportional to $C_{\ell}^{TT}$, and so with percent-level agreement between the power spectrum of our simulations we use to determine the mean-field and the data power spectrum  (discussed in Sect.~\ref{sec:simulations}), we can expect errors on the power spectra of the mask mean-field at the level of one part in $10^4$.
At $L=2$ for 143 and 217\,GHz this corresponds to a level of $[L(L+1)]^2 C_L^{\phi\phi} \approx 4\times 10^{-7}$, comparable to the lensing potential power spectrum.
There are several good ways to mitigate the effect of the mask mean-field.
By apodizing the mask before lens reconstruction, the gradients at the mask boundary can be greatly reduced, which significantly suppresses the mean-field \citep{vanEngelen:2012va,BenoitLevy:2013bc}.
At the map level, the only issue with such an apodization for our purposes is that it results in a position-dependent normalization which can complicate cross-correlation analyses.
A related approach, which we have taken in our power spectrum analysis, is to apodize the lens reconstruction \textit{after} performing the reconstruction.
This amounts to removing the regions of $\hat{\phi}$ near the mask boundary where the mean-field has all of its features.
This ``avoidance'' approach can also be used easily in cross-correlation analyses, as in \cite{Hirata:2004rp}.
A third approach is that of \cite{Namikawa:2012pe}, who show that it is possible to construct ``bias-hardened'' lensing estimators which have smaller mean-field contributions.
We have already shown consistency tests based on this approach in Sect.~\ref{sec:biashardened}.
In the low-$L$ bias-hardened comparisons of Fig.~\ref{fig:bh_lowl}, we saw discrepancies with the standard estimator of $[L(L+1)]^2 C_L^{\phi\phi} \approx 4 \times 10^{-7}$ and $7 \times 10^{-7}$, precisely the level expected above.

For our publicly released lensing map, we have chosen simply to high-pass filter our map to $L>10$, above which we believe that mean-field errors are subdominant to the reconstruction noise.

\subsection{Instrumental noise}
\label{sec:mf:noise}
The instrumental noise of \Planck\ is rich in statistically-anisotropic structure, as discussed in \cite{planck2013-p03}, which is sourced by the following.
\begin{enumerate}[(1)]
\item Striping effects, due to residual noise correlations on long time scales after destriping.
\item Correlations along the scan direction induced by the deconvolution of the detector time response and the low-pass rolloff filter.
\item Large variations in the pixel noise level due to the larger density of pixel hits near the Ecliptic poles.
\end{enumerate}
All of these effects can be avoided by forming estimators using pairs of data with independent noise realizations, as we have done in Sect.~\ref{sec:consistency:noisebias}, although our baseline results do not use such cross-correlations.
Here we discuss the size of the noise mean-field.
Items (1) and (2) are difficult to study analytically.
For white noise with a variance in each pixel given by $N(\hatn_p)$, 
item (3) can be identified with a statistical anisotropy ``$N$'' defined by \cite{Hanson:2009dr}
\begin{align}
N_{LM} &= \int d\hatn_p Y^*_{LM}(\hatn_p) N(\hatn_p) \nonumber \\
W_{\elt_1 \elt_2 L}^{N} &= 
 \sqrt{\frac{(2\elt_1+1)(2\elt_2+1)(2L+1)}{4\pi}}
\threej{\elt_1}{\elt_2}{L}{0}{0}{0} \frac{1}{B_{\elt_1}} \frac{1}{B_{\elt_2}},
\end{align}
Note that the shape of this bias is dictated by the hit count distribution, which is known quite well, rather than the noise level, which is more uncertain.
The shape of the noise mean-field is plotted in Fig.~\ref{fig:clpp_mf_beams_mask_noise}. 
As with the mask, it is large compared to our signal and noise but falls off quickly with multipole.
For $L<10$, given that there are known issues with our noise modelling at the $5\%$ level in power (as discussed in Sect.~\ref{sec:simulations}), it can be important to construct quadratic estimators from correlations of maps with independent noise realizations (apart from our bias-hardened estimator tests in Sect.~\ref{sec:biashardened}, this is also done for $L=1$ in \cite{planck2013-pipaberration}), however at the $L>10$ multipoles of our lensing map and power spectrum, uncertainties in the noise mean-field should be negligible.
\subsection{Beam asymmetry}
\label{sec:mf:beam}
\Planck\ observes the sky after convolution with a ``scanning beam'', which captures its effective response to the sky as a function of displacement from the nominal pointing direction.
Asymmetry in the shape of this beam (for example ellipticity) means that the fluctuations in the map are smeared depending on the beam orientation, which can mimic the shearing effects of gravitational lensing.
Here we present semi-analytical estimates of this beam asymmetry mean-field, and discuss how simulations of beam asymmetry are included in our simulations to correct for it.
As with the noise inhomogeneity, we will see that the large-scale coherency of the \Planck\ scan strategy means that beam asymmetries only mimic large-scale modes of $\phi_{LM}$.

Decomposing the scanning beam into harmonic coefficients $B_{{\elt}m}$, each time-ordered data (TOD) sample can be modelled as (neglecting the contribution from instrumental noise, which is independent of beam asymmetry)
\be
T_i = \sum_{{\elt}ms} e^{-i s \alpha_i} B_{{\elt}s} \tilde{T}_{{\elt}m} {}_s Y_{{\elt}m}(\theta_i, \phi_i),
\label{eqn:tod_beam}
\ee
where the TOD samples are indexed by $i$, and $\tilde{T}_{{\elt}m}$ is the underlying sky signal. 
The spin spherical harmonic ${}_s Y_{{\elt}m}$ rotates the scanning beam to the pointing location $(\theta, \phi)$, while the $e^{-i s \alpha_i}$ factor gives it the correct orientation.
On the small scales that lensing is sensitive to, it is a good approximation to assume that the procedure of mapmaking from TOD samples is just a process of binning:
\be
T(\hatn_p) = \sum_{i \in p} T_i / H_p,
\label{eqn:map_beam_full}
\ee
where $\hatn_p$ is the centroid of pixel $p$, and the sum is taken over all hits assigned to pixel $p$. 
$H_p$ is the number of hits in pixel $p$.
Here we will assume that the pointing for each observation of pixel $p$ is identical to the pixel center $\hatn_p$. 
The purpose of this assumption is simply to isolate the effect of beam asymmetry from the (small) effect of pixelization.
We will discuss the pixelization effect in Sect.~\ref{sec:pixelization}.
Placing all hits at pixel centers, combining Eq.~\eqref{eqn:tod_beam} and Eq.~\eqref{eqn:map_beam_full} we have
\be
T(\hatn) = \sum_{s} w(\hatn, -s) \left[ \sum_{{\elt}m} B_{{\elt}s} \tilde{T}_{{\elt}m} {}_s Y_{{\elt}m} (\hatn) \right],
\label{eqn:map_beam_centered}
\ee
where the ``scan strategy maps'' $w(\hatn, s)$ are given by
\be
w(\hatn_p, -s) = \sum_{i \in p} e^{-i s \alpha_i} / H_p.
\ee
The scan strategy maps are spin-$s$ objects, with corresponding harmonic decomposition
\be
{}_s w_{LM} = \int d^2 \hatn {}_s Y_{LM}^*(\hatn) w(\hatn, s).
\ee
We use Eq.~\eqref{eqn:map_beam_centered} to include the effect of asymmetric beams in the simulations that we use to debias our lensing estimates.
We can also propagate the effect of beam asymmetry to the lensing mean-field directly. 
Each mode $s$ of the scan strategy can be identified with a statistical anisotropy ``$b^{s}$'' of the type defined in Eq.~\eqref{eqn:tcov_aniso2}, with $b^{s}_{LM} = {}_s w_{LM}$ and
\begin{multline}
W^{b^s}_{\elt_1 \elt_2 L} = 
\sqrt{\frac{(2\elt_1+1)(2\elt_2+1)(2L+1)}{4\pi}} 
\\ \times
\Bigg[
\threej{\elt_1}{\elt_2}{L}{s}{0}{-s} 
\frac{ B_{\elt_1}^{(1)} B_{\elt_1s}^{(2) *} }{ B_{\elt_1}^{(1)} B_{\elt_2}^{(2)} } C_{\elt_1}^{TT} 
+
\threej{\elt_1}{\elt_2}{L}{0}{s}{-s} 
\frac{ B_{\elt_2s}^{(1) *} B_{\elt_2}^{(2)} }{ B_{\elt_1}^{(1)} B_{\elt_2}^{(2)} } C_{\elt_2}^{TT} 
 \Bigg].
\end{multline}
In Fig.~\ref{fig:clpp_mf_beams_mask_noise} we have plotted the expected full-sky bias for the best-fit \Planck\ beams, which are measured from planet observations in \cite{planck2013-p03c}.
There are considerable differences in the mean-field between frequencies, as well as the steep scale dependence, which is dictated by the large-scale coherency of the scan strategy.


\subsection{Pixelization} 
\label{sec:pixelization}
\label{sec:consistency:pixelization}
In the process of mapmaking, TOD samples are accumulated into bins defined by \HealpixPixelization\ pixels.
The \Planck\ HFI maps are generated at \HealpixPixelization\ resolution 11, with a typical size of 
$1\parcm7$, which is sufficiently close to the $2\parcm4$ scale of deflections expected from gravitational lensing effects that this remapping of the observed CMB is a source of concern.
To model this effect, we begin from Eq.~\eqref{eqn:tod_beam}, but treat the beam as symmetric to isolate the effect of pixelization from that of beam asymmetry. 
Binning hits into pixels, we then have
\begin{eqnarray}
T(\hatn) 
&=& \sum_{i \in p}  \sum_{{\elt}m} B_{{\elt}0} \tilde{T}_{{\elt}m} Y_{{\elt}m}(\theta_i, \phi_i) / H(\hatn) \nonumber \\
&\approx& \sum_{{\elt}m} B_{{\elt}0} \tilde{T}_{{\elt}m} Y_{{\elt}m}(\hatn + \vecd^{\rm pix}),
\label{eqn:t_deflection_approx}
\end{eqnarray}
where $\hatn$ is the center of pixel $p$ and
in the second line we have made a gradient approximation to $Y_{{\elt}m}(\hatn)$ about the pixel center, and used the effective deflection vector
\be
\vecd^{\rm pix}(\hatn_p) = \sum_{i \in p} \left( \hatn_i - \hatn_p \right) / H_p,
\ee
where $\hatn_i$ is the pointing for observation $i$, which is assigned to pixel $p$ (with center at $\hatn_p$ and number of hits $H_p$).
Implicit in this model is an assumption that on the arcminute pixelization scale the beam-convolved sky fluctuations are quite smooth, and can be modelled as a gradient.
Pixelization therefore introduces an effective deflection field, given by $\vecd^{\rm pix}$, which biases our reconstruction of the lensing potential.
We could in principle correct for the spurious lensing field which this introduces by including the remapping of Eq.~\eqref{eqn:t_deflection_approx} in the simulations used to estimate the mean-field correction, however as we will see it is quite small.
We can decompose the spin-$1$ deflection field as
\be
d^{\rm pix}_{LM} = \int d^2\hatn\, \sylm{1}{LM}{*}(\hatn) \left[ d_\theta^{\rm pix} + i d_{\phi}^{\rm pix} \right](\hatn),
\ee
where $d_\theta$ and $d_{\phi}$ are the components of $\vecd^{\rm pix}$ along the $\theta$ and $\phi$ directions respectively. 
The associated weight function, following Eq.~\eqref{eqn:tcov_aniso2}, is given by
\begin{multline}
W^{d}_{\elt_1 \elt_2 L} 
= -  \sqrt{\frac{(2\elt_1+1)(2\elt_2+1)(2L+1)}{4\pi}}  C_{\elt_1}^{TT} \\
\times 
\sqrt{ \frac{\elt_1(\elt_1+1)}{ L(L+1) } }
\frac{B_{\elt_1}}{B_{\elt_2}} \threej{\elt_1}{\elt_2}{L}{1}{0}{-1} + (\elt_1 \leftrightarrow  \elt_2).
\end{multline}
\begin{figure}[!t]
\centerline{\includegraphics[width=0.95\columnwidth]{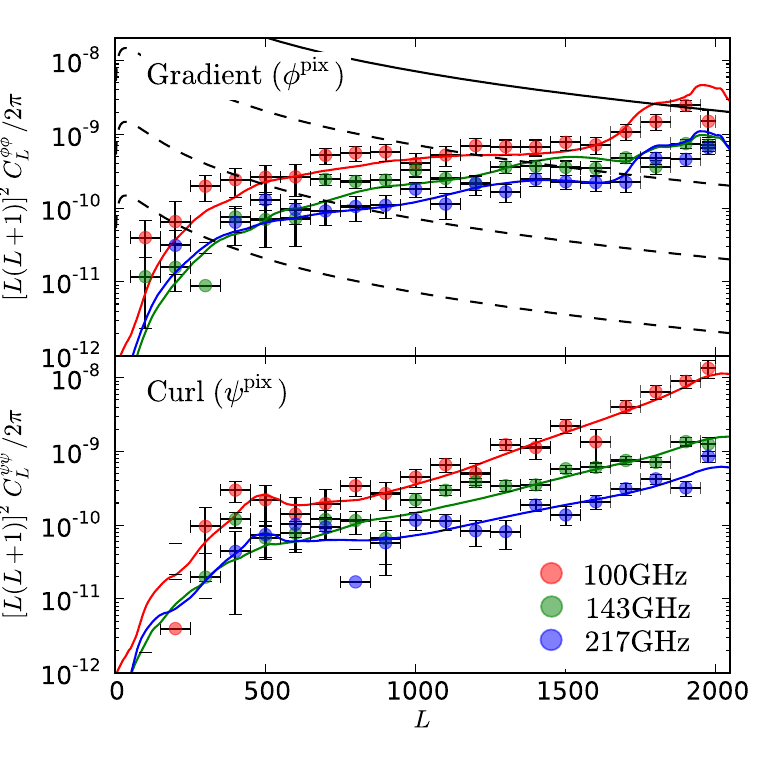}}
\caption{
Power spectra of the expected bias to our lensing estimators due to pixelization $\left. \Delta \hat{C}_L^{\phi\phi} \right|_{\rm pix}$ (thick solid lines),
as well as cross-correlation of the pixelization field with the estimators themselves $\resp_L^{\phi ({\rm pix})} \hat{C}_L^{\phi d}$ (binned points).
The theoretical lensing potential power spectrum $C_L^{\phi\phi}$ is plotted for gradient modes as solid black, while the lower dashed black curves give $(0.1, 0.01, 0.001)\times C_L^{\phi\phi}$ respectively.
The pixelization bias is always small compared to both signal and reconstruction noise.
\label{fig:cldd_pix_pp_xx}
}
\end{figure}
The pixelization deflection field can also be usefully split into gradient- and curl-type parts, as
\be
d_{LM}^{\rm pix} = d_{LM}^{G,\, {\rm pix}} + i d_{LM}^{C,\, {\rm pix}},
\ee
where the gradient and curl parts both satisfy the reality condition 
\be
d_{LM}^{G/C,\, {\rm pix}} = (-1)^{M} d_{L -M}^{* G/C,\, {\rm pix}}.
\label{eqn:dgcpix}
\ee
Only the gradient mode couples to the $\hat{\phi}$ lensing estimator, while the curl part only couples to the curl-mode lensing estimator $\hat{\psi}$ (discussed in Sect.~\ref{sec:consistency:curl}).
The statistical anisotropy introduced by $d_{}^{G,\, {\rm pix}}$ is almost completely degenerate with that of CMB lensing, differing only by the fact that it operates on the beam-convolved CMB sky rather than the primordial one.
In Fig.~\ref{fig:cldd_pix_pp_xx} we plot the expected bias to the lensing power spectrum estimator, which is given by
\begin{align}
\left. \Delta \hat{C}_L^{\phi\phi}  \right|_{\rm pix} &= \frac{1}{2L+1} \left( \resp_L^{\phi d} \right)^2 \sum_M \left| d_{LM}^{G} \right|^2,
\end{align}
as well as the analogous bias $\left. \Delta \hat{C}_L^{\psi\psi}  \right|_{\rm pix}$ for the curl null-test estimator.
To test for the presence of the pixelization lensing field in our reconstructions, we cross-correlate the deflection fields $d_{LM}^{G/C,\, {\rm pix}}$ with our lensing estimates $\hat{\phi}$, $\hat{\psi}$.
In Fig.~\ref{fig:cldd_pix_pp_xx} we plot
\be
\hat{C}_L^{\phi d} = \frac{f_{\rm sky}^{-1}}{2L+1} \sum_M \hat{\phi}_{LM} d_{LM}^{G*},
\ee
and similarly for $\psi$. 
We detect both $d^{G}_{LM}$  and $d^{C}_{LM}$ at greater than $15\sigma$ significance for all of our three main frequency bands, at the expected level.
This detection is only possible because we know $d_{LM}$ precisely from the satellite pointing.
Except at the very highest multipoles, the gradient mode of the deflection lensing field is always significantly smaller than the CMB lensing potential.
At $L<250$, which provides most of the weight for the \Planck\ lensing likelihood, the bias to the lensing potential is significantly less than $0.1\%$, while our statistical measurement uncertainty is at the $4 \%$ level.

\section{Likelihood validation}
\label{app:likelihood}
In the construction of our lensing likelihood in Sect.~\ref{sec:methodology:likelihood} we have used several simplifying approximations to the statistics of the reconstructed lensing potential, which we discuss briefly here.

\begin{figure}[!htpb]
\begin{overpic}[width=0.49\columnwidth]{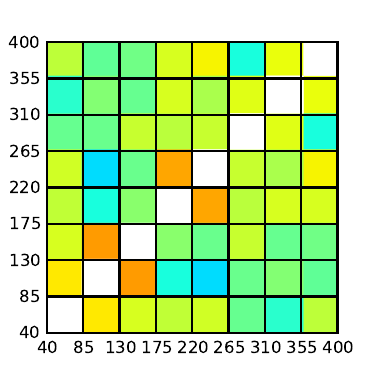} \put(12,92){100\,GHz:} \end{overpic}
\begin{overpic}[width=0.49\columnwidth]{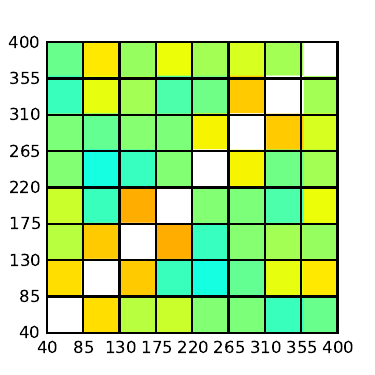} \put(12,92){143\,GHz:} \end{overpic}
\begin{overpic}[width=0.49\columnwidth]{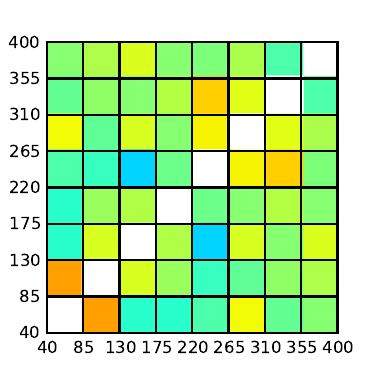} \put(12,92){217\,GHz:} \end{overpic}
\begin{overpic}[width=0.49\columnwidth]{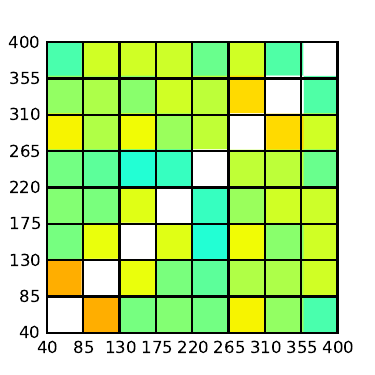} \put(12,92){MV:} \end{overpic}
\centering \includegraphics[width=0.95\columnwidth]{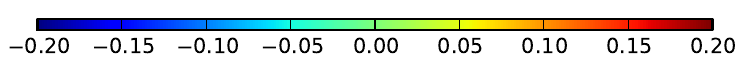}
\caption{
Measured correlation coefficients for the bins of Table~\ref{tab:bandpowers}, estimated from 1000 simulated lens reconstructions.
There is some evidence for correlations along the diagonal at the $10\%$ level, which is sufficiently small that we have chosen to neglect bin-bin correlations in our likelihood.
\label{fig:analysis_lens_bins_correlation}
}

\vspace{0.2in}
\begin{overpic}[width=0.49\columnwidth]{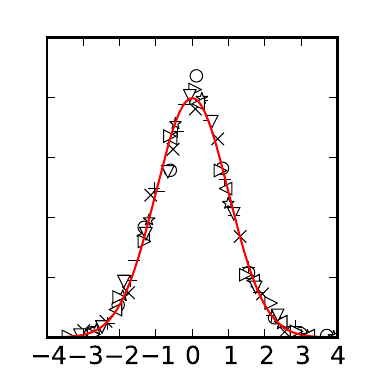} \put(12,92){100\,GHz:} \end{overpic}
\begin{overpic}[width=0.49\columnwidth]{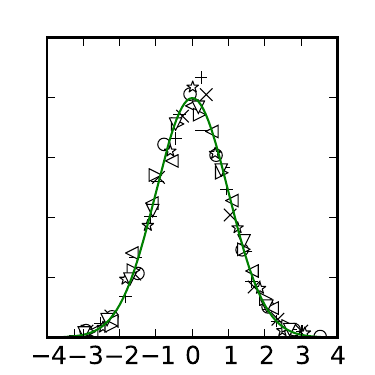} \put(12,92){143\,GHz:} \end{overpic}
\begin{overpic}[width=0.49\columnwidth]{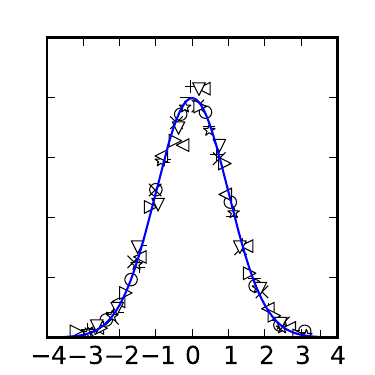} \put(12,92){217\,GHz:} \end{overpic}
\begin{overpic}[width=0.49\columnwidth]{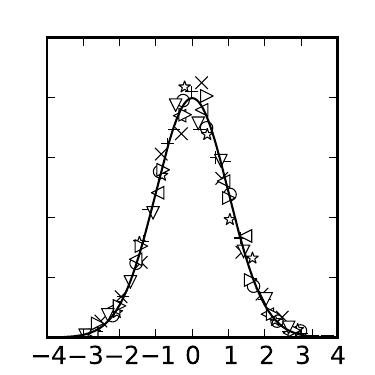} \put(12,92){MV:} \end{overpic}
\centering \includegraphics[width=0.95\columnwidth]{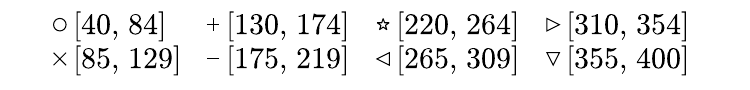}
\caption{
Histograms of amplitude coefficients for the bandpowers of Table.~\ref{tab:bandpowers}, taken from 1000 simulated lens reconstructions.
Each coefficient has been rescaled by its standard deviation so that it can be directly compared to a unit-variance Gaussian distribution (solid lines).
\label{fig:analysis_lens_bins_hist}
}
\end{figure}

\textit{Uncorrelated binned power spectrum estimates}: 
In Fig.~\ref{fig:analysis_lens_bins_correlation} we plot Monte-Carlo estimates of the bandpower correlation. These plots do not include the effect of beam uncertainties, errors in the point-source correction or cosmological uncertainty in the subtraction of the $N^{(1)}$ bias, which we include in our likelihood covariance matrix analytically.
Even for adjacent bandpowers, the correlation coefficient is constrained to be less than $10\%$ and we ignore such correlations in constructing the likelihood. As discussed in Sect.~\ref{sec:error_budget}, however, we do include bandpower correlations from beam uncertainties and the point-source and $N^{(1)}$ correction in the likelihood.

\textit{Gaussianity of the binned power spectrum estimates}:
Our estimator for the lensing potential power spectrum is a sum over the trispectrum of the observed CMB.
There is no simple, rigorous argument to expect our bandpower estimates to be Gaussian distributed, however keeping in mind that we are collapsing the information in a multi-million pixel map down to eight eigenmodes, it is quite reasonable to expect that the central limit theorem will drive their distribution in this direction.
With a set of simulated lens reconstructions, it is straightforward to assess the Gaussianity of our amplitude estimates.
In Fig.~\ref{fig:analysis_lens_bins_hist} we plot histograms of the bandpower coefficients from our set of fiducial simulations showing that the Gaussian approximation is good for the marginal distributions. Further tests of the assumption of a Gaussian likelihood are made in~\citet{Schmittfull:2013uea} for isotropic surveys. There it is shown that parameter constraints for a simple two-parameter model of $C_\elp^{\phi\phi}$ scatter across simulations in a manner consistent with the likelihood width in any realization.

\textit{No correlations between the lensing potential and temperature power spectra}:
Lensing may be detected at a significance of $10\sigma$ as a smearing of the acoustic peaks in the temperature power spectrum; see \cite{planck2013-p11}.
A potential worry when simply combining the temperature likelihood with the lensing power spectrum likelihood is that we are double counting the same information \citep{Lewis:2005tp,Perotto:2006rj,Smith:2006nk} since we do not account for the correlated effect of sample variance of the lenses in the estimated $\hat{C}_{\elp}^{\phi\phi}$ and $\hat{C}_\elt^{TT}$ spectra. This effect is explored in detail in~\citet{Schmittfull:2013uea} where it is shown to derive from the connected part of the 6-point function for the lensed temperature anisotropies. They estimate for \planck\ that the maximum correlation between $\hat{C}_{\elp}^{\phi\phi}$ and $\hat{C}_\elt^{TT}$
for unbinned spectra is around $0.05\%$, and that ignoring this leads to a misestimation of the errors on the lensing amplitude $A_{\mathrm L}$ (see Sect.~\ref{sec:results:cosmo}) in a joint analysis of only $2\%$.
A more pressing concern is that the statistical errors on the reconstruction power are correlated with those for the temperature power spectrum due to sample variance of the CMB anisotropies. This effect is also discussed in detail in~\citet{Schmittfull:2013uea}. The correlation derives from the fully disconnected part of the CMB 6-point function and leads to correlations of at most $0.2\%$ between the unbinned spectra. However, these correlations are removed exactly by the realization-dependent aspect of the $N^{(0)}$ debiasing procedure discussed in Sect.~\ref{sec:methodology:power_spectrum_reconstruction}.

To see that the disconnected contribution to $\mathrm{cov}(\hat{C}_\elp^{\phi\phi},
\hat{C}_\ell^{TT})$ vanishes, we note that with our data-dependent $N^{(0)}$ subtraction
the data enters the $\hat{C}_{\elp}^{\phi\phi}$ in the form\footnote{For compactness of presentation, we consider estimates built from a single temperature map here.}
\begin{multline}
\hat{C}_{\elp}^{\phi\phi} \sim \frac{\partial}{\partial C_{\elp}^{\phi\phi}} \langle
T_i T_j T_k T_l \rangle_c \biggl(\bar{T}_i \bar{T}_j \bar{T}_k \bar{T}_l \\
- \left[
C^{-1}_{ij} \bar{T}_k \bar{T}_l + C^{-1}_{kl} \bar{T}_i \bar{T}_j - \bar{C}^{-1}_{ij}
\bar{C}^{-1}_{kl} \right] \\
-\left[C^{-1}_{ik} \bar{T}_j \bar{T}_l + C^{-1}_{jl} \bar{T}_i \bar{T}_k - \bar{C}^{-1}_{ik}
\bar{C}^{-1}_{jl} \right] \\
- \left[C^{-1}_{il} \bar{T}_j \bar{T}_k + C^{-1}_{jk} \bar{T}_i \bar{T}_l - \bar{C}^{-1}_{jk}
\bar{C}^{-1}_{il} \right] \biggr)\, .
\label{eq:opttri1}
\end{multline}
Here, $\bar{T}_i$ is the inverse-variance filtered temperature, with covariance $C^{-1}_{ij}$, and summation over repeated indices is assumed.  Equation~(\ref{eq:opttri1}) arises naturally for optimal trispectrum estimation of weakly non-Gaussian fields (e.g.,~\citealt{Regan:2010cn}). The expectation value of the data combination on the right-hand side is simply the connected 4-point function, $\langle
T_i T_j T_k T_l \rangle_c$, but the data-dependent quadratic terms improve the variance of the estimator. The right-hand side of Eq.~(\ref{eq:opttri1}) is explicitly symmetric under regroupings of pairs of indices so that only the primary coupling of the trispectrum need be included. For lensing, the primary coupling factors such that
\begin{equation}
\frac{\partial}{\partial C_{\elp}^{\phi\phi}} \langle
T_i T_j T_k T_l \rangle_c^{\rm primary} = \sum_{M} (-1)^M X^{ij}_{LM} X^{kl}_{L-M} ;
\end{equation}
see~\citet{Hu:2001fa,Hanson:2010rp} for explicit expressions. The first set of terms in Eq.~(\ref{eq:opttri1}) combine to give
\begin{multline}
\frac{\partial}{\partial C_{\elp}^{\phi\phi}} \langle
T_i T_j T_k T_l \rangle_c^{\rm primary} \\
\times \left(\bar{T}_i \bar{T}_j \bar{T}_k \bar{T}_l 
- \left[
C^{-1}_{ij} \bar{T}_k \bar{T}_l + C^{-1}_{kl} \bar{T}_i \bar{T}_j - \bar{C}^{-1}_{ij}
\bar{C}^{-1}_{kl} \right]\right) \\
= \sum_M (-1)^M \left[X^{ij}_{LM} \left(\bar{T}_i \bar{T}_j - C^{-1}_{ij} \right)\right] \left[X^{kl}_{L-M} \left(\bar{T}_k \bar{T}_k - C^{-1}_{kl} \right)\right] , 
\end{multline}
where the terms in square brackets are exactly of the form of mean-field-subtracted,
unnormalized estimates of $\phi_{LM}$ and $\phi_{L -M}$.
The remaining terms are equivalent to the $N^{(0)}$ subtraction in Eq.~(\ref{eqn:cln0}).
Any quadratic estimate of the temperature power spectrum will involve a linear combination of terms of the form $T^{(1)}_p T^{(2)}_q$, where $T^{(1)}_p$ and $T^{(2)}_q$ may be different noisy estimates of the temperature anisotropies but are correlated with themselves and $T_i$ through (at least) the common temperature anisotropies. If we form the covariance between $T^{(1)}_p T^{(2)}_q$ and the data combination in Eq.~(\ref{eq:opttri1}) that enters the $C_\elp^{\phi\phi}$ estimate, the disconnected terms give the contribution to the covariance between the estimated lensing and temperature power spectra due to sample variance of the temperature anisotropies. However, it is straightforward to show that by including the additional terms in Eq.~(\ref{eq:opttri1}) that are quadratic in $\bar{T}$ the disconnected contribution vanishes. We see that, as well as mitigating against errors in the CMB and noise covariance matrices (see Sect.~\ref{sec:methodology}), and reducing the covariance between $C_\elp^{\phi\phi}$ estimates~\citep{Hanson:2010rp}, the data-dependent $N^{(0)}$ subtraction that we adopt removes much of the covariance with the temperature power spectrum.

The final effect that correlates the lensing and temperature power spectrum estimates is the correlation between $T$ and $\phi$ due to the late-time ISW effect in the former (Sect.~\ref{sec:results:iswlensing}). This produces only local correlations between the spectra and they fall rapidly with $\elp$ being less than $0.5\%$ at $\elp=40$. Their impact on the errors on the lensing amplitude $A_{\rm L}$ in a joint estimate from the temperature and lensing likelihoods is negligible.

\bibliographystyle{aat} 

\bibliography{planck_lensing,Planck_bib}

\raggedright
\end{document}

%% file: Planck.tex
\def\setsymbol#1#2{\expandafter\def\csname #1\endcsname{#2}}
\def\getsymbol#1{\csname #1\endcsname}

\def\Planck{\textit{Planck}}

\def\all2013resultspapers{\nocite{planck2013-p01, planck2013-p02, planck2013-p02a, planck2013-p02d, planck2013-p02b, planck2013-p03, planck2013-p03c, planck2013-p03f, planck2013-p03d, planck2013-p03e, planck2013-p01a, planck2013-p06, planck2013-p03a, planck2013-pip88, planck2013-p08, planck2013-p11, planck2013-p12, planck2013-p13, planck2013-p14, planck2013-p15, planck2013-p05b, planck2013-p17, planck2013-p09, planck2013-p09a, planck2013-p20, planck2013-p19, planck2013-pipaberration, planck2013-p05, planck2013-p05a, planck2013-pip56, planck2013-p06b}}

\newbox\tablebox    \newdimen\tablewidth
\def\leaderfil{\leaders\hbox to 5pt{\hss.\hss}\hfil}
\def\endPlancktable{\tablewidth=\columnwidth 
    $$\hss\copy\tablebox\hss$$
    \vskip-\lastskip\vskip -2pt}
\def\endPlancktablewide{\tablewidth=\textwidth 
    $$\hss\copy\tablebox\hss$$
    \vskip-\lastskip\vskip -2pt}
\def\tablenote#1 #2\par{\begingroup \parindent=0.8em
    \abovedisplayshortskip=0pt\belowdisplayshortskip=0pt
    \noindent
    $$\hss\vbox{\hsize\tablewidth \hangindent=\parindent \hangafter=1 \noindent
    \hbox to \parindent{$^#1$\hss}\strut#2\strut\par}\hss$$
    \endgroup}
\def\doubleline{\vskip 3pt\hrule \vskip 1.5pt \hrule \vskip 5pt}

\def\L2{\ifmmode L_2\else $L_2$\fi}

\def\DeltaT{\ifmmode \Delta T\else $\Delta T$\fi}
\def\deltat{\ifmmode \Delta t\else $\Delta t$\fi}
\def\fknee{\ifmmode f_{\rm knee}\else $f_{\rm knee}$\fi}
\def\Fmax{\ifmmode F_{\rm max}\else $F_{\rm max}$\fi}
\def\solar{\ifmmode{\rm M}_{\mathord\odot}\else${\rm M}_{\mathord\odot}$\fi}
\def\Msolar{\ifmmode{\rm M}_{\mathord\odot}\else${\rm M}_{\mathord\odot}$\fi}
\def\Lsolar{\ifmmode{\rm L}_{\mathord\odot}\else${\rm L}_{\mathord\odot}$\fi}

\def\inv{\ifmmode^{-1}\else$^{-1}$\fi}
\def\mo{\ifmmode^{-1}\else$^{-1}$\fi}
\def\sup#1{\ifmmode ^{\rm #1}\else $^{\rm #1}$\fi}
\def\expo#1{\ifmmode \times 10^{#1}\else $\times 10^{#1}$\fi}
\def\,{\thinspace}
\def\lsim{\mathrel{\raise .4ex\hbox{\rlap{$<$}\lower 1.2ex\hbox{$\sim$}}}}
\def\gsim{\mathrel{\raise .4ex\hbox{\rlap{$>$}\lower 1.2ex\hbox{$\sim$}}}}
\let\lea=\lsim

\def\simprop{\mathrel{\raise .4ex\hbox{\rlap{$\propto$}\lower 1.2ex\hbox{$\sim$}}}}
\def\deg{\ifmmode^\circ\else$^\circ$\fi}
\def\pdeg{\ifmmode $\setbox0=\hbox{$^{\circ}$}\rlap{\hskip.11\wd0 .}$^{\circ}
          \else \setbox0=\hbox{$^{\circ}$}\rlap{\hskip.11\wd0 .}$^{\circ}$\fi}
\def\arcs{\ifmmode {^{\scriptstyle\prime\prime}}
          \else $^{\scriptstyle\prime\prime}$\fi}
\def\arcm{\ifmmode {^{\scriptstyle\prime}}
          \else $^{\scriptstyle\prime}$\fi}
\newdimen\sa  \newdimen\sb
\def\parcs{\sa=.07em \sb=.03em
     \ifmmode \hbox{\rlap{.}}^{\scriptstyle\prime\kern -\sb\prime}\hbox{\kern -\sa}
     \else \rlap{.}$^{\scriptstyle\prime\kern -\sb\prime}$\kern -\sa\fi}
\def\parcm{\sa=.08em \sb=.03em
     \ifmmode \hbox{\rlap{.}\kern\sa}^{\scriptstyle\prime}\hbox{\kern-\sb}
     \else \rlap{.}\kern\sa$^{\scriptstyle\prime}$\kern-\sb\fi}
\def\ra[#1 #2 #3.#4]{#1\sup{h}#2\sup{m}#3\sup{s}\llap.#4}
\def\dec[#1 #2 #3.#4]{#1\deg#2\arcm#3\arcs\llap.#4}
\def\deco[#1 #2 #3]{#1\deg#2\arcm#3\arcs}
\def\rra[#1 #2]{#1\sup{h}#2\sup{m}}

\def\dots{\relax\ifmmode \ldots\else $\ldots$\fi}
\def\WHzsr{\ifmmode $W\,Hz\mo\,sr\mo$\else W\,Hz\mo\,sr\mo\fi}
\def\mHz{\ifmmode $\,mHz$\else \,mHz\fi}
\def\GHz{\ifmmode $\,GHz$\else \,GHz\fi}
\def\mKs{\ifmmode $\,mK\,s$^{1/2}\else \,mK\,s$^{1/2}$\fi}
\def\muKs{\ifmmode \,\mu$K\,s$^{1/2}\else \,$\mu$K\,s$^{1/2}$\fi}
\def\muKRJs{\ifmmode \,\mu$K$_{\rm RJ}$\,s$^{1/2}\else \,$\mu$K$_{\rm RJ}$\,s$^{1/2}$\fi}
\def\muKHz{\ifmmode \,\mu$K\,Hz$^{-1/2}\else \,$\mu$K\,Hz$^{-1/2}$\fi}
\def\MJysr{\ifmmode \,$MJy\,sr\mo$\else \,MJy\,sr\mo\fi}
\def\MJysrmK{\ifmmode \,$MJy\,sr\mo$\,mK$_{\rm CMB}\mo\else \,MJy\,sr\mo\,mK$_{\rm CMB}\mo$\fi}
\def\microns{\ifmmode \,\mu$m$\else \,$\mu$m\fi}

\def\muK{\ifmmode \,\mu$K$\else \,$\mu$\hbox{K}\fi}
\def\microK{\ifmmode \,\mu$K$\else \,$\mu$\hbox{K}\fi}
\def\muW{\ifmmode \,\mu$W$\else \,$\mu$\hbox{W}\fi}
\def\kms{\ifmmode $\,km\,s$^{-1}\else \,km\,s$^{-1}$\fi}
\def\kmsMpc{\ifmmode $\,\kms\,Mpc\mo$\else \,\kms\,Mpc\mo\fi}
\setsymbol{LFI:center:frequency:70GHz:units}{70.3\,GHz}
\setsymbol{LFI:center:frequency:44GHz:units}{44.1\,GHz}
\setsymbol{LFI:center:frequency:30GHz:units}{28.5\,GHz}

\setsymbol{LFI:center:frequency:70GHz}{70.3}
\setsymbol{LFI:center:frequency:44GHz}{44.1}
\setsymbol{LFI:center:frequency:30GHz}{28.5}

\setsymbol{LFI:center:frequency:LFI18:Rad:M:units}{71.7\GHz}
\setsymbol{LFI:center:frequency:LFI19:Rad:M:units}{67.5\GHz}
\setsymbol{LFI:center:frequency:LFI20:Rad:M:units}{69.2\GHz}
\setsymbol{LFI:center:frequency:LFI21:Rad:M:units}{70.4\GHz}
\setsymbol{LFI:center:frequency:LFI22:Rad:M:units}{71.5\GHz}
\setsymbol{LFI:center:frequency:LFI23:Rad:M:units}{70.8\GHz}
\setsymbol{LFI:center:frequency:LFI24:Rad:M:units}{44.4\GHz}
\setsymbol{LFI:center:frequency:LFI25:Rad:M:units}{44.0\GHz}
\setsymbol{LFI:center:frequency:LFI26:Rad:M:units}{43.9\GHz}
\setsymbol{LFI:center:frequency:LFI27:Rad:M:units}{28.3\GHz}
\setsymbol{LFI:center:frequency:LFI28:Rad:M:units}{28.8\GHz}
\setsymbol{LFI:center:frequency:LFI18:Rad:S:units}{70.1\GHz}
\setsymbol{LFI:center:frequency:LFI19:Rad:S:units}{69.6\GHz}
\setsymbol{LFI:center:frequency:LFI20:Rad:S:units}{69.5\GHz}
\setsymbol{LFI:center:frequency:LFI21:Rad:S:units}{69.5\GHz}
\setsymbol{LFI:center:frequency:LFI22:Rad:S:units}{72.8\GHz}
\setsymbol{LFI:center:frequency:LFI23:Rad:S:units}{71.3\GHz}
\setsymbol{LFI:center:frequency:LFI24:Rad:S:units}{44.1\GHz}
\setsymbol{LFI:center:frequency:LFI25:Rad:S:units}{44.1\GHz}
\setsymbol{LFI:center:frequency:LFI26:Rad:S:units}{44.1\GHz}
\setsymbol{LFI:center:frequency:LFI27:Rad:S:units}{28.5\GHz}
\setsymbol{LFI:center:frequency:LFI28:Rad:S:units}{28.2\GHz}

\setsymbol{LFI:center:frequency:LFI18:Rad:M}{71.7}
\setsymbol{LFI:center:frequency:LFI19:Rad:M}{67.5}
\setsymbol{LFI:center:frequency:LFI20:Rad:M}{69.2}
\setsymbol{LFI:center:frequency:LFI21:Rad:M}{70.4}
\setsymbol{LFI:center:frequency:LFI22:Rad:M}{71.5}
\setsymbol{LFI:center:frequency:LFI23:Rad:M}{70.8}
\setsymbol{LFI:center:frequency:LFI24:Rad:M}{44.4}
\setsymbol{LFI:center:frequency:LFI25:Rad:M}{44.0}
\setsymbol{LFI:center:frequency:LFI26:Rad:M}{43.9}
\setsymbol{LFI:center:frequency:LFI27:Rad:M}{28.3}
\setsymbol{LFI:center:frequency:LFI28:Rad:M}{28.8}
\setsymbol{LFI:center:frequency:LFI18:Rad:S}{70.1}
\setsymbol{LFI:center:frequency:LFI19:Rad:S}{69.6}
\setsymbol{LFI:center:frequency:LFI20:Rad:S}{69.5}
\setsymbol{LFI:center:frequency:LFI21:Rad:S}{69.5}
\setsymbol{LFI:center:frequency:LFI22:Rad:S}{72.8}
\setsymbol{LFI:center:frequency:LFI23:Rad:S}{71.3}
\setsymbol{LFI:center:frequency:LFI24:Rad:S}{44.1}
\setsymbol{LFI:center:frequency:LFI25:Rad:S}{44.1}
\setsymbol{LFI:center:frequency:LFI26:Rad:S}{44.1}
\setsymbol{LFI:center:frequency:LFI27:Rad:S}{28.5}
\setsymbol{LFI:center:frequency:LFI28:Rad:S}{28.2}

\setsymbol{LFI:white:noise:sensitivity:70GHz:units}{134.7\muKs}
\setsymbol{LFI:white:noise:sensitivity:44GHz:units}{164.7\muKs}
\setsymbol{LFI:white:noise:sensitivity:30GHz:units}{143.4\muKs}

\setsymbol{LFI:white:noise:sensitivity:70GHz}{134.7}
\setsymbol{LFI:white:noise:sensitivity:44GHz}{164.7}
\setsymbol{LFI:white:noise:sensitivity:30GHz}{143.4}

\setsymbol{LFI:white:noise:sensitivity:LFI18:Rad:M:units}{512.0\muKs}
\setsymbol{LFI:white:noise:sensitivity:LFI19:Rad:M:units}{581.4\muKs}
\setsymbol{LFI:white:noise:sensitivity:LFI20:Rad:M:units}{590.8\muKs}
\setsymbol{LFI:white:noise:sensitivity:LFI21:Rad:M:units}{455.2\muKs}
\setsymbol{LFI:white:noise:sensitivity:LFI22:Rad:M:units}{492.0\muKs}
\setsymbol{LFI:white:noise:sensitivity:LFI23:Rad:M:units}{507.7\muKs}
\setsymbol{LFI:white:noise:sensitivity:LFI24:Rad:M:units}{462.2\muKs}
\setsymbol{LFI:white:noise:sensitivity:LFI25:Rad:M:units}{413.6\muKs}
\setsymbol{LFI:white:noise:sensitivity:LFI26:Rad:M:units}{478.6\muKs}
\setsymbol{LFI:white:noise:sensitivity:LFI27:Rad:M:units}{277.7\muKs}
\setsymbol{LFI:white:noise:sensitivity:LFI28:Rad:M:units}{312.3\muKs}
\setsymbol{LFI:white:noise:sensitivity:LFI18:Rad:S:units}{465.7\muKs}
\setsymbol{LFI:white:noise:sensitivity:LFI19:Rad:S:units}{555.6\muKs}
\setsymbol{LFI:white:noise:sensitivity:LFI20:Rad:S:units}{623.2\muKs}
\setsymbol{LFI:white:noise:sensitivity:LFI21:Rad:S:units}{564.1\muKs}
\setsymbol{LFI:white:noise:sensitivity:LFI22:Rad:S:units}{534.4\muKs}
\setsymbol{LFI:white:noise:sensitivity:LFI23:Rad:S:units}{542.4\muKs}
\setsymbol{LFI:white:noise:sensitivity:LFI24:Rad:S:units}{399.2\muKs}
\setsymbol{LFI:white:noise:sensitivity:LFI25:Rad:S:units}{392.6\muKs}
\setsymbol{LFI:white:noise:sensitivity:LFI26:Rad:S:units}{418.6\muKs}
\setsymbol{LFI:white:noise:sensitivity:LFI27:Rad:S:units}{302.9\muKs}
\setsymbol{LFI:white:noise:sensitivity:LFI28:Rad:S:units}{285.3\muKs}

\setsymbol{LFI:white:noise:sensitivity:LFI18:Rad:M}{512.0}
\setsymbol{LFI:white:noise:sensitivity:LFI19:Rad:M}{581.4}
\setsymbol{LFI:white:noise:sensitivity:LFI20:Rad:M}{590.8}
\setsymbol{LFI:white:noise:sensitivity:LFI21:Rad:M}{455.2}
\setsymbol{LFI:white:noise:sensitivity:LFI22:Rad:M}{492.0}
\setsymbol{LFI:white:noise:sensitivity:LFI23:Rad:M}{507.7}
\setsymbol{LFI:white:noise:sensitivity:LFI24:Rad:M}{462.2}
\setsymbol{LFI:white:noise:sensitivity:LFI25:Rad:M}{413.6}
\setsymbol{LFI:white:noise:sensitivity:LFI26:Rad:M}{478.6}
\setsymbol{LFI:white:noise:sensitivity:LFI27:Rad:M}{277.7}
\setsymbol{LFI:white:noise:sensitivity:LFI28:Rad:M}{312.3}
\setsymbol{LFI:white:noise:sensitivity:LFI18:Rad:S}{465.7}
\setsymbol{LFI:white:noise:sensitivity:LFI19:Rad:S}{555.6}
\setsymbol{LFI:white:noise:sensitivity:LFI20:Rad:S}{623.2}
\setsymbol{LFI:white:noise:sensitivity:LFI21:Rad:S}{564.1}
\setsymbol{LFI:white:noise:sensitivity:LFI22:Rad:S}{534.4}
\setsymbol{LFI:white:noise:sensitivity:LFI23:Rad:S}{542.4}
\setsymbol{LFI:white:noise:sensitivity:LFI24:Rad:S}{399.2}
\setsymbol{LFI:white:noise:sensitivity:LFI25:Rad:S}{392.6}
\setsymbol{LFI:white:noise:sensitivity:LFI26:Rad:S}{418.6}
\setsymbol{LFI:white:noise:sensitivity:LFI27:Rad:S}{302.9}
\setsymbol{LFI:white:noise:sensitivity:LFI28:Rad:S}{285.3}

\setsymbol{LFI:knee:frequency:70GHz:units}{29.5\mHz}
\setsymbol{LFI:knee:frequency:44GHz:units}{56.2\mHz}
\setsymbol{LFI:knee:frequency:30GHz:units}{113.7\mHz}

\setsymbol{LFI:knee:frequency:70GHz}{29.5}
\setsymbol{LFI:knee:frequency:44GHz}{56.2}
\setsymbol{LFI:knee:frequency:30GHz}{113.7}

\setsymbol{LFI:knee:frequency:LFI18:Rad:M:units}{16.3\mHz}
\setsymbol{LFI:knee:frequency:LFI19:Rad:M:units}{15.1\mHz}
\setsymbol{LFI:knee:frequency:LFI20:Rad:M:units}{18.7\mHz}
\setsymbol{LFI:knee:frequency:LFI21:Rad:M:units}{37.2\mHz}
\setsymbol{LFI:knee:frequency:LFI22:Rad:M:units}{12.7\mHz}
\setsymbol{LFI:knee:frequency:LFI23:Rad:M:units}{34.6\mHz}
\setsymbol{LFI:knee:frequency:LFI24:Rad:M:units}{46.2\mHz}
\setsymbol{LFI:knee:frequency:LFI25:Rad:M:units}{24.9\mHz}
\setsymbol{LFI:knee:frequency:LFI26:Rad:M:units}{67.6\mHz}
\setsymbol{LFI:knee:frequency:LFI27:Rad:M:units}{187.4\mHz}
\setsymbol{LFI:knee:frequency:LFI28:Rad:M:units}{122.2\mHz}
\setsymbol{LFI:knee:frequency:LFI18:Rad:S:units}{17.7\mHz}
\setsymbol{LFI:knee:frequency:LFI19:Rad:S:units}{22.0\mHz}
\setsymbol{LFI:knee:frequency:LFI20:Rad:S:units}{8.7\mHz}
\setsymbol{LFI:knee:frequency:LFI21:Rad:S:units}{25.9\mHz}
\setsymbol{LFI:knee:frequency:LFI22:Rad:S:units}{15.8\mHz}
\setsymbol{LFI:knee:frequency:LFI23:Rad:S:units}{129.8\mHz}
\setsymbol{LFI:knee:frequency:LFI24:Rad:S:units}{100.9\mHz}
\setsymbol{LFI:knee:frequency:LFI25:Rad:S:units}{38.9\mHz}
\setsymbol{LFI:knee:frequency:LFI26:Rad:S:units}{58.9\mHz}
\setsymbol{LFI:knee:frequency:LFI27:Rad:S:units}{104.4\mHz}
\setsymbol{LFI:knee:frequency:LFI28:Rad:S:units}{40.7\mHz}

\setsymbol{LFI:knee:frequency:LFI18:Rad:M}{16.3}
\setsymbol{LFI:knee:frequency:LFI19:Rad:M}{15.1}
\setsymbol{LFI:knee:frequency:LFI20:Rad:M}{18.7}
\setsymbol{LFI:knee:frequency:LFI21:Rad:M}{37.2}
\setsymbol{LFI:knee:frequency:LFI22:Rad:M}{12.7}
\setsymbol{LFI:knee:frequency:LFI23:Rad:M}{34.6}
\setsymbol{LFI:knee:frequency:LFI24:Rad:M}{46.2}
\setsymbol{LFI:knee:frequency:LFI25:Rad:M}{24.9}
\setsymbol{LFI:knee:frequency:LFI26:Rad:M}{67.6}
\setsymbol{LFI:knee:frequency:LFI27:Rad:M}{187.4}
\setsymbol{LFI:knee:frequency:LFI28:Rad:M}{122.2}
\setsymbol{LFI:knee:frequency:LFI18:Rad:S}{17.7}
\setsymbol{LFI:knee:frequency:LFI19:Rad:S}{22.0}
\setsymbol{LFI:knee:frequency:LFI20:Rad:S}{8.7}
\setsymbol{LFI:knee:frequency:LFI21:Rad:S}{25.9}
\setsymbol{LFI:knee:frequency:LFI22:Rad:S}{15.8}
\setsymbol{LFI:knee:frequency:LFI23:Rad:S}{129.8}
\setsymbol{LFI:knee:frequency:LFI24:Rad:S}{100.9}
\setsymbol{LFI:knee:frequency:LFI25:Rad:S}{38.9}
\setsymbol{LFI:knee:frequency:LFI26:Rad:S}{58.9}
\setsymbol{LFI:knee:frequency:LFI27:Rad:S}{104.4}
\setsymbol{LFI:knee:frequency:LFI28:Rad:S}{40.7}

\setsymbol{LFI:slope:70GHz:units}{$-1.03$\mHz}
\setsymbol{LFI:slope:44GHz:units}{$-0.89$\mHz}
\setsymbol{LFI:slope:30GHz:units}{$-0.87$\mHz}

\setsymbol{LFI:slope:70GHz}{$-1.03$}
\setsymbol{LFI:slope:44GHz}{$-0.89$}
\setsymbol{LFI:slope:30GHz}{$-0.87$}

\setsymbol{LFI:slope:LFI18:Rad:M:units}{$-1.04$\mHz}
\setsymbol{LFI:slope:LFI19:Rad:M:units}{$-1.09$\mHz}
\setsymbol{LFI:slope:LFI20:Rad:M:units}{$-0.69$\mHz}
\setsymbol{LFI:slope:LFI21:Rad:M:units}{$-1.56$\mHz}
\setsymbol{LFI:slope:LFI22:Rad:M:units}{$-1.01$\mHz}
\setsymbol{LFI:slope:LFI23:Rad:M:units}{$-0.96$\mHz}
\setsymbol{LFI:slope:LFI24:Rad:M:units}{$-0.83$\mHz}
\setsymbol{LFI:slope:LFI25:Rad:M:units}{$-0.91$\mHz}
\setsymbol{LFI:slope:LFI26:Rad:M:units}{$-0.95$\mHz}
\setsymbol{LFI:slope:LFI27:Rad:M:units}{$-0.87$\mHz}
\setsymbol{LFI:slope:LFI28:Rad:M:units}{$-0.88$\mHz}
\setsymbol{LFI:slope:LFI18:Rad:S:units}{$-1.15$\mHz}
\setsymbol{LFI:slope:LFI19:Rad:S:units}{$-1.00$\mHz}
\setsymbol{LFI:slope:LFI20:Rad:S:units}{$-0.95$\mHz}
\setsymbol{LFI:slope:LFI21:Rad:S:units}{$-0.92$\mHz}
\setsymbol{LFI:slope:LFI22:Rad:S:units}{$-1.01$\mHz}
\setsymbol{LFI:slope:LFI23:Rad:S:units}{$-0.95$\mHz}
\setsymbol{LFI:slope:LFI24:Rad:S:units}{$-0.73$\mHz}
\setsymbol{LFI:slope:LFI25:Rad:S:units}{$-1.16$\mHz}
\setsymbol{LFI:slope:LFI26:Rad:S:units}{$-0.79$\mHz}
\setsymbol{LFI:slope:LFI27:Rad:S:units}{$-0.82$\mHz}
\setsymbol{LFI:slope:LFI28:Rad:S:units}{$-0.91$\mHz}

\setsymbol{LFI:slope:LFI18:Rad:M}{$-1.04$}
\setsymbol{LFI:slope:LFI19:Rad:M}{$-1.09$}
\setsymbol{LFI:slope:LFI20:Rad:M}{$-0.69$}
\setsymbol{LFI:slope:LFI21:Rad:M}{$-1.56$}
\setsymbol{LFI:slope:LFI22:Rad:M}{$-1.01$}
\setsymbol{LFI:slope:LFI23:Rad:M}{$-0.96$}
\setsymbol{LFI:slope:LFI24:Rad:M}{$-0.83$}
\setsymbol{LFI:slope:LFI25:Rad:M}{$-0.91$}
\setsymbol{LFI:slope:LFI26:Rad:M}{$-0.95$}
\setsymbol{LFI:slope:LFI27:Rad:M}{$-0.87$}
\setsymbol{LFI:slope:LFI28:Rad:M}{$-0.88$}
\setsymbol{LFI:slope:LFI18:Rad:S}{$-1.15$}
\setsymbol{LFI:slope:LFI19:Rad:S}{$-1.00$}
\setsymbol{LFI:slope:LFI20:Rad:S}{$-0.95$}
\setsymbol{LFI:slope:LFI21:Rad:S}{$-0.92$}
\setsymbol{LFI:slope:LFI22:Rad:S}{$-1.01$}
\setsymbol{LFI:slope:LFI23:Rad:S}{$-0.95$}
\setsymbol{LFI:slope:LFI24:Rad:S}{$-0.73$}
\setsymbol{LFI:slope:LFI25:Rad:S}{$-1.16$}
\setsymbol{LFI:slope:LFI26:Rad:S}{$-0.79$}
\setsymbol{LFI:slope:LFI27:Rad:S}{$-0.82$}
\setsymbol{LFI:slope:LFI28:Rad:S}{$-0.91$}

\setsymbol{LFI:FWHM:70GHz:units}{13\parcm01}
\setsymbol{LFI:FWHM:44GHz:units}{27\parcm92}
\setsymbol{LFI:FWHM:30GHz:units}{32\parcm65}

\setsymbol{LFI:FWHM:70GHz}{13.01}
\setsymbol{LFI:FWHM:44GHz}{27.92}
\setsymbol{LFI:FWHM:30GHz}{32.65}

\setsymbol{LFI:FWHM:LFI18:units}{13\parcm39}
\setsymbol{LFI:FWHM:LFI19:units}{13\parcm01}
\setsymbol{LFI:FWHM:LFI20:units}{12\parcm75}
\setsymbol{LFI:FWHM:LFI21:units}{12\parcm74}
\setsymbol{LFI:FWHM:LFI22:units}{12\parcm87}
\setsymbol{LFI:FWHM:LFI23:units}{13\parcm27}
\setsymbol{LFI:FWHM:LFI24:units}{22\parcm98}
\setsymbol{LFI:FWHM:LFI25:units}{30\parcm46}
\setsymbol{LFI:FWHM:LFI26:units}{30\parcm31}
\setsymbol{LFI:FWHM:LFI27:units}{32\parcm65}
\setsymbol{LFI:FWHM:LFI28:units}{32\parcm66}

\setsymbol{LFI:FWHM:LFI18}{13.39}
\setsymbol{LFI:FWHM:LFI19}{13.01}
\setsymbol{LFI:FWHM:LFI20}{12.75}
\setsymbol{LFI:FWHM:LFI21}{12.74}
\setsymbol{LFI:FWHM:LFI22}{12.87}
\setsymbol{LFI:FWHM:LFI23}{13.27}
\setsymbol{LFI:FWHM:LFI24}{22.98}
\setsymbol{LFI:FWHM:LFI25}{30.46}
\setsymbol{LFI:FWHM:LFI26}{30.31}
\setsymbol{LFI:FWHM:LFI27}{32.65}
\setsymbol{LFI:FWHM:LFI28}{32.66}

\setsymbol{LFI:FWHM:uncertainty:LFI18:units}{0.170\arcm}
\setsymbol{LFI:FWHM:uncertainty:LFI19:units}{0.174\arcm}
\setsymbol{LFI:FWHM:uncertainty:LFI20:units}{0.170\arcm}
\setsymbol{LFI:FWHM:uncertainty:LFI21:units}{0.156\arcm}
\setsymbol{LFI:FWHM:uncertainty:LFI22:units}{0.164\arcm}
\setsymbol{LFI:FWHM:uncertainty:LFI23:units}{0.171\arcm}
\setsymbol{LFI:FWHM:uncertainty:LFI24:units}{0.652\arcm}
\setsymbol{LFI:FWHM:uncertainty:LFI25:units}{1.075\arcm}
\setsymbol{LFI:FWHM:uncertainty:LFI26:units}{1.131\arcm}
\setsymbol{LFI:FWHM:uncertainty:LFI27:units}{1.266\arcm}
\setsymbol{LFI:FWHM:uncertainty:LFI28:units}{1.287\arcm}

\setsymbol{LFI:FWHM:uncertainty:LFI18}{0.170}
\setsymbol{LFI:FWHM:uncertainty:LFI19}{0.174}
\setsymbol{LFI:FWHM:uncertainty:LFI20}{0.170}
\setsymbol{LFI:FWHM:uncertainty:LFI21}{0.156}
\setsymbol{LFI:FWHM:uncertainty:LFI22}{0.164}
\setsymbol{LFI:FWHM:uncertainty:LFI23}{0.171}
\setsymbol{LFI:FWHM:uncertainty:LFI24}{0.652}
\setsymbol{LFI:FWHM:uncertainty:LFI25}{1.075}
\setsymbol{LFI:FWHM:uncertainty:LFI26}{1.131}
\setsymbol{LFI:FWHM:uncertainty:LFI27}{1.266}
\setsymbol{LFI:FWHM:uncertainty:LFI28}{1.287}

\setsymbol{HFI:center:frequency:100GHz:units}{100\,GHz}
\setsymbol{HFI:center:frequency:143GHz:units}{143\,GHz}
\setsymbol{HFI:center:frequency:217GHz:units}{217\,GHz}
\setsymbol{HFI:center:frequency:353GHz:units}{353\,GHz}
\setsymbol{HFI:center:frequency:545GHz:units}{545\,GHz}
\setsymbol{HFI:center:frequency:857GHz:units}{857\,GHz}

\setsymbol{HFI:center:frequency:100GHz}{100}
\setsymbol{HFI:center:frequency:143GHz}{143}
\setsymbol{HFI:center:frequency:217GHz}{217}
\setsymbol{HFI:center:frequency:353GHz}{353}
\setsymbol{HFI:center:frequency:545GHz}{545}
\setsymbol{HFI:center:frequency:857GHz}{857}

\setsymbol{HFI:Ndetectors:100GHz}{8}
\setsymbol{HFI:Ndetectors:143GHz}{11}
\setsymbol{HFI:Ndetectors:217GHz}{12}
\setsymbol{HFI:Ndetectors:353GHz}{12}
\setsymbol{HFI:Ndetectors:545GHz}{3}
\setsymbol{HFI:Ndetectors:857GHz}{4}

\setsymbol{HFI:FWHM:Maps:100GHz:units}{9\parcm88}
\setsymbol{HFI:FWHM:Maps:143GHz:units}{7\parcm18}
\setsymbol{HFI:FWHM:Maps:217GHz:units}{4\parcm87}
\setsymbol{HFI:FWHM:Maps:353GHz:units}{4\parcm65}
\setsymbol{HFI:FWHM:Maps:545GHz:units}{4\parcm72}
\setsymbol{HFI:FWHM:Maps:857GHz:units}{4\parcm39}
\setsymbol{HFI:FWHM:Maps:100GHz}{9.88}
\setsymbol{HFI:FWHM:Maps:143GHz}{7.18}
\setsymbol{HFI:FWHM:Maps:217GHz}{4.87}
\setsymbol{HFI:FWHM:Maps:353GHz}{4.65}
\setsymbol{HFI:FWHM:Maps:545GHz}{4.72}
\setsymbol{HFI:FWHM:Maps:857GHz}{4.39}

\setsymbol{HFI:beam:ellipticity:Maps:100GHz}{1.15}
\setsymbol{HFI:beam:ellipticity:Maps:143GHz}{1.01}
\setsymbol{HFI:beam:ellipticity:Maps:217GHz}{1.06}
\setsymbol{HFI:beam:ellipticity:Maps:353GHz}{1.05}
\setsymbol{HFI:beam:ellipticity:Maps:545GHz}{1.14}
\setsymbol{HFI:beam:ellipticity:Maps:857GHz}{1.19}

\setsymbol{HFI:FWHM:Mars:100GHz:units}{9\parcm37}
\setsymbol{HFI:FWHM:Mars:143GHz:units}{7\parcm04}
\setsymbol{HFI:FWHM:Mars:217GHz:units}{4\parcm68}
\setsymbol{HFI:FWHM:Mars:353GHz:units}{4\parcm43}
\setsymbol{HFI:FWHM:Mars:545GHz:units}{3\parcm80}
\setsymbol{HFI:FWHM:Mars:857GHz:units}{3\parcm67}

\setsymbol{HFI:FWHM:Mars:100GHz}{9.37}
\setsymbol{HFI:FWHM:Mars:143GHz}{7.04}
\setsymbol{HFI:FWHM:Mars:217GHz}{4.68}
\setsymbol{HFI:FWHM:Mars:353GHz}{4.43}
\setsymbol{HFI:FWHM:Mars:545GHz}{3.80}
\setsymbol{HFI:FWHM:Mars:857GHz}{3.67}

\setsymbol{HFI:beam:ellipticity:Mars:100GHz}{1.18}
\setsymbol{HFI:beam:ellipticity:Mars:143GHz}{1.03}
\setsymbol{HFI:beam:ellipticity:Mars:217GHz}{1.14}
\setsymbol{HFI:beam:ellipticity:Mars:353GHz}{1.09}
\setsymbol{HFI:beam:ellipticity:Mars:545GHz}{1.25}
\setsymbol{HFI:beam:ellipticity:Mars:857GHz}{1.03}

\setsymbol{HFI:CMB:relative:calibration:100GHz}{$\lsim 1\%$}
\setsymbol{HFI:CMB:relative:calibration:143GHz}{$\lsim 1\%$}
\setsymbol{HFI:CMB:relative:calibration:217GHz}{$\lsim 1\%$}
\setsymbol{HFI:CMB:relative:calibration:353GHz}{$\lsim 1\%$}
\setsymbol{HFI:CMB:relative:calibration:545GHz}{}
\setsymbol{HFI:CMB:relative:calibration:857GHz}{}

\setsymbol{HFI:CMB:absolute:calibration:100GHz}{$\lsim 2\%$}
\setsymbol{HFI:CMB:absolute:calibration:143GHz}{$\lsim 2\%$}
\setsymbol{HFI:CMB:absolute:calibration:217GHz}{$\lsim 2\%$}
\setsymbol{HFI:CMB:absolute:calibration:353GHz}{$\lsim 2\%$}
\setsymbol{HFI:CMB:absolute:calibration:545GHz}{}
\setsymbol{HFI:CMB:absolute:calibration:857GHz}{}

\setsymbol{HFI:FIRAS:gain:calibration:accuracy:statistical:100GHz}{}
\setsymbol{HFI:FIRAS:gain:calibration:accuracy:statistical:143GHz}{}
\setsymbol{HFI:FIRAS:gain:calibration:accuracy:statistical:217GHz}{}
\setsymbol{HFI:FIRAS:gain:calibration:accuracy:statistical:353GHz}{2.5\%}
\setsymbol{HFI:FIRAS:gain:calibration:accuracy:statistical:545GHz}{1\%}
\setsymbol{HFI:FIRAS:gain:calibration:accuracy:statistical:857GHz}{0.5\%}

\setsymbol{HFI:FIRAS:gain:calibration:accuracy:systematic:100GHz}{}
\setsymbol{HFI:FIRAS:gain:calibration:accuracy:systematic:143GHz}{}
\setsymbol{HFI:FIRAS:gain:calibration:accuracy:systematic:217GHz}{}
\setsymbol{HFI:FIRAS:gain:calibration:accuracy:systematic:353GHz}{}
\setsymbol{HFI:FIRAS:gain:calibration:accuracy:systematic:545GHz}{7\%}
\setsymbol{HFI:FIRAS:gain:calibration:accuracy:systematic:857GHz}{7\%}

\setsymbol{HFI:FIRAS:zero:point:accuracy:100GHz:units}{0.8\MJysr}
\setsymbol{HFI:FIRAS:zero:point:accuracy:143GHz:units}{}
\setsymbol{HFI:FIRAS:zero:point:accuracy:217GHz:units}{}
\setsymbol{HFI:FIRAS:zero:point:accuracy:353GHz:units}{1.4\MJysr}
\setsymbol{HFI:FIRAS:zero:point:accuracy:545GHz:units}{2.2\MJysr}
\setsymbol{HFI:FIRAS:zero:point:accuracy:857GHz:units}{1.7\MJysr}

\setsymbol{HFI:FIRAS:zero:point:accuracy:100GHz}{0.8}
\setsymbol{HFI:FIRAS:zero:point:accuracy:143GHz}{}
\setsymbol{HFI:FIRAS:zero:point:accuracy:217GHz}{}
\setsymbol{HFI:FIRAS:zero:point:accuracy:353GHz}{1.4}
\setsymbol{HFI:FIRAS:zero:point:accuracy:545GHz}{2.2}
\setsymbol{HFI:FIRAS:zero:point:accuracy:857GHz}{1.7}

\setsymbol{HFI:unit:conversion:100GHz:units}{0.2415\MJysrmK}
\setsymbol{HFI:unit:conversion:143GHz:units}{0.3694\MJysrmK}
\setsymbol{HFI:unit:conversion:217GHz:units}{0.4811\MJysrmK}
\setsymbol{HFI:unit:conversion:353GHz:units}{0.2883\MJysrmK}
\setsymbol{HFI:unit:conversion:545GHz:units}{0.05826\MJysrmK}
\setsymbol{HFI:unit:conversion:857GHz:units}{0.002238\MJysrmK}

\setsymbol{HFI:unit:conversion:100GHz}{0.2415}
\setsymbol{HFI:unit:conversion:143GHz}{0.3694}
\setsymbol{HFI:unit:conversion:217GHz}{0.4811}
\setsymbol{HFI:unit:conversion:353GHz}{0.2883}
\setsymbol{HFI:unit:conversion:545GHz}{0.05826}
\setsymbol{HFI:unit:conversion:857GHz}{0.002238}

\setsymbol{HFI:colour:correction:alpha=-2:V1.01:100GHz}{0.9893}
\setsymbol{HFI:colour:correction:alpha=-2:V1.01:143GHz}{0.9759}
\setsymbol{HFI:colour:correction:alpha=-2:V1.01:217GHz}{1.0007}
\setsymbol{HFI:colour:correction:alpha=-2:V1.01:353GHz}{1.0028}
\setsymbol{HFI:colour:correction:alpha=-2:V1.01:545GHz}{1.0019}
\setsymbol{HFI:colour:correction:alpha=-2:V1.01:857GHz}{0.9889}

\setsymbol{HFI:colour:correction:alpha=0:V1.01:100GHz}{1.0008}
\setsymbol{HFI:colour:correction:alpha=0:V1.01:143GHz}{1.0148}
\setsymbol{HFI:colour:correction:alpha=0:V1.01:217GHz}{0.9909}
\setsymbol{HFI:colour:correction:alpha=0:V1.01:353GHz}{0.9888}
\setsymbol{HFI:colour:correction:alpha=0:V1.01:545GHz}{0.9878}
\setsymbol{HFI:colour:correction:alpha=0:V1.01:857GHz}{1.0014}

\providecommand{\sorthelp}[1]{}

%% file: AuthorList_P12_Lensing_authors_and_institutes.tex
\author{\small
Planck Collaboration:
P.~A.~R.~Ade\inst{92}
\and
N.~Aghanim\inst{65}
\and
C.~Armitage-Caplan\inst{98}
\and
M.~Arnaud\inst{78}
\and
M.~Ashdown\inst{75, 6}
\and
F.~Atrio-Barandela\inst{19}
\and
J.~Aumont\inst{65}
\and
C.~Baccigalupi\inst{91}
\and
A.~J.~Banday\inst{101, 10}
\and
R.~B.~Barreiro\inst{72}
% ...
}
\institute{\small
APC, AstroParticule et Cosmologie, Universit\'{e} Paris Diderot, CNRS/IN2P3, CEA/lrfu, Observatoire de Paris, Sorbonne Paris Cit\'{e}, 10, rue Alice Domon et L\'{e}onie Duquet, 75205 Paris Cedex 13, France\\
\and
Aalto University Mets\"{a}hovi Radio Observatory, Mets\"{a}hovintie 114, FIN-02540 Kylm\"{a}l\"{a}, Finland\\
\and
African Institute for Mathematical Sciences, 6-8 Melrose Road, Muizenberg, Cape Town, South Africa\\
\and
Agenzia Spaziale Italiana Science Data Center, Via del Politecnico snc, 00133, Roma, Italy\\
\and
Agenzia Spaziale Italiana, Viale Liegi 26, Roma, Italy\\
\and
Astrophysics Group, Cavendish Laboratory, University of Cambridge, J J Thomson Avenue, Cambridge CB3 0HE, U.K.\\
\and
Astrophysics \& Cosmology Research Unit, School of Mathematics, Statistics \& Computer Science, University of KwaZulu-Natal, Westville Campus, Private Bag X54001, Durban 4000, South Africa\\
\and
Atacama Large Millimeter/submillimeter Array, ALMA Santiago Central Offices, Alonso de Cordova 3107, Vitacura, Casilla 763 0355, Santiago, Chile\\
\and
CITA, University of Toronto, 60 St. George St., Toronto, ON M5S 3H8, Canada\\
\and
CNRS, IRAP, 9 Av. colonel Roche, BP 44346, F-31028 Toulouse cedex 4, France\\
\and
California Institute of Technology, Pasadena, California, U.S.A.\\
\and
Centre for Theoretical Cosmology, DAMTP, University of Cambridge, Wilberforce Road, Cambridge CB3 0WA, U.K.\\
\and
Centro de Estudios de F\'{i}sica del Cosmos de Arag\'{o}n (CEFCA), Plaza San Juan, 1, planta 2, E-44001, Teruel, Spain\\
\and
Computational Cosmology Center, Lawrence Berkeley National Laboratory, Berkeley, California, U.S.A.\\
\and
Consejo Superior de Investigaciones Cient\'{\i}ficas (CSIC), Madrid, Spain\\
\and
DSM/Irfu/SPP, CEA-Saclay, F-91191 Gif-sur-Yvette Cedex, France\\
\and
DTU Space, National Space Institute, Technical University of Denmark, Elektrovej 327, DK-2800 Kgs. Lyngby, Denmark\\
\and
D\'{e}partement de Physique Th\'{e}orique, Universit\'{e} de Gen\`{e}ve, 24, Quai E. Ansermet,1211 Gen\`{e}ve 4, Switzerland\\
\and
Departamento de F\'{\i}sica Fundamental, Facultad de Ciencias, Universidad de Salamanca, 37008 Salamanca, Spain\\
\and
Departamento de F\'{\i}sica, Universidad de Oviedo, Avda. Calvo Sotelo s/n, Oviedo, Spain\\
\and
Department of Astronomy and Astrophysics, University of Toronto, 50 Saint George Street, Toronto, Ontario, Canada\\
\and
Department of Astrophysics/IMAPP, Radboud University Nijmegen, P.O. Box 9010, 6500 GL Nijmegen, The Netherlands\\
\and
Department of Electrical Engineering and Computer Sciences, University of California, Berkeley, California, U.S.A.\\
\and
Department of Physics \& Astronomy, University of British Columbia, 6224 Agricultural Road, Vancouver, British Columbia, Canada\\
\and
Department of Physics and Astronomy, Dana and David Dornsife College of Letter, Arts and Sciences, University of Southern California, Los Angeles, CA 90089, U.S.A.\\
\and
Department of Physics and Astronomy, University College London, London WC1E 6BT, U.K.\\
\and
Department of Physics and Astronomy, University of Sussex, Brighton BN1 9QH, U.K.\\
\and
Department of Physics, Carnegie Mellon University, 5000 Forbes Ave, Pittsburgh, PA 15213, U.S.A.\\
\and
Department of Physics, Florida State University, Keen Physics Building, 77 Chieftan Way, Tallahassee, Florida, U.S.A.\\
\and
Department of Physics, Gustaf H\"{a}llstr\"{o}min katu 2a, University of Helsinki, Helsinki, Finland\\
\and
Department of Physics, Princeton University, Princeton, New Jersey, U.S.A.\\
\and
Department of Physics, University of California, Berkeley, California, U.S.A.\\
\and
Department of Physics, University of California, One Shields Avenue, Davis, California, U.S.A.\\
\and
Department of Physics, University of California, Santa Barbara, California, U.S.A.\\
\and
Department of Physics, University of Illinois at Urbana-Champaign, 1110 West Green Street, Urbana, Illinois, U.S.A.\\
\and
Dipartimento di Fisica e Astronomia G. Galilei, Universit\`{a} degli Studi di Padova, via Marzolo 8, 35131 Padova, Italy\\
\and
Dipartimento di Fisica e Scienze della Terra, Universit\`{a} di Ferrara, Via Saragat 1, 44122 Ferrara, Italy\\
\and
Dipartimento di Fisica, Universit\`{a} La Sapienza, P. le A. Moro 2, Roma, Italy\\
\and
Dipartimento di Fisica, Universit\`{a} degli Studi di Milano, Via Celoria, 16, Milano, Italy\\
\and
Dipartimento di Fisica, Universit\`{a} degli Studi di Trieste, via A. Valerio 2, Trieste, Italy\\
\and
Dipartimento di Fisica, Universit\`{a} di Roma Tor Vergata, Via della Ricerca Scientifica, 1, Roma, Italy\\
\and
Discovery Center, Niels Bohr Institute, Blegdamsvej 17, Copenhagen, Denmark\\
\and
Dpto. Astrof\'{i}sica, Universidad de La Laguna (ULL), E-38206 La Laguna, Tenerife, Spain\\
\and
European Southern Observatory, ESO Vitacura, Alonso de Cordova 3107, Vitacura, Casilla 19001, Santiago, Chile\\
\and
European Space Agency, ESAC, Planck Science Office, Camino bajo del Castillo, s/n, Urbanizaci\'{o}n Villafranca del Castillo, Villanueva de la Ca\~{n}ada, Madrid, Spain\\
\and
European Space Agency, ESTEC, Keplerlaan 1, 2201 AZ Noordwijk, The Netherlands\\
\and
Finnish Centre for Astronomy with ESO (FINCA), University of Turku, V\"{a}is\"{a}l\"{a}ntie 20, FIN-21500, Piikki\"{o}, Finland\\
\and
Haverford College Astronomy Department, 370 Lancaster Avenue, Haverford, Pennsylvania, U.S.A.\\
\and
Helsinki Institute of Physics, Gustaf H\"{a}llstr\"{o}min katu 2, University of Helsinki, Helsinki, Finland\\
\and
INAF - Osservatorio Astrofisico di Catania, Via S. Sofia 78, Catania, Italy\\
\and
INAF - Osservatorio Astronomico di Padova, Vicolo dell'Osservatorio 5, Padova, Italy\\
\and
INAF - Osservatorio Astronomico di Roma, via di Frascati 33, Monte Porzio Catone, Italy\\
\and
INAF - Osservatorio Astronomico di Trieste, Via G.B. Tiepolo 11, Trieste, Italy\\
\and
INAF Istituto di Radioastronomia, Via P. Gobetti 101, 40129 Bologna, Italy\\
\and
INAF/IASF Bologna, Via Gobetti 101, Bologna, Italy\\
\and
INAF/IASF Milano, Via E. Bassini 15, Milano, Italy\\
\and
INFN, Sezione di Bologna, Via Irnerio 46, I-40126, Bologna, Italy\\
\and
INFN, Sezione di Roma 1, Universit\`{a} di Roma Sapienza, Piazzale Aldo Moro 2, 00185, Roma, Italy\\
\and
IPAG: Institut de Plan\'{e}tologie et d'Astrophysique de Grenoble, Universit\'{e} Joseph Fourier, Grenoble 1 / CNRS-INSU, UMR 5274, Grenoble, F-38041, France\\
\and
IUCAA, Post Bag 4, Ganeshkhind, Pune University Campus, Pune 411 007, India\\
\and
Imperial College London, Astrophysics group, Blackett Laboratory, Prince Consort Road, London, SW7 2AZ, U.K.\\
\and
Infrared Processing and Analysis Center, California Institute of Technology, Pasadena, CA 91125, U.S.A.\\
\and
Institut N\'{e}el, CNRS, Universit\'{e} Joseph Fourier Grenoble I, 25 rue des Martyrs, Grenoble, France\\
\and
Institut Universitaire de France, 103, bd Saint-Michel, 75005, Paris, France\\
\and
Institut d'Astrophysique Spatiale, CNRS (UMR8617) Universit\'{e} Paris-Sud 11, B\^{a}timent 121, Orsay, France\\
\and
Institut d'Astrophysique de Paris, CNRS (UMR7095), 98 bis Boulevard Arago, F-75014, Paris, France\\
\and
Institute for Space Sciences, Bucharest-Magurale, Romania\\
\and
Institute of Astronomy and Astrophysics, Academia Sinica, Taipei, Taiwan\\
\and
Institute of Astronomy, University of Cambridge, Madingley Road, Cambridge CB3 0HA, U.K.\\
\and
Institute of Theoretical Astrophysics, University of Oslo, Blindern, Oslo, Norway\\
\and
Instituto de Astrof\'{\i}sica de Canarias, C/V\'{\i}a L\'{a}ctea s/n, La Laguna, Tenerife, Spain\\
\and
Instituto de F\'{\i}sica de Cantabria (CSIC-Universidad de Cantabria), Avda. de los Castros s/n, Santander, Spain\\
\and
Jet Propulsion Laboratory, California Institute of Technology, 4800 Oak Grove Drive, Pasadena, California, U.S.A.\\
\and
Jodrell Bank Centre for Astrophysics, Alan Turing Building, School of Physics and Astronomy, The University of Manchester, Oxford Road, Manchester, M13 9PL, U.K.\\
\and
Kavli Institute for Cosmology Cambridge, Madingley Road, Cambridge, CB3 0HA, U.K.\\
\and
LAL, Universit\'{e} Paris-Sud, CNRS/IN2P3, Orsay, France\\
\and
LERMA, CNRS, Observatoire de Paris, 61 Avenue de l'Observatoire, Paris, France\\
\and
Laboratoire AIM, IRFU/Service d'Astrophysique - CEA/DSM - CNRS - Universit\'{e} Paris Diderot, B\^{a}t. 709, CEA-Saclay, F-91191 Gif-sur-Yvette Cedex, France\\
\and
Laboratoire Traitement et Communication de l'Information, CNRS (UMR 5141) and T\'{e}l\'{e}com ParisTech, 46 rue Barrault F-75634 Paris Cedex 13, France\\
\and
Laboratoire de Physique Subatomique et de Cosmologie, Universit\'{e} Joseph Fourier Grenoble I, CNRS/IN2P3, Institut National Polytechnique de Grenoble, 53 rue des Martyrs, 38026 Grenoble cedex, France\\
\and
Laboratoire de Physique Th\'{e}orique, Universit\'{e} Paris-Sud 11 \& CNRS, B\^{a}timent 210, 91405 Orsay, France\\
\and
Lawrence Berkeley National Laboratory, Berkeley, California, U.S.A.\\
\and
Max-Planck-Institut f\"{u}r Astrophysik, Karl-Schwarzschild-Str. 1, 85741 Garching, Germany\\
\and
McGill Physics, Ernest Rutherford Physics Building, McGill University, 3600 rue University, Montr\'{e}al, QC, H3A 2T8, Canada\\
\and
MilliLab, VTT Technical Research Centre of Finland, Tietotie 3, Espoo, Finland\\
\and
National University of Ireland, Department of Experimental Physics, Maynooth, Co. Kildare, Ireland\\
\and
Niels Bohr Institute, Blegdamsvej 17, Copenhagen, Denmark\\
\and
Observational Cosmology, Mail Stop 367-17, California Institute of Technology, Pasadena, CA, 91125, U.S.A.\\
\and
Optical Science Laboratory, University College London, Gower Street, London, U.K.\\
\and
SB-ITP-LPPC, EPFL, CH-1015, Lausanne, Switzerland\\
\and
SISSA, Astrophysics Sector, via Bonomea 265, 34136, Trieste, Italy\\
\and
School of Physics and Astronomy, Cardiff University, Queens Buildings, The Parade, Cardiff, CF24 3AA, U.K.\\
\and
School of Physics and Astronomy, University of Nottingham, Nottingham NG7 2RD, U.K.\\
\and
Space Research Institute (IKI), Russian Academy of Sciences, Profsoyuznaya Str, 84/32, Moscow, 117997, Russia\\
\and
Space Sciences Laboratory, University of California, Berkeley, California, U.S.A.\\
\and
Special Astrophysical Observatory, Russian Academy of Sciences, Nizhnij Arkhyz, Zelenchukskiy region, Karachai-Cherkessian Republic, 369167, Russia\\
\and
Stanford University, Dept of Physics, Varian Physics Bldg, 382 Via Pueblo Mall, Stanford, California, U.S.A.\\
\and
Sub-Department of Astrophysics, University of Oxford, Keble Road, Oxford OX1 3RH, U.K.\\
\and
Theory Division, PH-TH, CERN, CH-1211, Geneva 23, Switzerland\\
\and
UPMC Univ Paris 06, UMR7095, 98 bis Boulevard Arago, F-75014, Paris, France\\
\and
Universit\'{e} de Toulouse, UPS-OMP, IRAP, F-31028 Toulouse cedex 4, France\\
% ...
}

%% file: planck_lensing.bbl
\def\eprinttmppp@#1arXiv:@{#1}
\providecommand{\arxivlink[1]}{\href{http://arxiv.org/abs/#1}{arXiv:#1}}
\def\eprinttmp@#1arXiv:#2 [#3]#4@{\ifthenelse{\equal{#3}{x}}{\ifthenelse{
\equal{#1}{}}{\arxivlink{\eprinttmppp@#2@}}{\arxivlink{#1}}}{\arxivlink{#2}
  [#3]}}
\providecommand{\eprintlink}[1]{\eprinttmp@#1arXiv: [x]@}
\providecommand{\eprint}[1]{\eprintlink{#1}}
\providecommand{\adsurl}[1]{\href{#1}{ADS}}
\begin{thebibliography}{122}
\expandafter\ifx\csname natexlab\endcsname\relax\def\natexlab#1{#1}\fi

\bibitem[{{Abrial} {et~al.}(2007){Abrial}, {Moudden}, {Starck}, {Bobin},
  {Fadili}, {Afeyan}, \& {Nguyen}}]{inpainting:abrial06}
{Abrial}, P., {Moudden}, Y., {Starck}, J., {et~al.}, {Morphological Component
  Analysis and Inpainting on the sphere: Application in Physics and
  Astrophysics}. 2007, Journal of Fourier Analysis and Applications, 13, 729

\bibitem[{{Abrial} {et~al.}(2008){Abrial}, {Moudden}, {Starck}, {Fadili},
  {Delabrouille}, \& {Nguyen}}]{Abrial:2008}
{Abrial}, P., {Moudden}, Y., {Starck}, J.-L., {et~al.}, {CMB data analysis and
  sparsity}. 2008, Statistical Methodology, 5, 289, arXiv:0804.1295

\bibitem[{Basak {et~al.}(2009)Basak, Prunet, \& Benabed}]{Basak:2008pq}
Basak, S., Prunet, S., \& Benabed, K., {Simulating weak lensing on CMB maps}.
  2009, \aap, 508, 2213, \eprint{0811.1677}

\bibitem[{Bauer {et~al.}(2012)Bauer, Baltay, Ellman, Jerke, Rabinowitz,
  {et~al.}}]{Bauer:2012rs}
Bauer, A.~H., Baltay, C., Ellman, N., {et~al.}, {The Mass-Richness Relation of
  MaxBCG Clusters from Quasar Lensing Magnification using Variability}. 2012,
  \apj, 749, 56, \eprint{1202.1371}

\bibitem[{Bennett {et~al.}(1996)Bennett, Banday, Gorski, Hinshaw, Jackson,
  {et~al.}}]{Bennett:1996ce}
Bennett, C., Banday, A., Gorski, K., {et~al.}, {Four year COBE DMR cosmic
  microwave background observations: Maps and basic results}. 1996, \apj, 464,
  L1, \eprint{astro-ph/9601067}

\bibitem[{Bennett {et~al.}(2011)Bennett, Hill, Hinshaw, Larson, Smith,
  {et~al.}}]{Bennett:2010jb}
Bennett, C., Hill, R., Hinshaw, G., {et~al.}, {Seven-Year Wilkinson Microwave
  Anisotropy Probe (WMAP) Observations: Are There Cosmic Microwave Background
  Anomalies?} 2011, \apjs, 192, 17, \eprint{1001.4758}

\bibitem[{Bennett {et~al.}(2013)}]{Bennett:2012zja}
Bennett, C. {et~al.}, {Nine-Year Wilkinson Microwave Anisotropy Probe (WMAP)
  Observations: Final Maps and Results}. 2013, \apjs, 208, 20,
  \eprint{1212.5225}

\bibitem[{Benoit-Levy {et~al.}(2013)Benoit-Levy, Dechelette, Benabed, Cardoso,
  Hanson, {et~al.}}]{BenoitLevy:2013bc}
Benoit-Levy, A., Dechelette, T., Benabed, K., {et~al.}, {Full-sky CMB lensing
  reconstruction in presence of sky-cuts}. 2013, \aap, 555, A37,
  \eprint{1301.4145}

\bibitem[{Bernardeau(1997)}]{Bernardeau:1996aa}
Bernardeau, F., {Weak lensing detection in CMB maps}. 1997, \aap, 324, 15,
  \eprint{astro-ph/9611012}

\bibitem[{{Blanchard} \& {Schneider}(1987)}]{1987A&A...184....1B}
{Blanchard}, A. \& {Schneider}, J., {Gravitational lensing effect on the
  fluctuations of the cosmic background radiation}. 1987, \aap, 184, 1

\bibitem[{Bleem {et~al.}(2012)Bleem, van Engelen, Holder, Aird, Armstrong,
  {et~al.}}]{Bleem:2012gm}
Bleem, L., van Engelen, A., Holder, G., {et~al.}, {A Measurement of the
  Correlation of Galaxy Surveys with CMB Lensing Convergence Maps from the
  South Pole Telescope}. 2012, \apjl, 753, L9, \eprint{1203.4808}

\bibitem[{Bucher {et~al.}(2012)Bucher, Carvalho, Moodley, \&
  Remazeilles}]{Bucher:2010iv}
Bucher, M., Carvalho, C.~S., Moodley, K., \& Remazeilles, M., {CMB Lensing
  Reconstruction in Real Space}. 2012, \prd, 85, 043016, \eprint{1004.3285}

\bibitem[{Challinor \& Lewis(2005)}]{Challinor:2005jy}
Challinor, A. \& Lewis, A., {Lensed CMB power spectra from all-sky correlation
  functions}. 2005, \prd, 71, 103010, \eprint{astro-ph/0502425}

\bibitem[{Challinor \& van Leeuwen(2002)}]{Challinor:2002zh}
Challinor, A. \& van Leeuwen, F., {Peculiar velocity effects in high resolution
  microwave background experiments}. 2002, \prd, 65, 103001,
  \eprint{astro-ph/0112457}

\bibitem[{{Cole} \& {Efstathiou}(1989)}]{1989MNRAS.239..195C}
{Cole}, S. \& {Efstathiou}, G., {Gravitational lensing of fluctuations in the
  microwave background radiation}. 1989, \mnras, 239, 195

\bibitem[{{Condon} {et~al.}(1998){Condon}, {Cotton}, {Greisen}, {Yin},
  {Perley}, {Taylor}, \& {Broderick}}]{NVSS1998}
{Condon}, J.~J., {Cotton}, W.~D., {Greisen}, E.~W., {et~al.}, {The NRAO VLA Sky
  Survey}. 1998, \aj, 115, 1693

\bibitem[{Creminelli {et~al.}(2006)Creminelli, Nicolis, Senatore, Tegmark, \&
  Zaldarriaga}]{Creminelli:2005hu}
Creminelli, P., Nicolis, A., Senatore, L., Tegmark, M., \& Zaldarriaga, M.,
  {Limits on non-gaussianities from wmap data}. 2006, JCAP, 0605, 004,
  \eprint{astro-ph/0509029}

\bibitem[{Das {et~al.}(2013)Das, Louis, Nolta, Addison, Battistelli,
  {et~al.}}]{Das:2013zf}
Das, S., Louis, T., Nolta, M.~R., {et~al.}, {The Atacama Cosmology Telescope:
  Temperature and Gravitational Lensing Power Spectrum Measurements from Three
  Seasons of Data}. 2013, JCAP, submitted, \eprint{1301.1037}

\bibitem[{Das {et~al.}(2011)Das, Sherwin, Aguirre, Appel, Bond,
  {et~al.}}]{Das:2011ak}
Das, S., Sherwin, B.~D., Aguirre, P., {et~al.}, {Detection of the Power
  Spectrum of Cosmic Microwave Background Lensing by the Atacama Cosmology
  Telescope}. 2011, \prl, 107, 021301, \eprint{1103.2124}

\bibitem[{De~Zotti {et~al.}(2010)De~Zotti, Massardi, Negrello, \&
  Wall}]{DeZotti:2009an}
De~Zotti, G., Massardi, M., Negrello, M., \& Wall, J., {Radio and Millimeter
  Continuum Surveys and their Astrophysical Implications}. 2010,
  Astron.Astrophys.Rev., 18, 1, \eprint{0908.1896}

\bibitem[{{Delabrouille} {et~al.}(2012){Delabrouille}, {Betoule}, {Melin},
  {Miville-Desch{\^e}nes}, {Gonzalez-Nuevo}, {Le Jeune}, {Castex}, {de Zotti},
  {Basak}, {Ashdown}, {Aumont}, {Baccigalupi}, {Banday}, {Bernard}, {Bouchet},
  {Clements}, {da Silva}, {Dickinson}, {Dodu}, {Dolag}, {Elsner}, {Fauvet},
  {Fa{\"y}}, {Giardino}, {Leach}, {Lesgourgues}, {Liguori}, {Macias-Perez},
  {Massardi}, {Matarrese}, {Mazzotta}, {Montier}, {Mottet}, {Paladini},
  {Partridge}, {Piffaretti}, {Prezeau}, {Prunet}, {Ricciardi}, {Roman},
  {Schaefer}, \& {Toffolatti}}]{delabrouille2012}
{Delabrouille}, J., {Betoule}, M., {Melin}, J.-B., {et~al.}, {The pre-launch
  Planck Sky Model: a model of sky emission at submillimetre to centimetre
  wavelengths}. 2012, ArXiv e-prints, \eprint{1207.3675}

\bibitem[{{Efstathiou} \& {Bond}(1999)}]{1999MNRAS.304...75E}
{Efstathiou}, G. \& {Bond}, J.~R., {Cosmic confusion: degeneracies among
  cosmological parameters derived from measurements of microwave background
  anisotropies}. 1999, \mnras, 304, 75, \eprint{arXiv:astro-ph/9807103}

\bibitem[{Feng {et~al.}(2012{\natexlab{a}})Feng, Aslanyan, Manohar, Keating,
  Paar, {et~al.}}]{Feng:2012uf}
Feng, C., Aslanyan, G., Manohar, A.~V., {et~al.}, {Measuring Gravitational
  Lensing of the Cosmic Microwave Background using cross-correlation with large
  scale structure}. 2012{\natexlab{a}}, \prd, 86, 063519, \eprint{1207.3326}

\bibitem[{Feng {et~al.}(2012{\natexlab{b}})Feng, Keating, Paar, \&
  Zahn}]{Feng:2011jx}
Feng, C., Keating, B., Paar, H.~P., \& Zahn, O., {Reconstruction of
  Gravitational Lensing Using WMAP 7-Year Data}. 2012{\natexlab{b}}, \prd, 85,
  043513, \eprint{1111.2371}

\bibitem[{{G{\'o}rski} {et~al.}(2005){G{\'o}rski}, {Hivon}, {Banday},
  {Wandelt}, {Hansen}, {Reinecke}, \& {Bartelmann}}]{gorski2005}
{G{\'o}rski}, K.~M., {Hivon}, E., {Banday}, A.~J., {et~al.}, {HEALPix: A
  Framework for High-Resolution Discretization and Fast Analysis of Data
  Distributed on the Sphere}. 2005, \apj, 622, 759,
  \eprint{arXiv:astro-ph/0409513}

\bibitem[{Goto {et~al.}(2012)Goto, Szapudi, \& Granett}]{Goto:2012yc}
Goto, T., Szapudi, I., \& Granett, B.~R., {Cross-correlation of WISE Galaxies
  with the Cosmic Microwave Background}. 2012, \mnras\ Letters, 422,
  \eprint{1202.5306}

\bibitem[{Gruppuso {et~al.}(2011)Gruppuso, Finelli, Natoli, Paci, Cabella,
  {et~al.}}]{Gruppuso:2010nd}
Gruppuso, A., Finelli, F., Natoli, P., {et~al.}, {New constraints on Parity
  Symmetry from a re-analysis of the WMAP-7 low resolution power spectra}.
  2011, \mnras, 411, 1445, \eprint{1006.1979}

\bibitem[{Hall {et~al.}(2010)Hall, Knox, Reichardt, Ade, Aird,
  {et~al.}}]{Hall:2009rv}
Hall, N., Knox, L., Reichardt, C., {et~al.}, {Angular Power Spectra of the
  Millimeter Wavelength Background Light from Dusty Star-forming Galaxies with
  the South Pole Telescope}. 2010, \apj, 718, 632, \eprint{0912.4315}

\bibitem[{Hanson {et~al.}(2011)Hanson, Challinor, Efstathiou, \&
  Bielewicz}]{Hanson:2010rp}
Hanson, D., Challinor, A., Efstathiou, G., \& Bielewicz, P., {CMB temperature
  lensing power reconstruction}. 2011, \prd, 83, 043005, \eprint{1008.4403}

\bibitem[{Hanson {et~al.}(2010)Hanson, Lewis, \& Challinor}]{Hanson:2010gu}
Hanson, D., Lewis, A., \& Challinor, A., {Asymmetric Beams and CMB Statistical
  Anisotropy}. 2010, \prd, 81, 103003, \eprint{1003.0198}

\bibitem[{Hanson {et~al.}(2009)Hanson, Rocha, \& Gorski}]{Hanson:2009dr}
Hanson, D., Rocha, G., \& Gorski, K., {Lensing reconstruction from PLANCK sky
  maps: inhomogeneous noise}. 2009, \mnras, 400, 2169, \eprint{0907.1927}

\bibitem[{Hinshaw {et~al.}(2009)}]{Hinshaw:2008kr}
Hinshaw, G. {et~al.}, {Five-Year Wilkinson Microwave Anisotropy Probe (WMAP)
  Observations: Data Processing, Sky Maps, and Basic Results}. 2009, \apjs,
  180, 225, \eprint{0803.0732}

\bibitem[{Hirata {et~al.}(2008)Hirata, Ho, Padmanabhan, Seljak, \&
  Bahcall}]{Hirata:2008cb}
Hirata, C.~M., Ho, S., Padmanabhan, N., Seljak, U., \& Bahcall, N.~A.,
  {Correlation of CMB with large-scale structure: II. Weak lensing}. 2008,
  \prd, 78, 043520, \eprint{0801.0644}

\bibitem[{Hirata {et~al.}(2004)Hirata, Padmanabhan, Seljak, Schlegel, \&
  Brinkmann}]{Hirata:2004rp}
Hirata, C.~M., Padmanabhan, N., Seljak, U., Schlegel, D., \& Brinkmann, J.,
  {Cross-correlation of CMB with large-scale structure: weak gravitational
  lensing}. 2004, \prd, 70, 103501, \eprint{astro-ph/0406004}

\bibitem[{Hirata \& Seljak(2003{\natexlab{a}})}]{Hirata:2002jy}
Hirata, C.~M. \& Seljak, U., {Analyzing weak lensing of the cosmic microwave
  background using the likelihood function}. 2003{\natexlab{a}}, \prd, 67,
  043001, \eprint{astro-ph/0209489}

\bibitem[{Hirata \& Seljak(2003{\natexlab{b}})}]{Hirata:2003ka}
Hirata, C.~M. \& Seljak, U., {Reconstruction of lensing from the cosmic
  microwave background polarization}. 2003{\natexlab{b}}, \prd, 68, 083002,
  \eprint{astro-ph/0306354}

\bibitem[{Ho {et~al.}(2012)Ho, Cuesta, Seo, de~Putter, Ross,
  {et~al.}}]{Ho:2012vy}
Ho, S., Cuesta, A., Seo, H.-J., {et~al.}, {Clustering of Sloan Digital Sky
  Survey III Photometric Luminous Galaxies: The Measurement, Systematics and
  Cosmological Implications}. 2012, \apj, 761, 14, \eprint{1201.2137}

\bibitem[{{Howlett} {et~al.}(2012){Howlett}, {Lewis}, {Hall}, \&
  {Challinor}}]{2012JCAP...04..027H}
{Howlett}, C., {Lewis}, A., {Hall}, A., \& {Challinor}, A., {CMB power spectrum
  parameter degeneracies in the era of precision cosmology}. 2012, JCAP, 4, 27,
  \eprint{1201.3654}

\bibitem[{Hu(2001)}]{Hu:2001fa}
Hu, W., {Angular trispectrum of the cosmic microwave background}. 2001, \prd,
  64, 083005, \eprint{astro-ph/0105117}

\bibitem[{Hu \& Okamoto(2002)}]{Hu:2001kj}
Hu, W. \& Okamoto, T., {Mass Reconstruction with CMB Polarization}. 2002, \apj,
  574, 566, \eprint{astro-ph/0111606}

\bibitem[{H{\"u}tsi(2009)}]{Huetsi:2009zq}
H{\"u}tsi, G., {Power spectrum of the maxBCG sample: detection of acoustic
  oscillations using galaxy clusters}. 2009, \mnras, 401, 2477,
  \eprint{0910.0492}

\bibitem[{Kamionkowski \& Knox(2003)}]{Kamionkowski:2002nd}
Kamionkowski, M. \& Knox, L., {Aspects of the cosmic microwave background
  dipole}. 2003, \prd, 67, 063001, \eprint{astro-ph/0210165}

\bibitem[{Kaplinghat {et~al.}(2003)Kaplinghat, Knox, \&
  Song}]{Kaplinghat:2003bh}
Kaplinghat, M., Knox, L., \& Song, Y.-S., {Determining neutrino mass from the
  CMB alone}. 2003, \prl, 91, 241301, \eprint{astro-ph/0303344}

\bibitem[{Keisler {et~al.}(2011)Keisler, Reichardt, Aird, Benson, Bleem,
  {et~al.}}]{Keisler:2011aw}
Keisler, R., Reichardt, C., Aird, K., {et~al.}, {A Measurement of the Damping
  Tail of the Cosmic Microwave Background Power Spectrum with the South Pole
  Telescope}. 2011, \apj, 743, 28, \eprint{1105.3182}

\bibitem[{Kesden {et~al.}(2003)Kesden, Cooray, \& Kamionkowski}]{Kesden:2003cc}
Kesden, M.~H., Cooray, A., \& Kamionkowski, M., {Lensing reconstruction with
  CMB temperature and polarization}. 2003, \prd, 67, 123507,
  \eprint{astro-ph/0302536}

\bibitem[{Kim \& Naselsky(2010)}]{Kim:2010gf}
Kim, J. \& Naselsky, P., {Anomalous parity asymmetry of the Wilkinson Microwave
  Anisotropy Probe power spectrum data at low multipoles}. 2010, \apj, 714,
  L265, \eprint{1001.4613}

\bibitem[{Knox \& Song(2002)}]{Knox:2002pe}
Knox, L. \& Song, Y.-S., {A limit on the detectability of the energy scale of
  inflation}. 2002, \prl, 89, 011303, \eprint{astro-ph/0202286}

\bibitem[{Koester {et~al.}(2007)}]{Koester:2007bg}
Koester, B. {et~al.}, {A MaxBCG Catalog of 13,823 Galaxy Clusters from the
  Sloan Digital Sky Survey}. 2007, \apj, 660, 239, \eprint{astro-ph/0701265}

\bibitem[{Komatsu {et~al.}(2005)Komatsu, Spergel, \& Wandelt}]{Komatsu:2003iq}
Komatsu, E., Spergel, D.~N., \& Wandelt, B.~D., {Measuring primordial
  non-Gaussianity in the cosmic microwave background}. 2005, \apj, 634, 14,
  \eprint{astro-ph/0305189}

\bibitem[{Kovacs {et~al.}(2013)Kovacs, Szapudi, Granett, \&
  Frei}]{Kovacs:2013rs}
Kovacs, A., Szapudi, I., Granett, B.~R., \& Frei, Z., {Cross-correlation of
  WMAP7 and the WISE Full Data Release}. 2013, \mnras\ Letters, 431,
  \eprint{1301.0475}

\bibitem[{Land \& Magueijo(2005)}]{Land:2005jq}
Land, K. \& Magueijo, J., {Is the Universe odd?} 2005, \prd, 72, 101302,
  \eprint{astro-ph/0507289}

\bibitem[{Lewis(2005)}]{Lewis:2005tp}
Lewis, A., {Lensed CMB simulation and parameter estimation}. 2005, \prd, 71,
  083008, \eprint{astro-ph/0502469}

\bibitem[{Lewis \& Challinor(2006)}]{Lewis:2006fu}
Lewis, A. \& Challinor, A., {Weak gravitational lensing of the cmb}. 2006,
  Phys.Rept., 429, 1, \eprint{astro-ph/0601594}

\bibitem[{Lewis {et~al.}(2011)Lewis, Challinor, \& Hanson}]{Lewis:2011fk}
Lewis, A., Challinor, A., \& Hanson, D., {The shape of the CMB lensing
  bispectrum}. 2011, JCAP, 1103, 018, \eprint{1101.2234}

\bibitem[{Limber(1954)}]{Limber1954}
Limber, D.~N. 1954, \apj, 119, 655

\bibitem[{{Linder}(1990)}]{1990MNRAS.243..353L}
{Linder}, E.~V., {Analysis of gravitationally lensed microwave background
  anisotropies}. 1990, \mnras, 243, 353

\bibitem[{Metcalf \& Silk(1997)}]{Metcalf:1997ih}
Metcalf, R.~B. \& Silk, J., {Gravitational Magnification of the Cosmic
  Microwave Background}. 1997, \apj, 489, 1, \eprint{astro-ph/9708059}

\bibitem[{Munshi {et~al.}(2011{\natexlab{a}})Munshi, Heavens, Cooray, Smidt,
  Coles, {et~al.}}]{Munshi:2009wy}
Munshi, D., Heavens, A., Cooray, A., {et~al.}, {New Optimised Estimators for
  the Primordial Trispectrum}. 2011{\natexlab{a}}, \mnras, 412, 1993,
  \eprint{0910.3693}

\bibitem[{Munshi {et~al.}(2011{\natexlab{b}})Munshi, Valageas, Cooray, \&
  Heavens}]{Munshi:2009fr}
Munshi, D., Valageas, P., Cooray, A., \& Heavens, A., {Secondary
  non-Gaussianity and Cross-Correlation Analysis}. 2011{\natexlab{b}}, \mnras,
  414, 3173, \eprint{0907.3229}

\bibitem[{Namikawa {et~al.}(2013)Namikawa, Hanson, \&
  Takahashi}]{Namikawa:2012pe}
Namikawa, T., Hanson, D., \& Takahashi, R., {Bias-Hardened CMB Lensing}. 2013,
  \mnras, 431, 609, \eprint{1209.0091}

\bibitem[{Namikawa {et~al.}(2012)Namikawa, Yamauchi, \&
  Taruya}]{Namikawa:2011cs}
Namikawa, T., Yamauchi, D., \& Taruya, A., {Full-sky lensing reconstruction of
  gradient and curl modes from CMB maps}. 2012, JCAP, 1201, 007,
  \eprint{1110.1718}

\bibitem[{Okamoto \& Hu(2003)}]{Okamoto:2003zw}
Okamoto, T. \& Hu, W., {CMB Lensing Reconstruction on the Full Sky}. 2003,
  \prd, 67, 083002, \eprint{astro-ph/0301031}

\bibitem[{{Osborne} {et~al.}(2014){Osborne}, {Hanson}, \&
  {Dor{\'e}}}]{2013arXiv1310.7547O}
{Osborne}, S.~J., {Hanson}, D., \& {Dor{\'e}}, O., {Extragalactic Foreground
  Contamination in Temperature-based CMB Lens Reconstruction}. 2014, JCAP, 03,
  \eprint{1310.7547}

\bibitem[{{Page} {et~al.}(2007){Page}, {Hinshaw}, {Komatsu}, {Nolta},
  {Spergel}, {Bennett}, {Barnes}, {Bean}, {Dor{\'e}}, {Dunkley}, {Halpern},
  {Hill}, {Jarosik}, {Kogut}, {Limon}, {Meyer}, {Odegard}, {Peiris}, {Tucker},
  {Verde}, {Weiland}, {Wollack}, \& {Wright}}]{2007ApJS..170..335P}
{Page}, L., {Hinshaw}, G., {Komatsu}, E., {et~al.}, {Three-Year Wilkinson
  Microwave Anisotropy Probe (WMAP) Observations: Polarization Analysis}. 2007,
  \apjs, 170, 335, \eprint{arXiv:astro-ph/0603450}

\bibitem[{Perotto {et~al.}(2010)Perotto, Bobin, Plaszczynski, Starck, \&
  Lavabre}]{Perotto:2009tv}
Perotto, L., Bobin, J., Plaszczynski, S., Starck, J.~L., \& Lavabre, A.,
  {Reconstruction of the CMB lensing for Planck}. 2010, \aap, 519, 14,
  \eprint{0903.1308}

\bibitem[{Perotto {et~al.}(2006)Perotto, Lesgourgues, Hannestad, Tu, \&
  Wong}]{Perotto:2006rj}
Perotto, L., Lesgourgues, J., Hannestad, S., Tu, H., \& Wong, Y. Y.~Y.,
  {Probing cosmological parameters with the CMB: Forecasts from full Monte
  Carlo simulations}. 2006, JCAP, 0610, 013, \eprint{astro-ph/0606227}

\bibitem[{{Planck Collaboration}(2013)}]{planck2013-p28}
{Planck Collaboration}. 2013, {The Explanatory Supplement to the Planck 2013
  results,
  http://www.sciops.esa.int/wikiSI/planckpla/index.php?title=Main\_Page}
  ({ESA})

\bibitem[{{\sorthelp{Planck Collaboration 2011G}}{Planck Collaboration
  VII}(2011)}]{planck2011-1.10}
{\sorthelp{Planck Collaboration 2011G}}{Planck Collaboration VII}, {Planck
  early results. VII. The Early Release Compact Source Catalogue}. 2011, \aap,
  536, A7

\bibitem[{{\sorthelp{Planck Collaboration 2011R}}{Planck Collaboration
  XVIII}(2011)}]{planck2011-6.6}
{\sorthelp{Planck Collaboration 2011R}}{Planck Collaboration XVIII}, {Planck
  early results. XVIII. The power spectrum of cosmic infrared background
  anisotropies}. 2011, \aap, 536, A18

\bibitem[{{\sorthelp{Planck Collaboration 2011X}}{Planck Collaboration
  XXIV}(2011)}]{planck2011-7.12}
{\sorthelp{Planck Collaboration 2011X}}{Planck Collaboration XXIV}, {Planck
  early results. XXIV. Dust in the diffuse interstellar medium and the Galactic
  halo}. 2011, \aap, 536, A24

\bibitem[{{\sorthelp{Planck Collaboration 2013A}}{Planck Collaboration
  I}(2014)}]{planck2013-p01}
{\sorthelp{Planck Collaboration 2013A}}{Planck Collaboration I},
  {\textit{Planck} 2013 results: Overview of Planck Products and Scientific
  Results}. 2014, \aap, in press, \eprint{arXiv:1303.5062}

\bibitem[{{\sorthelp{Planck Collaboration 2013B}}{Planck Collaboration
  II}(2014)}]{planck2013-p02}
{\sorthelp{Planck Collaboration 2013B}}{Planck Collaboration II},
  {\textit{Planck} 2013 results: The Low Frequency Instrument data processing}.
  2014, \aap, in press, \eprint{arXiv:1303.5063}

\bibitem[{{\sorthelp{Planck Collaboration 2013C}}{Planck Collaboration
  III}(2014)}]{planck2013-p02a}
{\sorthelp{Planck Collaboration 2013C}}{Planck Collaboration III},
  {\textit{Planck} 2013 results: LFI systematic uncertainties}. 2014, \aap, in
  press, \eprint{arXiv:1303.5064}

\bibitem[{{\sorthelp{Planck Collaboration 2013D}}{Planck Collaboration
  IV}(2014)}]{planck2013-p02d}
{\sorthelp{Planck Collaboration 2013D}}{Planck Collaboration IV},
  {\textit{Planck} 2013 results: LFI Beams}. 2014, \aap, in press,
  \eprint{arXiv:1303.5065}

\bibitem[{{\sorthelp{Planck Collaboration 2013E}}{Planck Collaboration
  V}(2014)}]{planck2013-p02b}
{\sorthelp{Planck Collaboration 2013E}}{Planck Collaboration V},
  {\textit{Planck} 2013 results: LFI Calibration}. 2014, \aap, in press,
  \eprint{arXiv:1303.5066}

\bibitem[{{\sorthelp{Planck Collaboration 2013F}}{Planck Collaboration
  VI}(2014)}]{planck2013-p03}
{\sorthelp{Planck Collaboration 2013F}}{Planck Collaboration VI},
  {\textit{Planck} 2013 results: High Frequency Instrument Data Processing}.
  2014, \aap, in press, \eprint{arXiv:1303.5067}

\bibitem[{{\sorthelp{Planck Collaboration 2013G}}{Planck Collaboration
  VII}(2014)}]{planck2013-p03c}
{\sorthelp{Planck Collaboration 2013G}}{Planck Collaboration VII},
  {\textit{Planck} 2013 results: HFI time response and beams}. 2014, \aap, in
  press, \eprint{arXiv:1303.5068}

\bibitem[{{\sorthelp{Planck Collaboration 2013H}}{Planck Collaboration
  VIII}(2014)}]{planck2013-p03f}
{\sorthelp{Planck Collaboration 2013H}}{Planck Collaboration VIII},
  {\textit{Planck} 2013 results: HFI calibration and Map-making}. 2014, \aap,
  in press, \eprint{arXiv:1303.5069}

\bibitem[{{\sorthelp{Planck Collaboration 2013I}}{Planck Collaboration
  IX}(2014)}]{planck2013-p03d}
{\sorthelp{Planck Collaboration 2013I}}{Planck Collaboration IX},
  {\textit{Planck} 2013 results: HFI spectral response}. 2014, \aap, in press,
  \eprint{arXiv:1303.5070}

\bibitem[{{\sorthelp{Planck Collaboration 2013J}}{Planck Collaboration
  X}(2014)}]{planck2013-p03e}
{\sorthelp{Planck Collaboration 2013J}}{Planck Collaboration X},
  {\textit{Planck} 2013 results: HFI energetic particle effects}. 2014, \aap,
  in press, \eprint{arXiv:1303.5071}

\bibitem[{{\sorthelp{Planck Collaboration 2013K}}{Planck Collaboration
  XI}(2014)}]{planck2013-p06b}
{\sorthelp{Planck Collaboration 2013K}}{Planck Collaboration XI},
  {\textit{Planck} 2013 results: All-sky model of thermal dust emission}. 2014,
  \aap, in press, \eprint{arXiv:1312.1300}

\bibitem[{{\sorthelp{Planck Collaboration 2013L}}{Planck Collaboration
  XII}(2014)}]{planck2013-p06}
{\sorthelp{Planck Collaboration 2013L}}{Planck Collaboration XII},
  {\textit{Planck} 2013 results: Component separation}. 2014, \aap, in press,
  \eprint{arXiv:1303.5072}

\bibitem[{{\sorthelp{Planck Collaboration 2013M}}{Planck Collaboration
  XIII}(2014)}]{planck2013-p03a}
{\sorthelp{Planck Collaboration 2013M}}{Planck Collaboration XIII},
  {\textit{Planck} 2013 results: Galactic CO emission as seen by Planck}. 2014,
  \aap, in press, \eprint{arXiv:1303.5073}

\bibitem[{{\sorthelp{Planck Collaboration 2013N}}{Planck Collaboration
  XIV}(2014)}]{planck2013-pip88}
{\sorthelp{Planck Collaboration 2013N}}{Planck Collaboration XIV},
  {\textit{Planck} 2013 results: Zodiacal emission}. 2014, \aap, in press,
  \eprint{arXiv:1303.5074}

\bibitem[{{\sorthelp{Planck Collaboration 2013O}}{Planck Collaboration
  XV}(2014)}]{planck2013-p08}
{\sorthelp{Planck Collaboration 2013O}}{Planck Collaboration XV},
  {\textit{Planck} 2013 results: CMB power spectra and likelihood}. 2014, \aap,
  in press, \eprint{arXiv:1303.5075}

\bibitem[{{\sorthelp{Planck Collaboration 2013P}}{Planck Collaboration
  XVI}(2014)}]{planck2013-p11}
{\sorthelp{Planck Collaboration 2013P}}{Planck Collaboration XVI},
  {\textit{Planck} 2013 results: Cosmological parameters}. 2014, \aap, in
  press, \eprint{arXiv:1303.5076}

\bibitem[{{\sorthelp{Planck Collaboration 2013Q}}{Planck Collaboration
  XVII}(2014)}]{planck2013-p12}
{\sorthelp{Planck Collaboration 2013Q}}{Planck Collaboration XVII},
  {\textit{Planck} 2013 results: Gravitational lensing by large-scale
  structure}. 2014, \aap, in press, \eprint{arXiv:1303.5077}

\bibitem[{{\sorthelp{Planck Collaboration 2013R}}{Planck Collaboration
  XVIII}(2014)}]{planck2013-p13}
{\sorthelp{Planck Collaboration 2013R}}{Planck Collaboration XVIII},
  {\textit{Planck} 2013 results: Gravitational lensing by star-forming
  galaxies}. 2014, \aap, in press, \eprint{arXiv:1303.5078}

\bibitem[{{\sorthelp{Planck Collaboration 2013S}}{Planck Collaboration
  XIX}(2014)}]{planck2013-p14}
{\sorthelp{Planck Collaboration 2013S}}{Planck Collaboration XIX},
  {\textit{Planck} 2013 results: The integrated Sachs-Wolfe effect}. 2014,
  \aap, in press, \eprint{arXiv:1303.5079}

\bibitem[{{\sorthelp{Planck Collaboration 2013T}}{Planck Collaboration
  XX}(2014)}]{planck2013-p15}
{\sorthelp{Planck Collaboration 2013T}}{Planck Collaboration XX},
  {\textit{Planck} 2013 results: Cosmology from Planck SZ cluster counts}.
  2014, \aap, in press, \eprint{arXiv:1303.5080}

\bibitem[{{\sorthelp{Planck Collaboration 2013U}}{Planck Collaboration
  XXI}(2014)}]{planck2013-p05b}
{\sorthelp{Planck Collaboration 2013U}}{Planck Collaboration XXI},
  {\textit{Planck} 2013 results: All-sky Compton parameter map and
  characterization}. 2014, \aap, in press, \eprint{arXiv:1303.5081}

\bibitem[{{\sorthelp{Planck Collaboration 2013V}}{Planck Collaboration
  XXII}(2014)}]{planck2013-p17}
{\sorthelp{Planck Collaboration 2013V}}{Planck Collaboration XXII},
  {\textit{Planck} 2013 results: Constraints on inflation}. 2014, \aap, in
  press, \eprint{arXiv:1303.5082}

\bibitem[{{\sorthelp{Planck Collaboration 2013W}}{Planck Collaboration
  XXIII}(2014)}]{planck2013-p09}
{\sorthelp{Planck Collaboration 2013W}}{Planck Collaboration XXIII},
  {\textit{Planck} 2013 results: Isotropy and statistics of the CMB}. 2014,
  \aap, in press, \eprint{arXiv:1303.5083}

\bibitem[{{\sorthelp{Planck Collaboration 2013X}}{Planck Collaboration
  XXIV}(2014)}]{planck2013-p09a}
{\sorthelp{Planck Collaboration 2013X}}{Planck Collaboration XXIV},
  {\textit{Planck} 2013 results: Constraints on primordial non-Gaussianity}.
  2014, \aap, in press, \eprint{arXiv:1303.5084}

\bibitem[{{\sorthelp{Planck Collaboration 2013Y}}{Planck Collaboration
  XXV}(2014)}]{planck2013-p20}
{\sorthelp{Planck Collaboration 2013Y}}{Planck Collaboration XXV},
  {\textit{Planck} 2013 results: Searches for cosmic strings and other
  topological defects}. 2014, \aap, in press, \eprint{arXiv:1303.5085}

\bibitem[{{\sorthelp{Planck Collaboration 2013ZA}}{Planck Collaboration
  XXVI}(2014)}]{planck2013-p19}
{\sorthelp{Planck Collaboration 2013ZA}}{Planck Collaboration XXVI},
  {\textit{Planck} 2013 results: Geometry and topology of the Universe}. 2014,
  \aap, in press, \eprint{arXiv:1303.5086}

\bibitem[{{\sorthelp{Planck Collaboration 2013ZB}}{Planck Collaboration
  XXVII}(2014)}]{planck2013-pipaberration}
{\sorthelp{Planck Collaboration 2013ZB}}{Planck Collaboration XXVII},
  {\textit{Planck} 2013 results: Special relativistic effects on the CMB
  dipole}. 2014, \aap, in press, \eprint{arXiv:1303.5087}

\bibitem[{{\sorthelp{Planck Collaboration 2013ZC}}{Planck Collaboration
  XXVIII}(2014)}]{planck2013-p05}
{\sorthelp{Planck Collaboration 2013ZC}}{Planck Collaboration XXVIII},
  {\textit{Planck} 2013 results: The Planck Catalogue of Compact Sources}.
  2014, \aap, in press, \eprint{arXiv:1303.5088}

\bibitem[{{\sorthelp{Planck Collaboration 2013ZD}}{Planck Collaboration
  XXIX}(2014)}]{planck2013-p05a}
{\sorthelp{Planck Collaboration 2013ZD}}{Planck Collaboration XXIX},
  {\textit{Planck} 2013 results: The Planck catalogue of Sunyaev-Zeldovich
  sources}. 2014, \aap, in press, \eprint{arXiv:1303.5089}

\bibitem[{{\sorthelp{Planck Collaboration 2013ZE}}{Planck Collaboration
  XXX}(2014)}]{planck2013-pip56}
{\sorthelp{Planck Collaboration 2013ZE}}{Planck Collaboration XXX},
  {\textit{Planck} 2013 results: Cosmic infrared background measurements and
  implications for star formation}. 2014, \aap, in press,
  \eprint{arXiv:1309.0382}

\bibitem[{{\sorthelp{Planck Collaboration 2013ZF}}{Planck Collaboration
  XXXI}(2014)}]{planck2013-p01a}
{\sorthelp{Planck Collaboration 2013ZF}}{Planck Collaboration XXXI},
  {\textit{Planck} 2013 results: Consistency of the data}. 2014, In preparation

\bibitem[{{\sorthelp{Planck Collaboration Int G}}{Planck Collaboration Int.
  VII}(2013)}]{planck2012-VII}
{\sorthelp{Planck Collaboration Int G}}{Planck Collaboration Int. VII}, {Planck
  intermediate results. VII. Statistical properties of infrared and radio
  extragalactic sources from the Planck Early Release Compact Source Catalogue
  at frequencies between 100 and 857 GHz}. 2013, \aap, 550, A133

\bibitem[{{Plaszczynski} {et~al.}(2012){Plaszczynski}, {Lavabre}, {Perotto}, \&
  {Starck}}]{plaszczynski}
{Plaszczynski}, S., {Lavabre}, A., {Perotto}, L., \& {Starck}, J.-L., {A hybrid
  approach to cosmic microwave background lensing reconstruction from all-sky
  intensity maps}. 2012, \aap, 544, A27, \eprint{1201.5779}

\bibitem[{{Rees} \& {Sciama}(1968)}]{ReesSciama1968}
{Rees}, M.~J. \& {Sciama}, D.~W., {Large-scale Density Inhomogeneities in the
  Universe}. 1968, \nat, 217, 511

\bibitem[{Regan {et~al.}(2010)Regan, Shellard, \& Fergusson}]{Regan:2010cn}
Regan, D., Shellard, E., \& Fergusson, J., {General CMB and Primordial
  Trispectrum Estimation}. 2010, \prd, 82, 023520, \eprint{1004.2915}

\bibitem[{{Reichardt} {et~al.}(2012){Reichardt}, {Shaw}, {Zahn}, {Aird},
  {Benson}, {Bleem}, {Carlstrom}, {Chang}, {Cho}, {Crawford}, {Crites}, {de
  Haan}, {Dobbs}, {Dudley}, {George}, {Halverson}, {Holder}, {Holzapfel},
  {Hoover}, {Hou}, {Hrubes}, {Joy}, {Keisler}, {Knox}, {Lee}, {Leitch},
  {Lueker}, {Luong-Van}, {McMahon}, {Mehl}, {Meyer}, {Millea}, {Mohr},
  {Montroy}, {Natoli}, {Padin}, {Plagge}, {Pryke}, {Ruhl}, {Schaffer},
  {Shirokoff}, {Spieler}, {Staniszewski}, {Stark}, {Story}, {van Engelen},
  {Vanderlinde}, {Vieira}, \& {Williamson}}]{2012ApJ...755...70R}
{Reichardt}, C.~L., {Shaw}, L., {Zahn}, O., {et~al.}, {A Measurement of
  Secondary Cosmic Microwave Background Anisotropies with Two Years of South
  Pole Telescope Observations}. 2012, \apj, 755, 70, \eprint{1111.0932}

\bibitem[{Ross {et~al.}(2011)Ross, Ho, Cuesta, Tojeiro, Percival,
  {et~al.}}]{Ross:2011cz}
Ross, A.~J., Ho, S., Cuesta, A.~J., {et~al.}, {Ameliorating Systematic
  Uncertainties in the Angular Clustering of Galaxies: A Study using SDSS-III}.
  2011, \mnras, 417, 1350, \eprint{1105.2320}

\bibitem[{{Sachs} \& {Wolfe}(1967)}]{SachsWolfe1967}
{Sachs}, R.~K. \& {Wolfe}, A.~M., {Perturbations of a Cosmological Model and
  Angular Variations of the Microwave Background}. 1967, \apj, 147, 73

\bibitem[{Schmittfull {et~al.}(2013)Schmittfull, Challinor, Hanson, \&
  Lewis}]{Schmittfull:2013uea}
Schmittfull, M.~M., Challinor, A., Hanson, D., \& Lewis, A., {On the joint
  analysis of CMB temperature and lensing-reconstruction power spectra}. 2013,
  \prd, 88, \eprint{1308.0286}

\bibitem[{Seljak(1996)}]{Seljak:1995ve}
Seljak, U., {Gravitational lensing effect on cosmic microwave background
  anisotropies: A Power spectrum approach}. 1996, \apj, 463, 1,
  \eprint{astro-ph/9505109}

\bibitem[{Sherwin {et~al.}(2012)Sherwin, Das, Hajian, Addison, Bond,
  {et~al.}}]{Sherwin:2012mr}
Sherwin, B.~D., Das, S., Hajian, A., {et~al.}, {The Atacama Cosmology
  Telescope: Cross-Correlation of CMB Lensing and Quasars}. 2012, \prd, 86,
  083006, \eprint{1207.4543}

\bibitem[{Sherwin {et~al.}(2011)}]{Sherwin:2011gv}
Sherwin, B.~D. {et~al.}, {Evidence for dark energy from the cosmic microwave
  background alone using the Atacama Cosmology Telescope lensing measurements}.
  2011, \prl, 107, 021302, \eprint{1105.0419}

\bibitem[{Smidt {et~al.}(2011)}]{Smidt:2010by}
Smidt, J. {et~al.}, {A Constraint On the Integrated Mass Power Spectrum out to
  z = 1100 from Lensing of the Cosmic Microwave Background}. 2011, \apj, 728,
  L1, \eprint{1012.1600}

\bibitem[{Smith {et~al.}(2006)Smith, Hu, \& Kaplinghat}]{Smith:2006nk}
Smith, K.~M., Hu, W., \& Kaplinghat, M., {Cosmological Information from Lensed
  CMB Power Spectra}. 2006, \prd, 74, 123002, \eprint{astro-ph/0607315}

\bibitem[{Smith {et~al.}(2007)Smith, Zahn, Dore, \& Nolta}]{Smith:2007rg}
Smith, K.~M., Zahn, O., Dore, O., \& Nolta, M.~R., {Detection of Gravitational
  Lensing in the Cosmic Microwave Background}. 2007, \prd, 76, 043510,
  \eprint{0705.3980}

\bibitem[{{Stompor} \& {Efstathiou}(1999)}]{1999MNRAS.302..735S}
{Stompor}, R. \& {Efstathiou}, G., {Gravitational lensing of cosmic microwave
  background anisotropies and cosmological parameter estimation}. 1999, \mnras,
  302, 735, \eprint{arXiv:astro-ph/9805294}

\bibitem[{Story {et~al.}(2013)Story, Reichardt, Hou, Keisler, Aird,
  {et~al.}}]{Story:2012wx}
Story, K., Reichardt, C., Hou, Z., {et~al.}, {A Measurement of the Cosmic
  Microwave Background Damping Tail from the 2500-square-degree SPT-SZ survey}.
  2013, \apj, 779, 86, \eprint{1210.7231}

\bibitem[{{Sunyaev} \& {Zeldovich}(1980)}]{SunyaevZeldovich1980}
{Sunyaev}, R.~A. \& {Zeldovich}, I.~B., {Microwave background radiation as a
  probe of the contemporary structure and history of the universe}. 1980,
  \araa, 18, 537

\bibitem[{Tinker {et~al.}(2010)Tinker, Robertson, Kravtsov, Klypin, Warren,
  {et~al.}}]{Tinker:2010my}
Tinker, J.~L., Robertson, B.~E., Kravtsov, A.~V., {et~al.}, {The Large Scale
  Bias of Dark Matter Halos: Numerical Calibration and Model Tests}. 2010,
  \apj, 724, 878, \eprint{1001.3162}

\bibitem[{van Engelen {et~al.}(2012)van Engelen, Keisler, Zahn, Aird, Benson,
  {et~al.}}]{vanEngelen:2012va}
van Engelen, A., Keisler, R., Zahn, O., {et~al.}, {A measurement of
  gravitational lensing of the microwave background using South Pole Telescope
  data}. 2012, \apj, 756, 142, \eprint{1202.0546}

\bibitem[{Wright {et~al.}(2010)Wright, Eisenhardt, Mainzer, Ressler, Cutri,
  {et~al.}}]{Wright:2010qw}
Wright, E.~L., Eisenhardt, P.~R., Mainzer, A., {et~al.}, {The Wide-field
  Infrared Survey Explorer (WISE): Mission Description and Initial On-orbit
  Performance}. 2010, Astron.J., 140, 1868, \eprint{1008.0031}

\bibitem[{Zaldarriaga \& Seljak(1998)}]{Zaldarriaga:1998ar}
Zaldarriaga, M. \& Seljak, U., {Gravitational Lensing Effect on Cosmic
  Microwave Background Polarization}. 1998, \prd, 58, 023003,
  \eprint{astro-ph/9803150}

\end{thebibliography}
